\documentclass[12pt]{report} 

\usepackage{amssymb}
\usepackage{amsmath}
\usepackage{graphicx}
\usepackage{natbib}
\usepackage{multicol}
\usepackage{float}
\usepackage[caption = false]{subfig}

\usepackage{hyperref,graphicx}

\def\mytitle{Simulation Scenarios in One-Dimensional Self-Gravitating System}
\def\myname{Suman Pramanick}

\def\mysupervisor{Somnath Bharadwaj}

\def\mydep{Department of Physics}

\begin{document}
 \baselineskip=18pt plus1pt
 \setcounter{secnumdepth}{3}
 \setcounter{tocdepth}{3}
 \pagenumbering{roman}
 
 %Front Matter
 \thispagestyle{empty}
\begin{center}
    { \large {\bfseries {\mytitle}} \par}
\vspace{3\baselineskip}
    {\textit{A guided study for the partial fulfillment of Ph.D. coursework}\\
    }\par

\vspace{\baselineskip}
    {\textit{by} \par}
\vspace{\baselineskip}
    {{\large {\bf \myname }} \par}
%\vspace{-0.1\baselineskip}
%    {{\large {\bf \myrollno}} \par}
\vspace{1.5\baselineskip}
    {Advisor \par}
\vspace{\baselineskip}
    {{\large \bf \mysupervisor} \par}

%{\large \vspace*{1ex}
\vspace{1.5\baselineskip}
    {{\mydep} \par}
\vspace*{1ex}
    { {Indian Institute of Technology Kharagpur\\ Kharagpur, West Bengal 721302, India} \par}
 \end{center}
 \thispagestyle{plain}
\begin{center}
    \Large \textbf{\uppercase{Abstract}}
\end{center}

\vspace{3\baselineskip}

\noindent

We study and compare different numerical differential equation solvers on the basis of numerical complexity, energy conservation, and stable solution in phase-space for the Simple Harmonic Oscillation (SHM) problem. We conclude and show that the Leapfrog method is the best for our problem. We solve Poisson's equation in gravity on the computation grid by the finite difference method and Fourier method. We solve Poisson's equation for source not in a grid point by cloud-in cell method. Finally, we simulate one-dimensional self-gravitating system and show the evolution of the system via phase-space trajectories.

\vspace{\baselineskip}

\noindent
%\textbf{Keywords}: CIC, Leapfrog, Fourier Transform, Inverse Fourier Transform
 \tableofcontents
 %\listoffigures
 
 %Main material
 \clearpage
 \pagenumbering{arabic}

 \chapter{Solutions by Euler, RK-2 and RK-4 Methods}
\section{Aim}
To Solve the Simple Harmonic Oscillation (SHM) problem by Euler method, Runge-Kutta second order (RK-2) method and Runge-Kutta fourth order (RK-4) method.

\section{Introduction}
There are very few real systems which can be solve analytically, safely we can say most of the real life systems we can not solve analytically. However we can formulate the system mathematically and can write the differential equations to solve the system. These systems can be solved using numerical methods with obviously some boundary conditions specified. The solution we get from numerical techniques are not error free, because in every numerical technique we assume some simple steps to reduce the complexity of the system. Those assumptions are good approximation of the real system but not exactly the real system. That leads to a accumulation of errors in each computational steps (iterations). For a long run these errors sum up and sometime make the numerical system way deviate from the real system. 

There are two things we need keep in mind. We have to take each infinitesimal steps such that they don not deviate much from the real system. The second point is to do such we cannot suggest an infinitesimal process which makes a huge computational complexity, because that will make more run time and sometimes less efficient. So, basically we have to balance between these two things, and have to choose the most efficient technique according to our requirements. 

\section{Problem}
In this assignment we will solve the simple harmonic oscillation (SHM) problem. The equations of the problem are:

\begin{equation}
    \frac{dx}{dt}=\frac{p}{m} \;\mathrm { and } \; \frac{dp}{dt}=-kx
\end{equation}

The parameters for our problem are:

\begin{equation}
    k=1; \: m=1; \: \omega=1 \;\mathrm { and } \; T=2\pi
\end{equation}

Before solving the equations with different numerical techniques we will briefly discuss about the details of some numerical techniques to solve differential equations.

\section{Euler Method}
Euler method is a first-order method of solving differential equation. The error per step in Euler method is proportional to the square of the step size. The global error of Euler method is however proportional to the step size. This is the simplest and basic method of solving differential equation. 

\begin{figure}[ht]
\centering
\subfloat[fig 1]{\includegraphics[width = 2.8in]{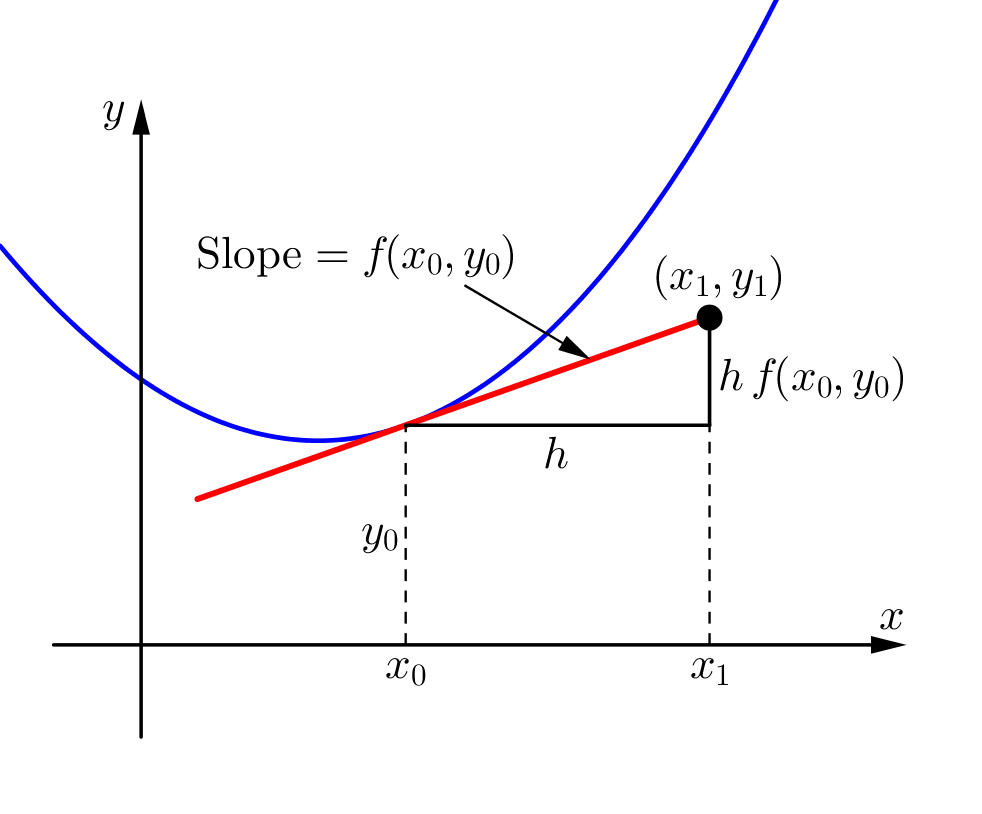}} 
\subfloat[fig 2]{\includegraphics[width = 2.8in]{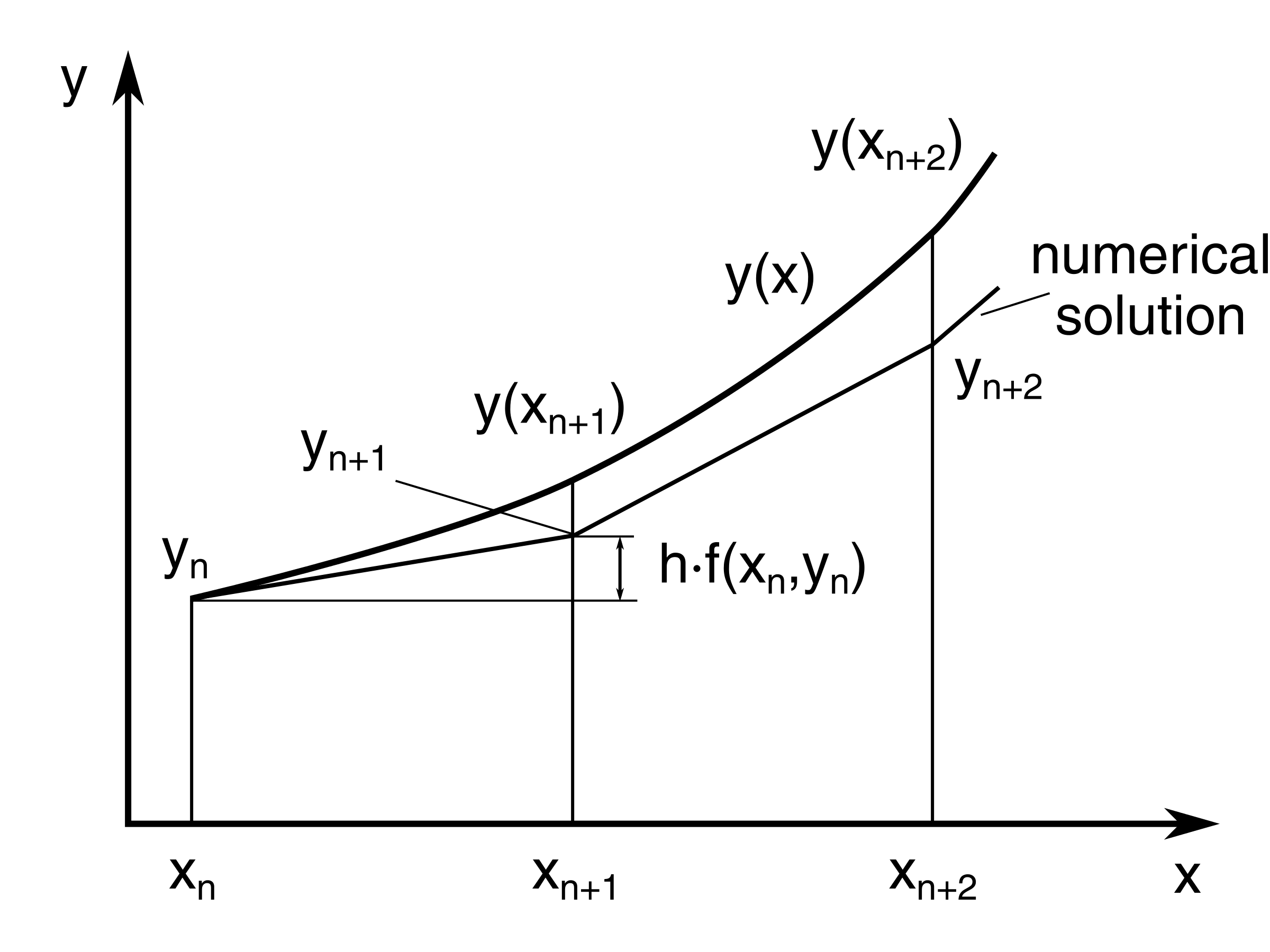}}
\caption{Euler method approximate solution}
\label{figa2_1}
\end{figure}

According to figure \ref{figa2_1}, we can see that at point $(x_0,y_0)$ the slope (slope of the tangent at that point) is $\frac{d y}{d x}$. lets say $\frac{d y}{d x}=f(x,y)$ then at point $(x_0,y_0)$ the slope is $f(x_0,y_0)$. for a small increment $h$ we can approximate $y_0+hf(x_0,y_0)$ as the new value of $y$ at point $x_0+h$. So, we can write the iterative formula as
\begin{equation}
    x_{n+1}=x_n+h
\end{equation}
\begin{equation}
    y_{n+1}=y_n+hf(x_n,y_n)
\end{equation}

So, we need the initial values of $x$ and $y$ to compute the whole function in iterative method. If we want other points along the path of the true solution, and yet we don't actually have the true solution, then it looks like using the tangent line as an approximation might be our best bet! After all, at least on this picture, it looks like the line stays pretty close to the curve if you don't move too far away from the initial point.

In this way according to figure \ref{figa2_1} we can see that we can at least trace the function and if the step size $h$ is very small then we can actually very accurately calculate the function. 

\subsection{Solution of SHM Problem}
Now let us discuss the SHM problem which is our main goal in this chapter. 

\begin{figure}[ht]
\centering
\subfloat[$N=10$]{\includegraphics[width = 2.9in]{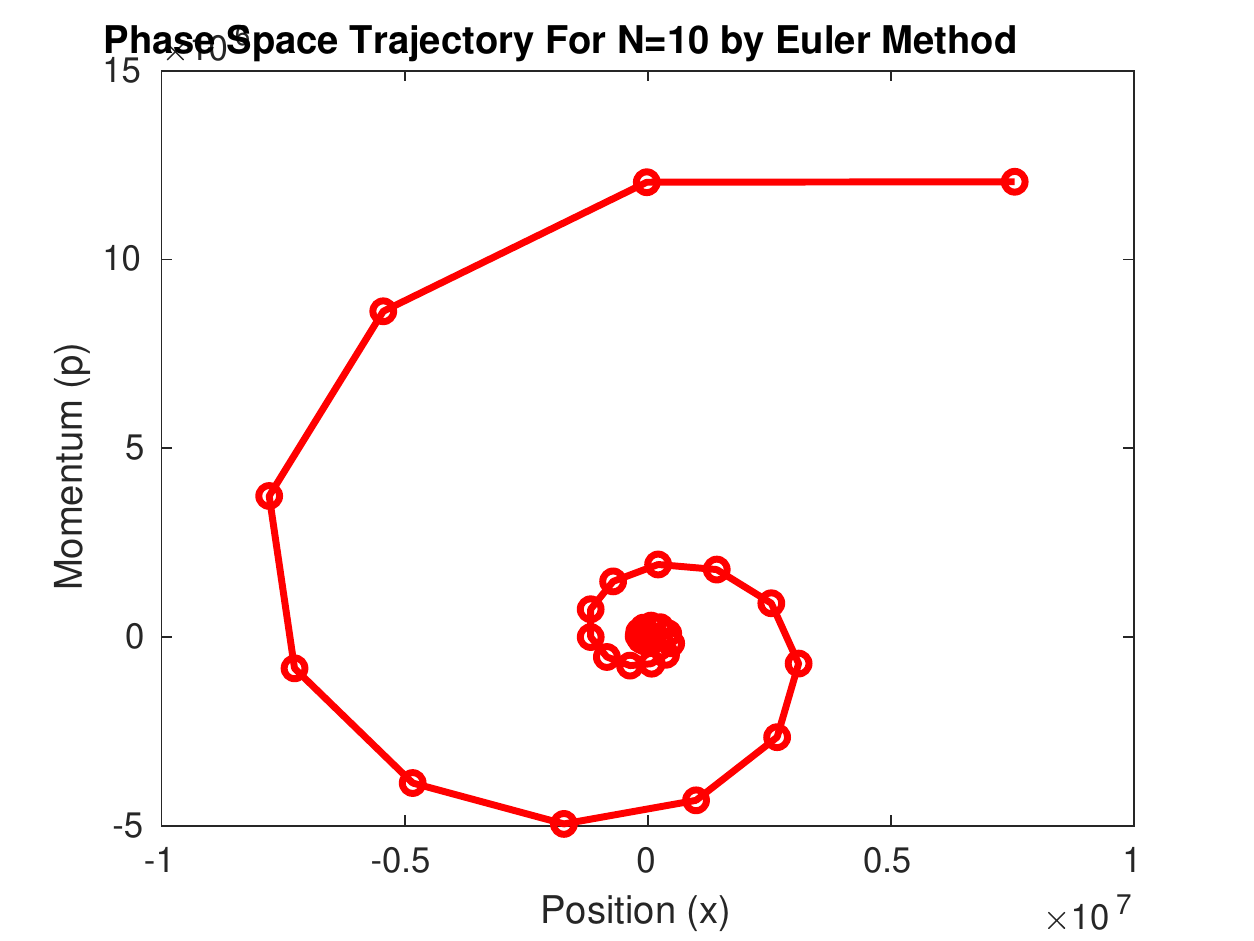}} 
\subfloat[$N=100$]{\includegraphics[width = 2.9in]{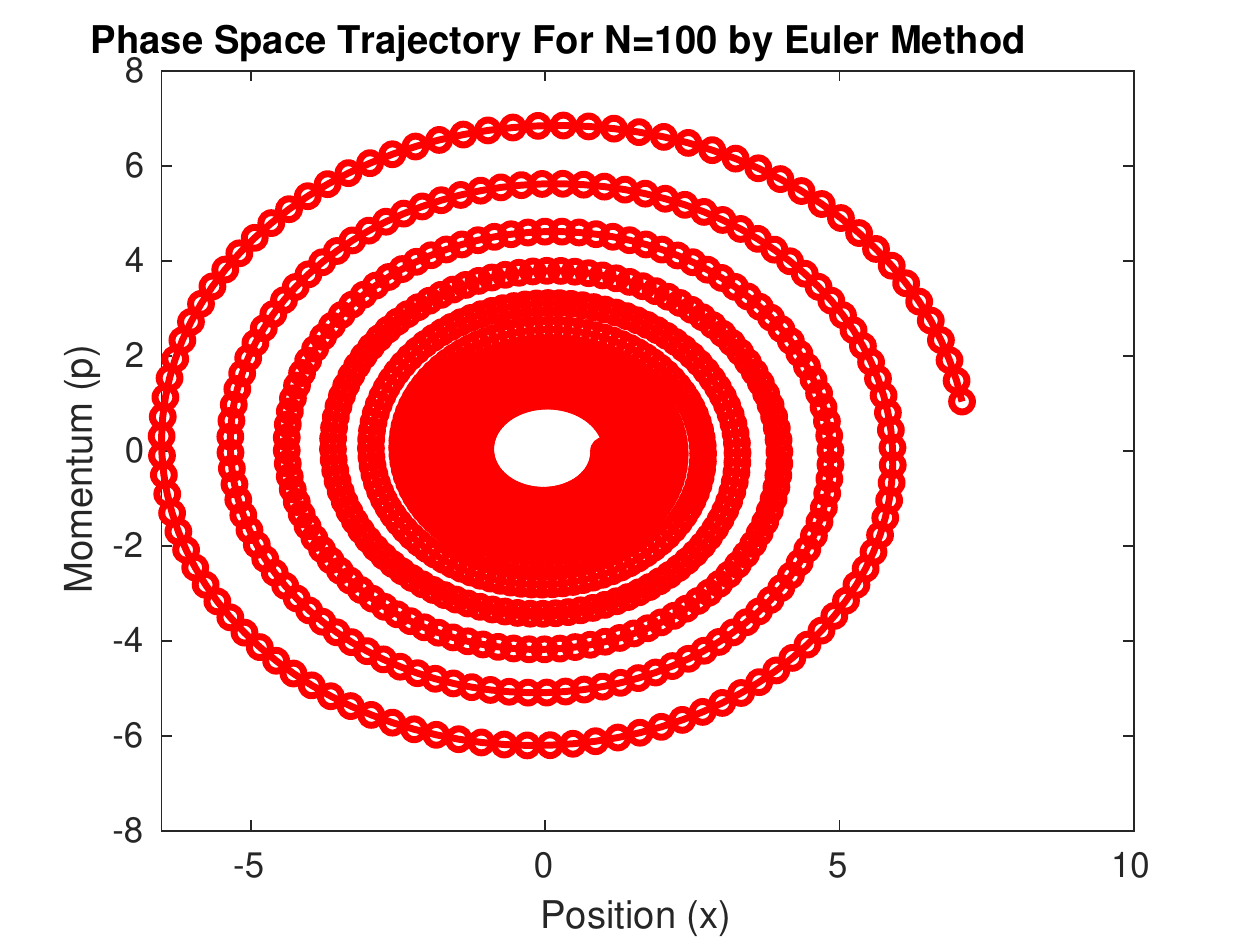}}\\
\subfloat[$N=1000$]{\includegraphics[width = 2.9in]{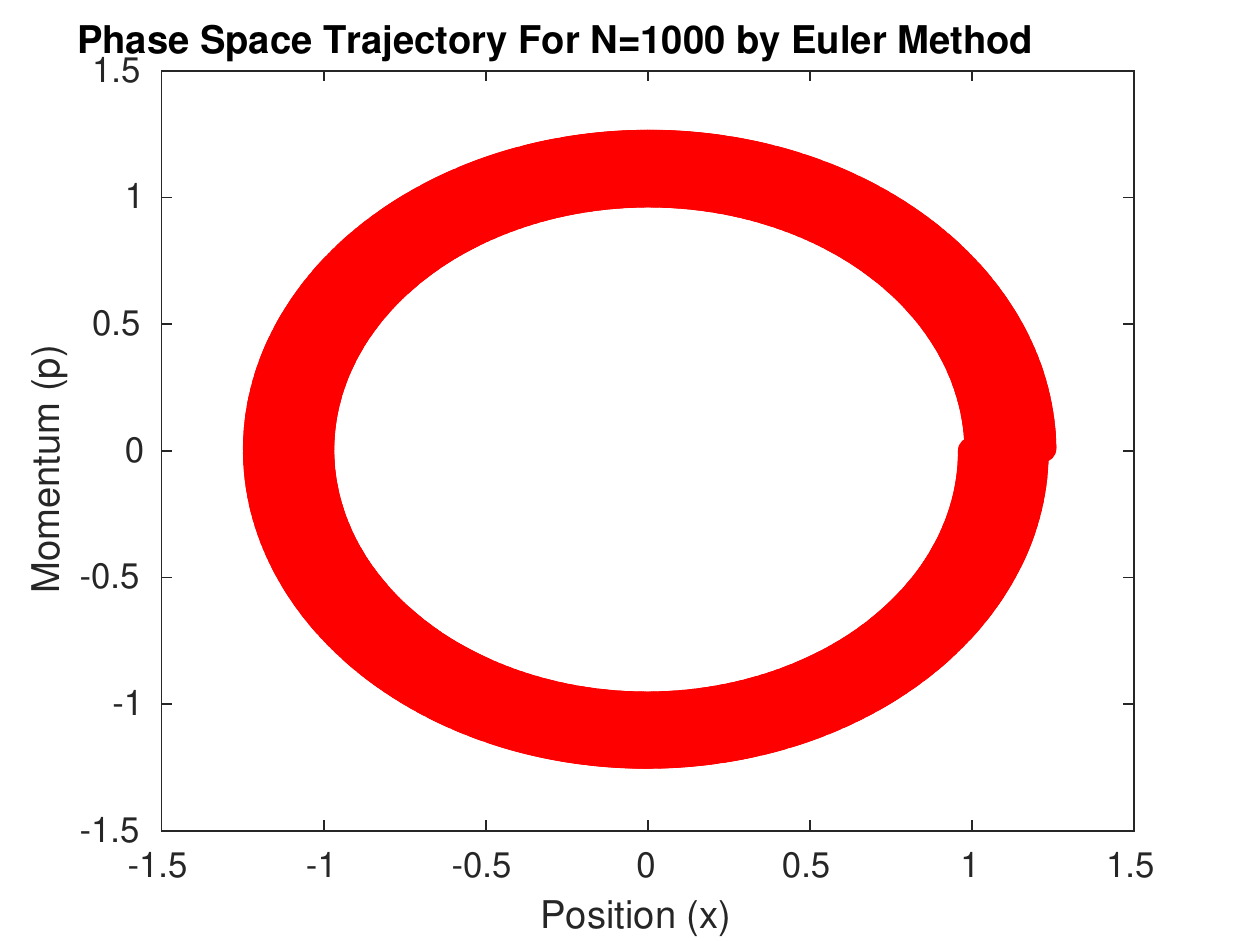}}
\subfloat[$N=10000$]{\includegraphics[width = 2.9in]{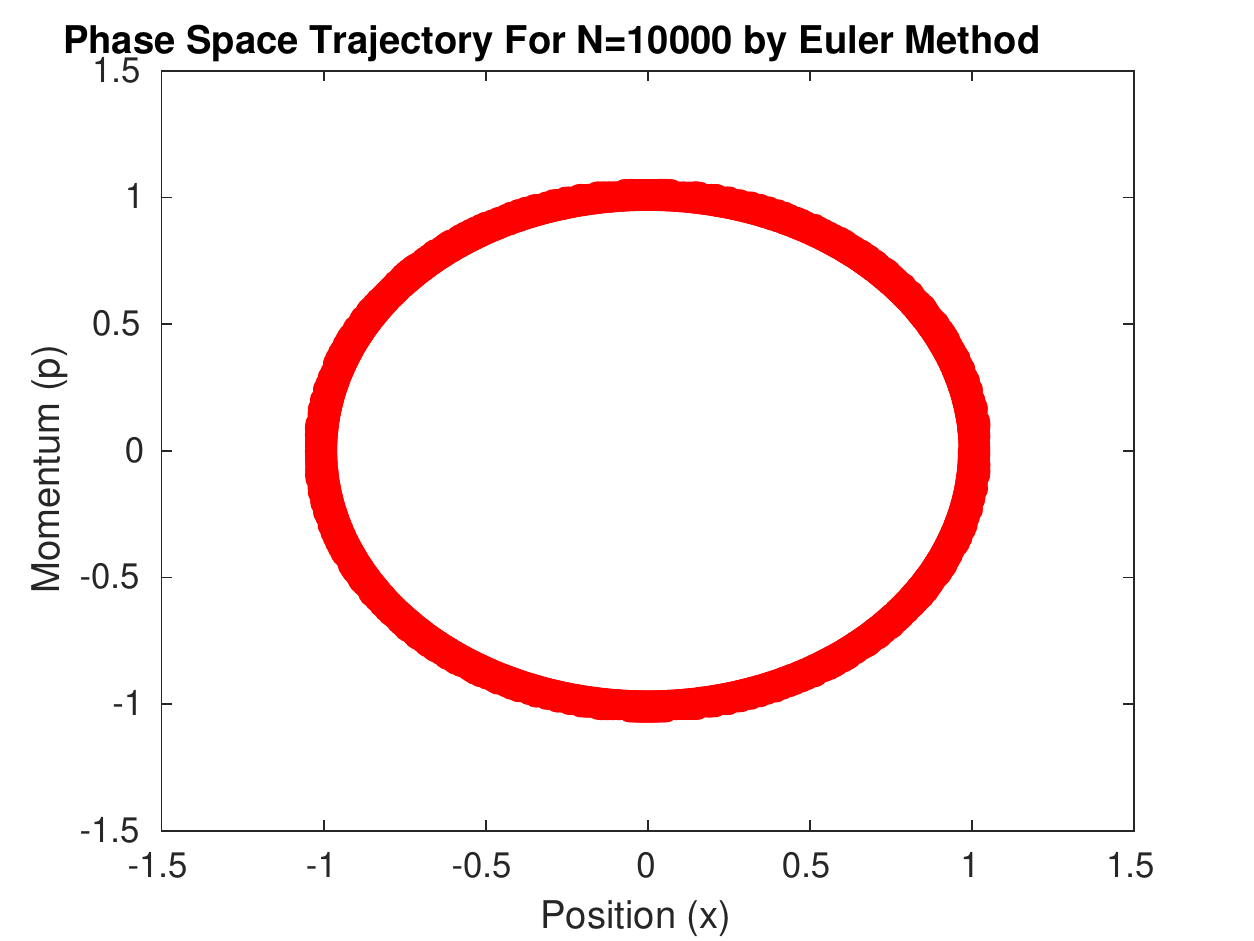}} 
\caption{Phase space trajectories for different number of steps ($N$) inside a full time period $T$ (For Euler method)}
\label{fig_eu1}
\end{figure}

Form figure \ref{fig_eu1} we can see that the solution is diverging for small $N$ values but for large $N$, like $N=10000$ the solution is very close to the actual elliptical solution. For small $N$ the step size $h$ is very small, so error accumulation is large in each step. 

\begin{figure}[ht]
\centering
\subfloat[$N=10$]{\includegraphics[width = 2.9in]{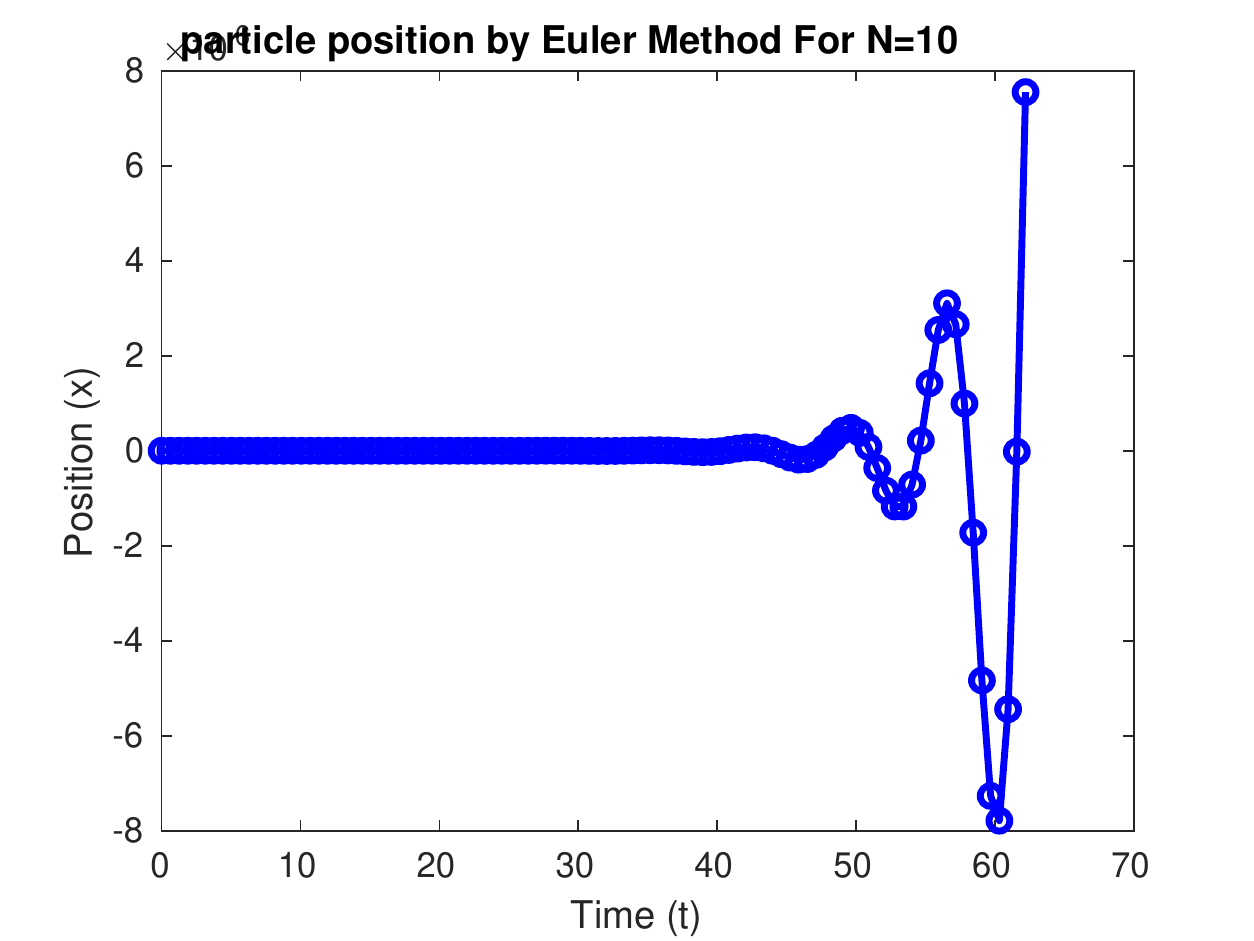}} 
\subfloat[$N=100$]{\includegraphics[width = 2.9in]{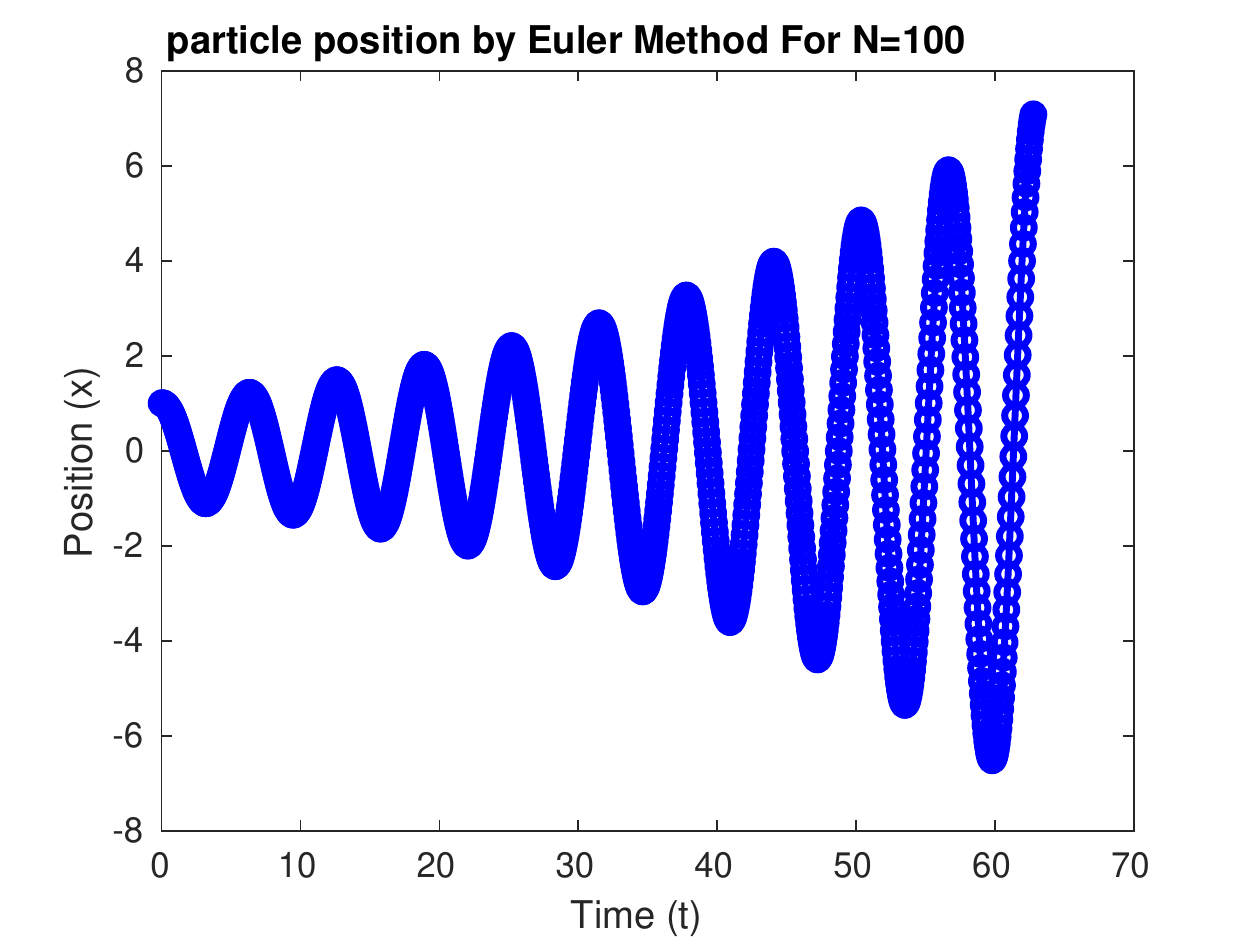}}\\
\subfloat[$N=1000$]{\includegraphics[width = 2.9in]{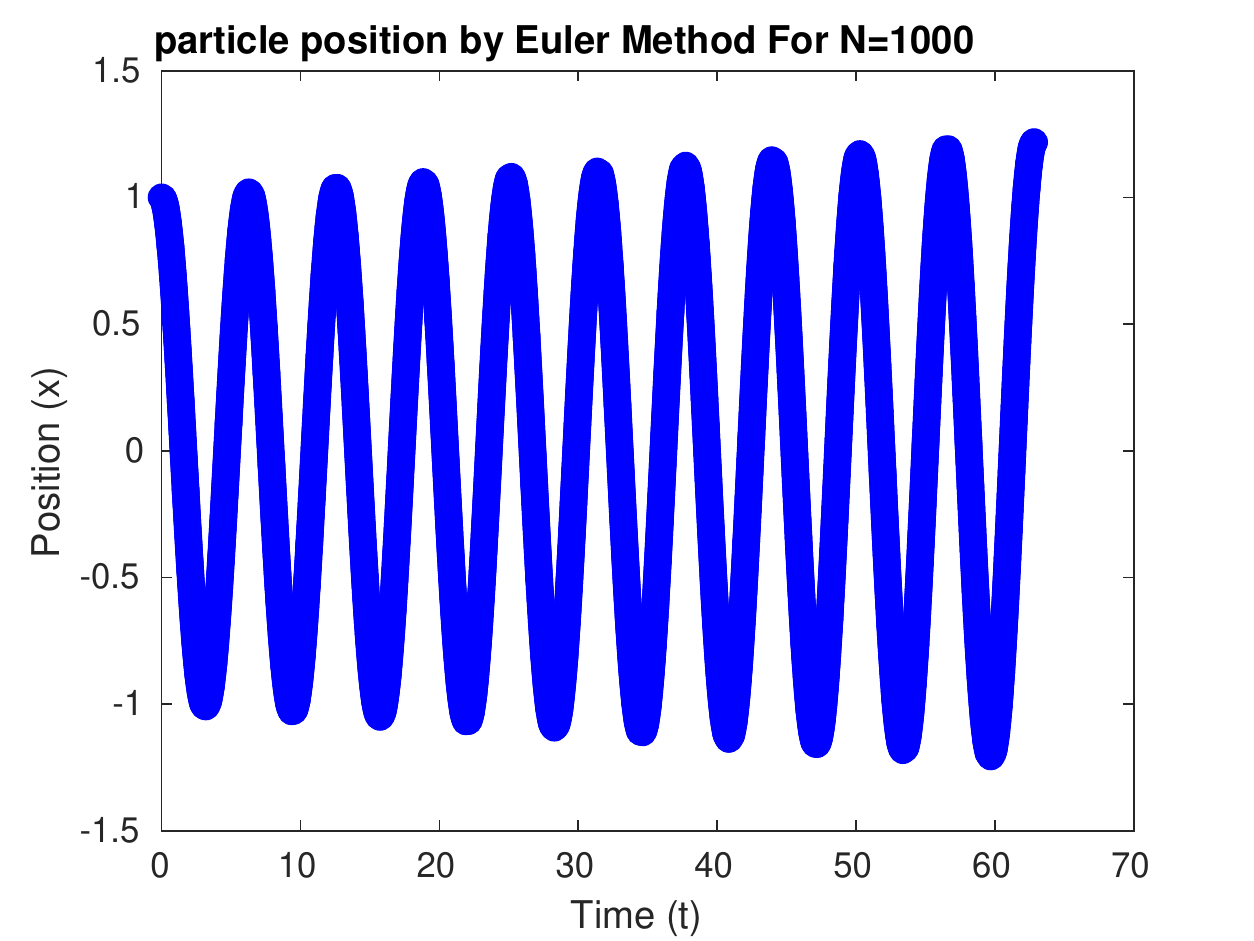}}
\subfloat[$N=10000$]{\includegraphics[width = 2.9in]{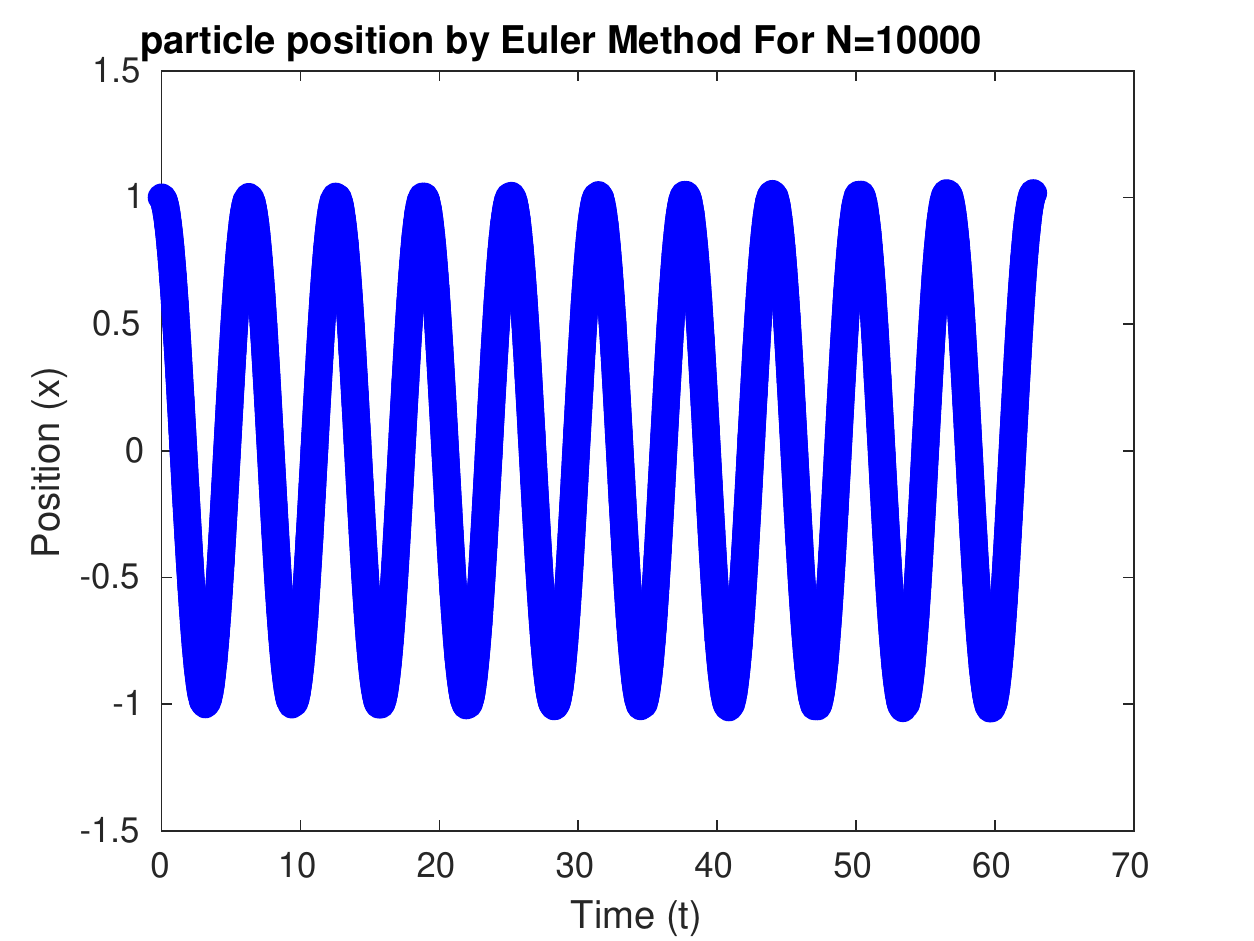}} 
\caption{Solution of $x(t)$ vs $t$ plot for different number of steps ($N$) inside a full time period $T$ (For Euler method)}
\label{fig_eu2}
\end{figure}

\begin{figure}[ht]
\centering
\subfloat[$N=10$]{\includegraphics[width = 2.9in]{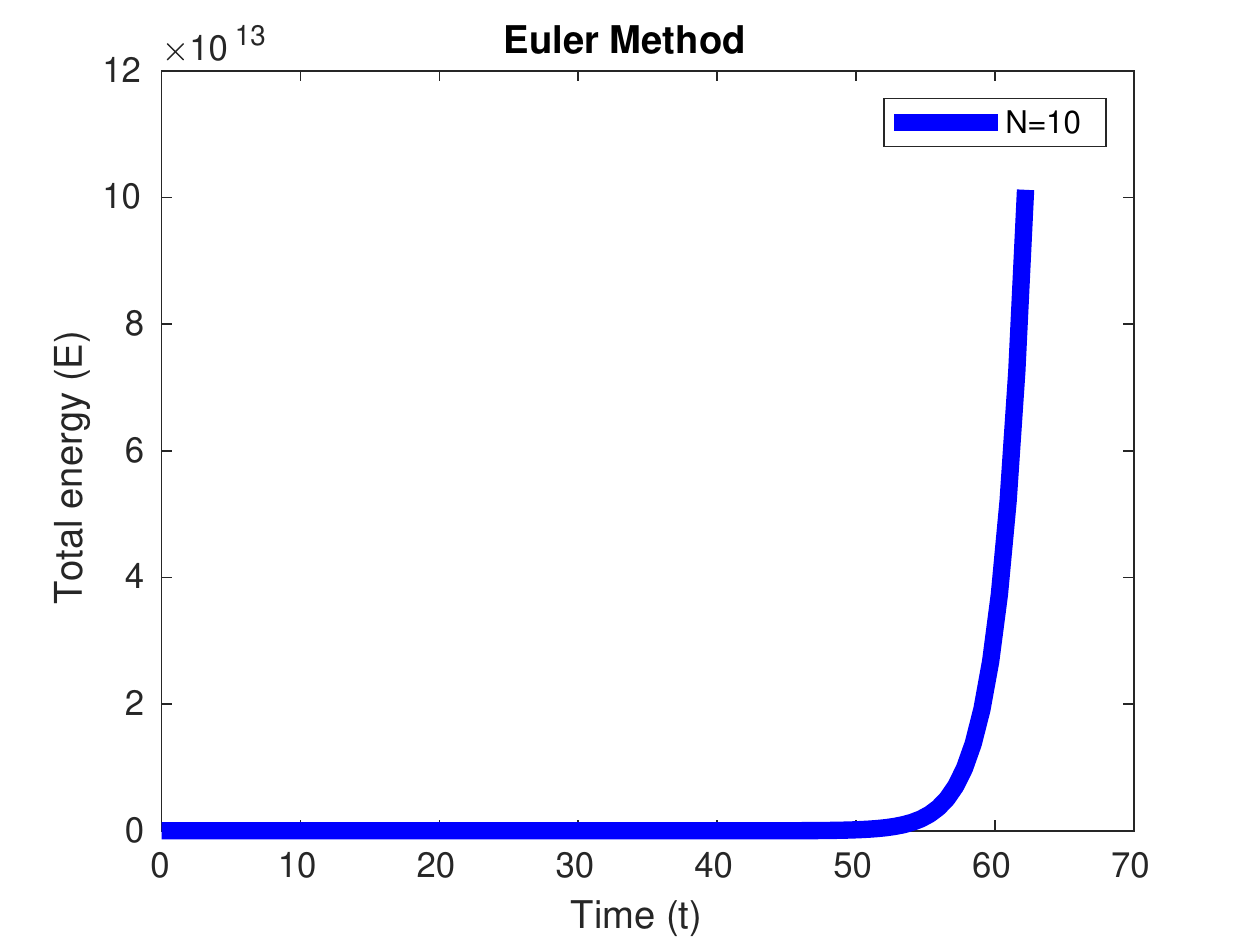}} 
\subfloat[$N=100$]{\includegraphics[width = 2.9in]{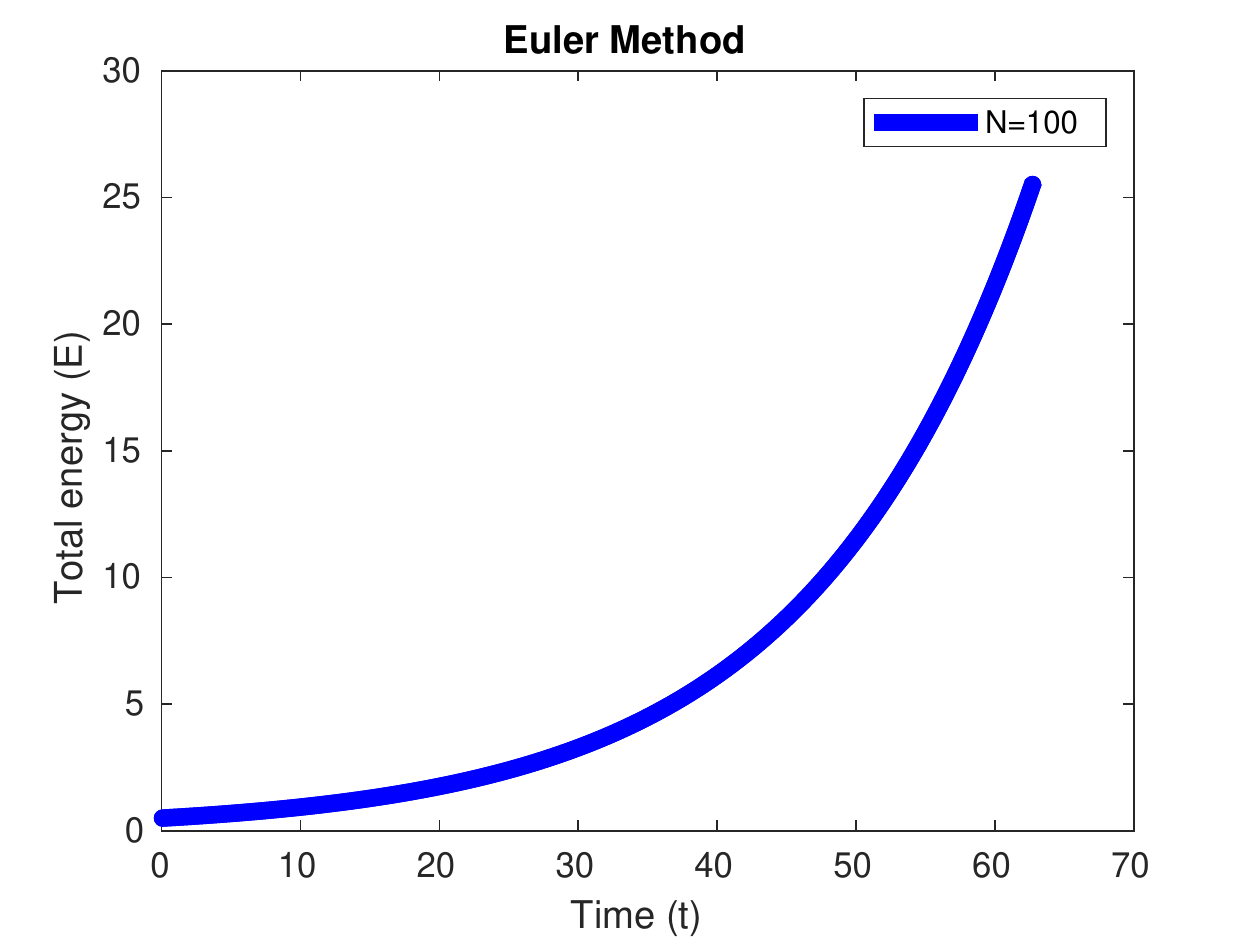}}\\
\subfloat[]{\includegraphics[width = 2.9in]{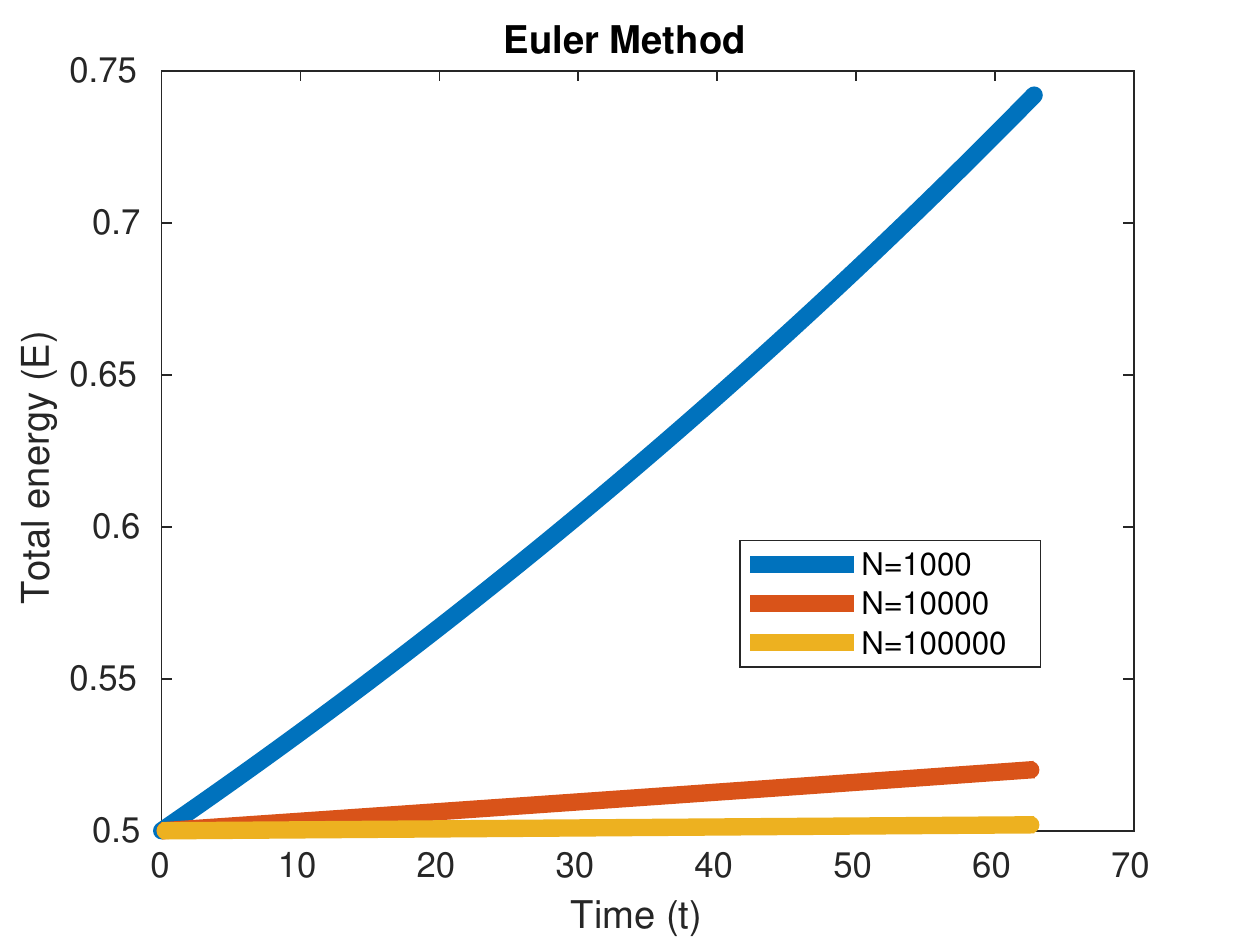}}
\caption{Constant energy value of the solution over time (For Euler method).}
\label{fig_eu3}
\end{figure}

The energy of the system should be constant as the system is conservative and has no damping. From figure \ref{fig_eu3} we can see the energy is diverging for small $N$. However for large $N$ the energy is comparably constant of the process.

\section{Second order Runge-Kutta method}
Runge-Kutta methods are among the
most popular ODE solvers. They were first studied by Carle Runge and Martin Kutta around 1900. In contrast to the multistep methods of the previous section, Runge-Kutta methods
are single-step methods — however, with multiple stages per step. They are motivated
by the dependence of the Taylor methods on the specific IVP. These new methods do
not require derivatives of the right-hand side function $f$ in the code, and are therefore
general-purpose initial value problem solvers.

\subsection{Derivation of RK-2}\label{secrk_2}
Let us consider 
\begin{equation}
    \frac{d y}{d x}=f(x,y)
\end{equation}

The Taylor expansion of $y(x)$ is

\begin{equation}\label{eq1}
    y(x+h)=y(x)+h y'(x)+\frac{h^2}{2} y''(x)+\mathcal{O}(h^3)
\end{equation}
The second derivative of $y$ in terms of $f$ is

\begin{equation}
    y''(x)=f_x(x,y)+f_y(x,y)f(x,y)
\end{equation}
$f_y$ is the Jacobian. the Taylor series expansion of $y$ becomes
\begin{equation}
    y(x+h)=y(x)+h f(x,y)+\frac{h^2}{2}[f_x(x,y)+f_y(x,y)f(x,y)]+\mathcal{O}(h^3)
\end{equation}
\begin{equation}
    y(x+h)=y(x)+\frac{h}{2} f(x,y)+\frac{h}{2}[f_x(x,y)+hf_x(x,y)+h f_y(x,y)f(x,y)]+\mathcal{O}(h^3)
\end{equation}
From multivariate Taylor series expansion 
\begin{equation}
    f(x+h,y+k)=f(x,y)+h f_x(x,y)+f_y(x,y)k+... 
\end{equation}
So,
\begin{equation}
    f(x+h,y+h f(x,y))=f(x,y)+h f_x(x,y)+h f_y(x,y) f(x,y) + \mathbf{O}(h^2)
\end{equation}
Therefore, we get 
\begin{equation}
    y(x+h)=y(x)+\frac{h}{2} f(x,y) + \frac{h}{2} \left[f(x+h,y+h f(x,y))\right]+\mathcal{O}(h^3)
\end{equation}
\[
 \boxed{y_{n+1}=y_n+h\left(\frac{1}{2} k_1+\frac{1}{2} k_2 \right)}
 \]
 Where,
 \begin{equation}
     k_1=f(x_n,y_n)
 \end{equation}
 \begin{equation}
     k_2=f(x_n+h,y_n+h k_1)
 \end{equation}
 
 \subsection{Explanations of plots}
\begin{figure}[ht]
\centering
\subfloat[$N=10$]{\includegraphics[width = 2.9in]{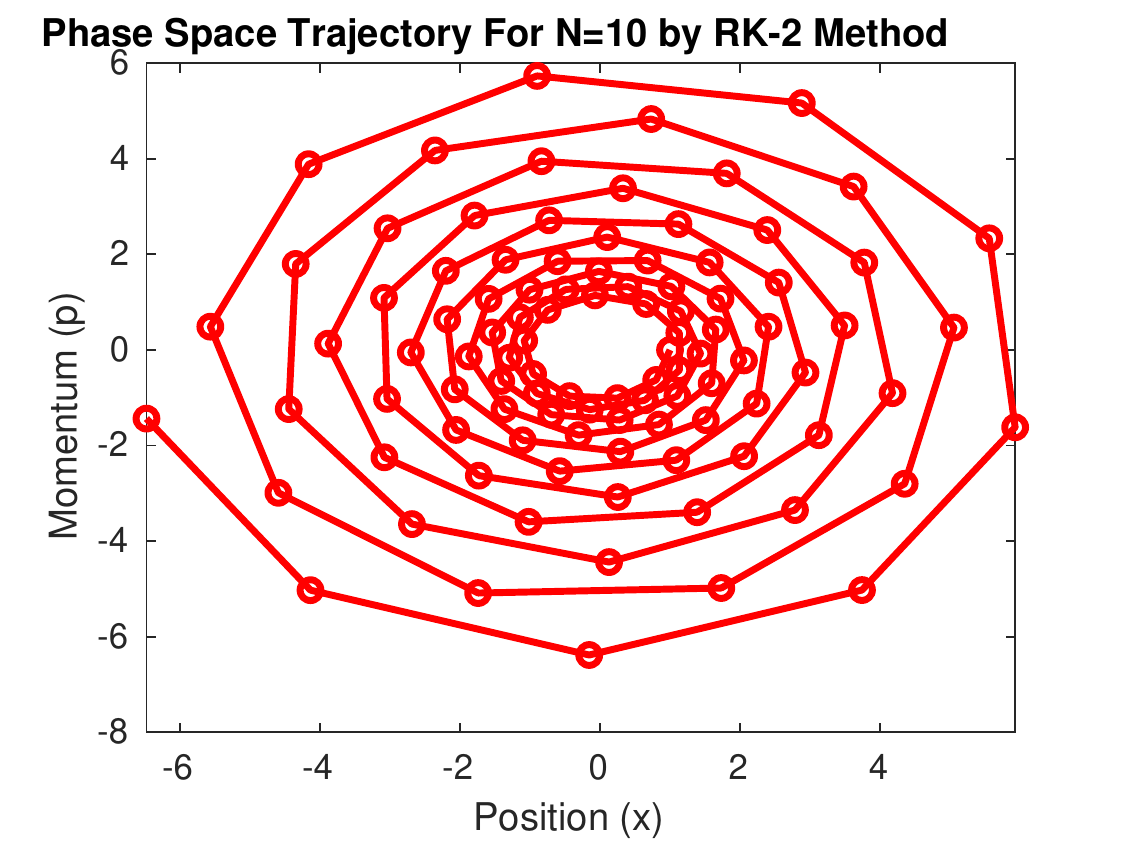}} 
\subfloat[$N=100$]{\includegraphics[width = 2.9in]{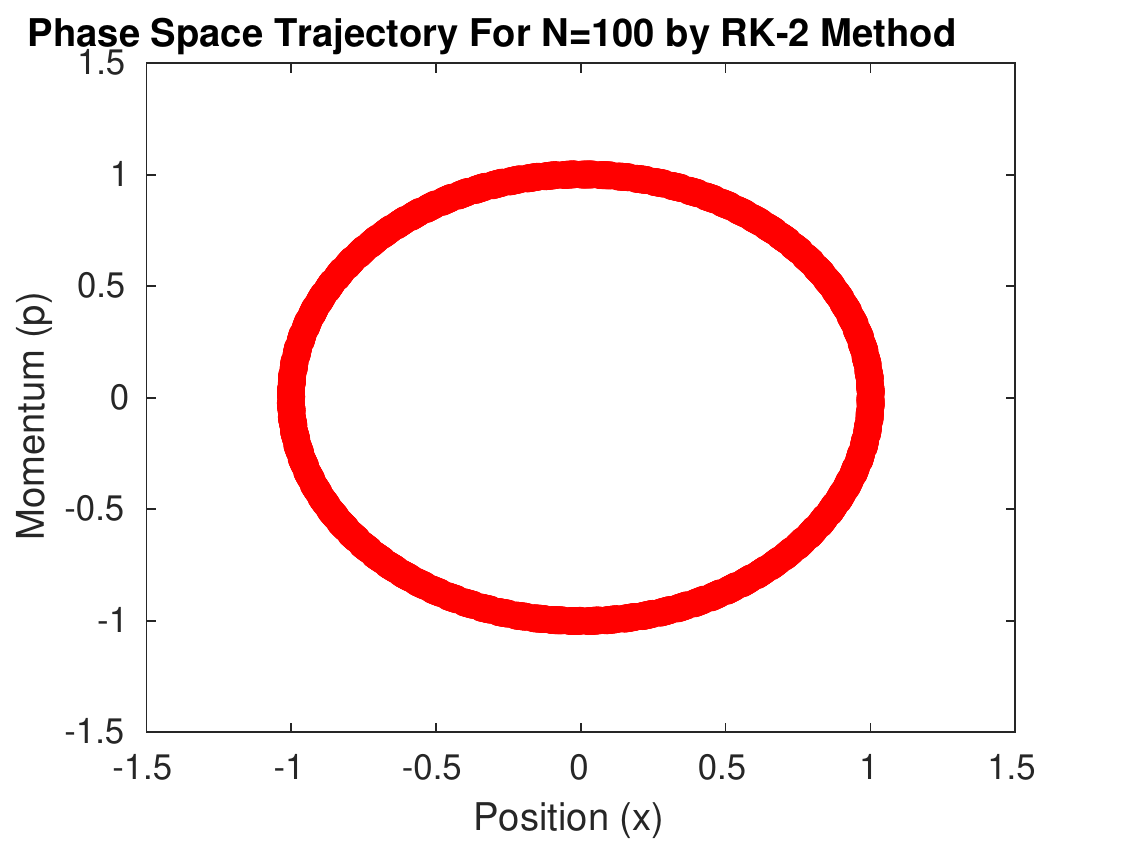}}\\
\subfloat[$N=1000$]{\includegraphics[width = 2.9in]{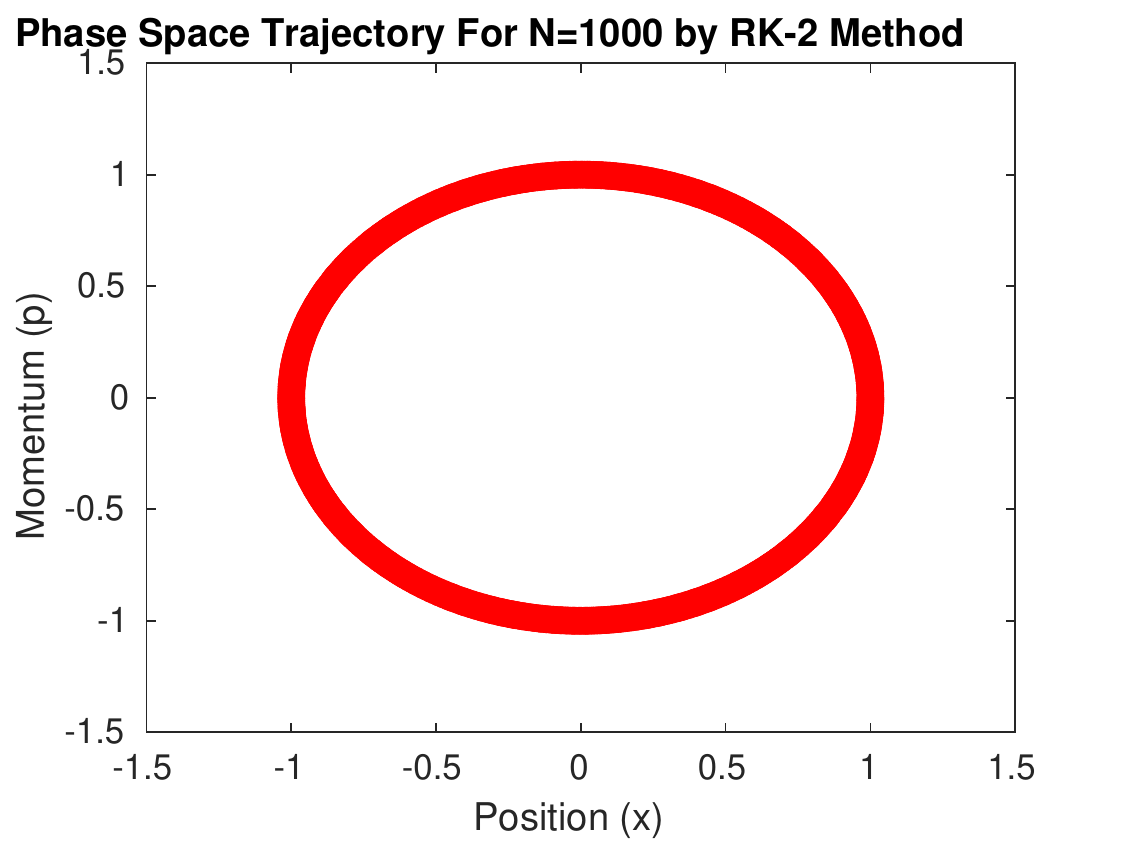}}
\subfloat[$N=10000$]{\includegraphics[width = 2.9in]{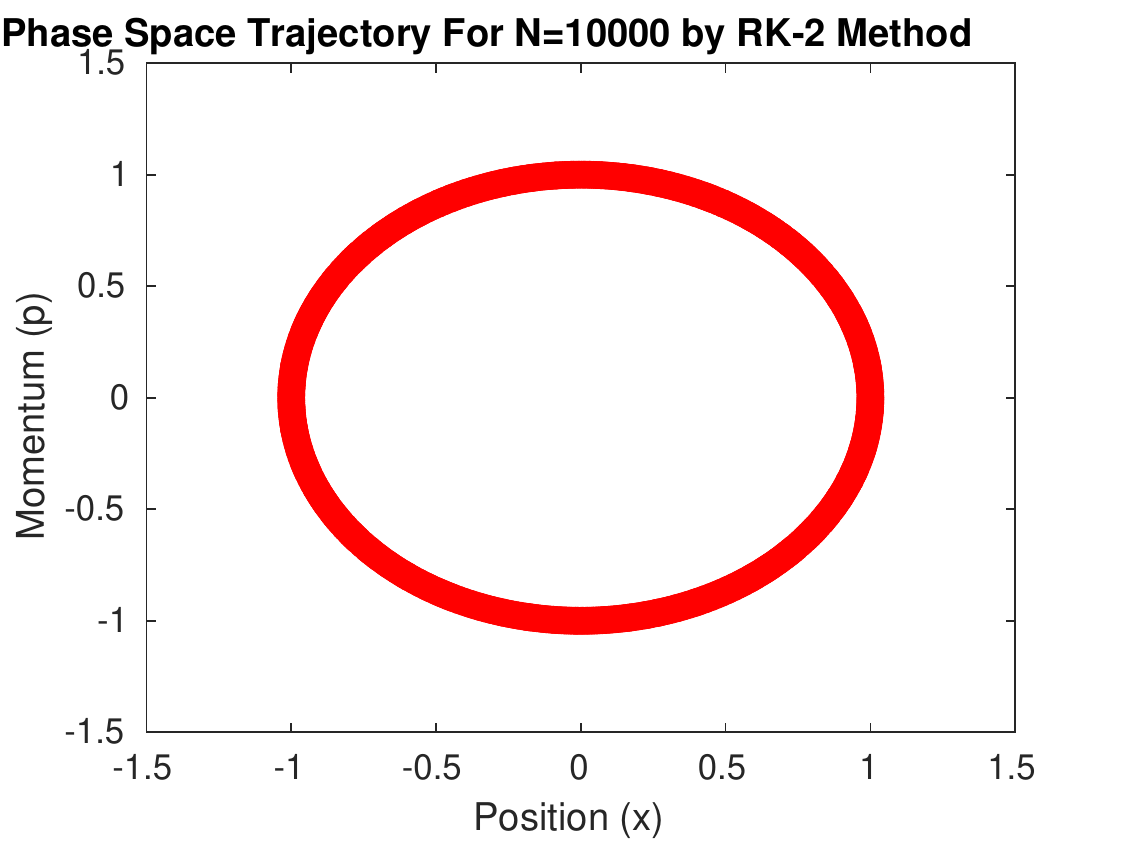}} 
\caption{Phase space trajectories for different number of steps ($N$) inside a full time period $T$ (For second-order Runge-Kutta Method)}
\label{fig_rk2_1}
\end{figure}

 From figure \ref{fig_rk2_1} we can see for small value of $N$ the phase space solution is diverging. However for larger values of $N$ we get very close solution to actual elliptical one.
 
\begin{figure}[ht]
\centering
\subfloat[$N=10$]{\includegraphics[width = 2.9in]{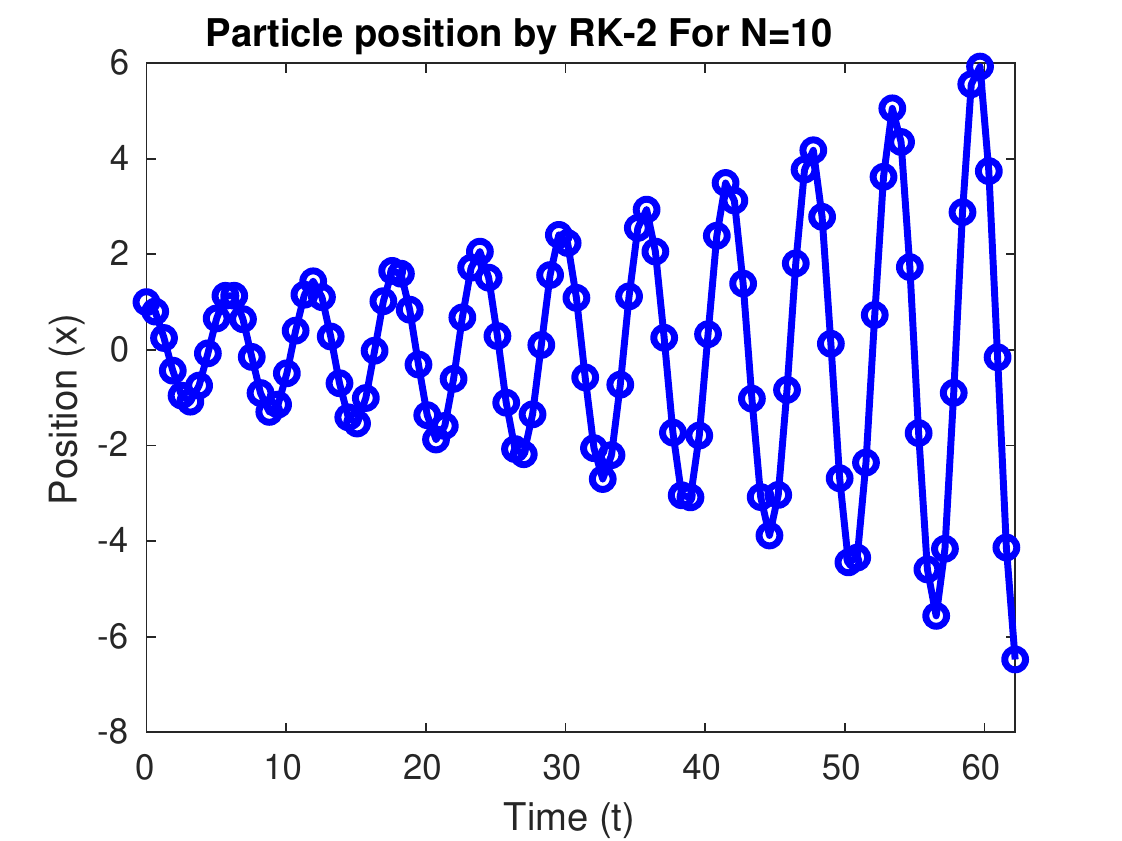}} 
\subfloat[$N=100$]{\includegraphics[width = 2.9in]{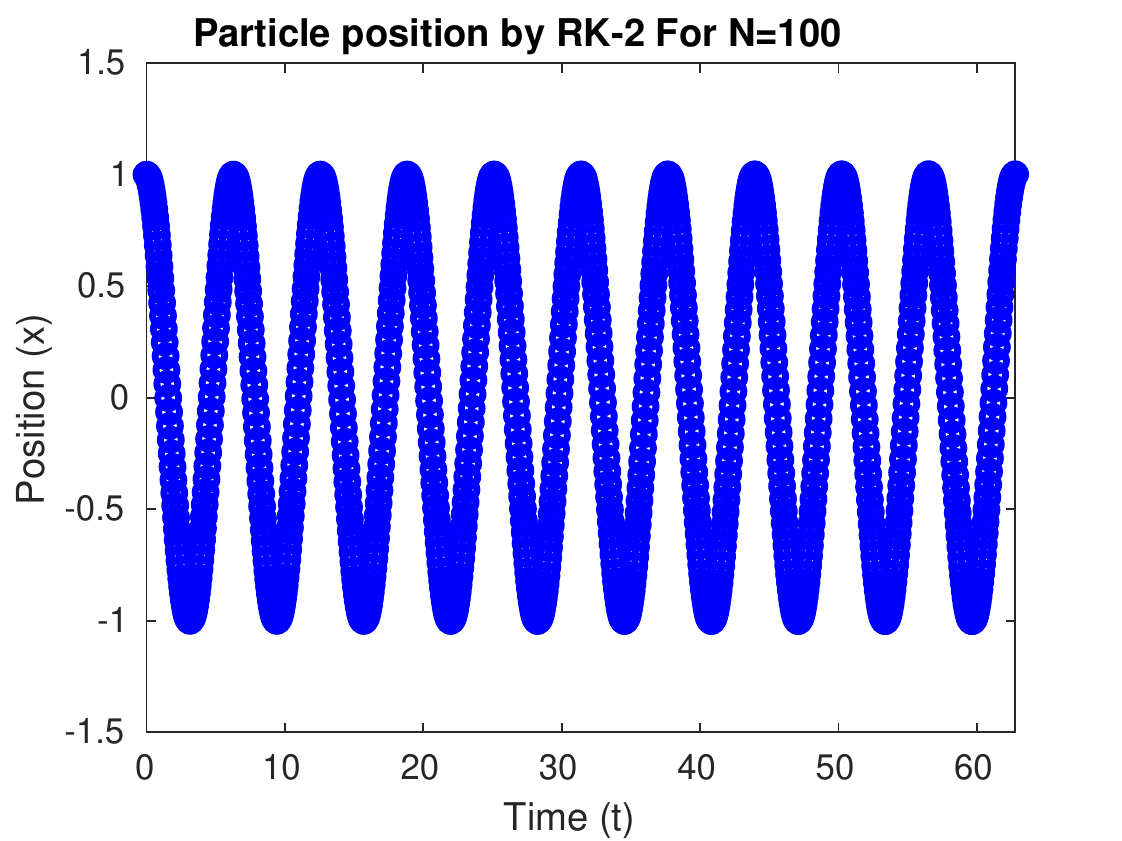}}\\
\subfloat[$N=1000$]{\includegraphics[width = 2.9in]{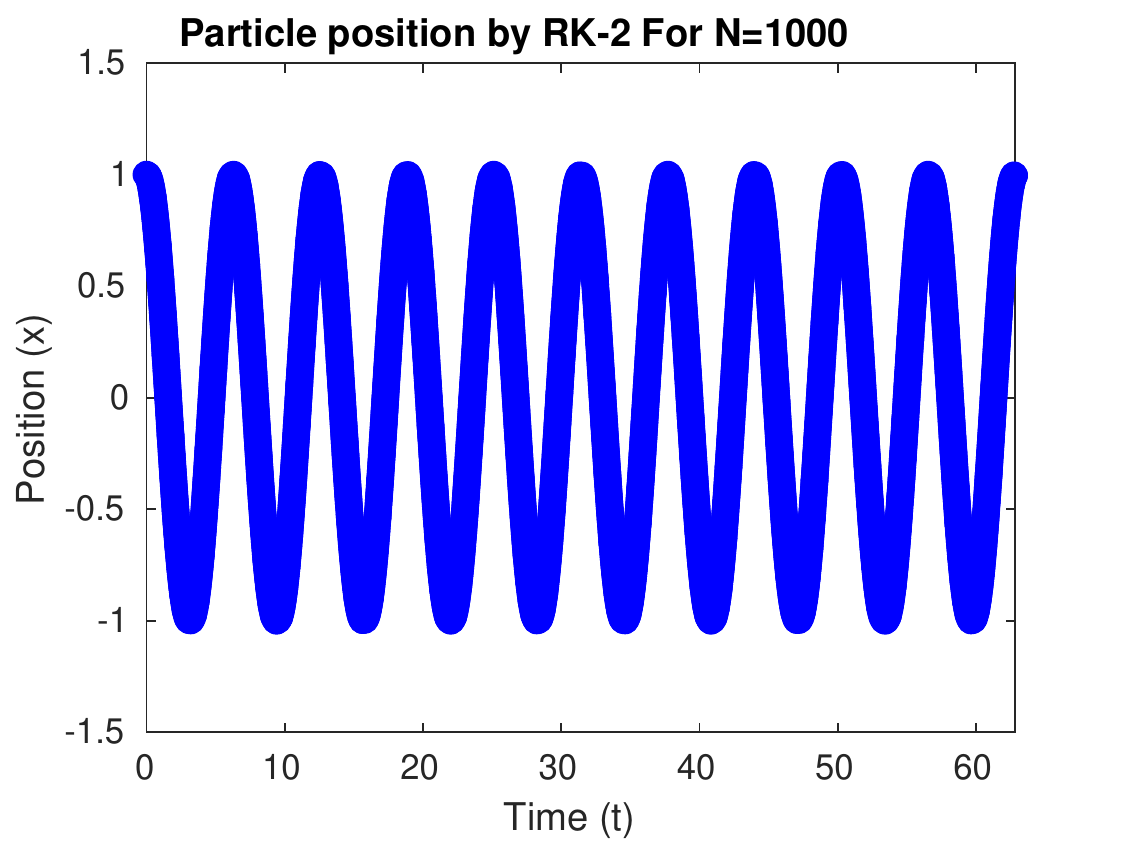}}
\subfloat[$N=10000$]{\includegraphics[width = 2.9in]{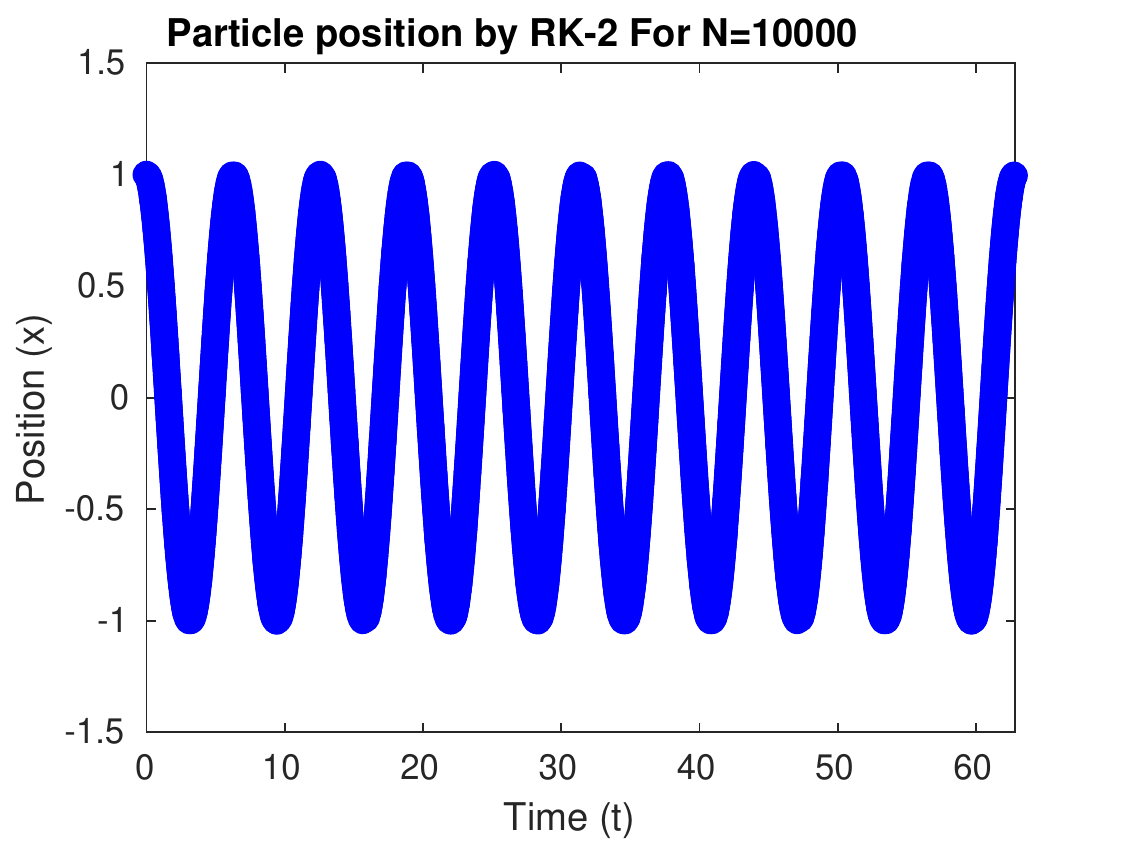}} 
\caption{Solution of $x(t)$ vs $t$ plot for different number of steps ($N$) inside a full time period $T$ (For second-order Runge-Kutta Method)}
\label{fig_rk2_2}
\end{figure}

 From figure \ref{fig_rk2_2} we can see that the solution of position $x(t)$ is diverging for small $N$, but for large $N$ we get very stable solution. 
 
\begin{figure}[ht]
\centering
\subfloat[]{\includegraphics[width = 2.9in]{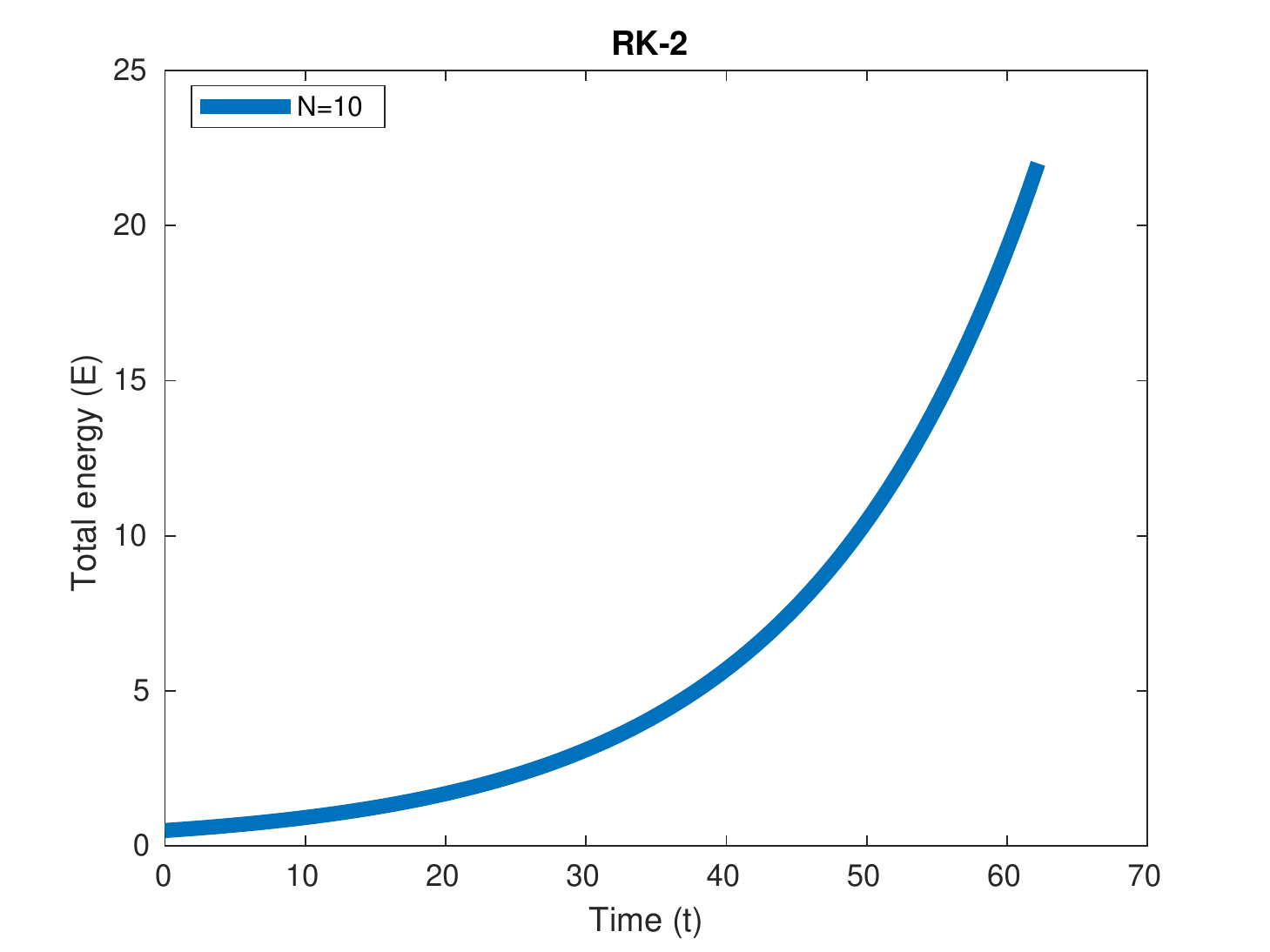}} 
\subfloat[]{\includegraphics[width = 2.9in]{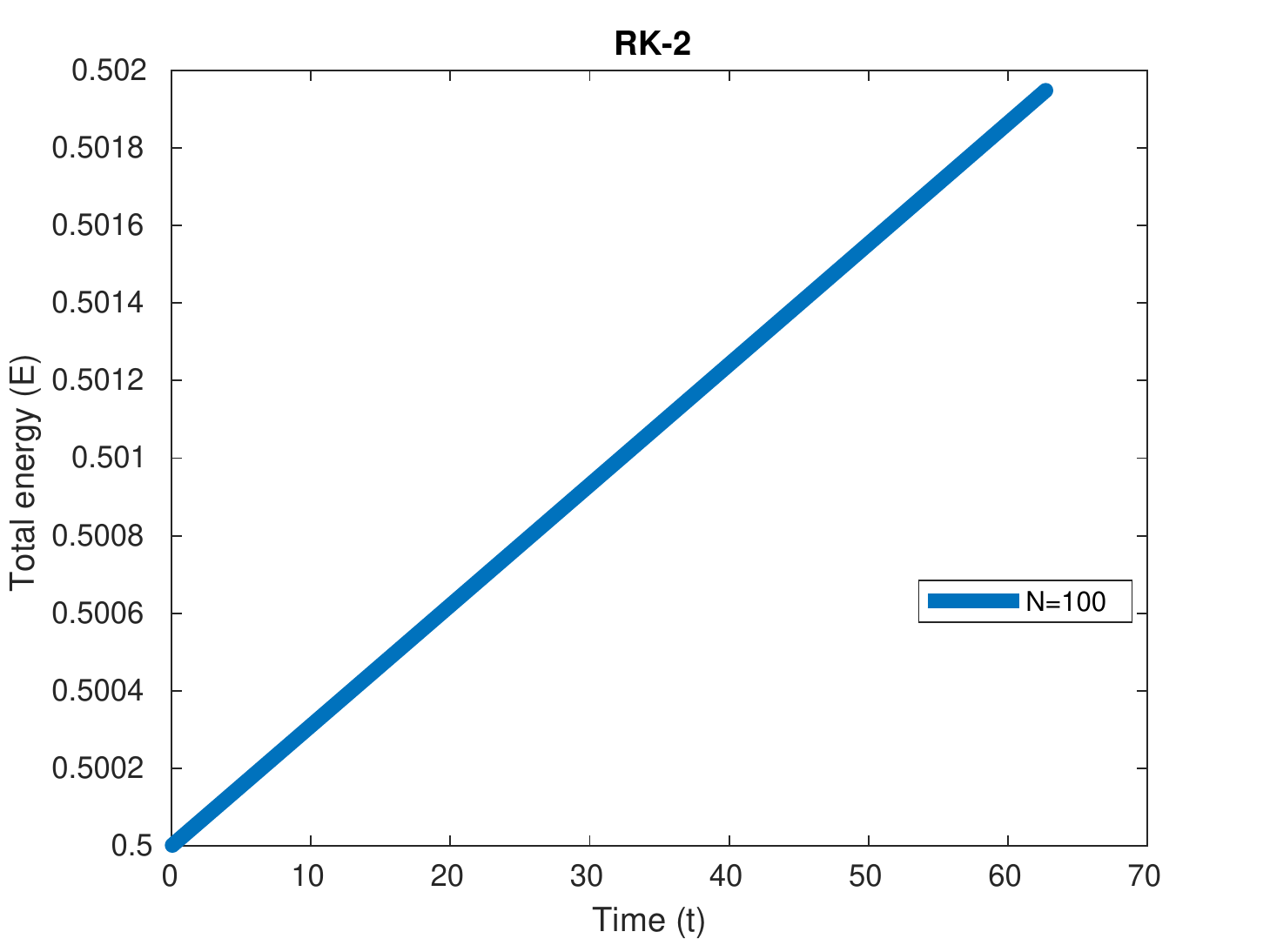}}\\
\subfloat[]{\includegraphics[width = 2.9in]{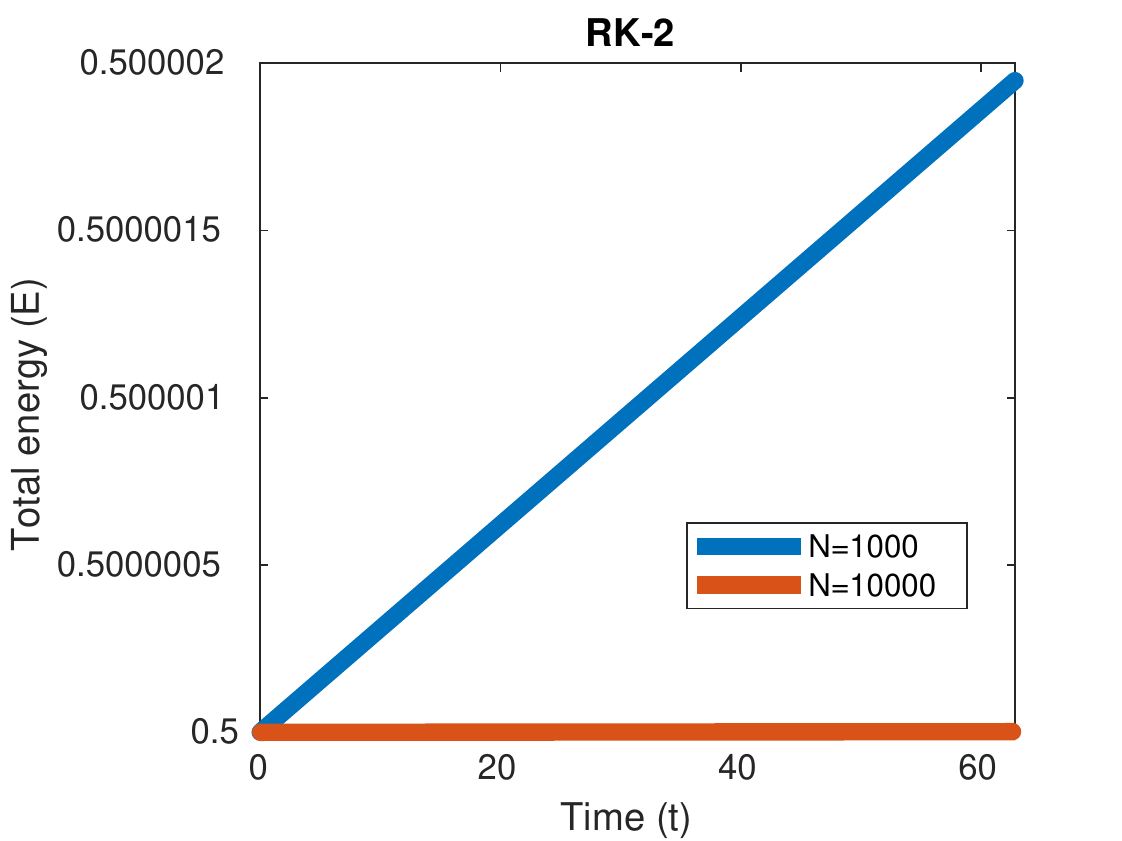}}
\caption{Constant energy value of the solution over time (For second-order Runge-Kutta Method)}
\label{fig_rk2_3}
\end{figure}

 The constancy of total energy value also can be observed from figure \ref{fig_rk2_3}. However for small $N$ energy value is not constant after some iteration, but for large $N$ values it is well constant.

\section{Fourth-Order Runge-Kutta method}
\subsection{Iteration steps}
The classical fourth-order Runge-Kutta Method is given by
\begin{equation}
    y_{n+1}=y_n+h\left[\frac{k_1}{6}+\frac{k_2}{3}=\frac{k_3}{3}+\frac{k_4}{6} \right]
\end{equation}
where,
\begin{equation}
    k_1=f(x_n,y_n)
\end{equation}
\begin{equation}
    k_2=f\left( x_n+\frac{h}{2},y_n+\frac{h}{2} k_1 \right)
\end{equation}
\begin{equation}
    k_3=f\left( x_n+\frac{h}{2},y_n+\frac{h}{2} k_2 \right)
\end{equation}
\begin{equation}
    k_4=f\left( x_n+h,y_n+h k_3 \right)
\end{equation}

\subsection{Explanations of plots}
\begin{figure}[ht]
\centering
\subfloat[$N=10$]{\includegraphics[width = 2.9in]{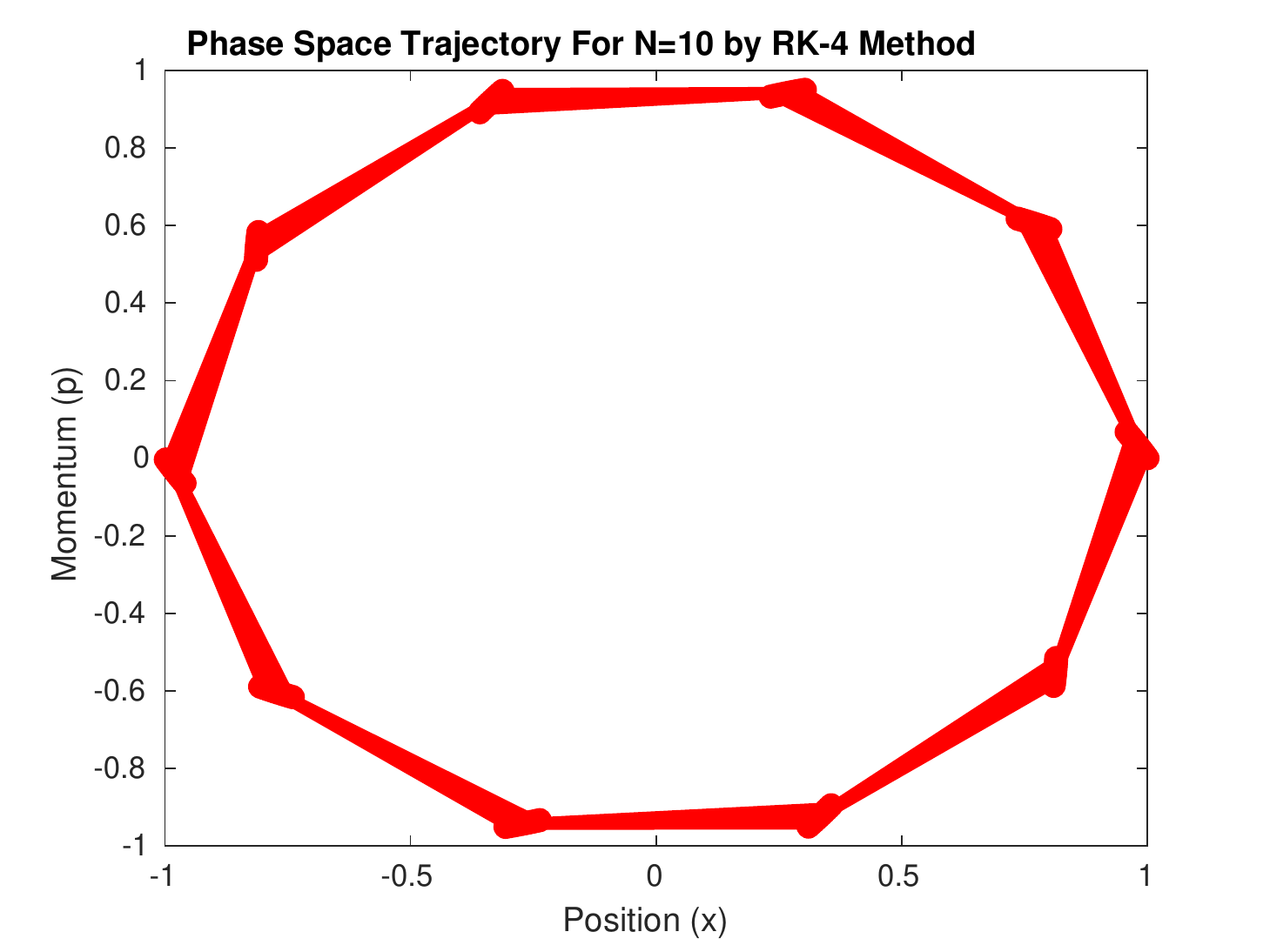}} 
\subfloat[$N=100$]{\includegraphics[width = 2.9in]{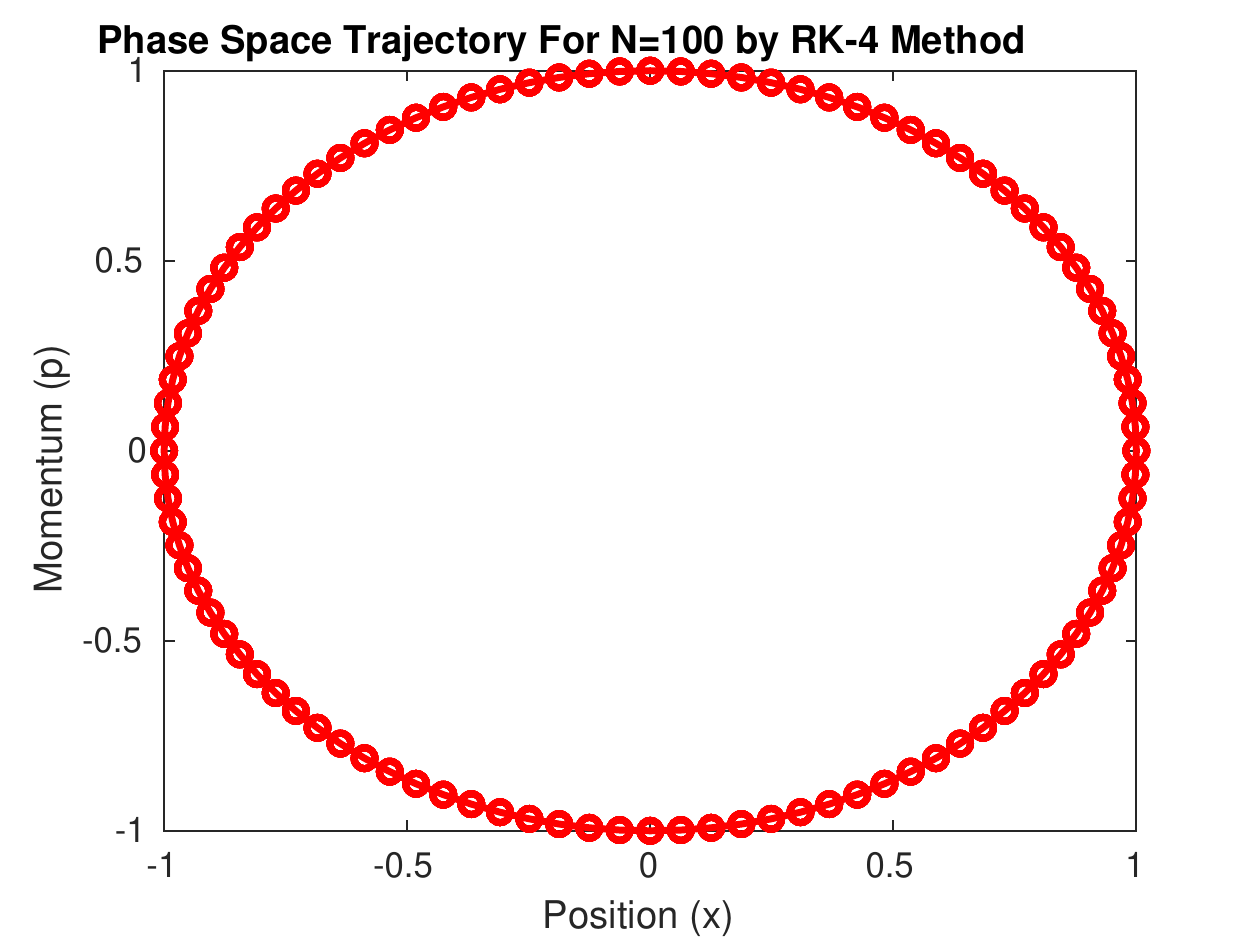}}\\
\subfloat[$N=1000$]{\includegraphics[width = 2.9in]{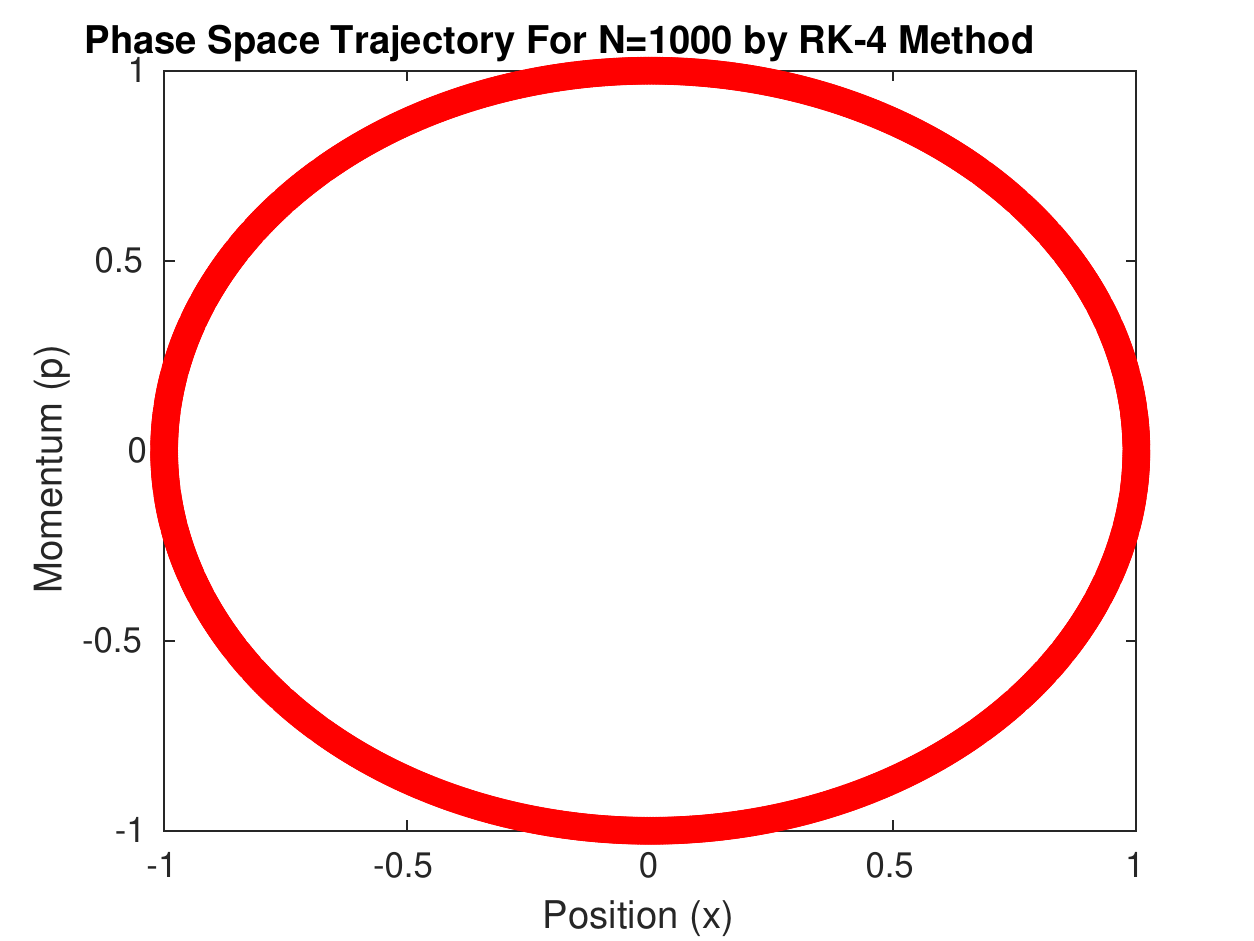}}
\subfloat[$N=10000$]{\includegraphics[width = 2.9in]{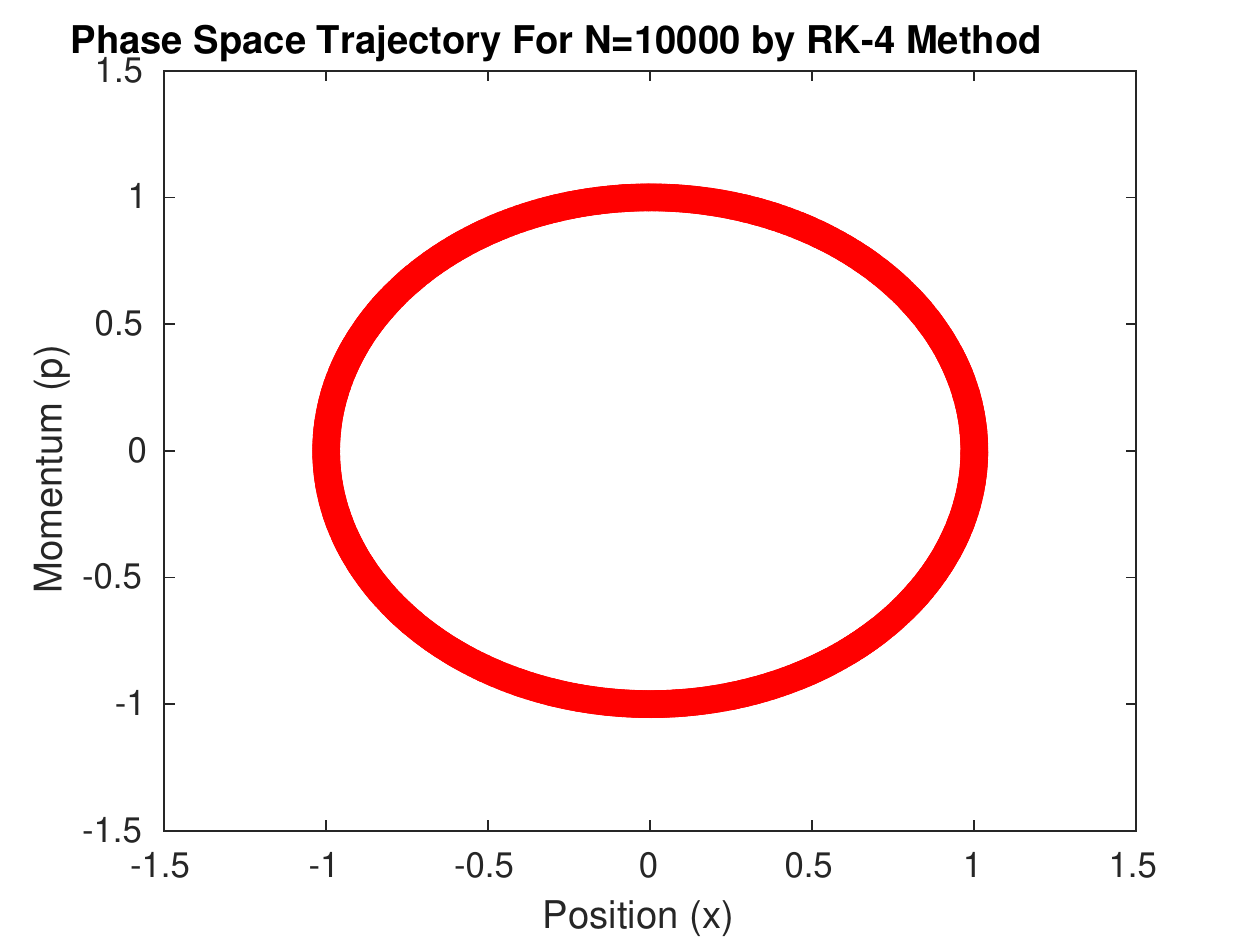}} 
\caption{Phase space trajectories for different number of steps ($N$) inside a full time period $T$ (For fourth-order Runge-Kutta Method)}
\label{fig_rk4_1}
\end{figure}

From figure \ref{fig_rk4_1} we can see that also for very small value of $N$ $(N=10)$ the phase space trajectory is stable and form a well ellipse.

\begin{figure}[ht]
\centering
\subfloat[$N=10$]{\includegraphics[width = 2.9in]{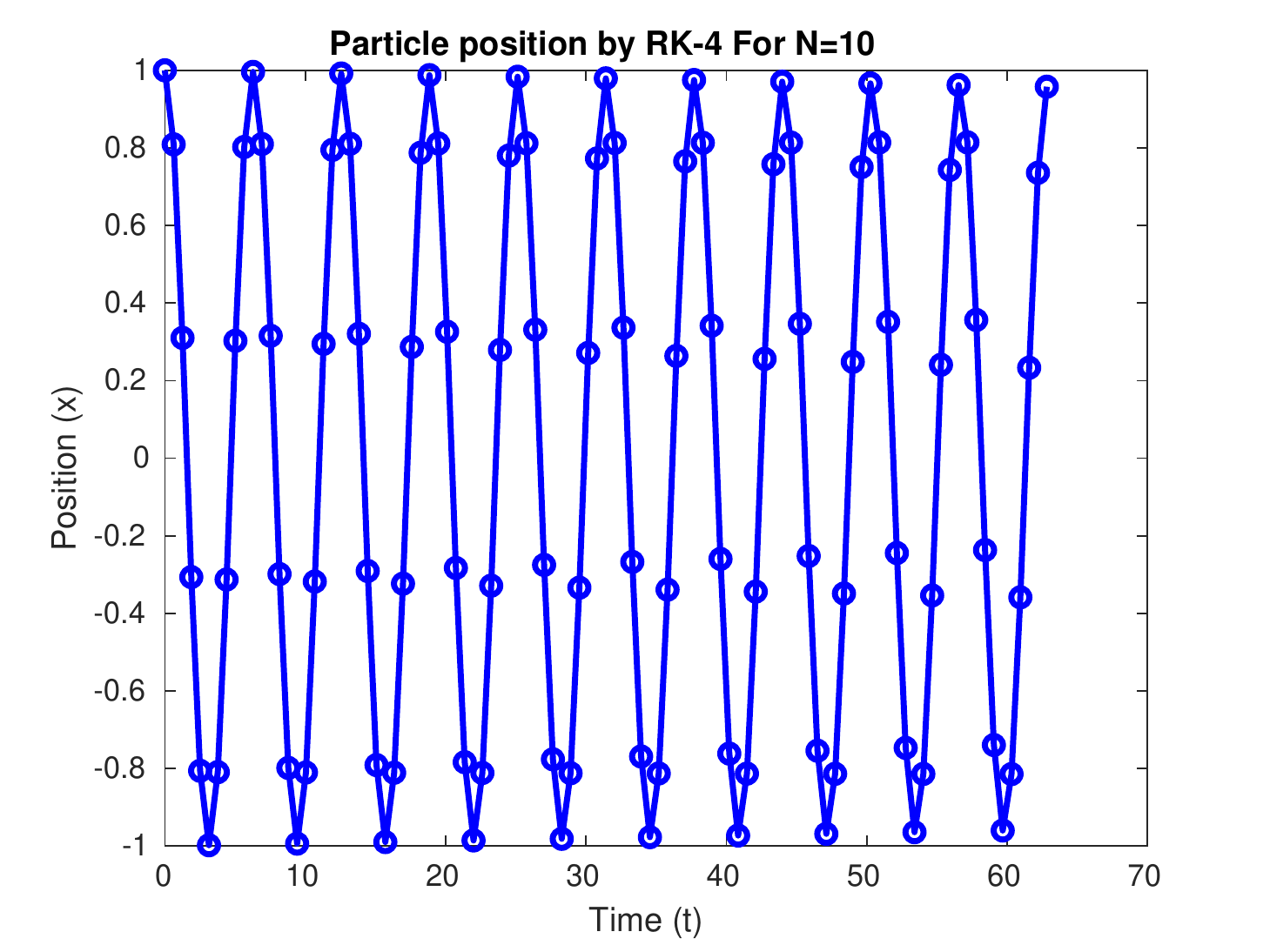}} 
\subfloat[$N=100$]{\includegraphics[width = 2.9in]{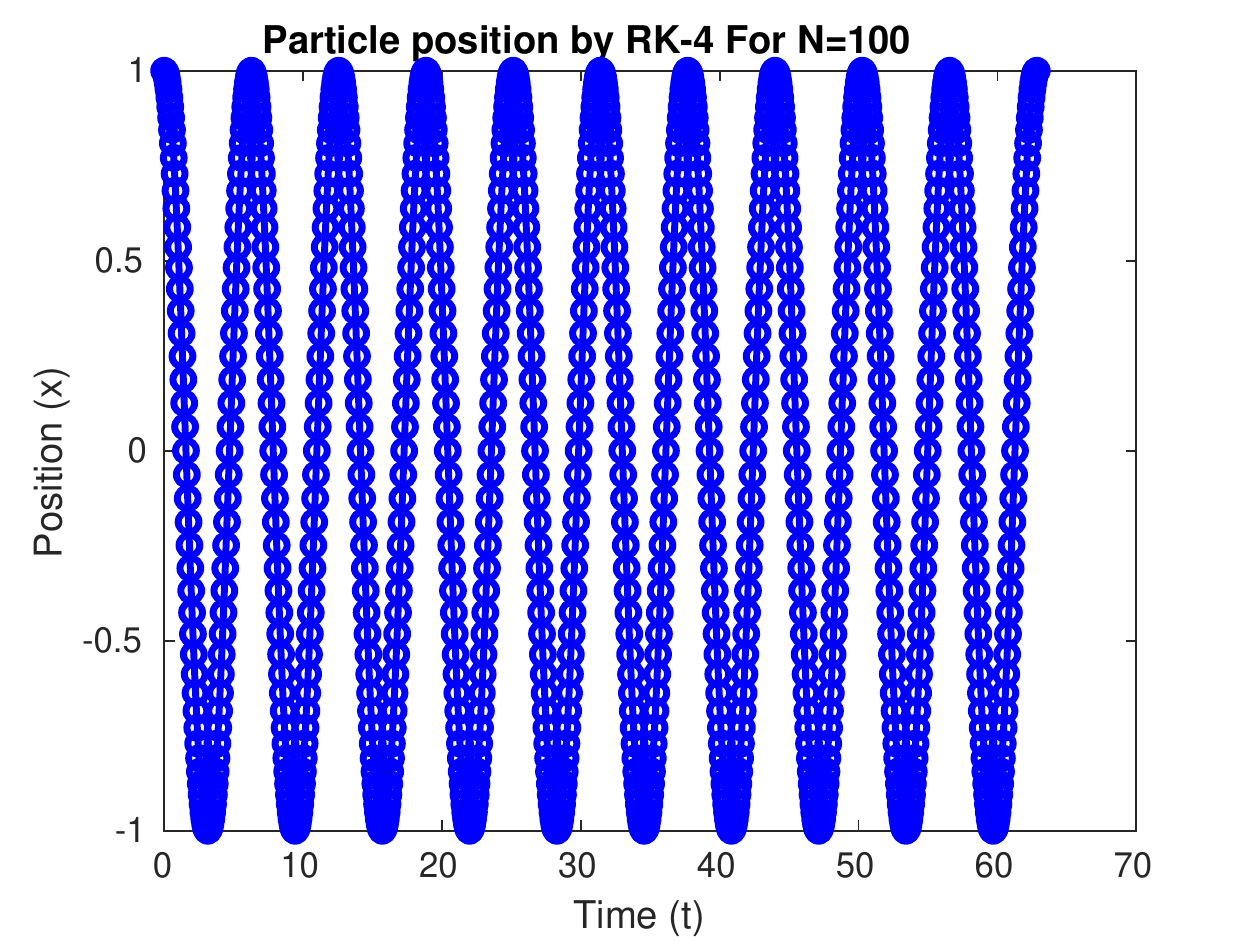}}\\
\subfloat[$N=1000$]{\includegraphics[width = 2.9in]{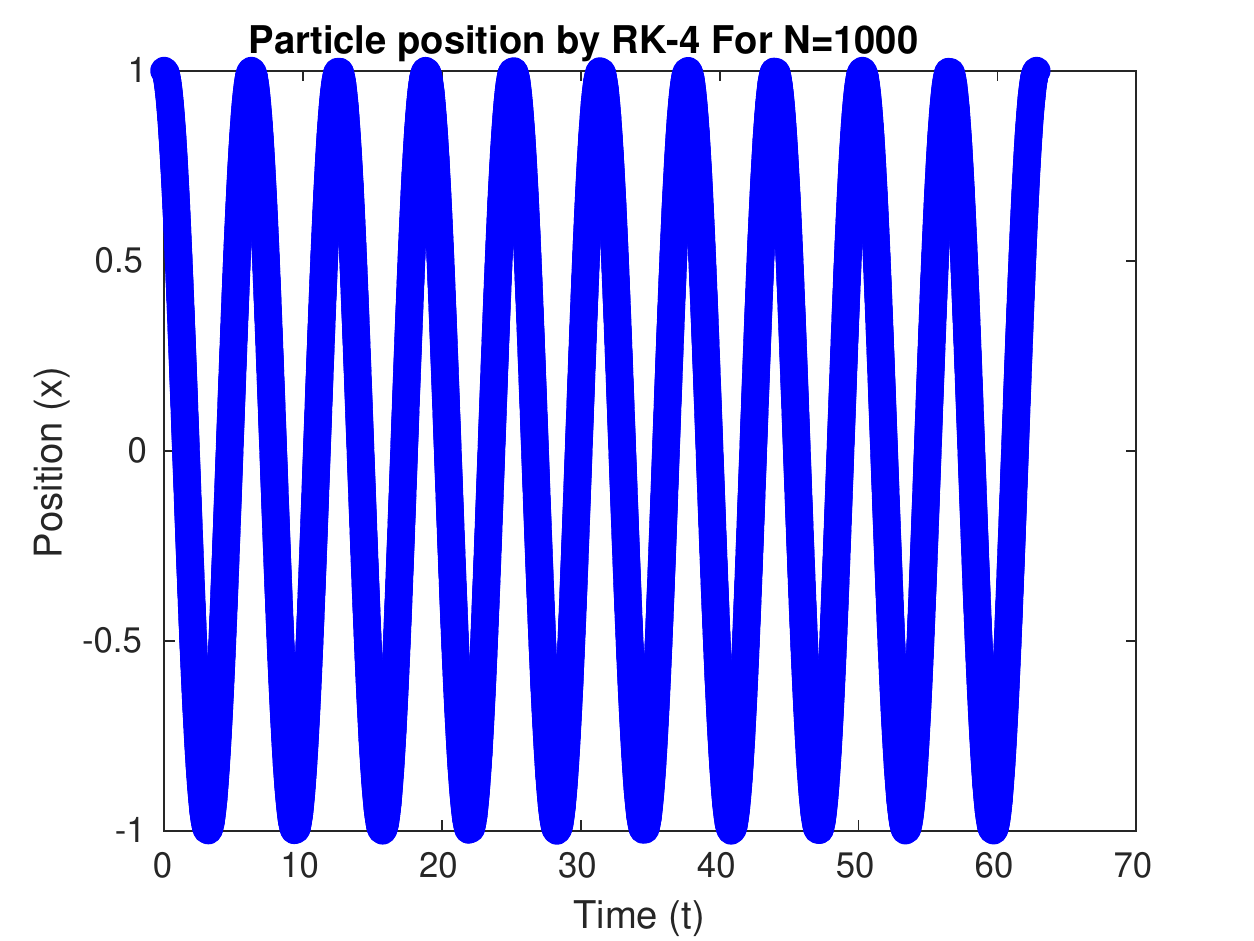}}
\subfloat[$N=10000$]{\includegraphics[width = 2.9in]{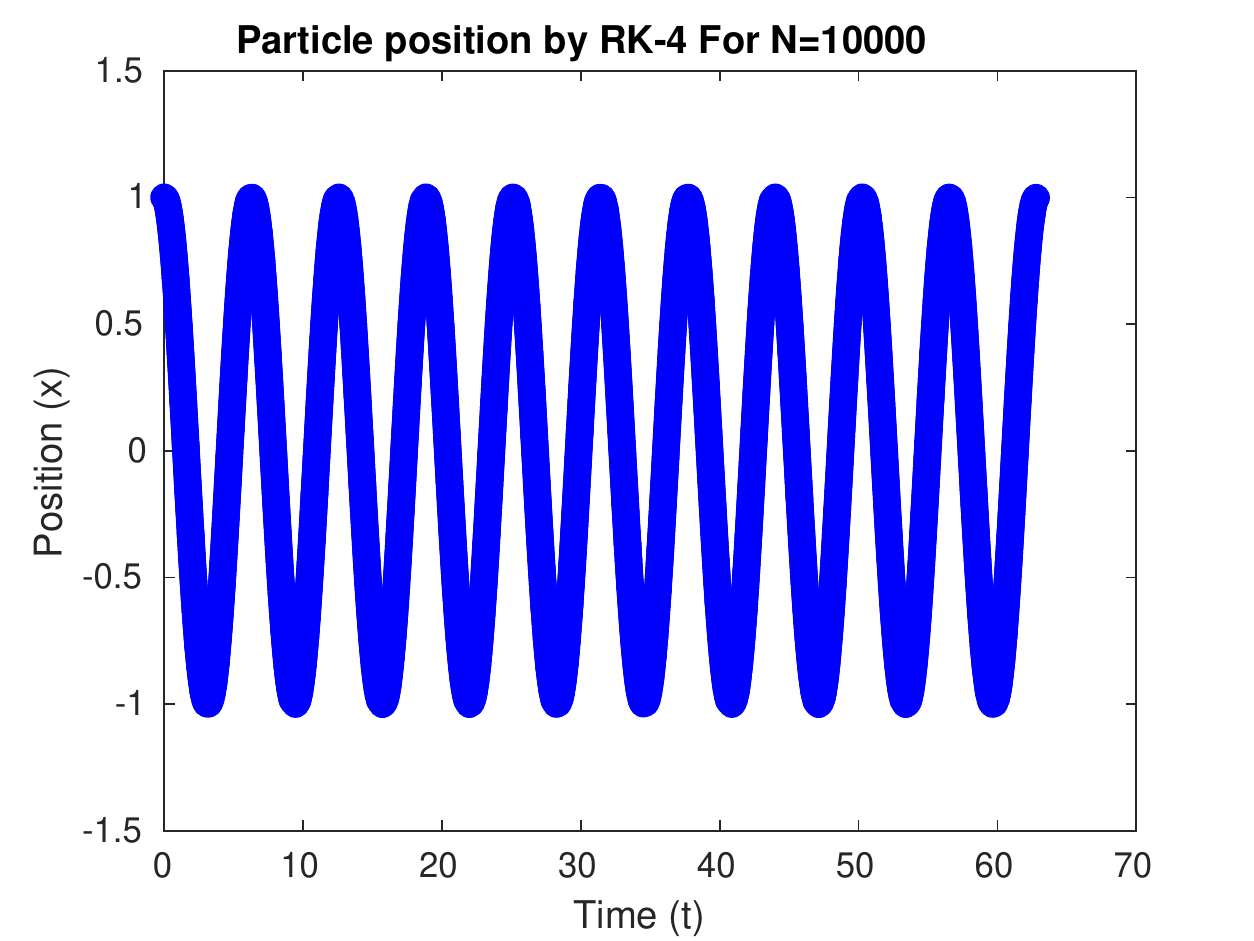}} 
\caption{Solution of $x(t)$ vs $t$ plot for different number of steps ($N$) inside a full time period $T$ (For Fourth-order Runge-Kutta Method)}
\label{fig_rk4_2}
\end{figure}

Same thing we can see in $x(t)$ plots (figure \ref{fig_rk4_2}). The solution is stable and well sinusoidal. 

\begin{figure}[ht]
\centering
\subfloat[]{\includegraphics[width = 2.9in]{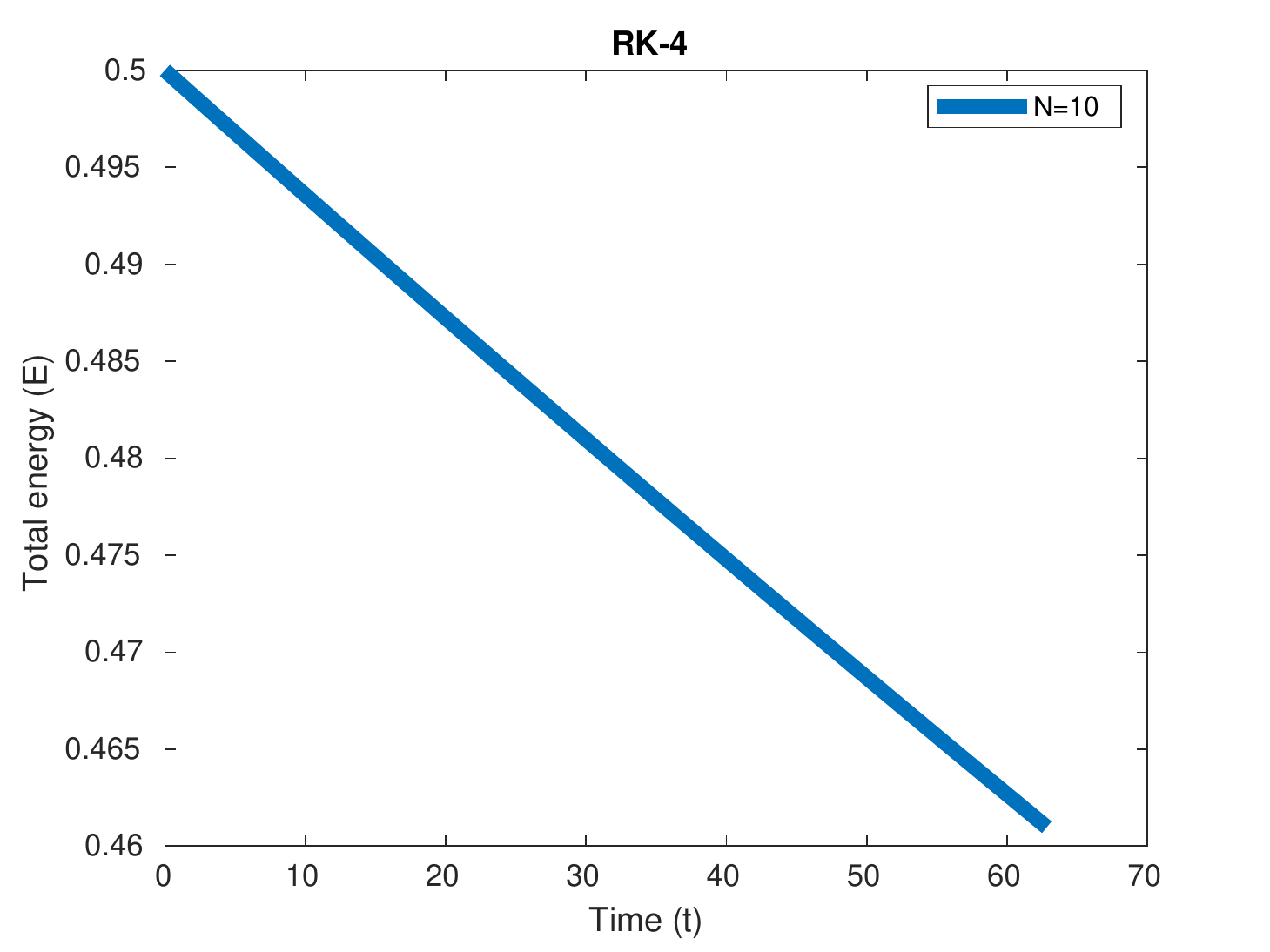}} 
\subfloat[]{\includegraphics[width = 2.9in]{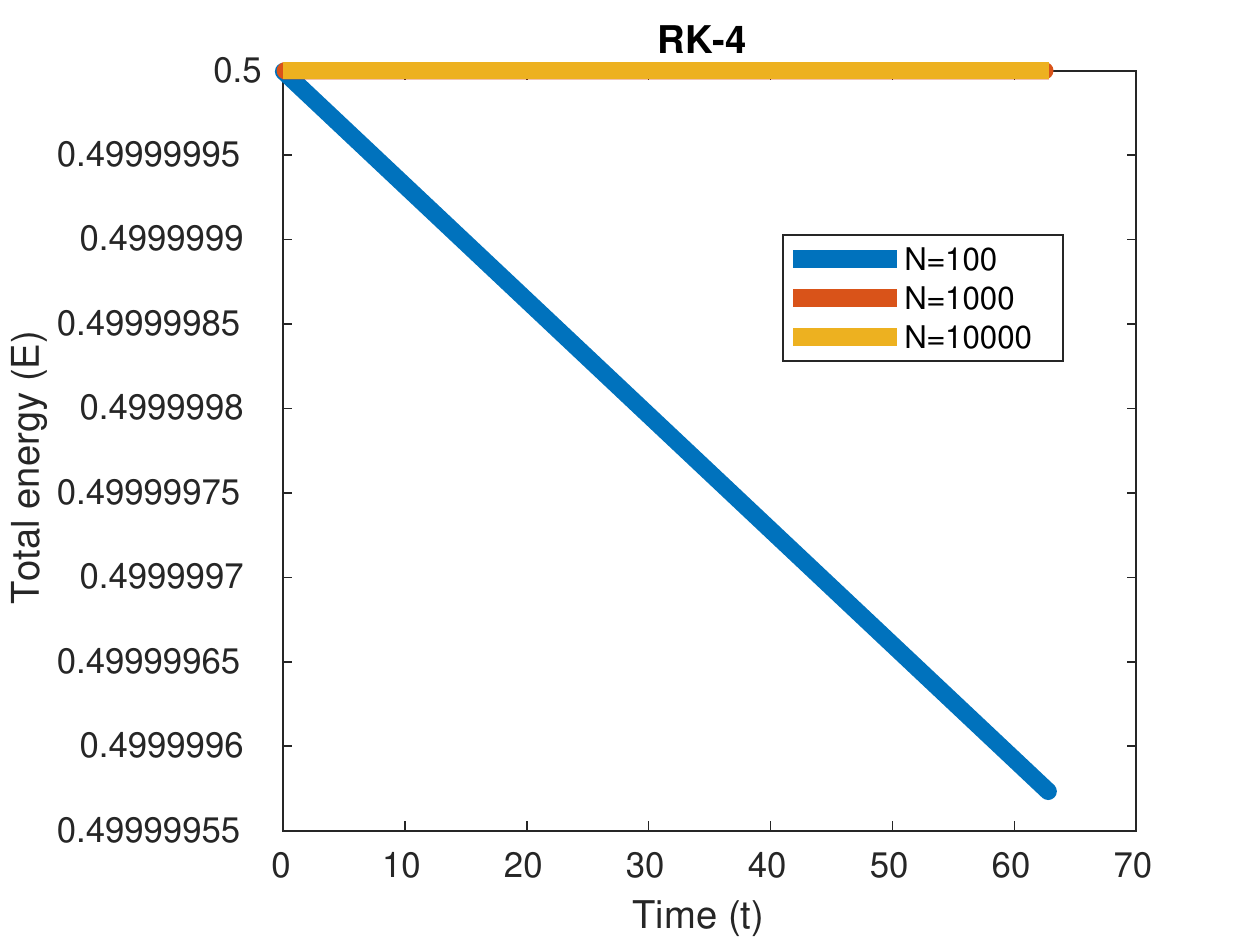}}
\caption{Constant energy value of the solution over time (For Fourth-order Runge-Kutta Method)}
\label{fig_rk4_3}
\end{figure}

The constancy of energy value is very good in this method. We can see that a very negligible energy change is there (figure \ref{fig_rk4_3}) which is very less compering to other methods.

\section{Error Analysis and Discussions }
From the plots we clearly can see that for a fixed number of steps $N$ the solution is better in RK-2 method than Euler method. Again RK-4 solution is much better than RK-2 solutions. Euler method is first order method, where RK-2 is a second order method. so just by increasing an order we see that result improve much. Same thing is happening in the case of RK-2 and Rk-4 methods. RK-4 is a Fourth-order approximation whereas RK-2 is second-order. So, just by increasing two orders the error of the solution minimizes a lot. In terms of Taylor series 
\begin{equation}
    y(x+h)=y(x)+h y'(x)+\frac{h^2}{2} y''(x)+...
\end{equation}

The taylor series is an infinite series. computationally it is impossible to take infinite terms. If we just take two terms in the RHS, then that is a first-order approximation, which is Euler method. If we take three terms then that is a second-order approximation and it has been shown in section \ref{secrk_2} that it is nothing but RK-2 method.
\begin{figure}[ht]
    \centering
    \includegraphics[width = 6in]{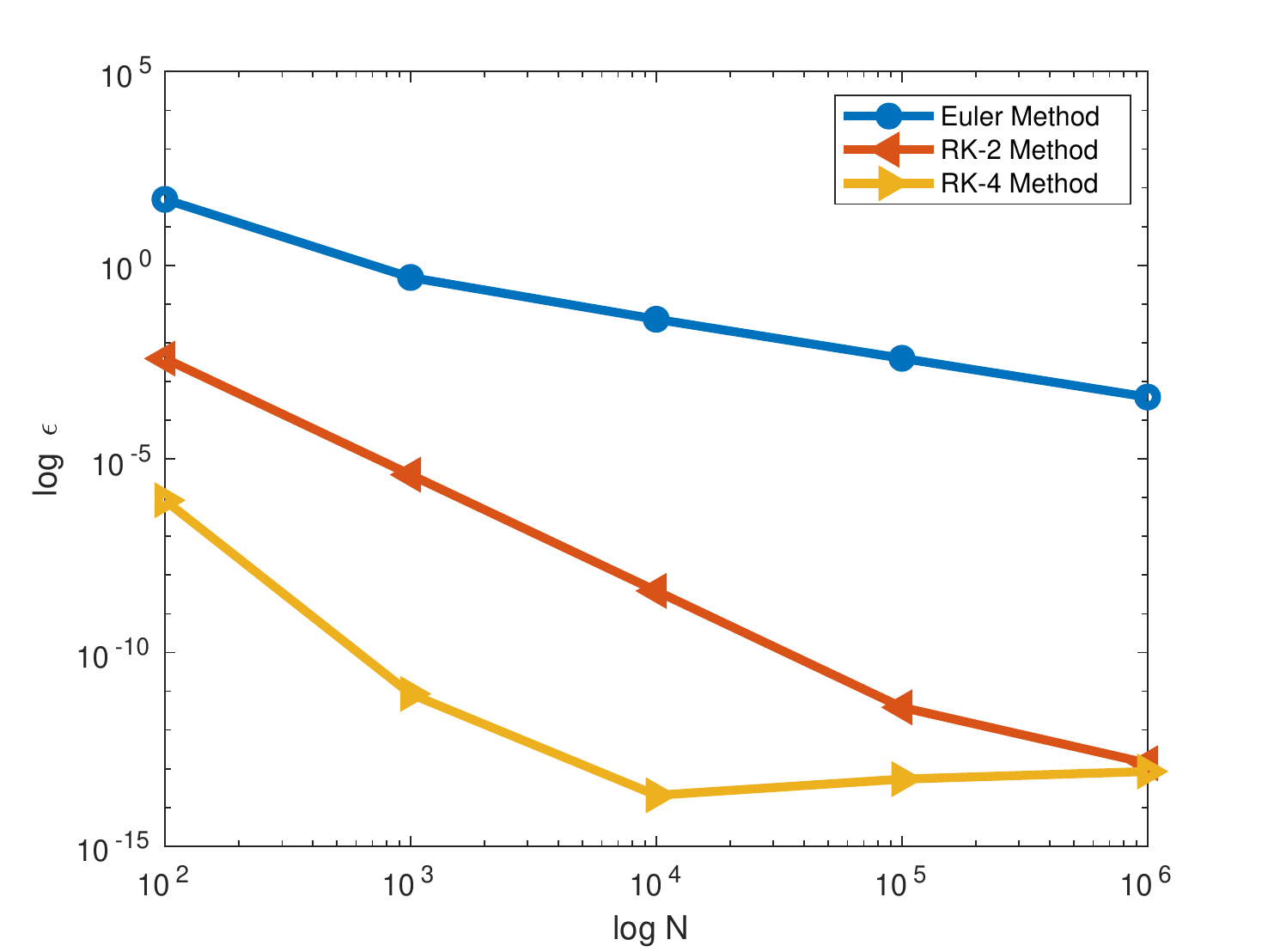}
    \caption{$\epsilon$ vs $N$ plot in log-log scale for different numerical methods}
    \label{fig_err}
\end{figure}
Now in figure \ref{fig_err} we plotted the relative error in energy, i.e. $\varepsilon=\frac{\Delta E}{E}$ where $E$ is the total energy of the system, which should be constant of the system and $\Delta E$ is the energy change in the process due to accumulation of error. We plotted this relative error with respect to number of steps $N$ in log-log scale. Theoretically $\Delta E =0$ so we should get a line parallel to $\varepsilon$ axis and passing through $0$. But due to error accumulation in each step, the actual plot that we get for different methods is shown in figure \ref{fig_err}.

From figure \ref{fig_err} it is clear for every process the error is less for larger number of steps $N$. The slop of the plot is however less for Euler method, RK-2 is in middle and RK-4 has maximum slop. Larger slope means the error is falling faster with $N$. 

\section{Conclusion}
From this assignment we conclude that obviously high order methods are less erratic and error per step is also less. But for computational simplicity we cannot choose higher order methods as much as we want. So, we have choose a simple less order method but again with less error. In next chapter we will talk about such a method, which is first order but still energy conserving.

 \chapter{Solution by Leapfrog Method}
\section{Aim}
To solve the SHM problem by Leap frog method and check its energy conserving and reversible properties. To compare the solution of SHM problem by Leapfrog method with solutions of the same by previously discussed methods (Euler, RK-2 and RK-4).

\section{Introduction}
Leapfrog method is a numerical differential equation solver of the form 
\begin{equation}
  \frac{d^2 x}{d t^2}=a(x)
\end{equation}

This second order differential equation can be written in form of a coupled first order differential equation as 
\begin{equation}
    \frac{d x}{d t}=v(x) \; \mathrm{and} \; \frac{d v}{d t}=a(x)
\end{equation}

This method is very much useful for dynamical mechanical systems. This is a first order method, so the computational complexity is less, at the same time it is less erratic process. The accumulation of errors in energy is very less, so this process is energy conserving. For our SHM problem, the force is conservative so the energy should be constant. In previous Assignment we see that this constancy of energy is no longer retained in first order methods like Euler method. Here we will solve the same using leapfrog method and will compare our results with previous assignment (Assignment-2).

\section{Algorithm of Leapfrog Method}
We are going to solve a second order differential equation, so to solve it we need two initial conditions. The initial conditions are initial position of the particle $(x_0)$ and initial momentum of the particle $(p_0)$ or velocity $(v_0)$. 

\begin{figure}[ht]
    \centering
    \includegraphics[width = 5in]{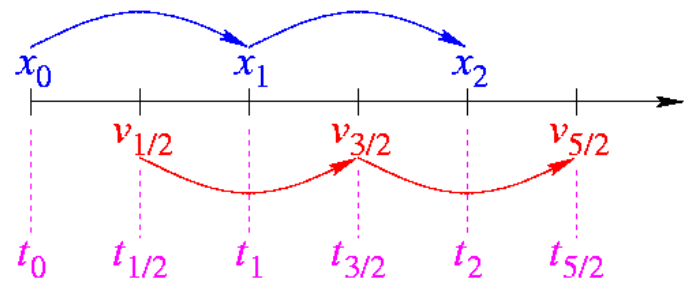}
    \caption{Pictorial representation of iteration steps in Leapfrog method}
    \label{fig_lp}
\end{figure}

\begin{figure}[ht]
\centering
\subfloat[$N=10$]{\includegraphics[width = 2.9in]{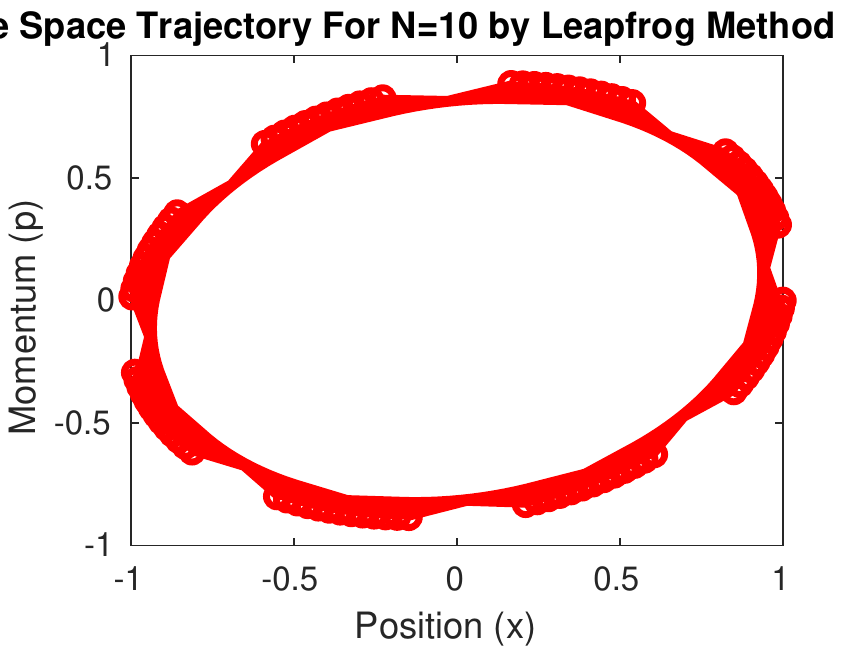}} 
\subfloat[$N=100$]{\includegraphics[width = 2.9in]{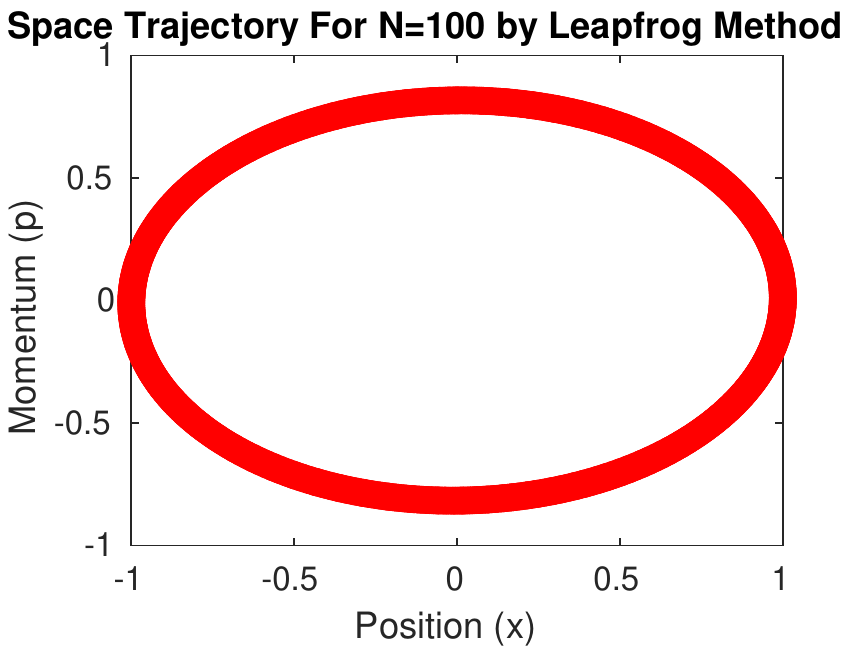}}\\
\subfloat[$N=1000$]{\includegraphics[width = 2.9in]{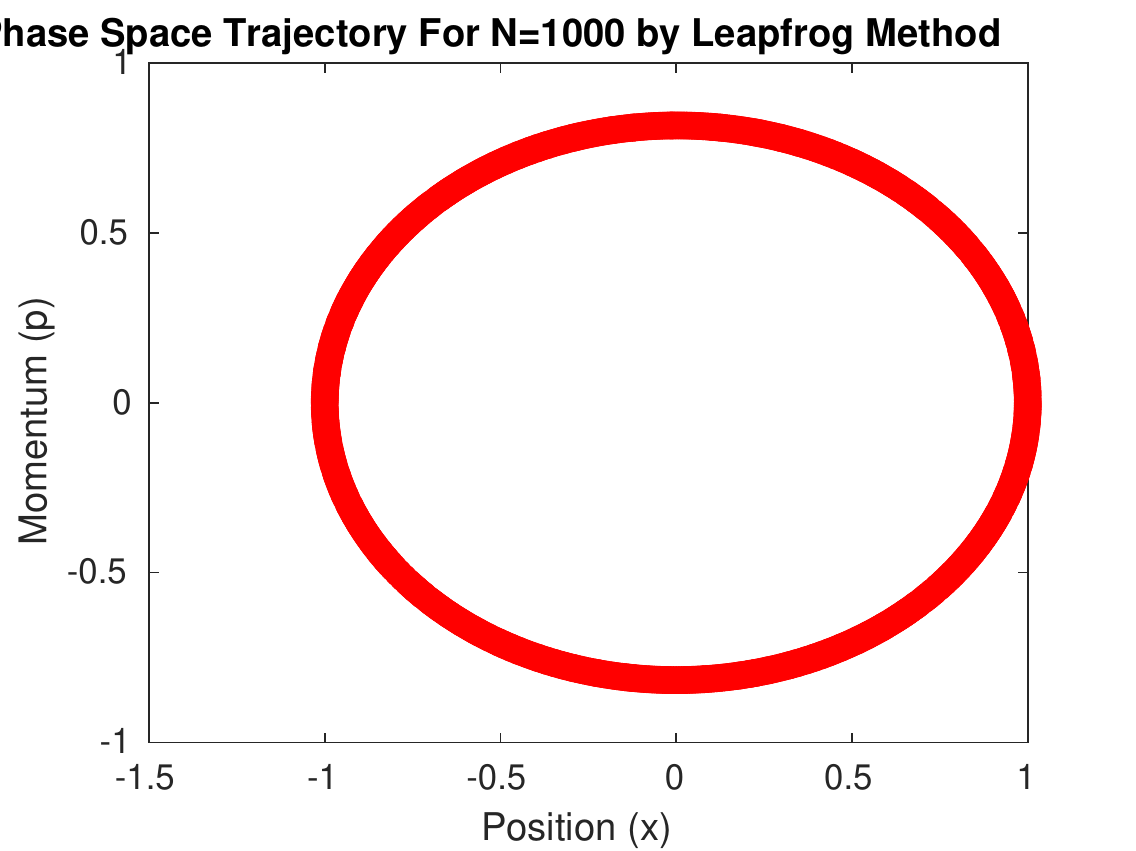}}
\subfloat[$N=10000$]{\includegraphics[width = 2.9in]{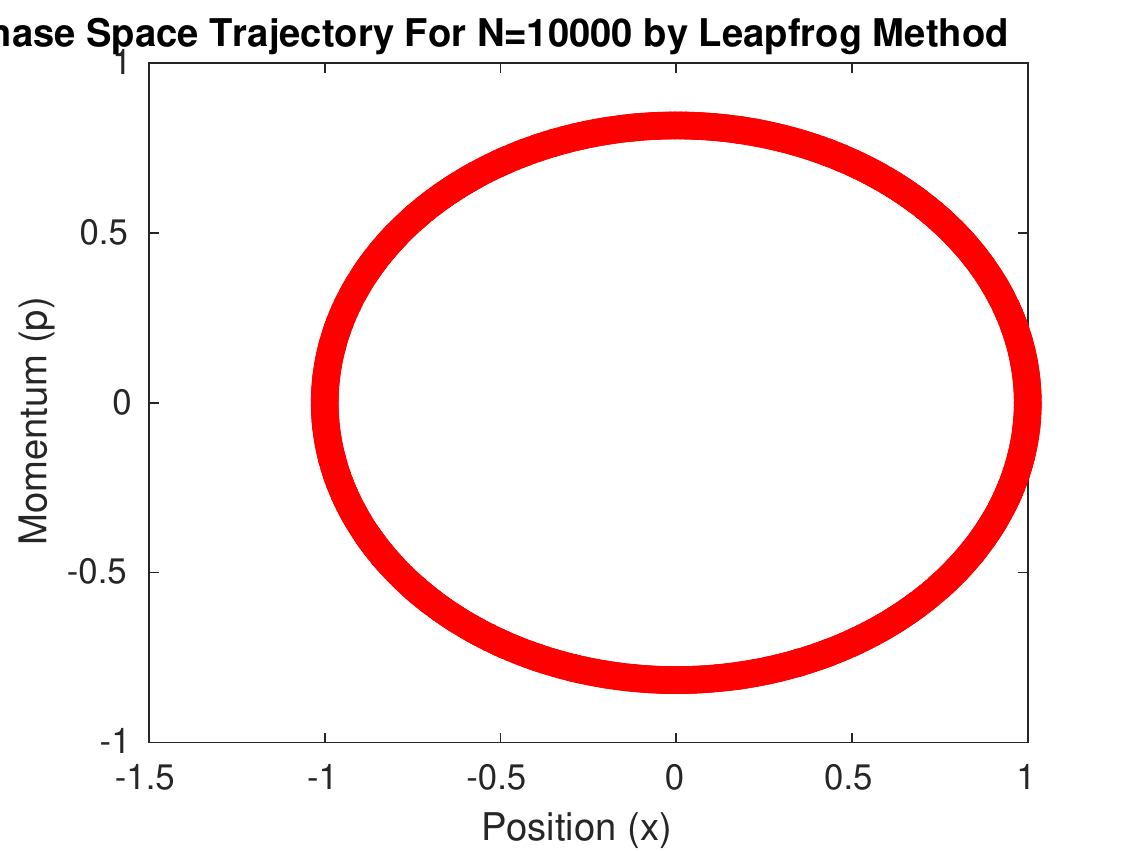}} 
\caption{Phase space trajectories for different number of steps ($N$) inside a full time period $T$ (By Leapfrog method)}
\label{fig_lf_1}
\end{figure}

\begin{figure}[ht]
\centering
\subfloat[$N=10$]{\includegraphics[width = 2.9in]{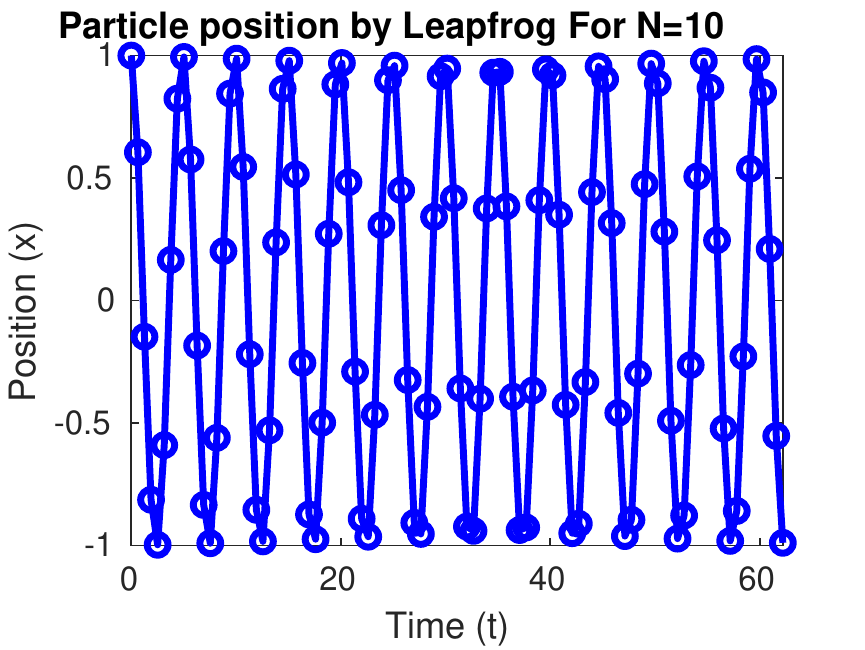}} 
\subfloat[$N=100$]{\includegraphics[width = 2.9in]{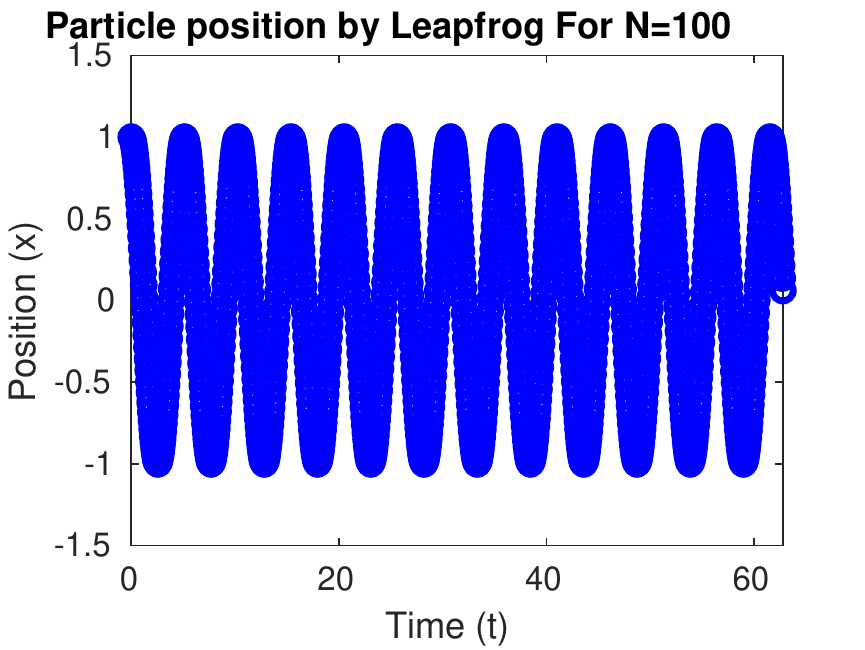}}\\
\subfloat[$N=1000$]{\includegraphics[width = 2.9in]{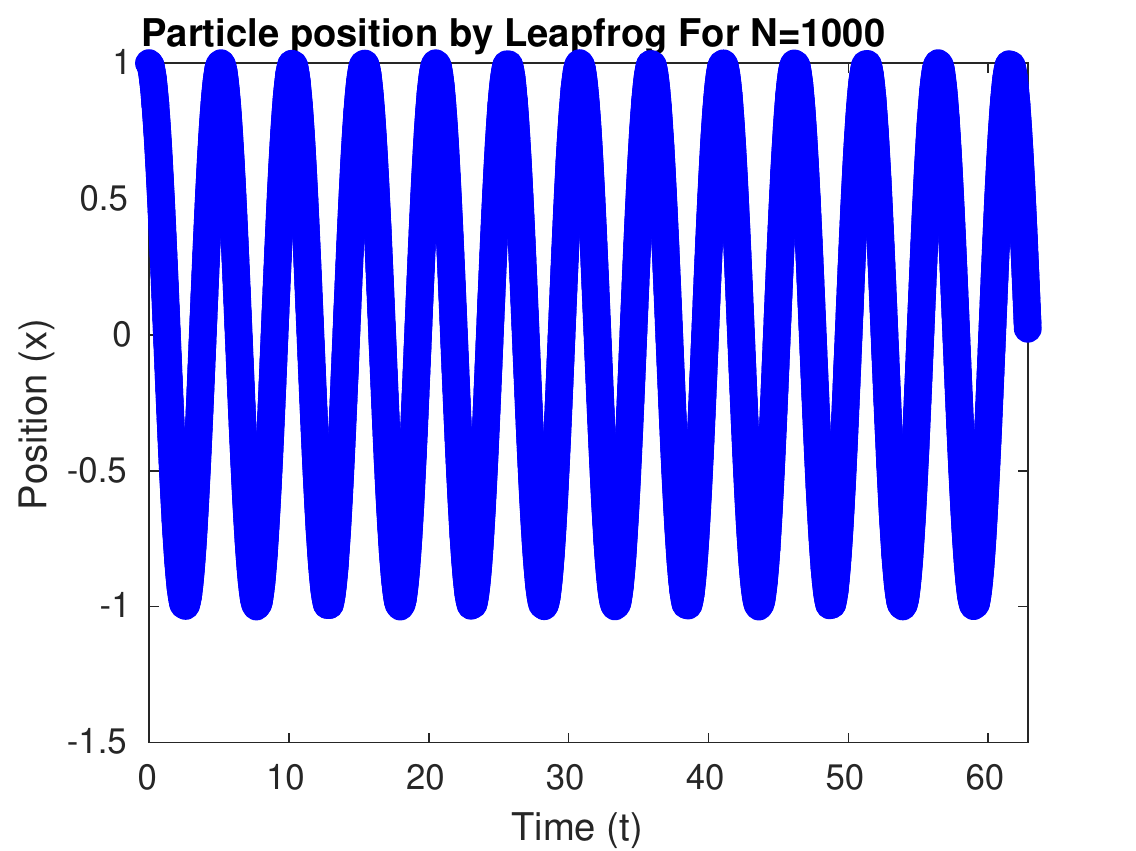}}
\subfloat[$N=10000$]{\includegraphics[width = 2.9in]{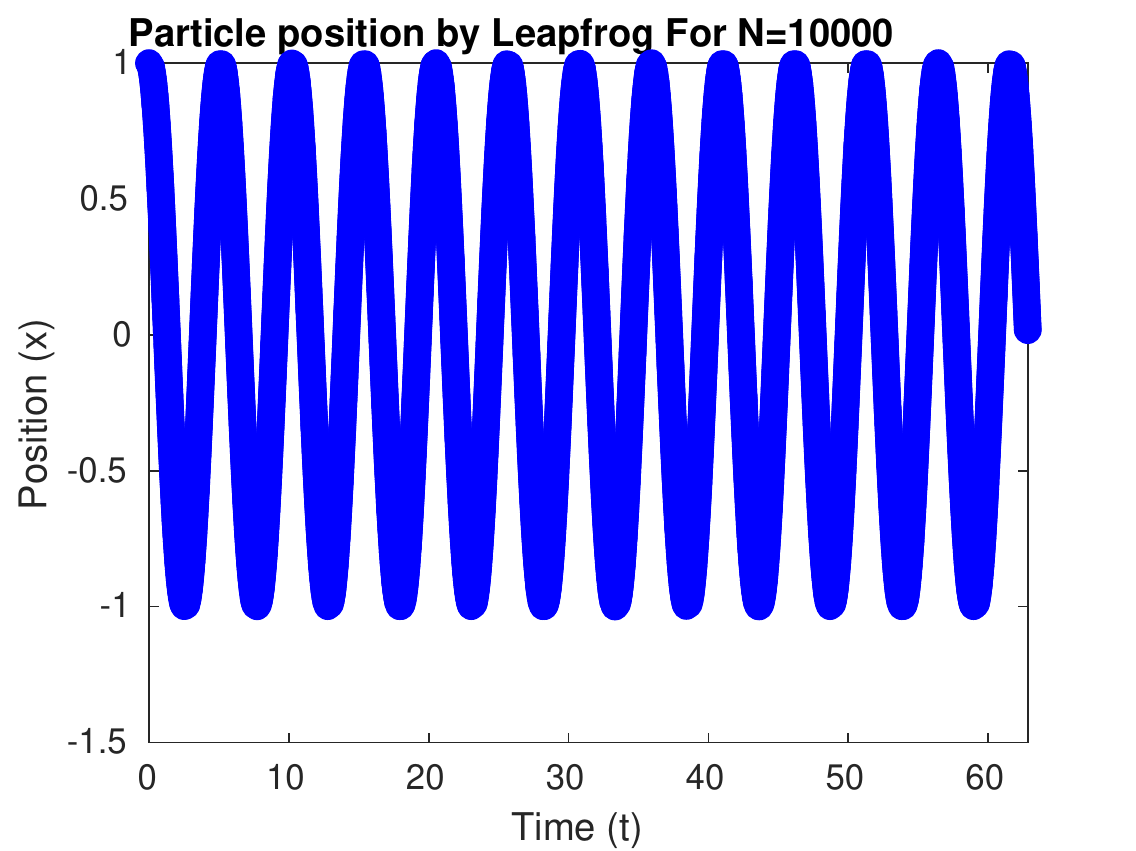}} 
\caption{Solution of $x(t)$ vs $t$ plot for different number of steps ($N$) inside a full time period $T$ (By Leapfrog method)}
\label{fig_lf_2}
\end{figure}

For our SHM problem 
\begin{equation}
    \frac{d x}{d t}=p(x)/m \; \mathrm{and} \; \frac{d p}{d t}=-k x
\end{equation}

From $x_0$ and $p_0$, the half step of $x$ by Leapfrog method is
\begin{equation}
    x_{\frac{1}{2}}=x_0+\frac{p_0}{m} \frac{\Delta t}{2}
\end{equation}
Now, by Leapfrog method full step on $p$ and $x$ can be calculated for rest of $n=0$ to $n=N$ by
\begin{equation}
    p_(n+1)=p_n-k x_{n+\frac{1}{2}} \Delta t
\end{equation}
again,
\begin{equation}
    x_{n+\frac{3}{2}}=x_{n+\frac{1}{2}}+\frac{p_{n+1}}{m} \Delta t
\end{equation}

There are two primary strengths to leapfrog integration when applied to mechanics problems. The first is the time-reversibility of the Leapfrog method. One can integrate forward n steps, and then reverse the direction of integration and integrate backwards n steps to arrive at the same starting position. The second strength is its symplectic nature, which implies that it conserves the (slightly modified) energy of dynamical systems. This is especially useful when computing orbital dynamics, as many other integration schemes, such as the (order-4) Runge–Kutta method, do not conserve energy and allow the system to drift substantially over time.

\begin{figure}[ht]
\centering
\subfloat[$N=10$]{\includegraphics[width = 2.9in]{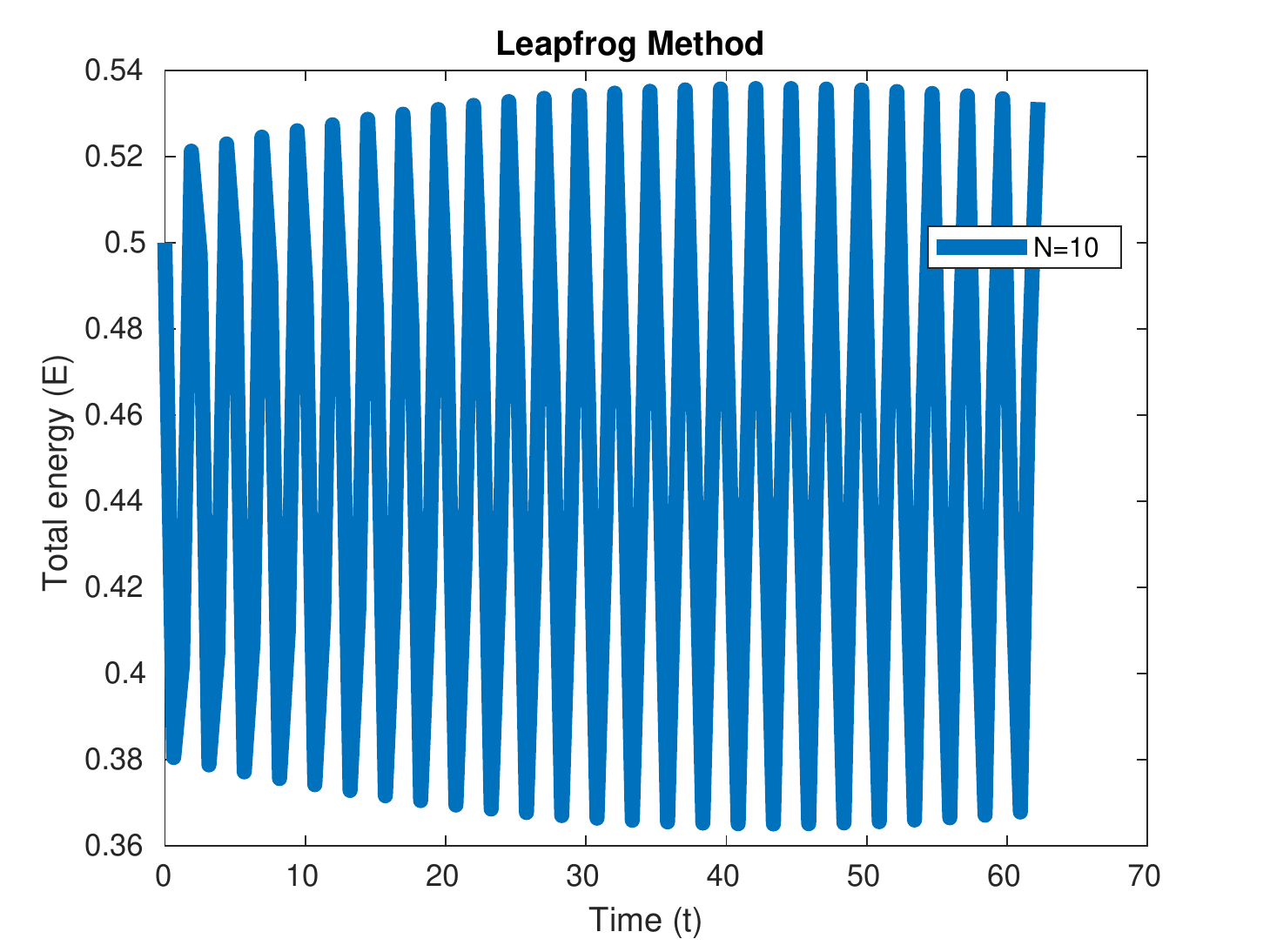}} 
\subfloat[$N=100$]{\includegraphics[width = 2.9in]{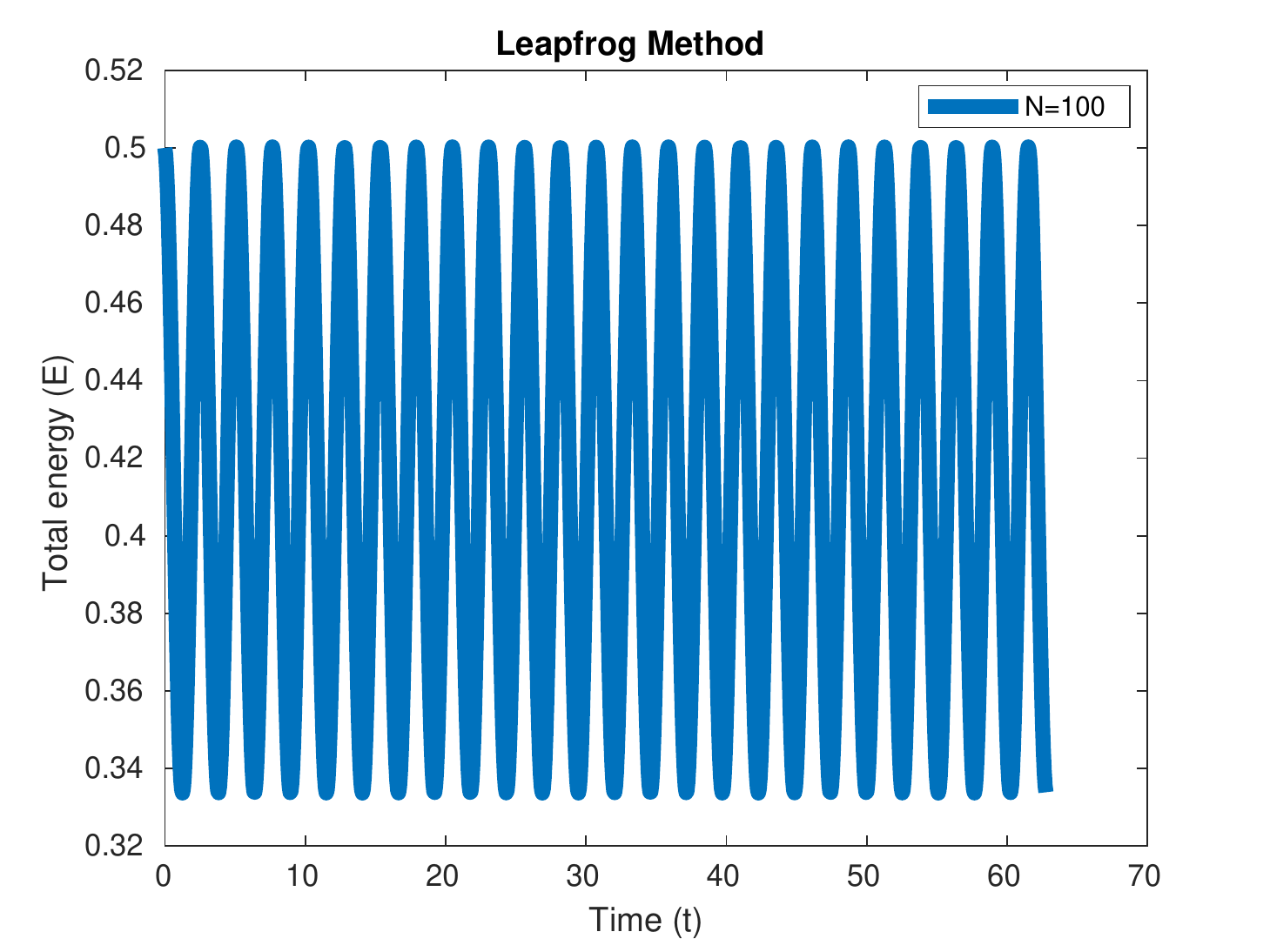}}\\
\subfloat[$N=1000$]{\includegraphics[width = 2.9in]{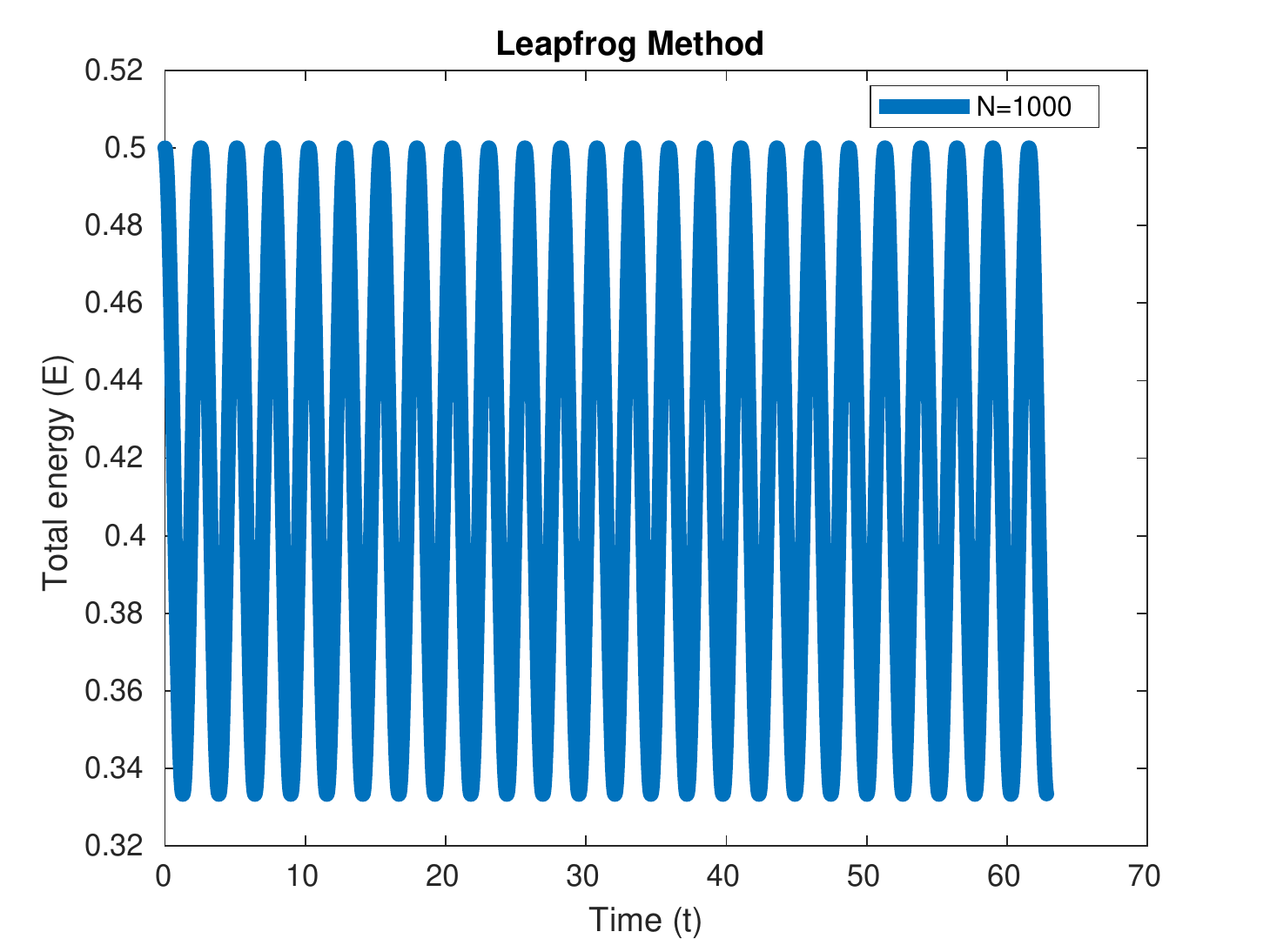}}
\subfloat[$N=10000$]{\includegraphics[width = 2.9in]{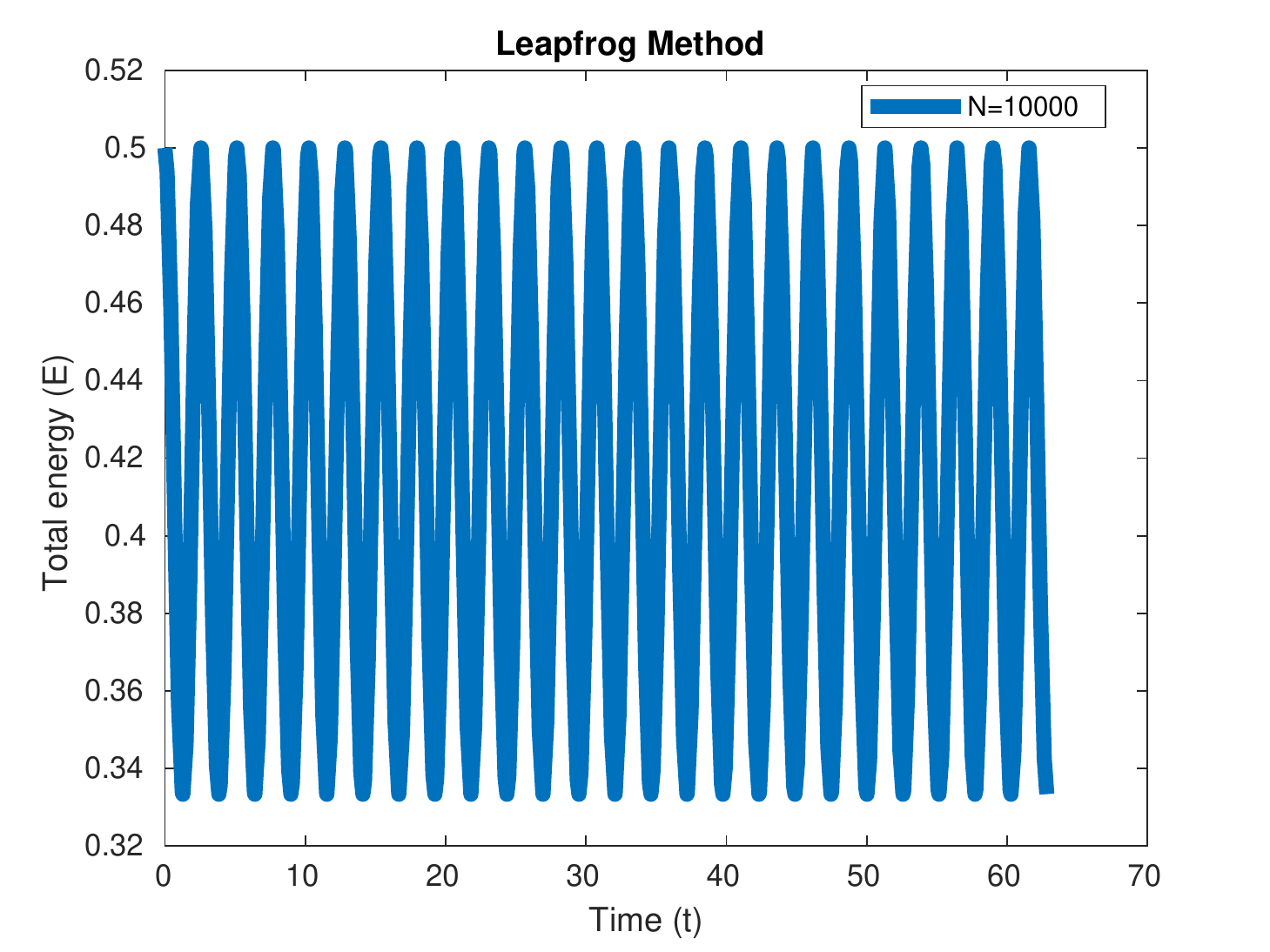}} 
\caption{Constant energy value of the solution over time (By Leapfrog method). We can see the energy is oscillating around a constant value with very small amplitude.}
\label{fig_lf_3}
\end{figure}

\section{Explanation of Plots}
In figure \ref{fig_lf_1} we plot the phase-space trajectories for different number of steps ($N$) inside a time period. We can see for $N=10$ also we get an ellipse (slightly of axis from the analytical solution). The best thing is that like Euler and RK method for small $N$, like the case of $N=10$ the solution is not diverging here. This clearly shows why the Leapfrog method is reversible. 

In figure \ref{fig_lf_2} we plot the $x(t)$ versus $t$. The solution is stable, unlike the solutions from Euler method and RK method, Leapfrog method gives a stable solution over time. So even for small $N$ the solution is not diverging. That ensures that in leapfrog method the error accumulation is very small. 

Figure \ref{fig_lf_3} shows the constancy of total energy with time. Unlike the Euler method and RK method, here we are not getting a ever growing or ever falling energy pattern. The total energy here oscillates around a constant energy value. The average line of this oscillation is the constant energy value of the system. 

\section{Discussions and Conclusions}

\begin{figure}[ht]
\centering
\subfloat[For Leapfrog only]{\includegraphics[width = 2.9in]{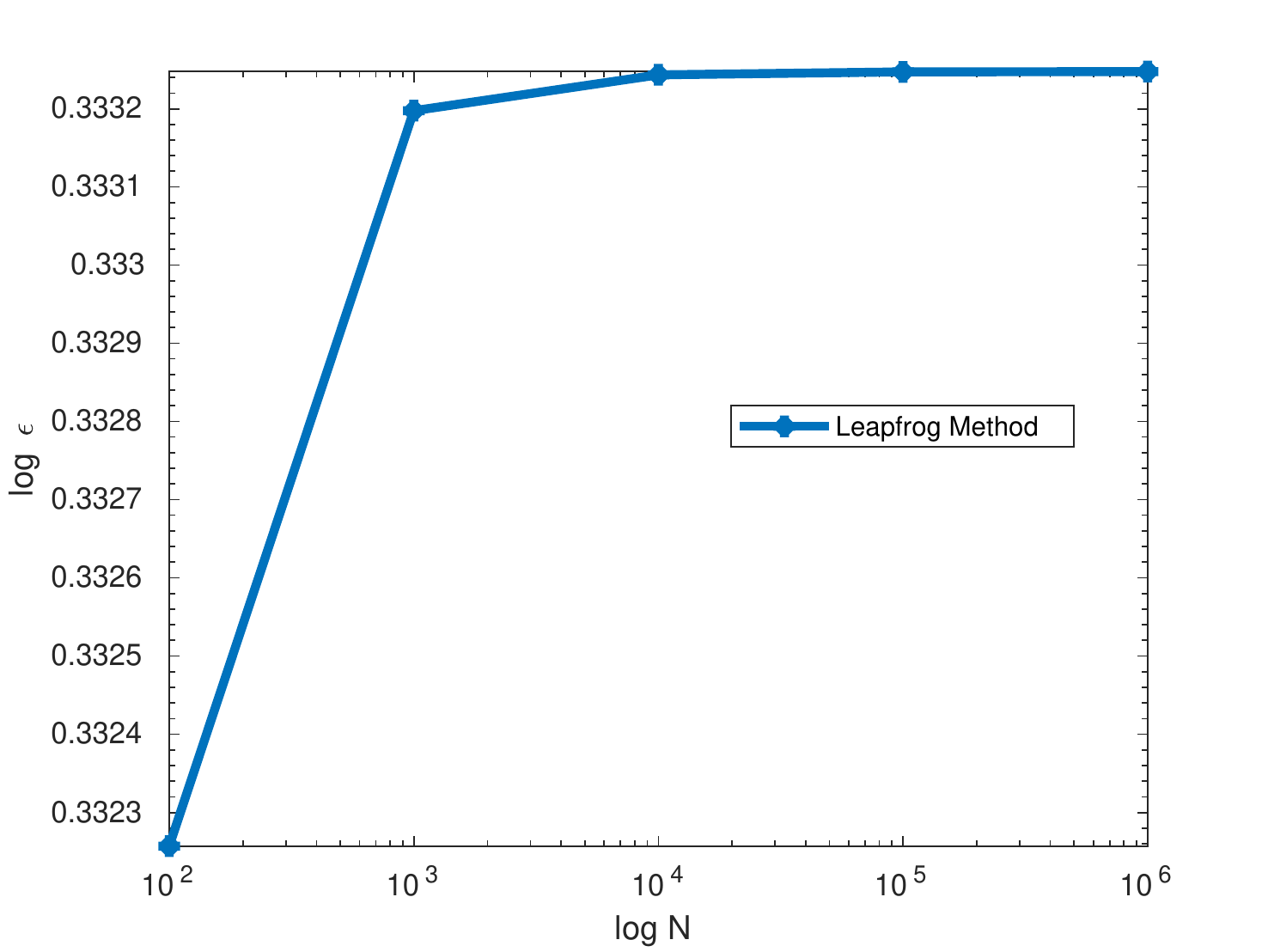}} 
\subfloat[Comparison with other methods]{\includegraphics[width = 2.9in]{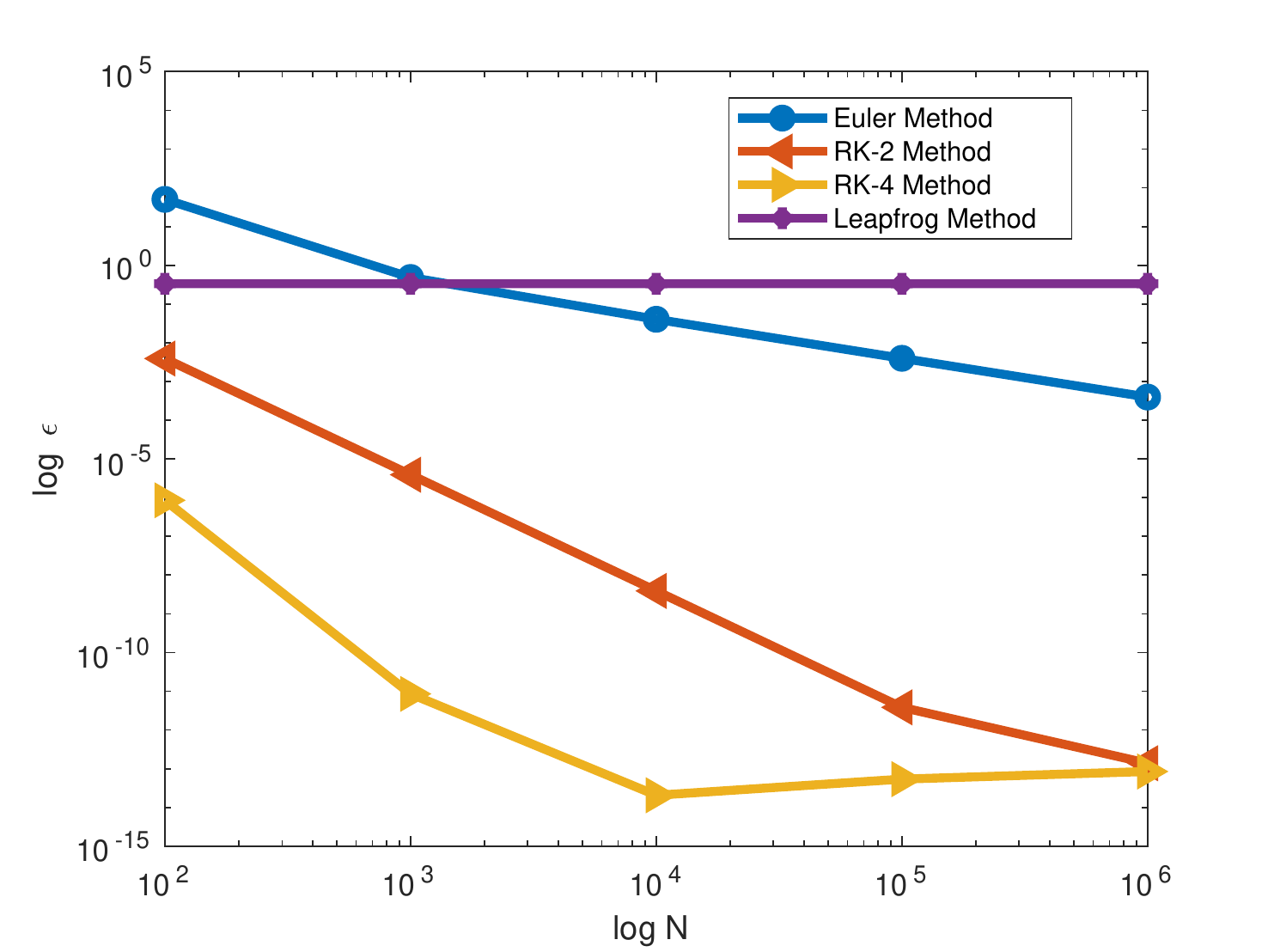}}
\caption{$\epsilon$ vs $N$ plot in log-log scale for different numerical methods}
\label{fig_lferr_3}
\end{figure}

As we discussed the error in energy in Leapfrog method is negligible. In figure \ref{fig_lf_3} we have plotted relative error ($\varepsilon=\frac{\Delta E}{E}$) with number of steps $N$ in log scale. In figure \ref{fig_lf_3} the first plot (plot (a)) shows the negligible amount of energy change for a change in $N$. The change is so small that when we plot the relative error in Leapfrog method with the relative errors from other methods like Euler, RK-2 and RK-4 (figure \ref{fig_lf_3}), the Leap frog method form a horizontal line close to $\varepsilon=0$ in $\varepsilon$ vs. $N$ plot. This plot confirms that Leapfrog method is best energy conserving method. 

So, whenever in a system we need energy conserving results, Leapfrog method is the best way. 

1) It is very simple to implement. 

2) It is first order so computational complexity is less. 

3) It gives stable solution.

4) The error accumulation in each step is negligible.

5) The solutions are reversible.

6) The solutions are energy conserving.

7) For small number of steps $N$ the solution is not diverging.

 \chapter{Solution of Poisson's Equation using Finite Difference Method}

\section{Aim}
To discuss the Finite Difference Method (FDM) and solve Poisson's equation using FDM. 

\section{Introduction}
Our goal is to solve Poisson equation. In previous assignments we learnt how to solve Poisson equation using Fourier method. In this assignment we will solve Poisson equation by more direct approach, Finite difference method.

\section{Brief Theory}
We want to solve 
\begin{equation}
     \nabla ^2 \phi =4\pi G \rho
\end{equation}
where $\rho$ is given and we want to solve for $\phi$. In 1D the Poisson equation is 
\begin{equation}\label{po1}
    \frac{d^2 \phi}{d x^2}=4 \pi G \rho
\end{equation}

Now, by Taylor series expansion we have

\begin{equation}\label{eq11}
    \phi (x+h) = \phi (x) + h \: \phi ' (x) +\frac{h^2}{2} \: \phi '' (x) +...
\end{equation}

\begin{equation}\label{eq22}
    \phi (x-h) = \phi (x) - h \: \phi ' (x) +\frac{h^2}{2} \: \phi '' (x) -...
\end{equation}

By $eq \: (\ref{eq11}) - eq \: (\ref{eq22})$ we have

\begin{equation}\label{fd1}
    \phi ' (x) = \frac{\phi (x+h) - \phi (x-h)}{2 h} + \frac{\mathcal{O} (h^3)}{h}
\end{equation}

and by $eq \: (\ref{eq11}) + eq \: (\ref{eq22})$ we have

\begin{equation}\label{fd2}
    \phi '' (x) = \frac{\phi (x+h)+ \phi (x-h) -2 \phi (x)}{h^2} + \frac{\mathcal{O} (h^3)}{h^2}
\end{equation}

Eq. \ref{fd1} and eq. \ref{fd2} shows the discrete version of first and second derivative of a function respectively. In the discrete form eq \ref{po1} can be written as

\begin{equation}
    \frac{\phi (x+h)+ \phi (x-h) -2 \phi (x)}{h^2} = 4 \pi G \rho (x) + \frac{\mathcal{O} (h^3)}{h^2}
\end{equation}

In discrete space (computational space) $x$ is replaced by discrete integer $a$. 
\begin{equation}
   \boxed{ \frac{\phi _{a+1}+ \phi _{a-1} -2 \phi _a}{L^2} = 4 \pi G \rho _a }
\end{equation}

neglecting higher order terms, $L$ is the grid spacing. 

\section{Simulation Parameters}
We done the simulation for a 2D grid, the parameters are:

1) square grid side length $=100$

2) grid spacing $L=1$

3) The value of potential $\phi$ at boundaries of the square grid is assumed to be zero (As with distance the potential value falls down, and for infinite distance it is zero. However we can not work with infinity, so it is a good approximation to set potential values zero at boundaries of the finite square grid over that we are simulating). 

\section{Potential Plots}
\begin{figure}[ht]
\centering
\subfloat[3D surface plot]{\includegraphics[width = 3in]{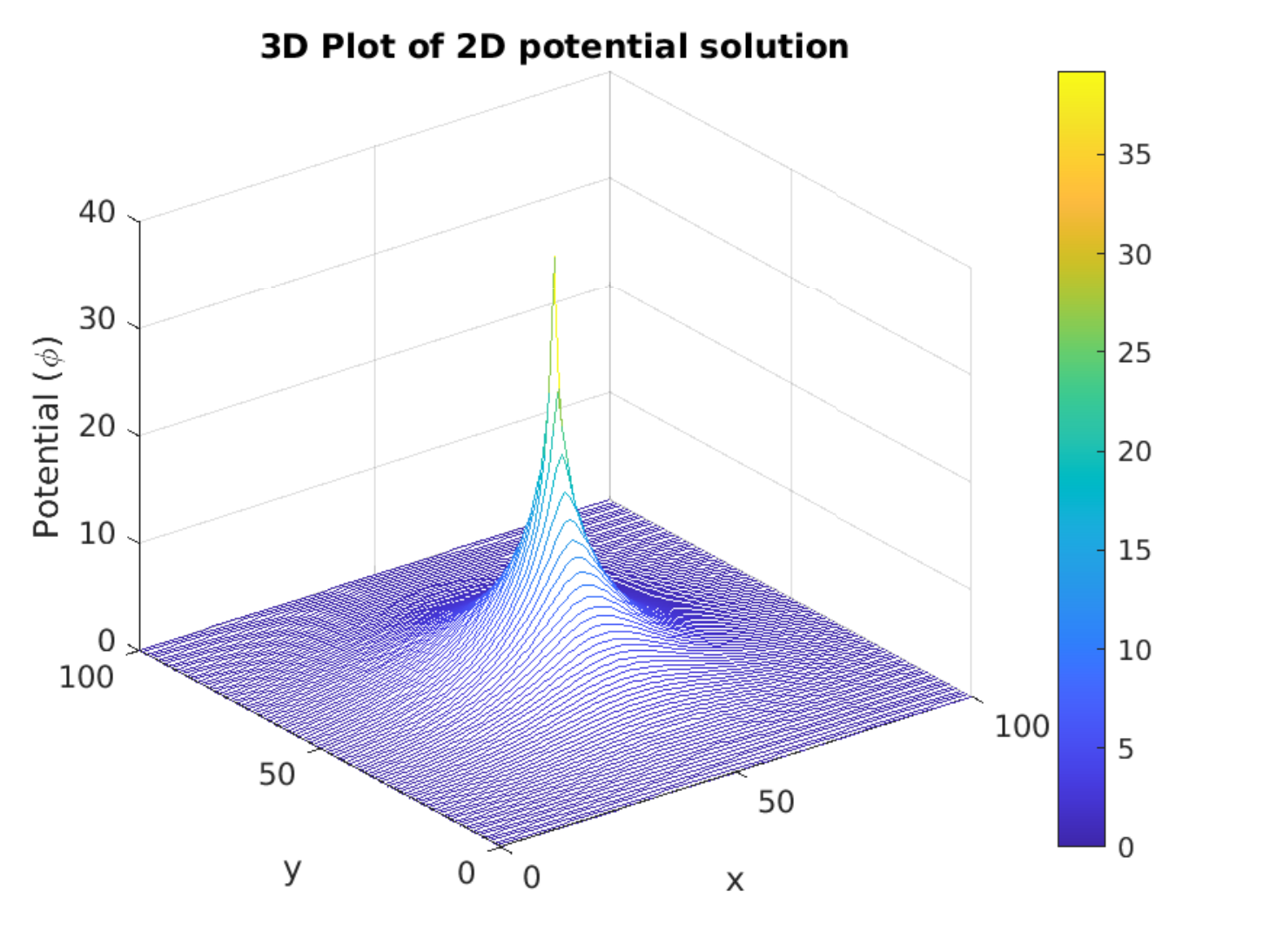}} 
\subfloat[Color plot]{\includegraphics[width = 3in]{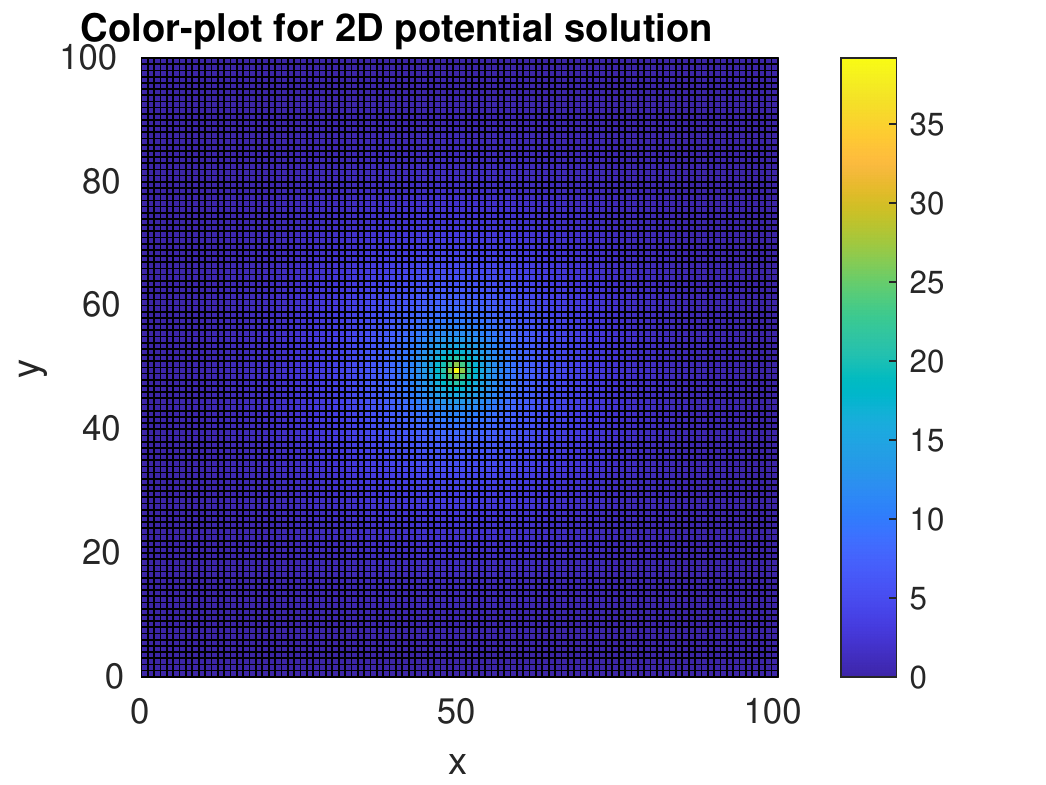}} 
\caption{Potential ($\phi$) plot for a point mass at the centre ($50,50$) of 2D grid ($100\times 100$)}
\label{fig_poi1}
\end{figure}

\begin{figure}[ht]
\centering
\subfloat[3D surface plot]{\includegraphics[width = 3in]{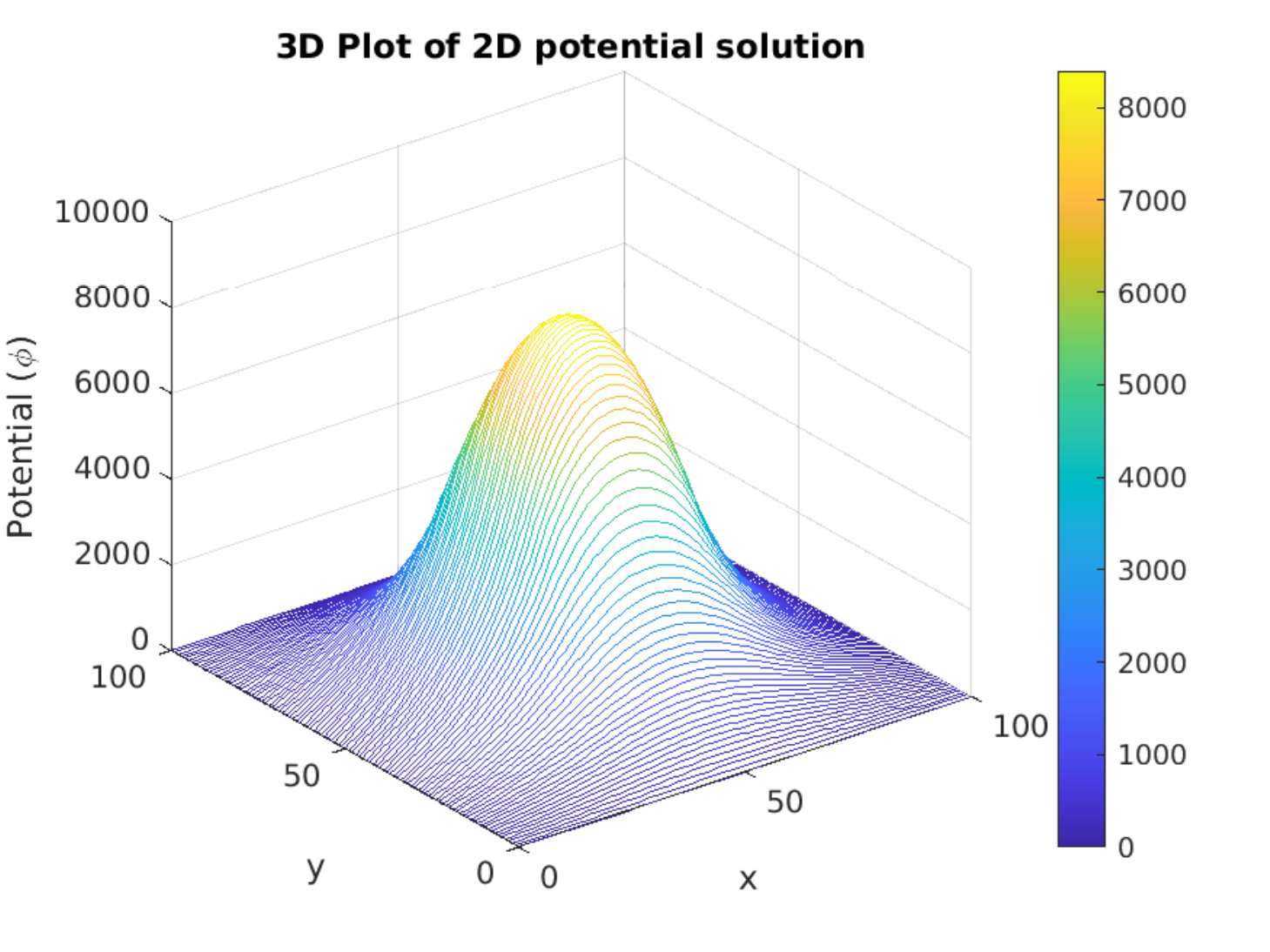}}
\subfloat[Color plot]{\includegraphics[width = 3in]{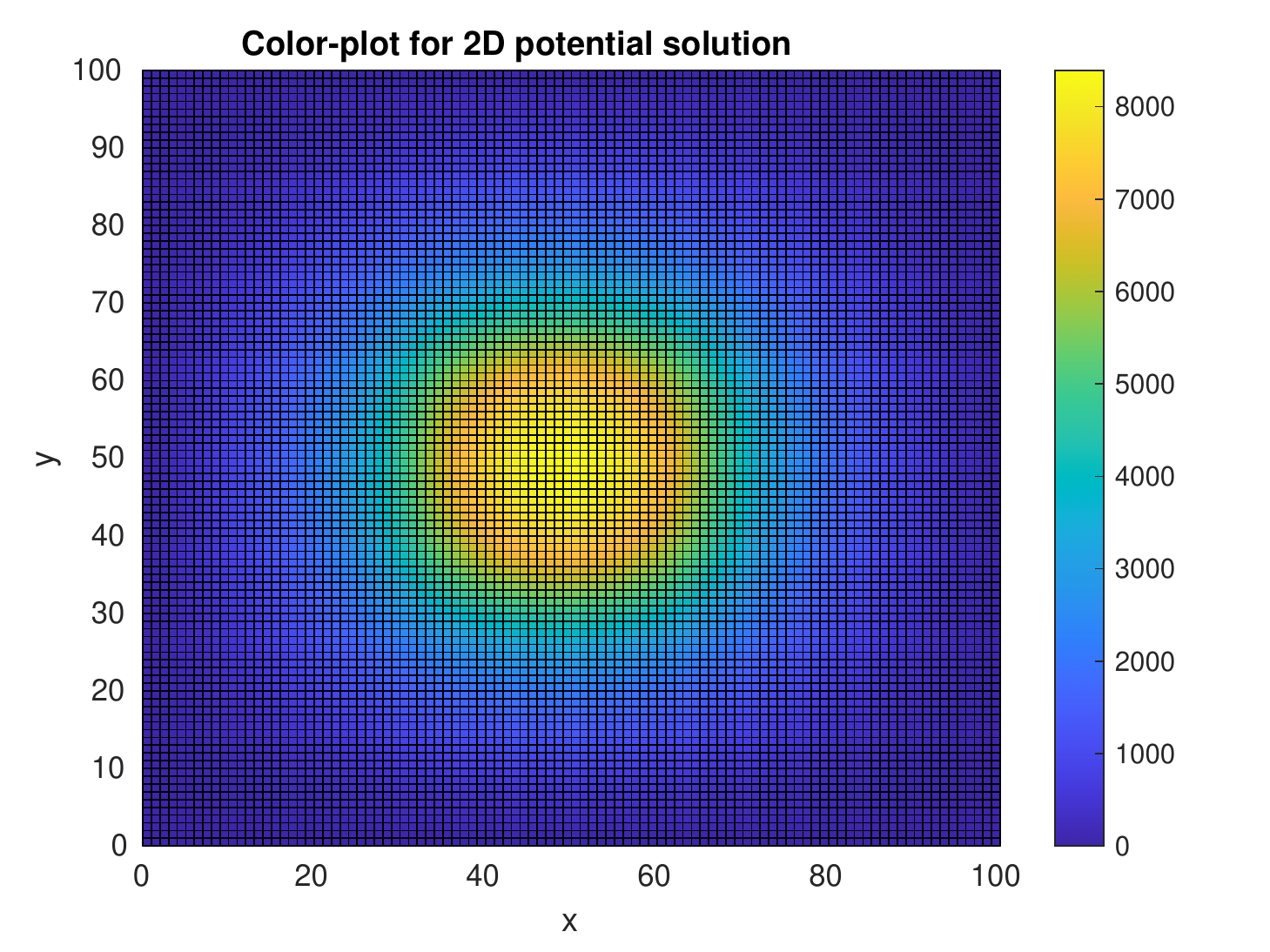}} 
\caption{Potential ($\phi$) plot for a square 2D uniform bulk mass distributed between $x=33$, $x=66$, $y=33$ and $y=66$ of a ($100\times 100$) 2D grid}
\label{fig_poi4}
\end{figure}

Figure \ref{fig_poi1} to figure \ref{fig_poi4} shows potential solutions for different source mass configurations. (For getting a clear plot I just reverse the sign of the potential before plotting).

\begin{figure}[ht]
\centering
\subfloat[3D surface plot]{\includegraphics[width = 3in]{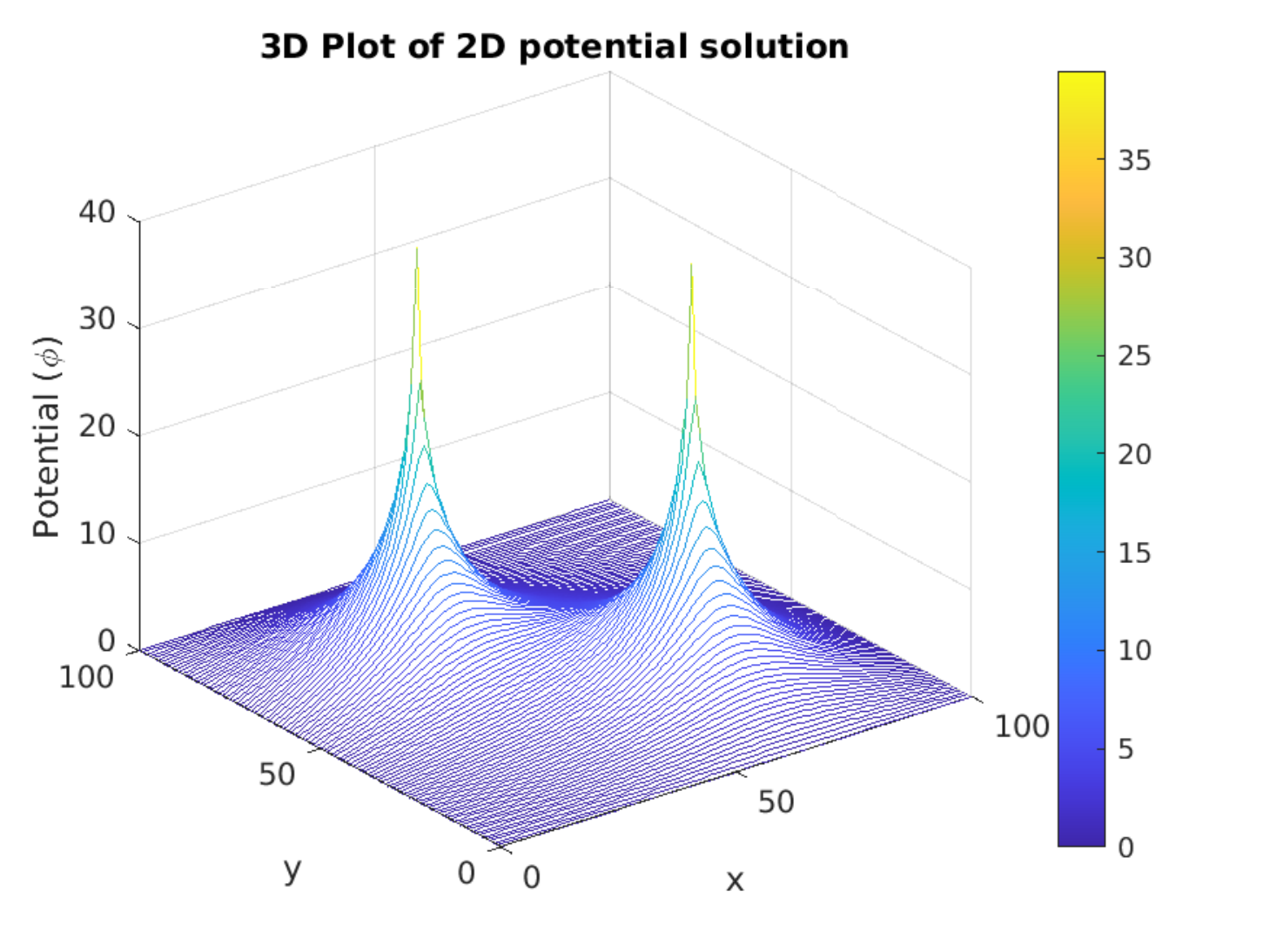}} 
\subfloat[Color plot]{\includegraphics[width = 3in]{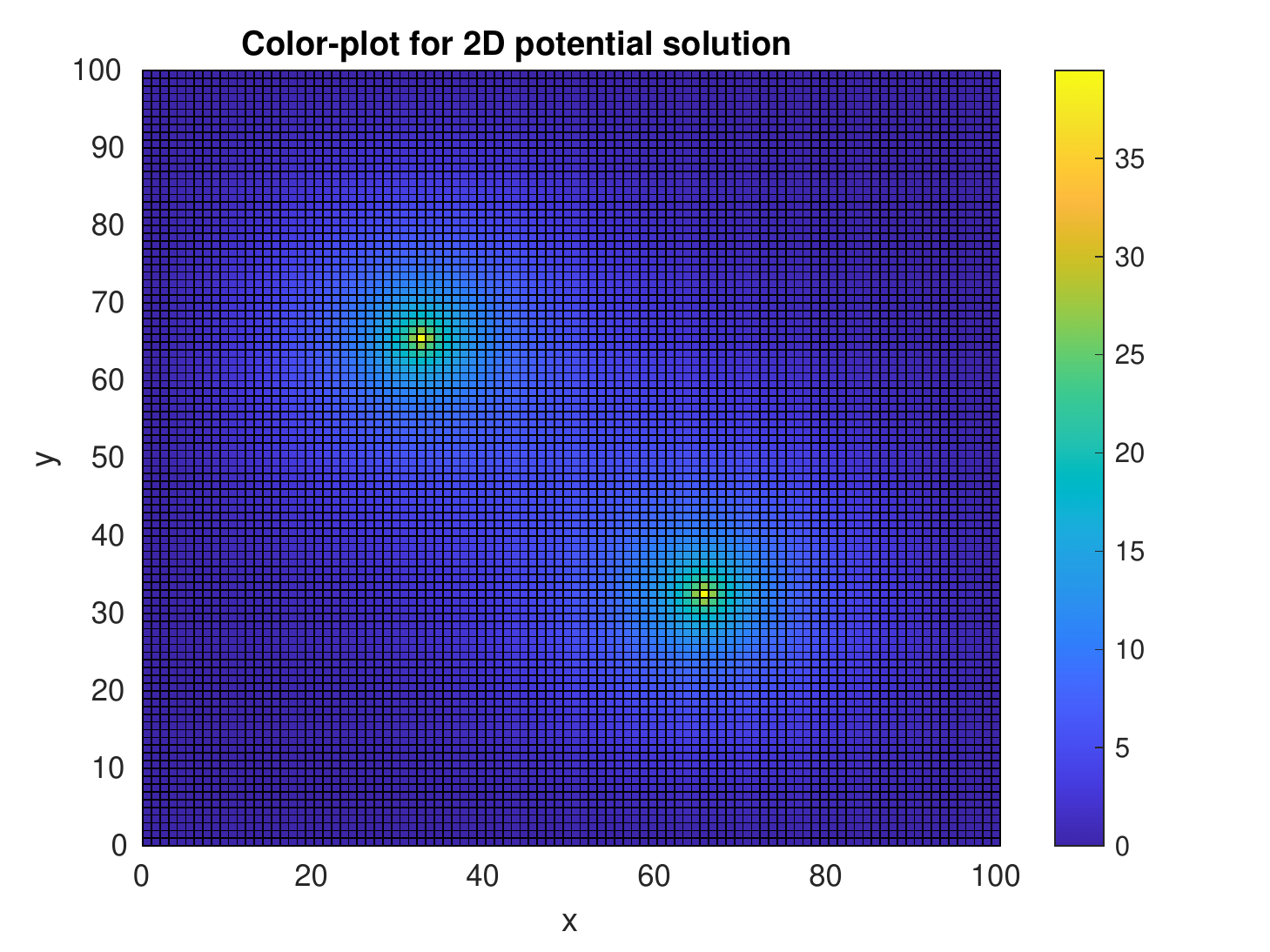}} 
\caption{Potential ($\phi$) plot for two point masses at ($33,66$) and ($66,33$) of a ($100\times 100$) 2D grid}
\label{fig_poi2}
\end{figure}

\begin{figure}[ht]
\centering
\subfloat[3D surface plot]{\includegraphics[width = 3in]{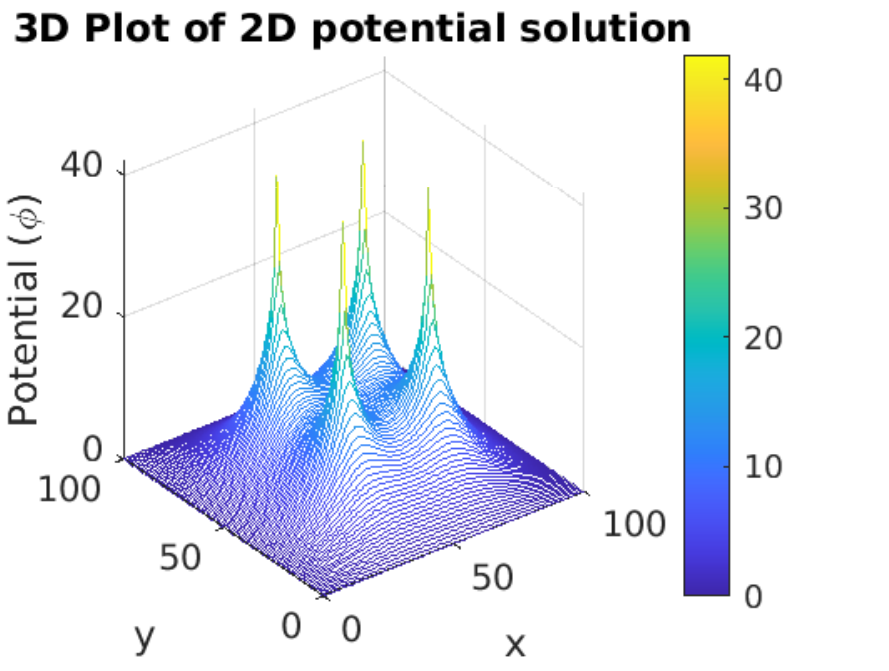}}
\subfloat[Color plot]{\includegraphics[width = 3in]{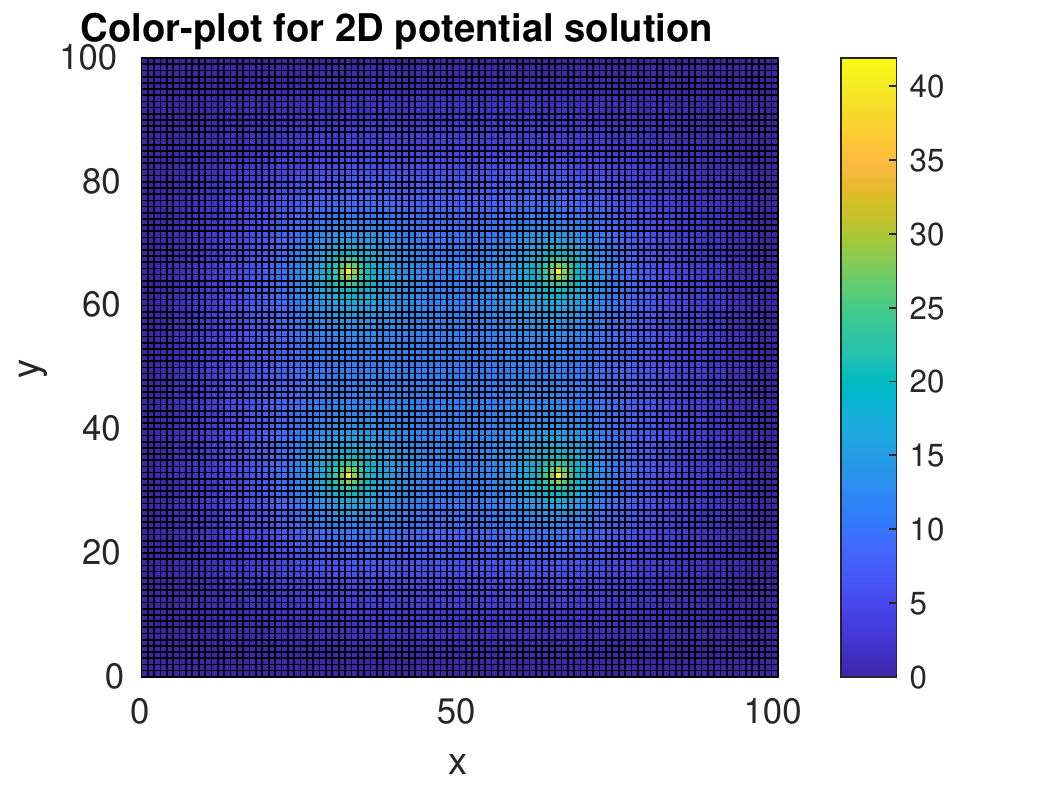}} 
\caption{Potential ($\phi$) plot for four point masses at ($33,66$). ($33,33$), ($66,33$) and ($66,66$) of a ($100\times 100$) 2D grid}
\label{fig_poi3}
\end{figure}

 \chapter{Fourier Method Solution of Poisson's Equation}
\section{Aim}
To find the potential-field and acceleration-field created by a known mass density distribution ($\rho$) for 1D system by Fourier transform and inverse Fourier transform method. 

\section{Introduction}
We can write any function in Fourier series expansion, as the terms in Fourier series forms a complete basis set. Same is true for Fourier transform. Fourier transform forms a complete continuous basis to expand any function. 

We want to solve the potential and finally acceleration field created by a mass density distribution $(\rho)$. So, the source mass distribution is $\rho$, which is known as function of distance. 

We cannot deal with continuous variables in computation. The distance $x$ is just a number $a$ in discrete system. So we have $\rho_a$. By discrete Fourier transform of this we have

\begin{equation}
    \Tilde{\rho_n}=\frac{1}{N} \sum_{a=0}^{N-1} \rho_a \exp{\left[- \frac{2 \pi i n a}{N}\right]}
\end{equation}

From this mass density in Fourier space we can calculate the potential in Fourier space via the Formula derived in the class:

\begin{equation}
    \Tilde{\phi_n}=-\frac{(4 \pi G) \Tilde{\rho_n} L^2}{4 \sin^2{\frac{\pi n}{N}}} \; \: \: \: \: \: \text{for $n > 0$}
\end{equation}

The acceleration-field in Fourier space can be calculated using the formula:

\begin{equation}
    \Tilde{A_n}=- i \sin{\left(\frac{2 \pi n}{N}\right)} \frac{\Tilde{\phi_n}}{L}
\end{equation}

Now, we have $\Tilde{\phi_n}$ and $\Tilde{A_n}$, so, we can just take inverse Fourier transform to calculate our requre potential field $\phi_a$ and acceleration field $A_a$ in real space:

\begin{equation}
    {\phi_a}= \sum_{n=0}^{N-1} \Tilde{\phi_n} \exp{\left[ \frac{2 \pi i n a}{N}\right]}
\end{equation}
and,
\begin{equation}
    {A_a}= \sum_{n=0}^{N-1} \Tilde{A_n} \exp{\left[ \frac{2 \pi i n a}{N}\right]}
\end{equation}

\section{Problem Parameters}
In this assignment we take a 1D grid of grid size $N=128$. We work with three types of mass density $\rho_a$.

\section{Explanation of Plots}
\begin{figure}[ht]
\centering
\subfloat[]{\includegraphics[width = 2.9in]{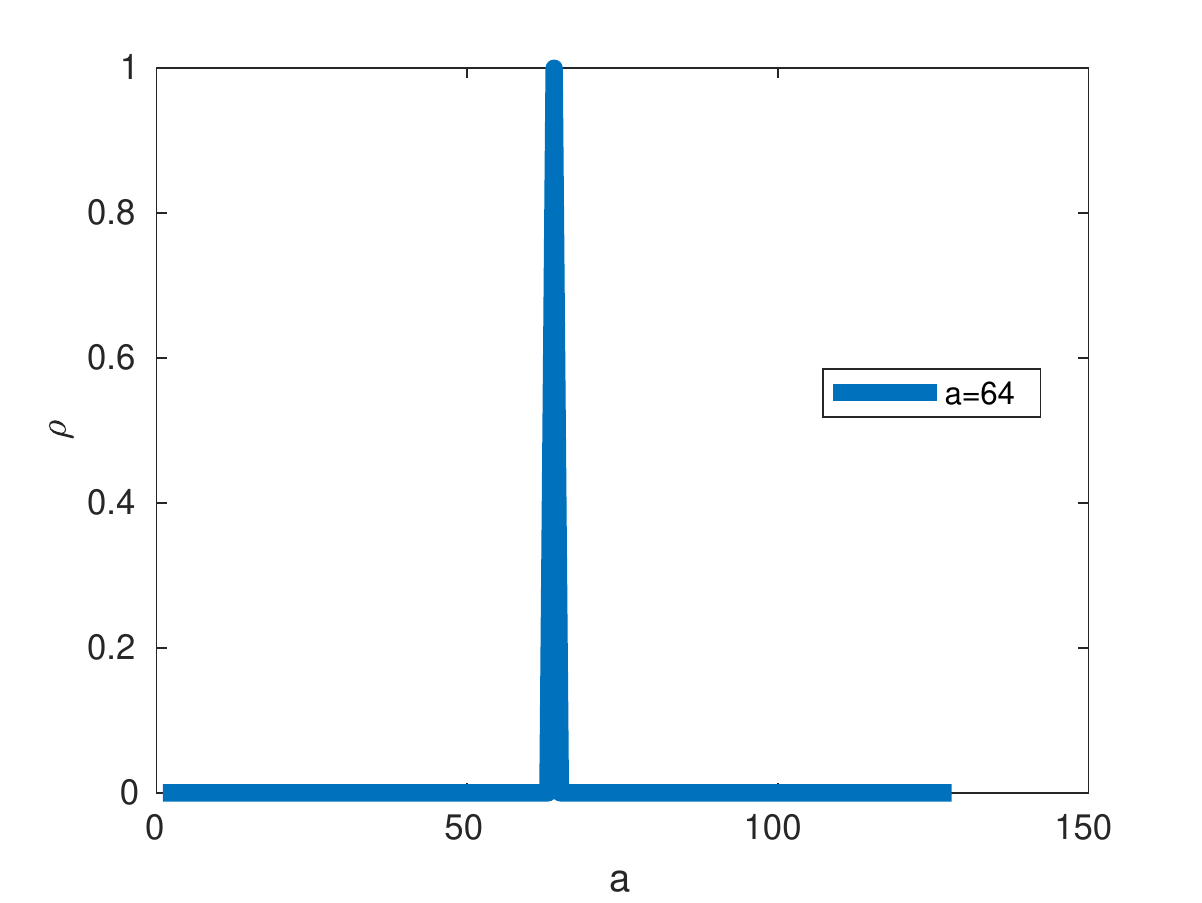}} 
\subfloat[]{\includegraphics[width = 2.9in]{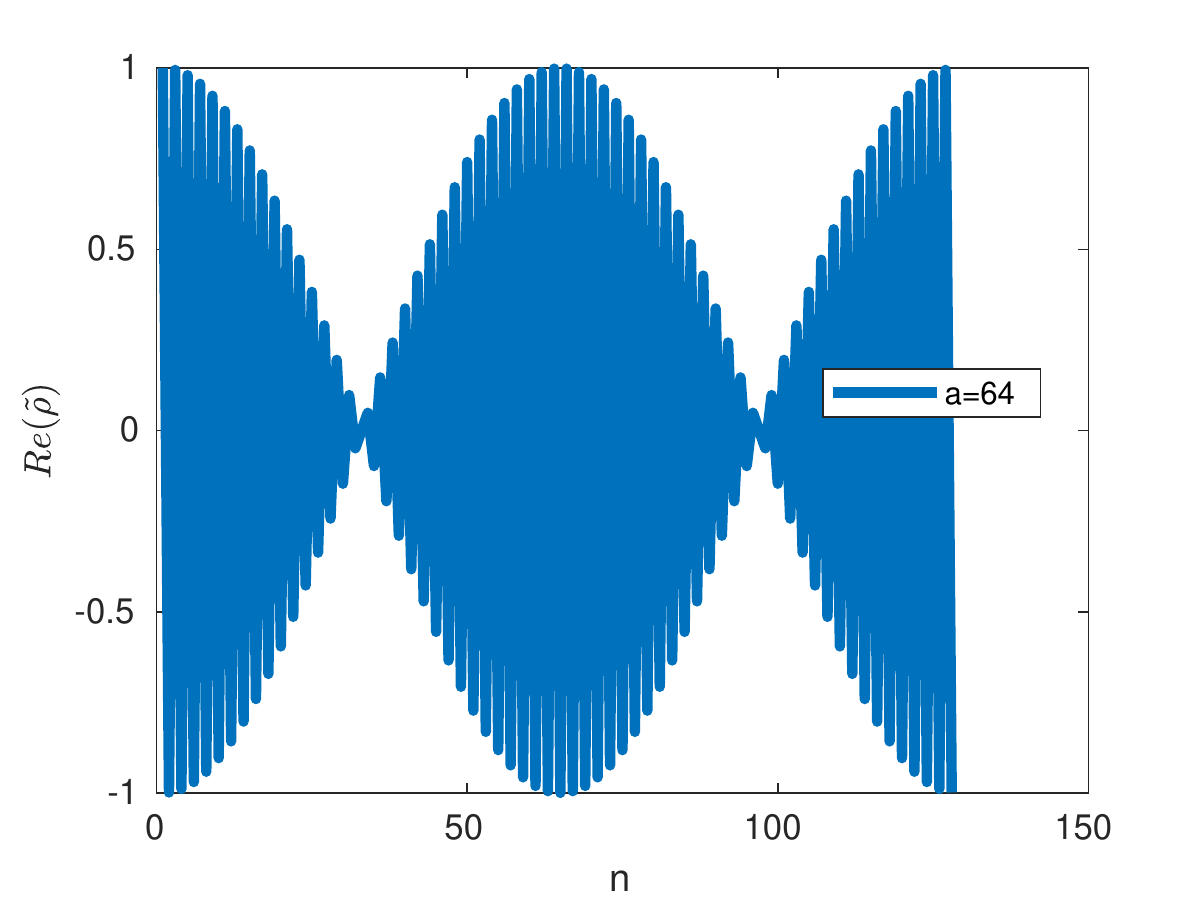}}\\
\subfloat[]{\includegraphics[width = 2.9in]{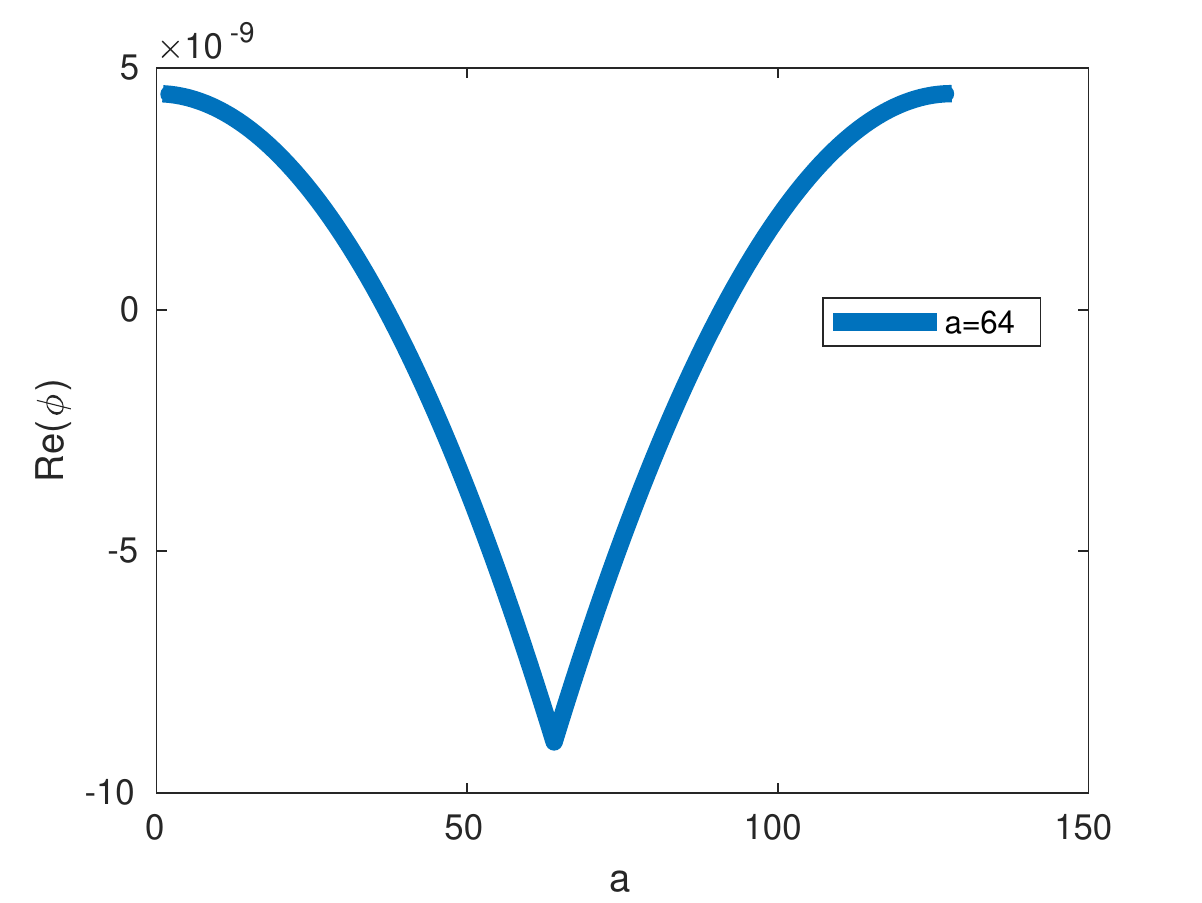}}
\subfloat[]{\includegraphics[width = 2.9in]{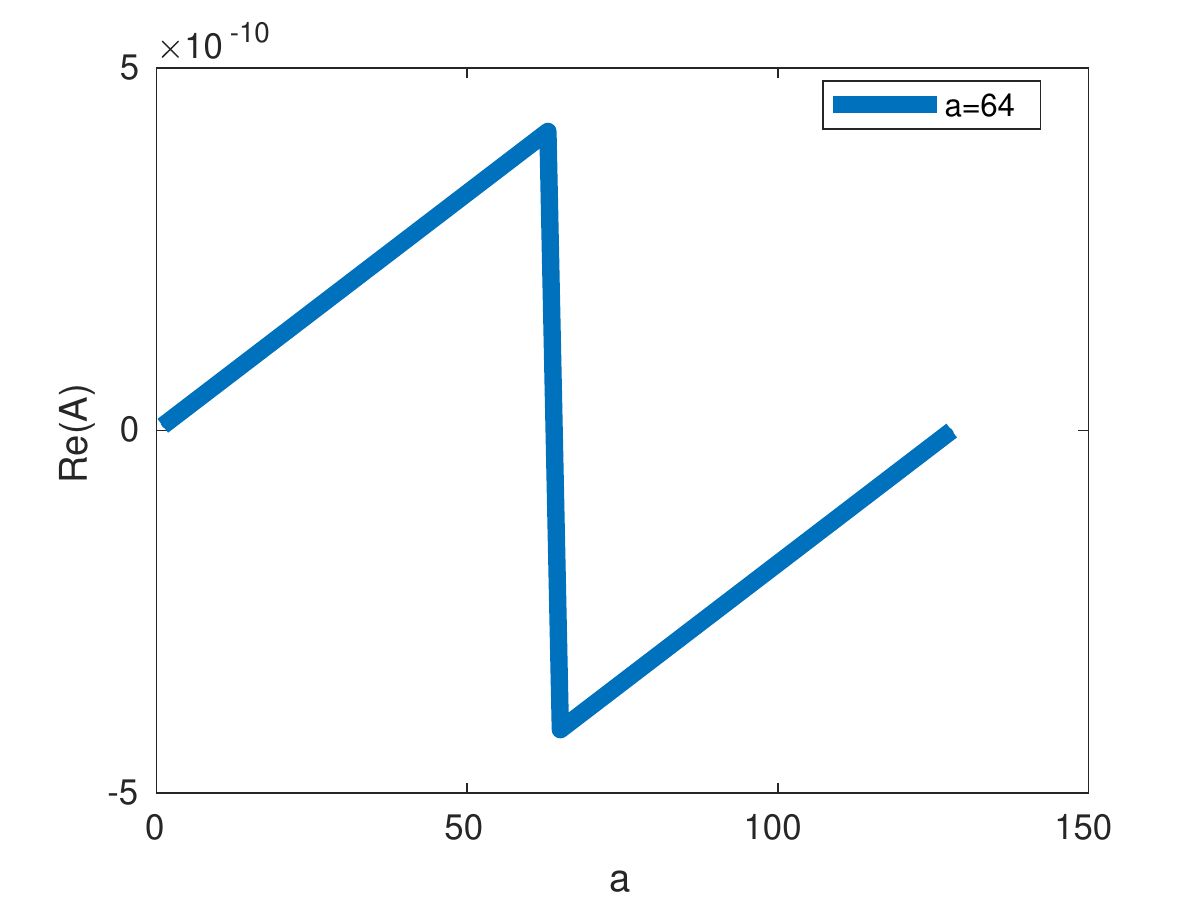}}\\
\subfloat[]{\includegraphics[width = 2.9in]{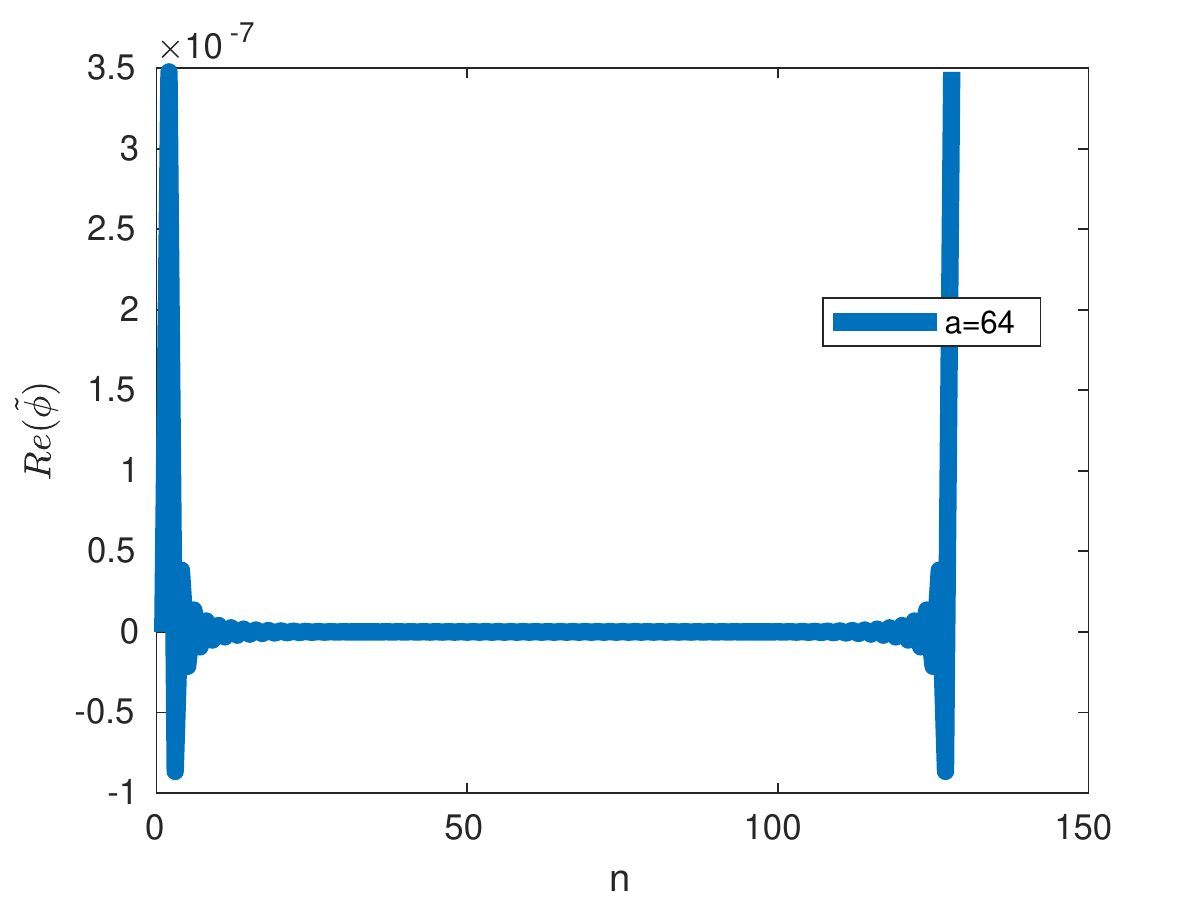}}
\subfloat[]{\includegraphics[width = 2.9in]{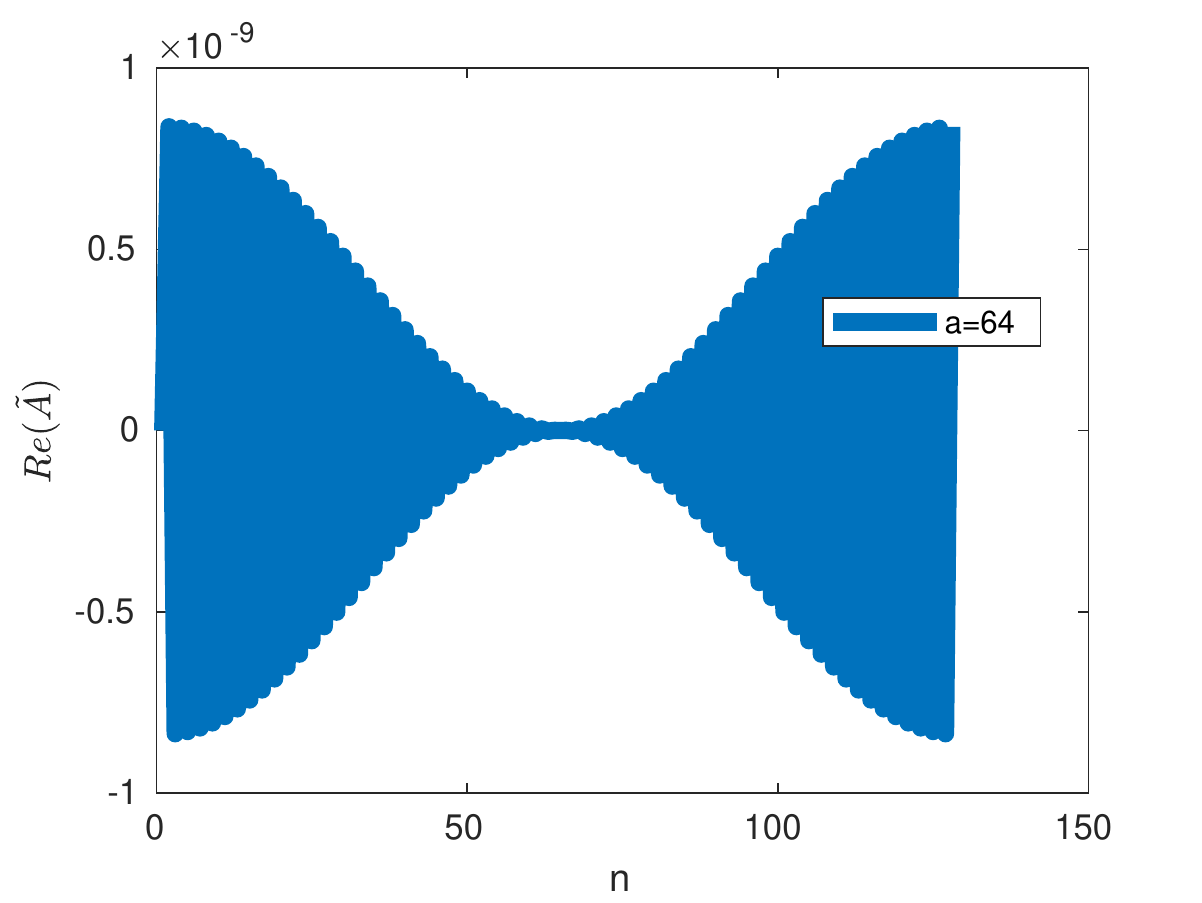}}
\caption{Real parts of all different complex parameters (a point mass at $a=64$ is the source mass)}
\label{fig_fr_1}
\end{figure}

\begin{figure}[ht]
\centering
\subfloat[]{\includegraphics[width = 2.9in]{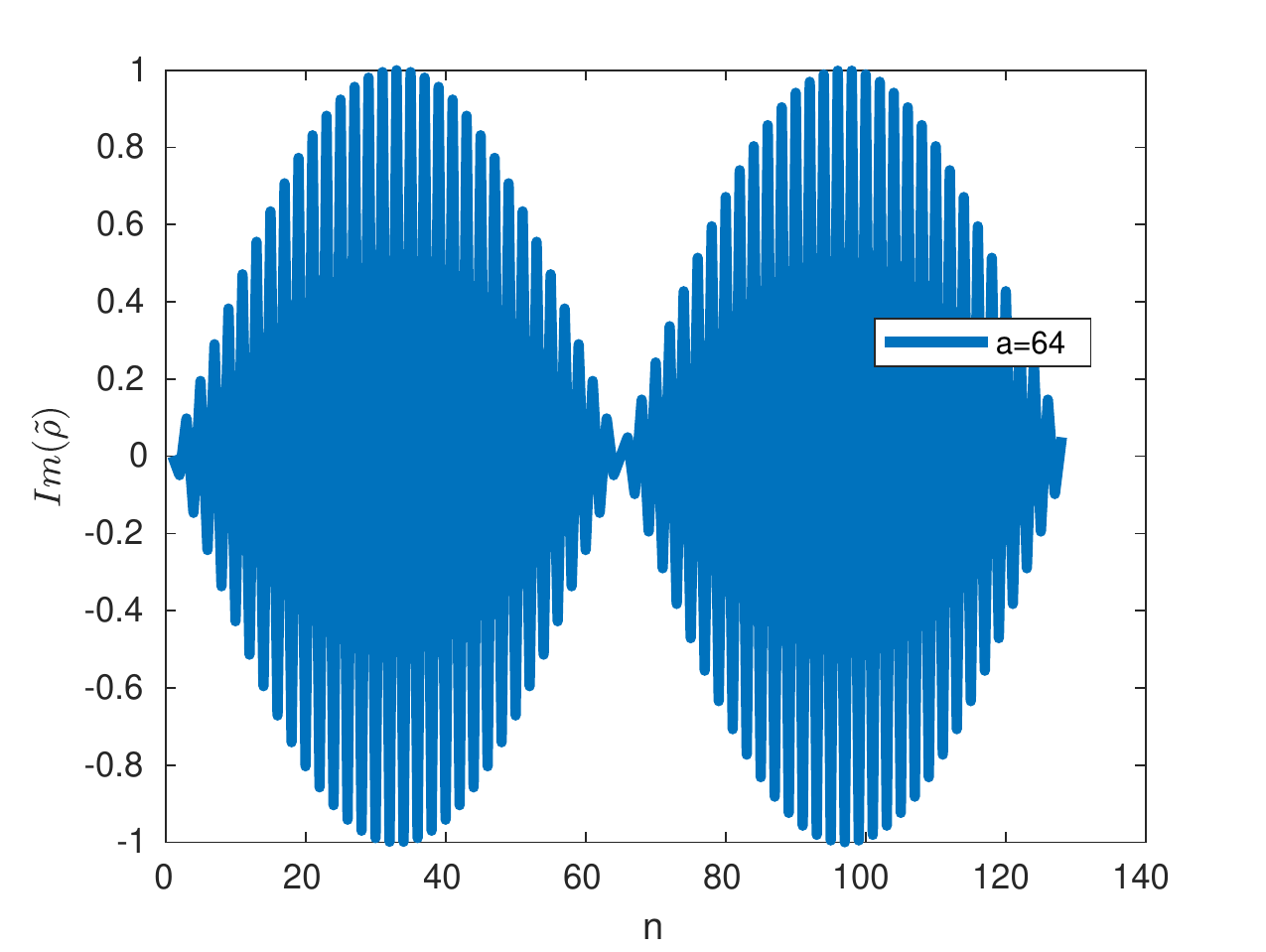}} 
\subfloat[]{\includegraphics[width = 2.9in]{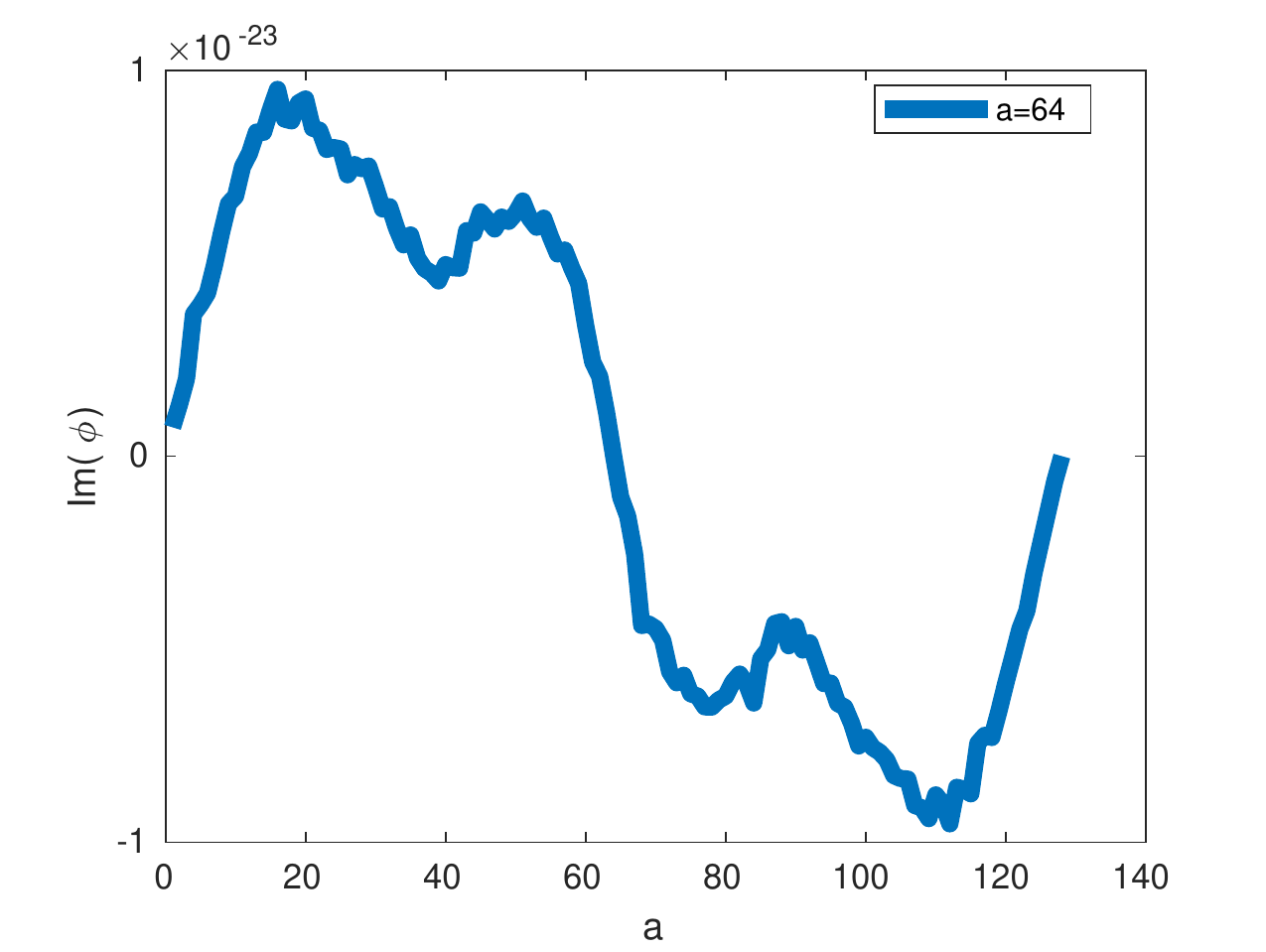}}\\
\subfloat[]{\includegraphics[width = 2.9in]{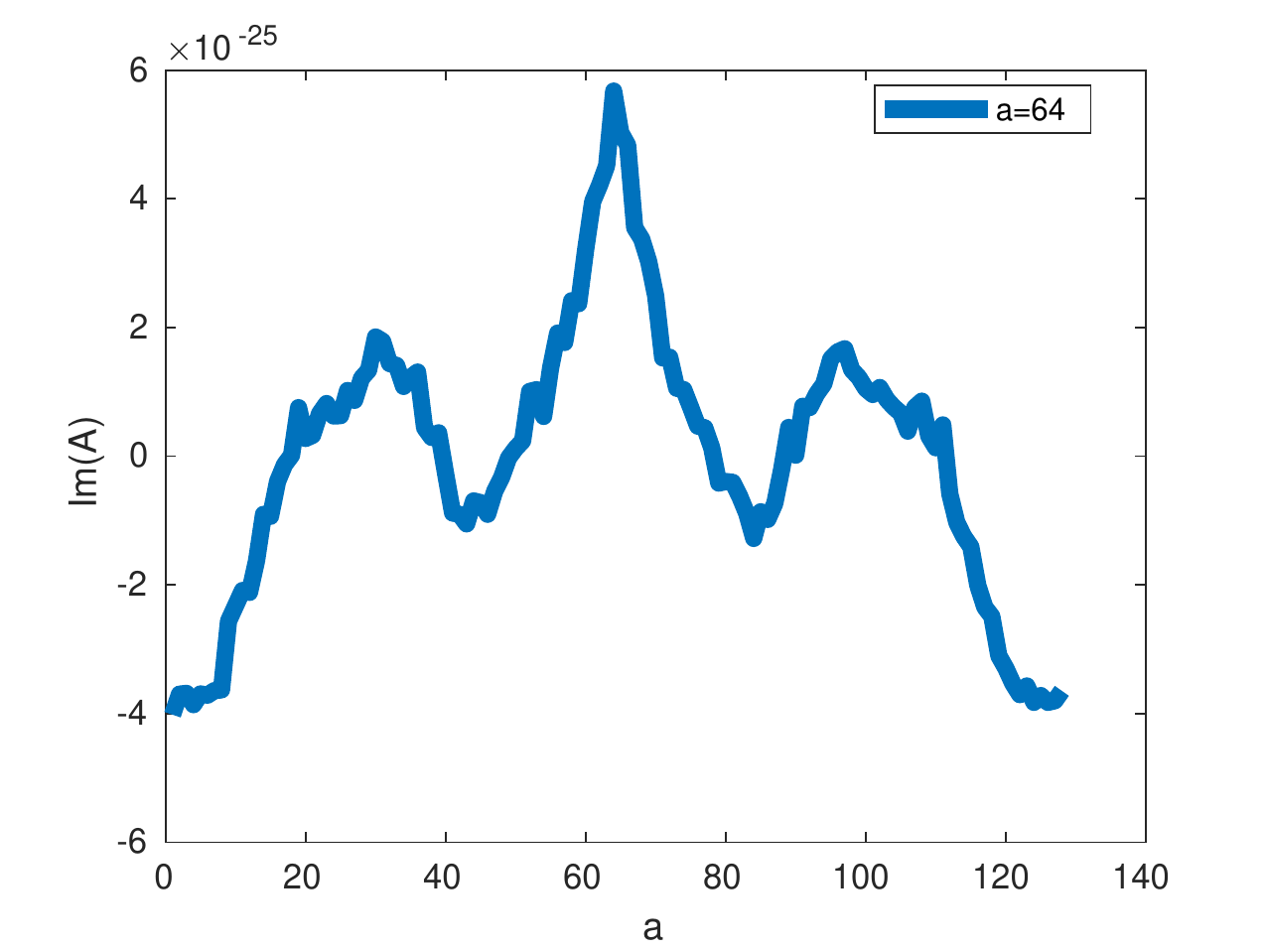}}
\subfloat[]{\includegraphics[width = 2.9in]{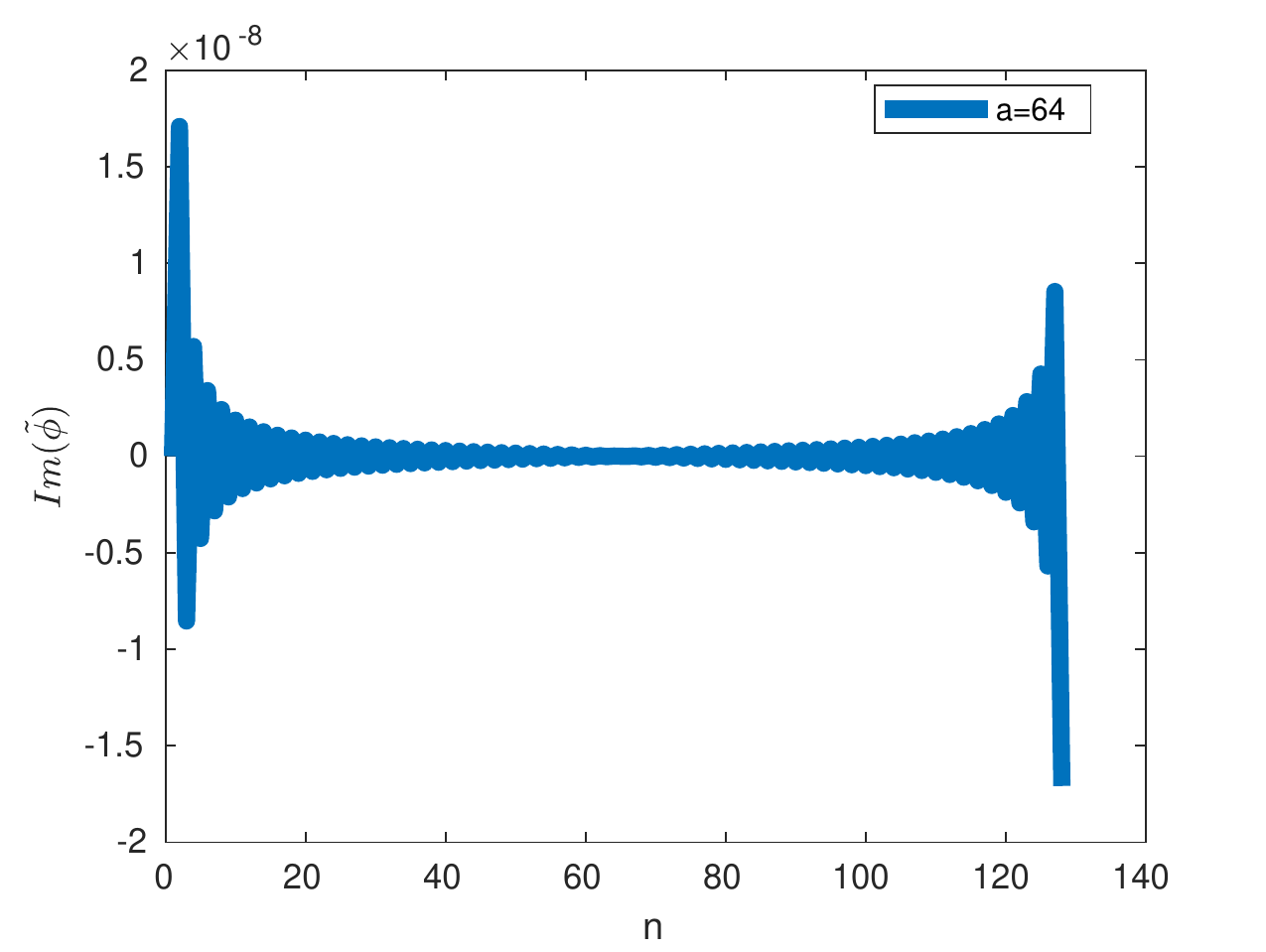}}\\
\subfloat[]{\includegraphics[width = 2.9in]{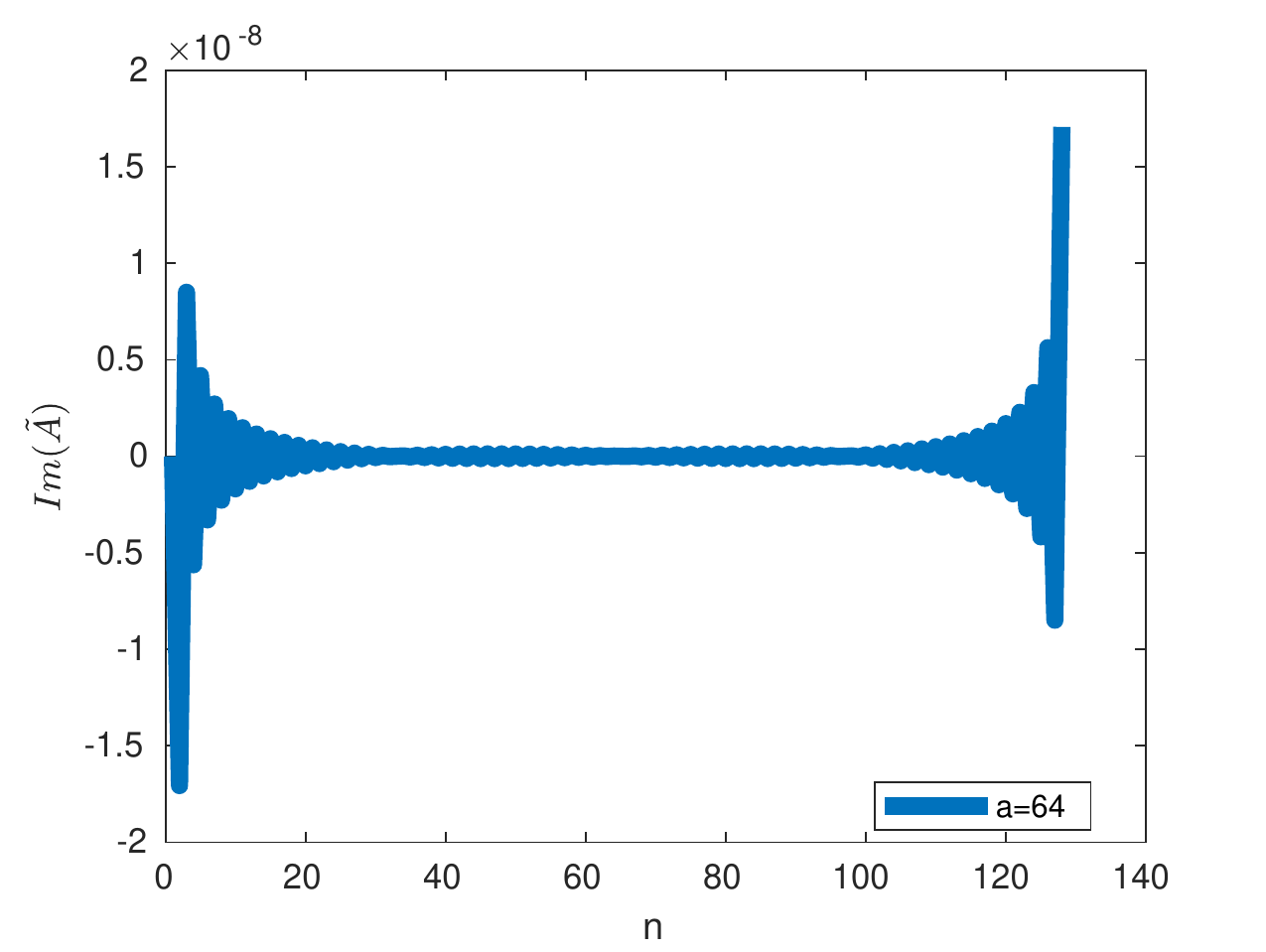}}
\caption{Imaginary parts of all different complex parameters (a point mass at $a=64$ is the source mass)}
\label{fig_fr_2}
\end{figure}

In figure \ref{fig_fr_1} and figure \ref{fig_fr_2} we have a point mass (source) at $a=64$. For this mass configuration these figure shows the plots for all different parameters in real and Fourier space. We plotted real and imaginary parts separately for all complex parameters. 

\begin{figure}[ht]
\centering
\subfloat[]{\includegraphics[width = 2.9in]{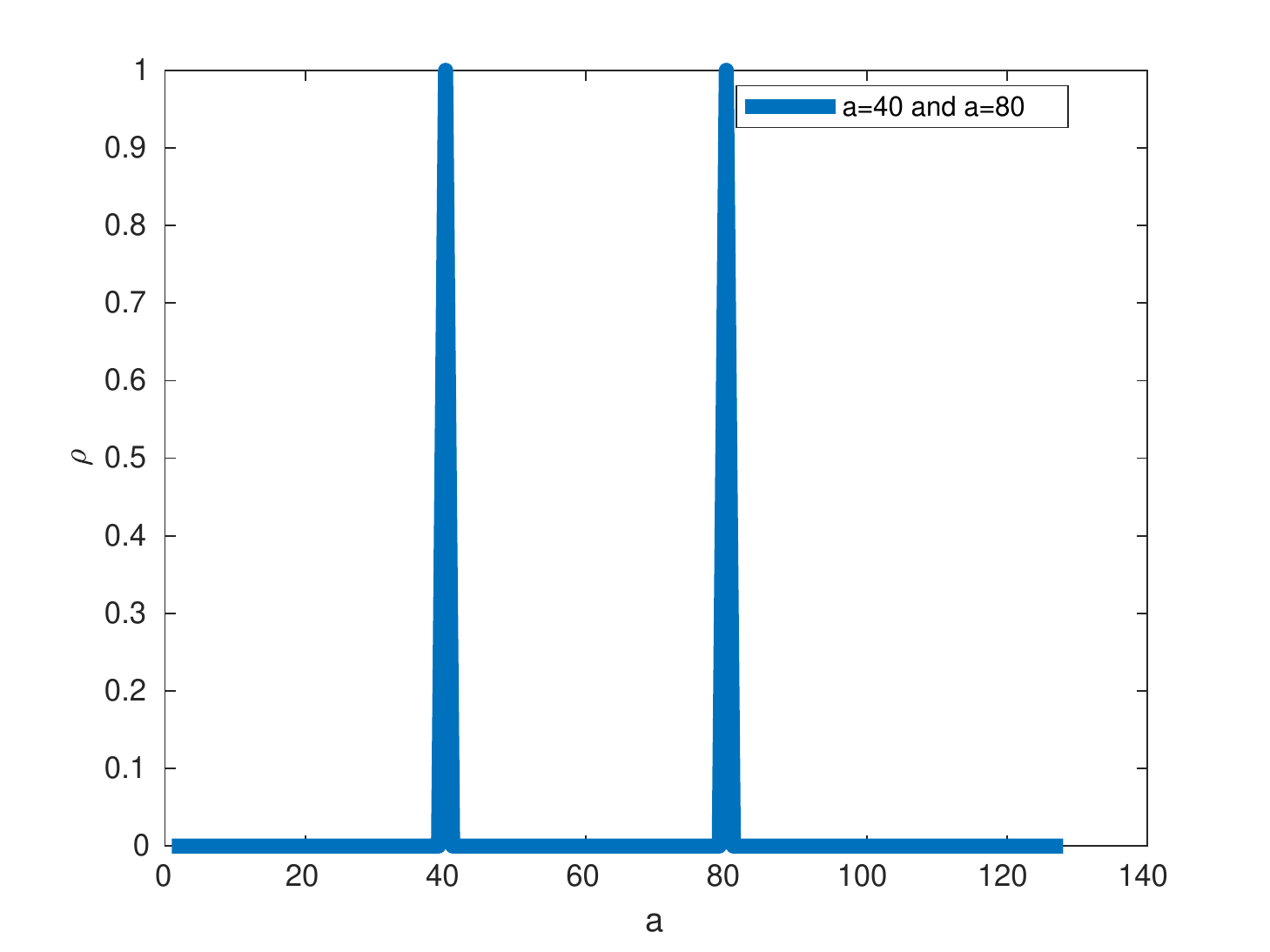}} 
\subfloat[]{\includegraphics[width = 2.9in]{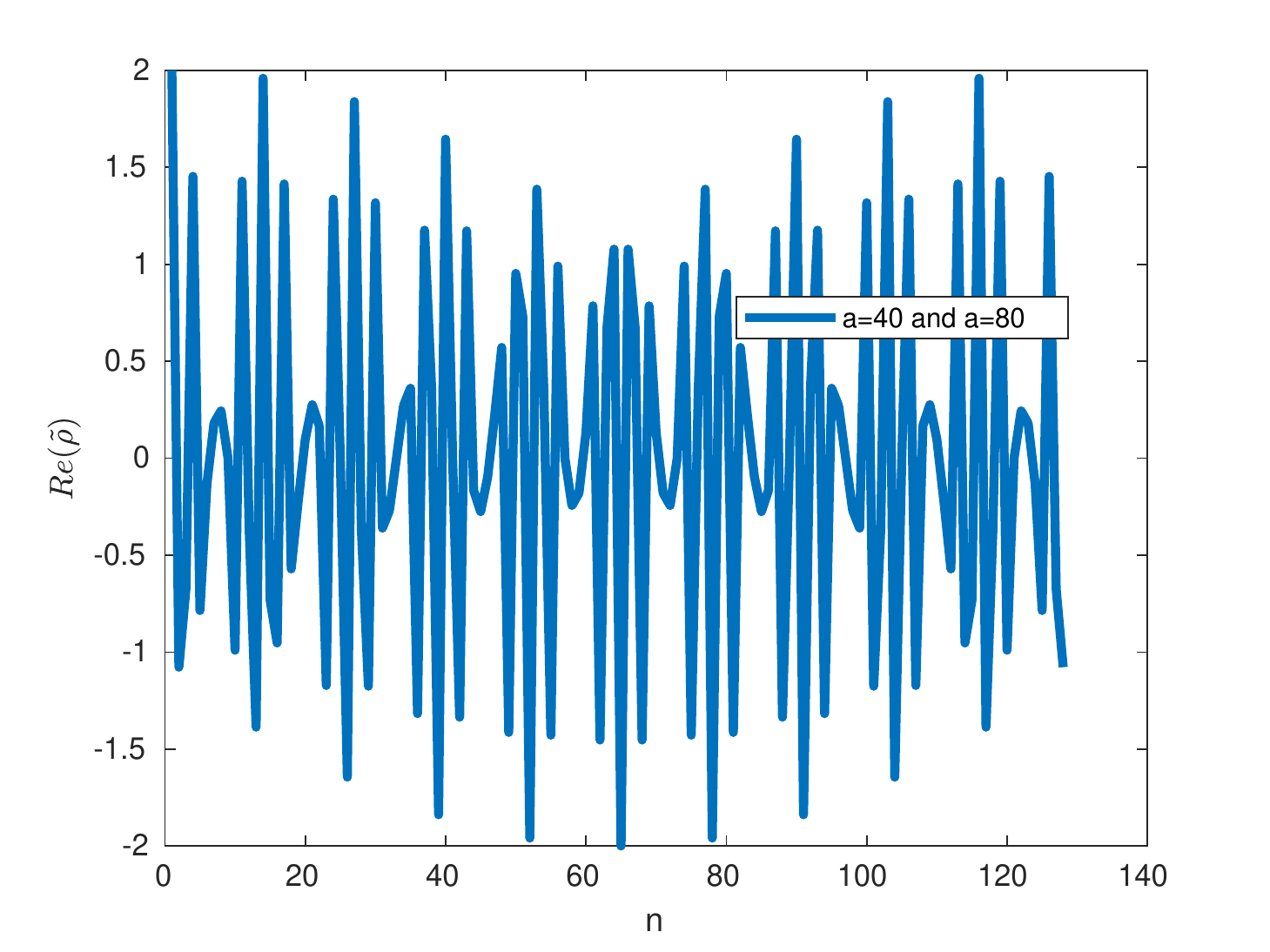}}\\
\subfloat[]{\includegraphics[width = 2.9in]{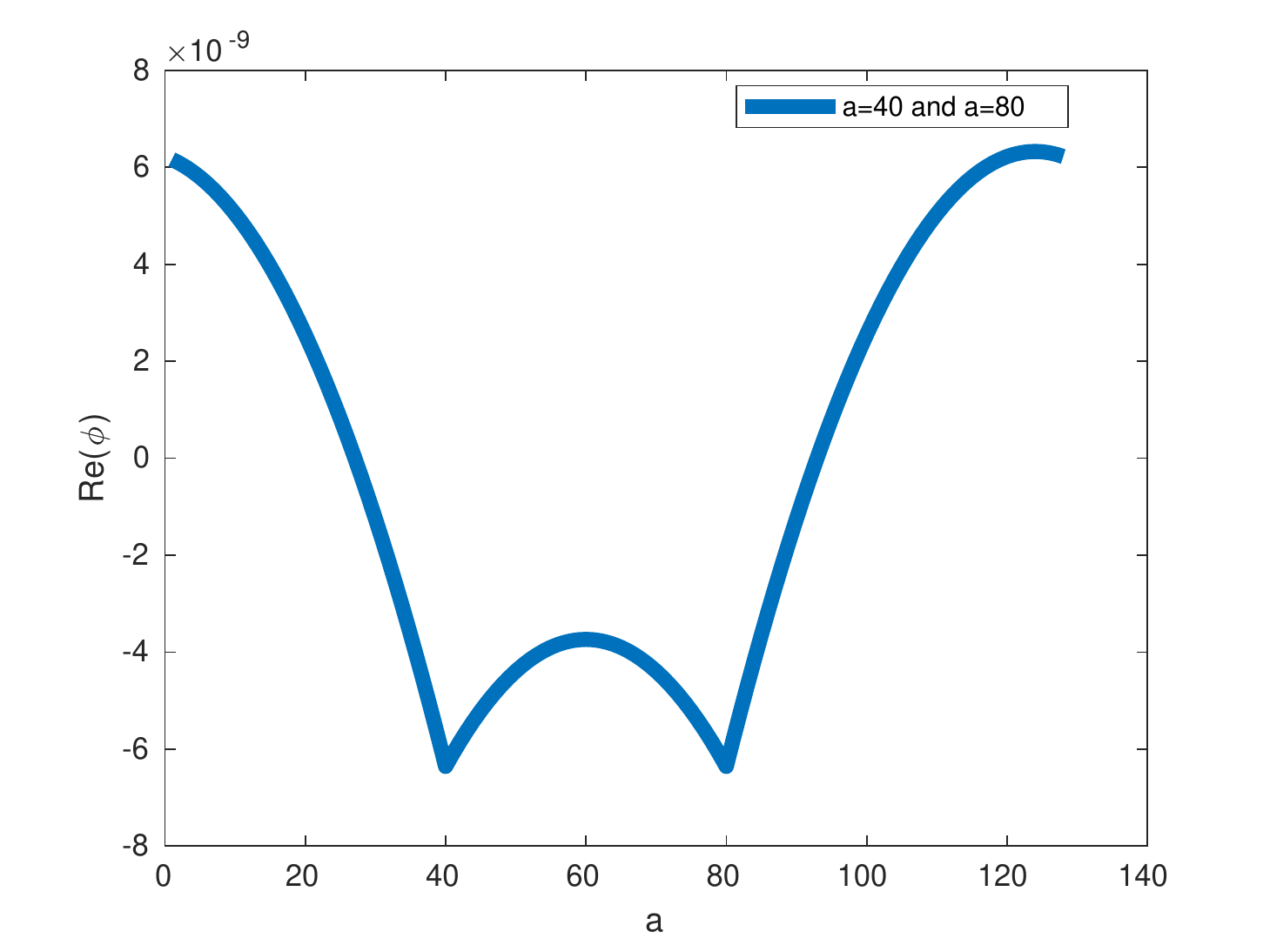}}
\subfloat[]{\includegraphics[width = 2.9in]{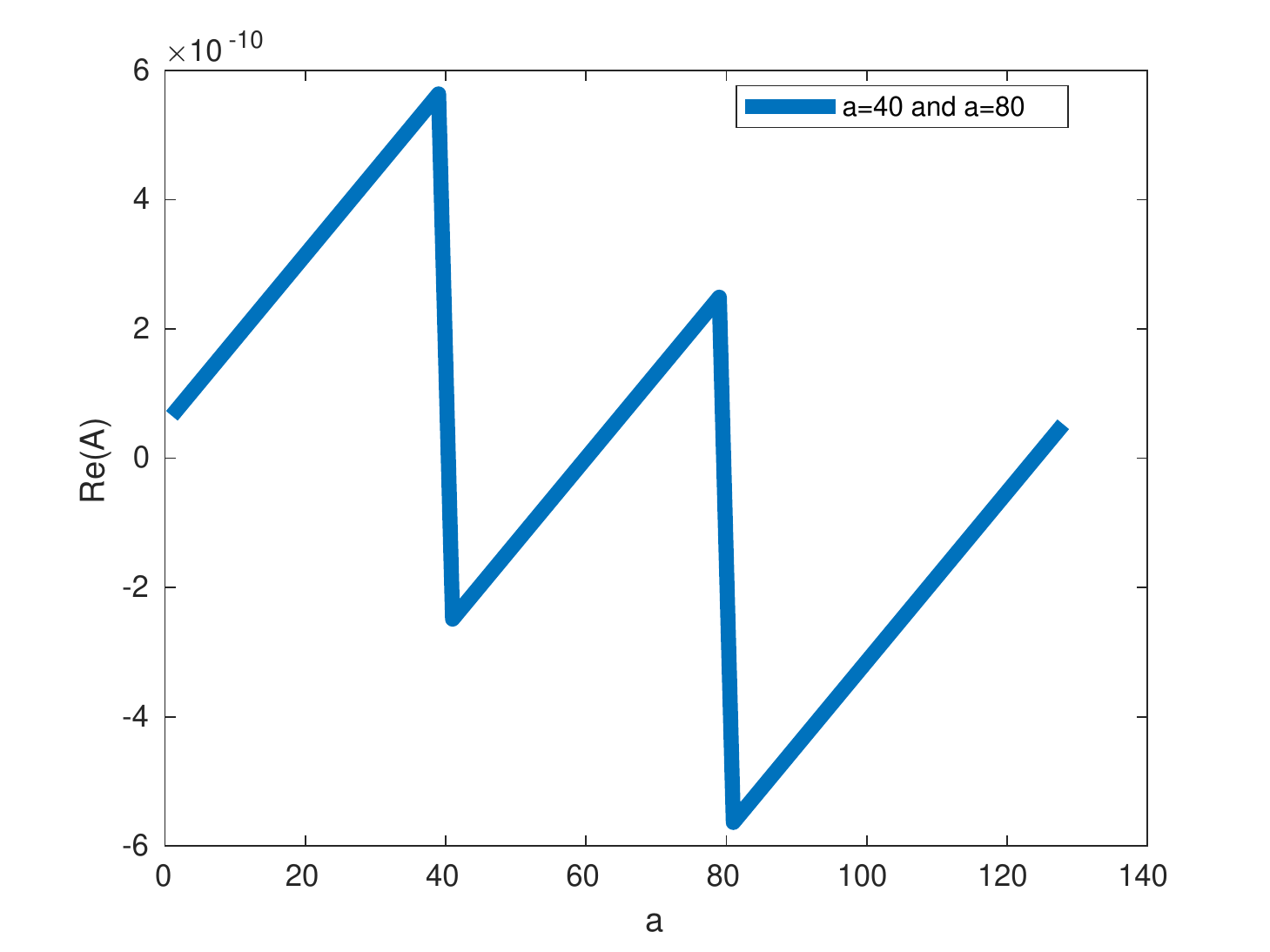}}\\
\subfloat[]{\includegraphics[width = 2.9in]{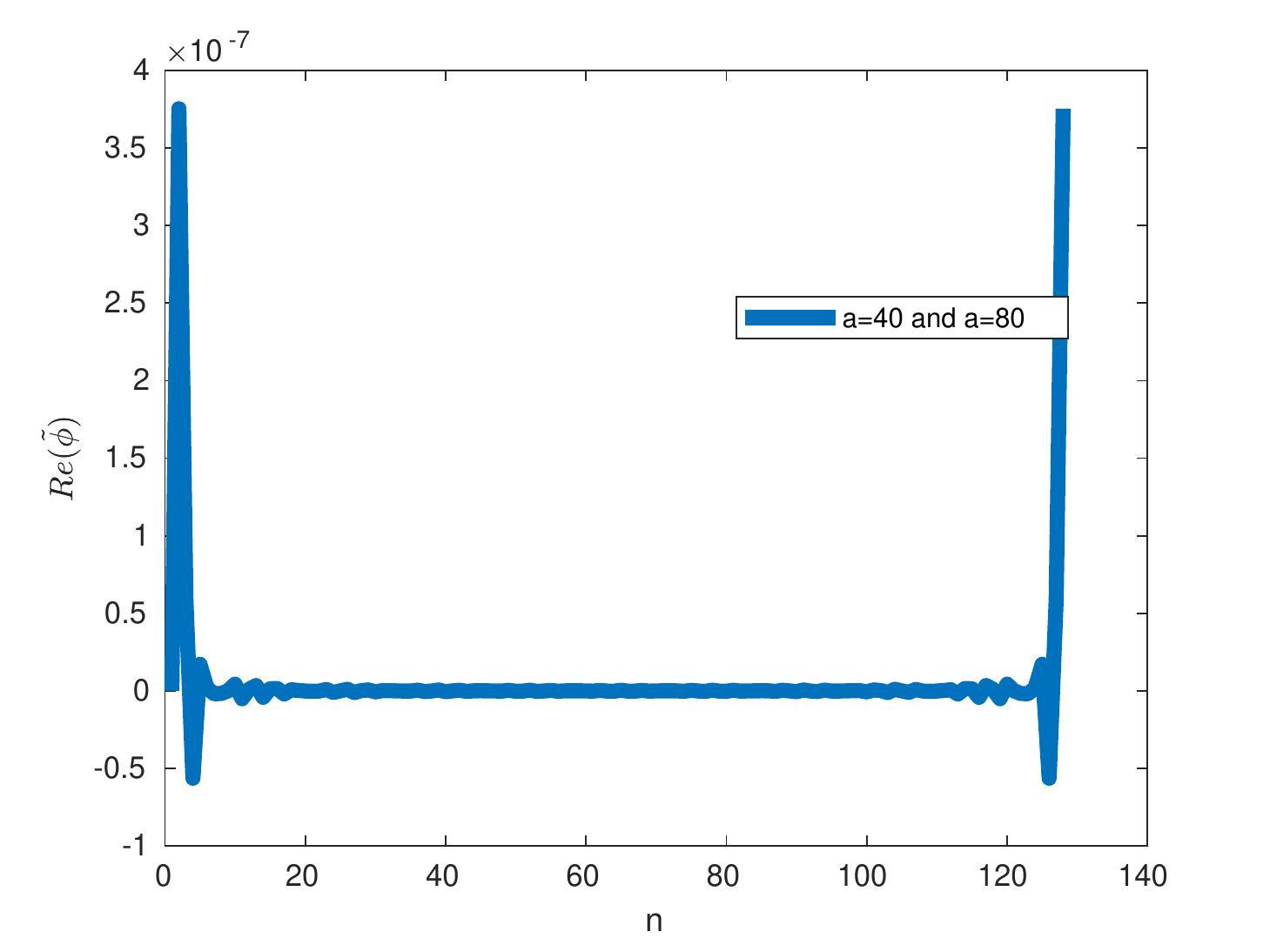}}
\subfloat[]{\includegraphics[width = 2.9in]{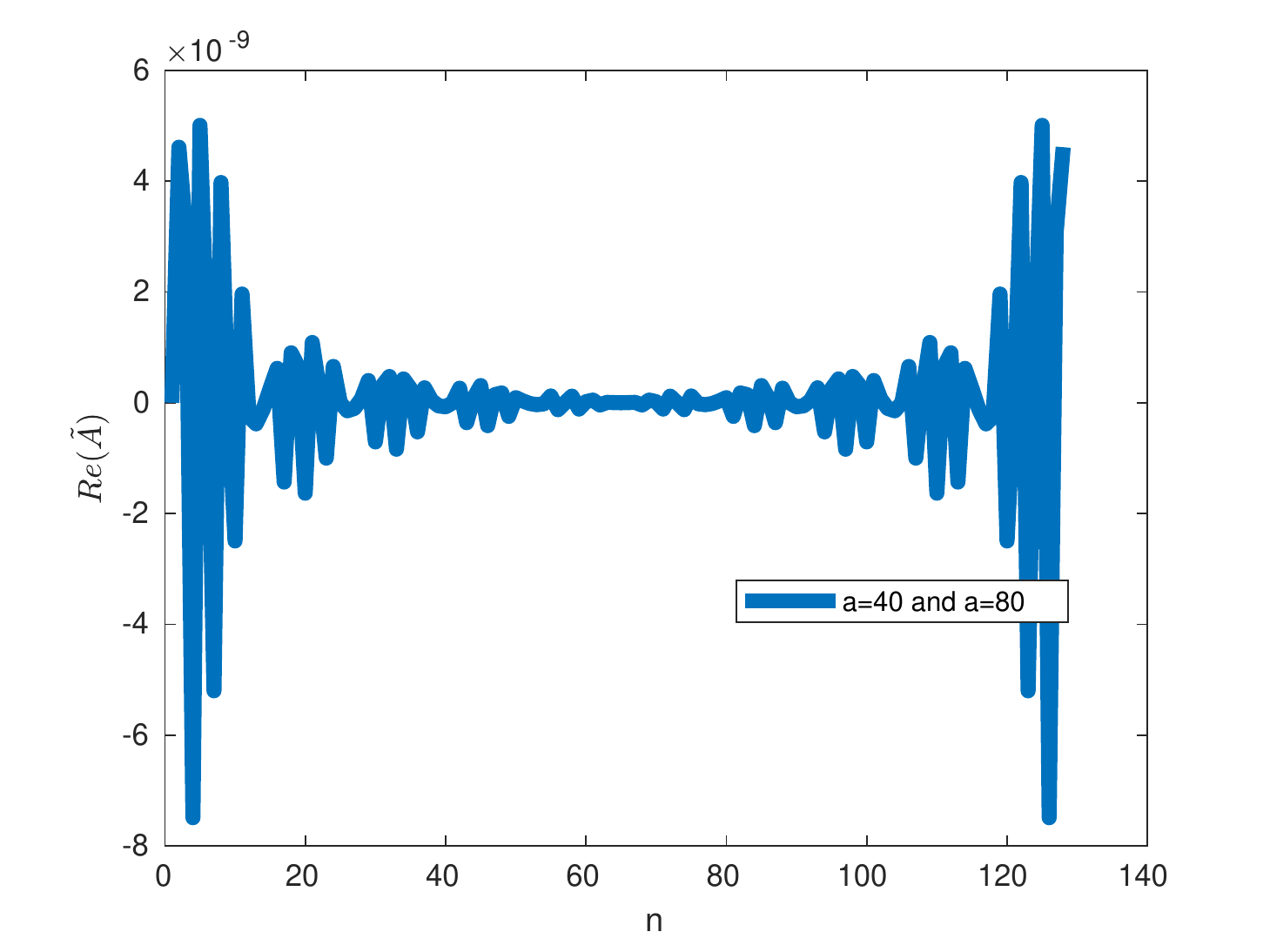}}
\caption{Real parts of all different complex parameters (two point masses at $a=40$ and $a=80$ are the source distribution)}
\label{fig_fr_3}
\end{figure}

\begin{figure}[ht]
\centering
\subfloat[]{\includegraphics[width = 2.9in]{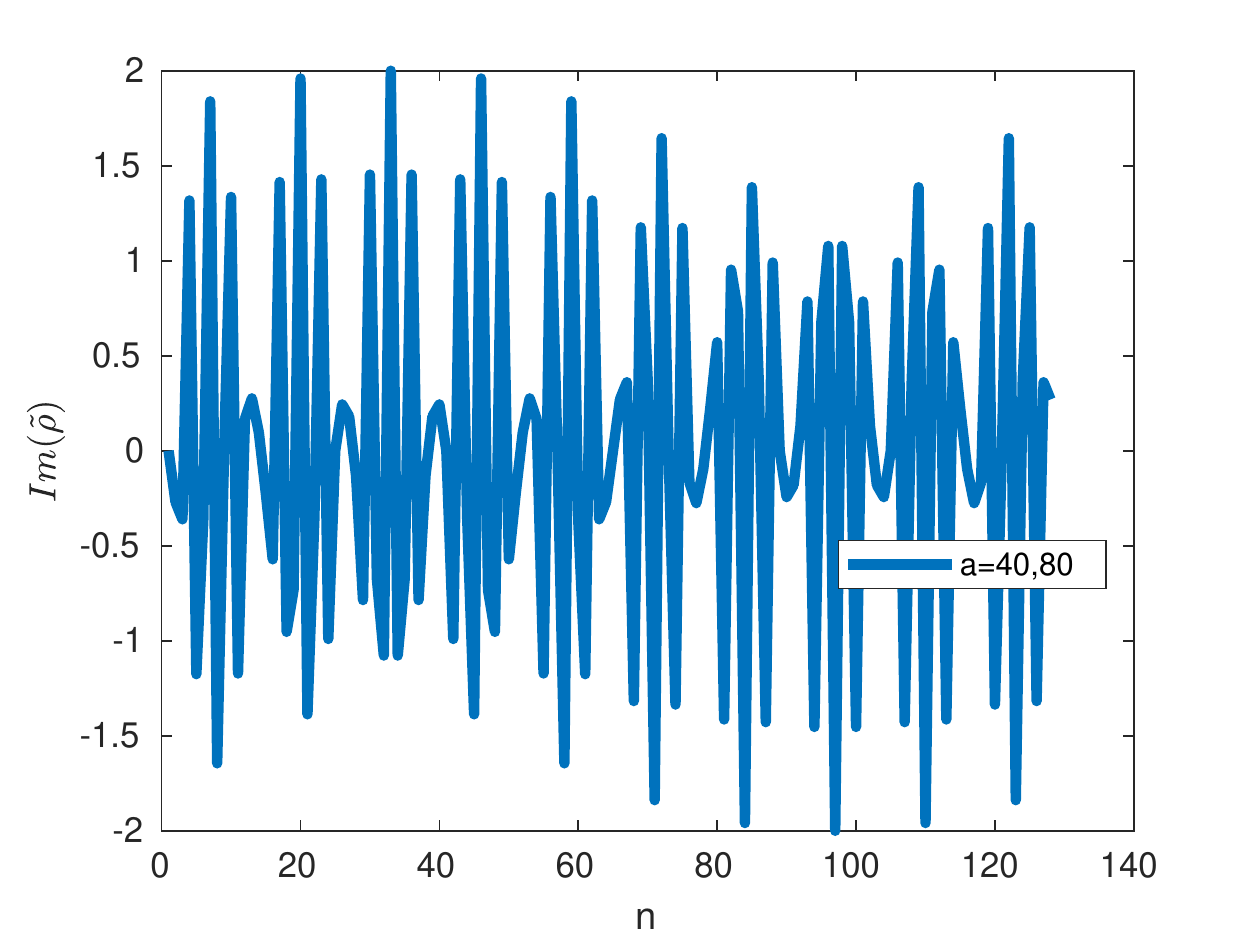}} 
\subfloat[]{\includegraphics[width = 2.9in]{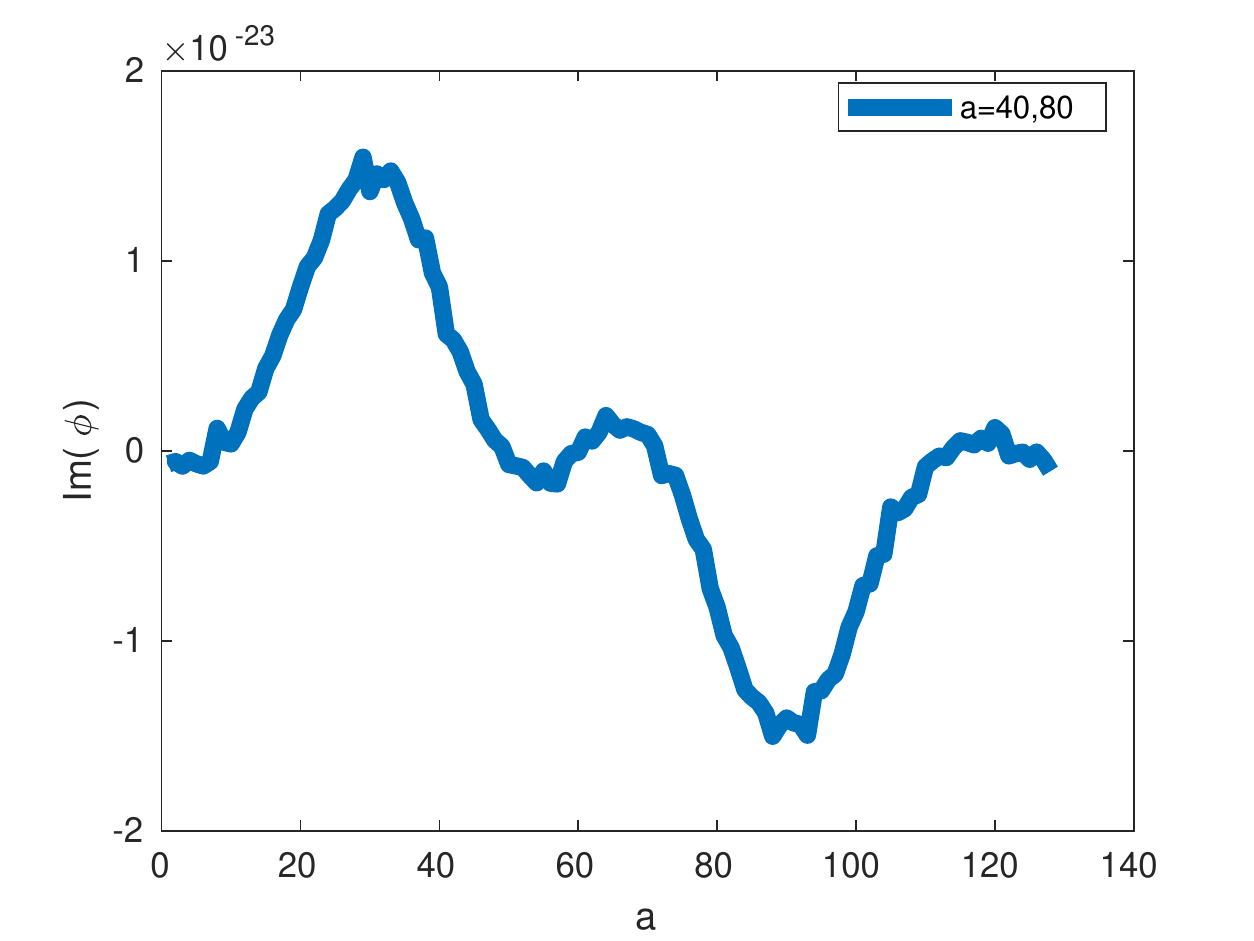}}\\
\subfloat[]{\includegraphics[width = 2.9in]{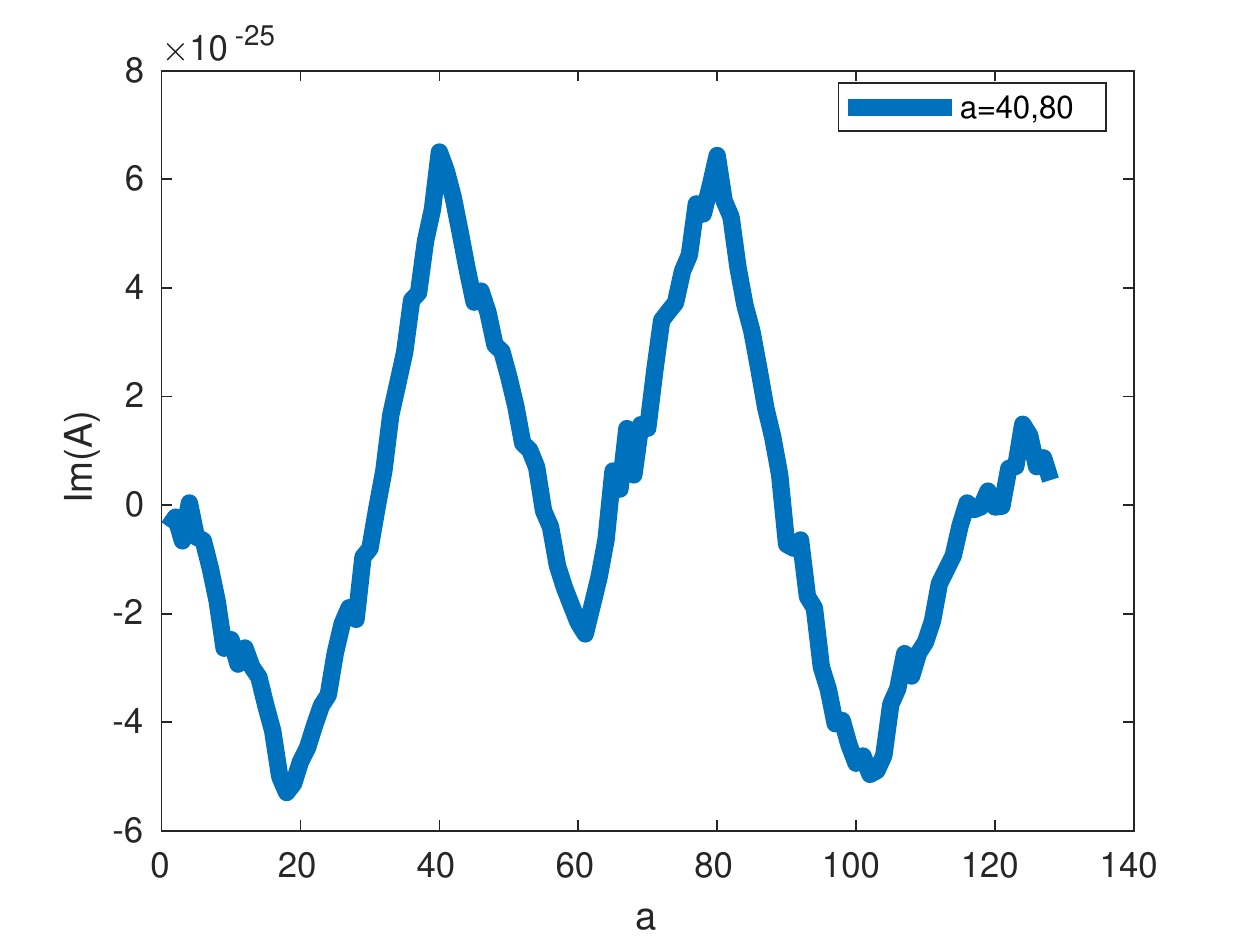}}
\subfloat[]{\includegraphics[width = 2.9in]{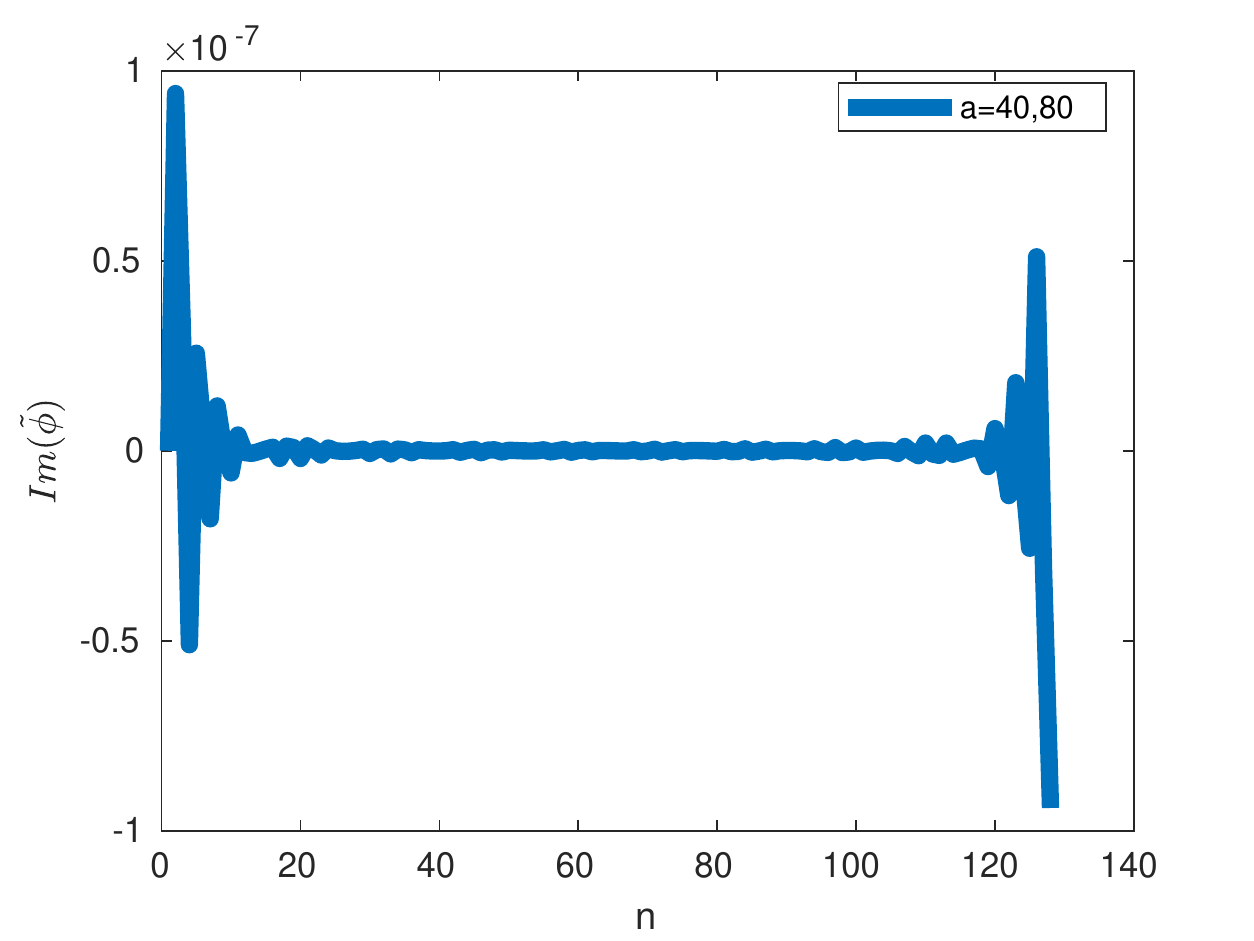}}\\
\subfloat[]{\includegraphics[width = 2.9in]{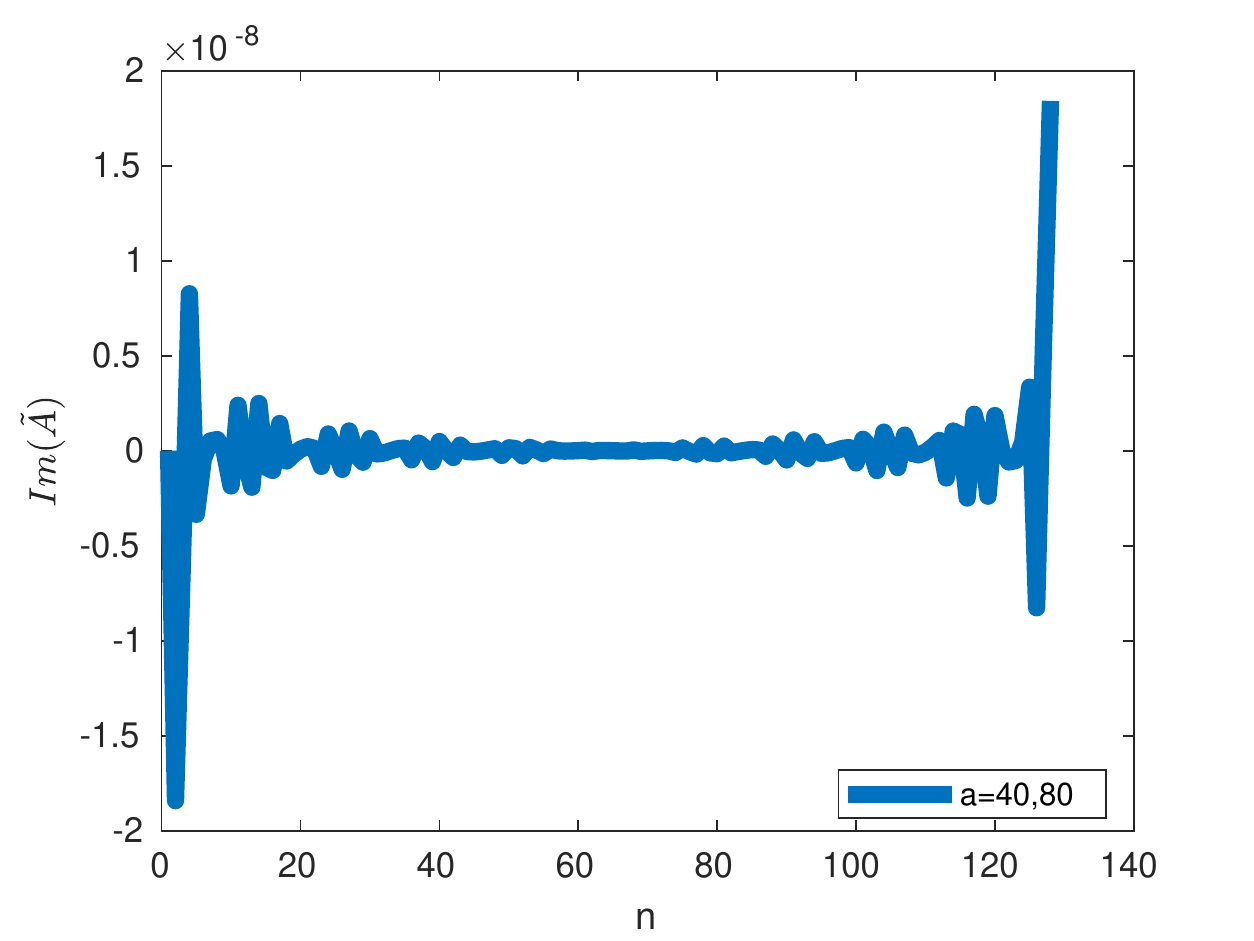}}
\caption{Imaginary parts of all different complex parameters (two point masses at $a=40$ and $a=80$ are the source distribution)}
\label{fig_fr_4}
\end{figure}

In figure \ref{fig_fr_3} and figure \ref{fig_fr_4} the source is two point masses situated at $a=40$ and $a=80$. 

\begin{figure}[ht]
\centering
\subfloat[]{\includegraphics[width = 2.9in]{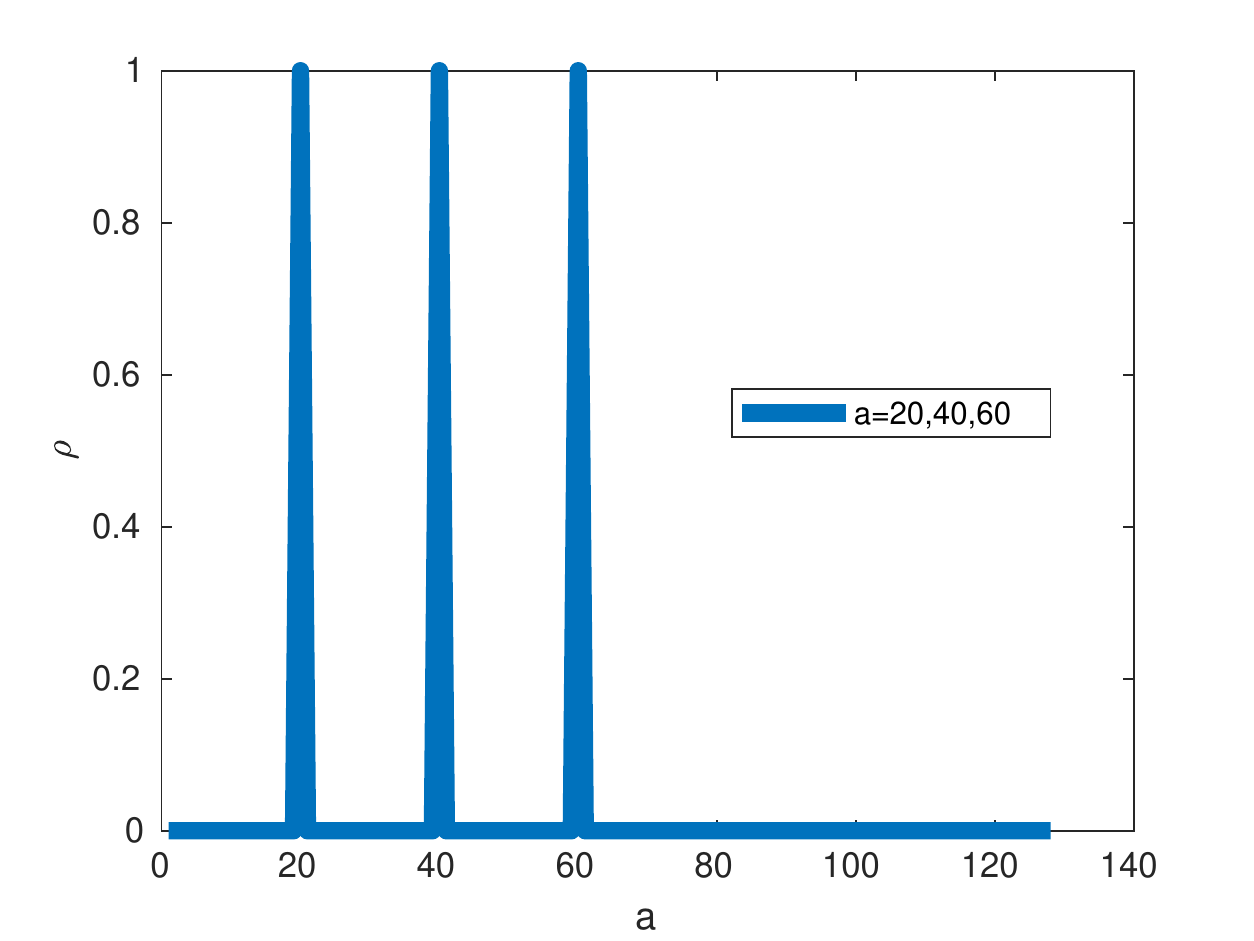}} 
\subfloat[]{\includegraphics[width = 2.9in]{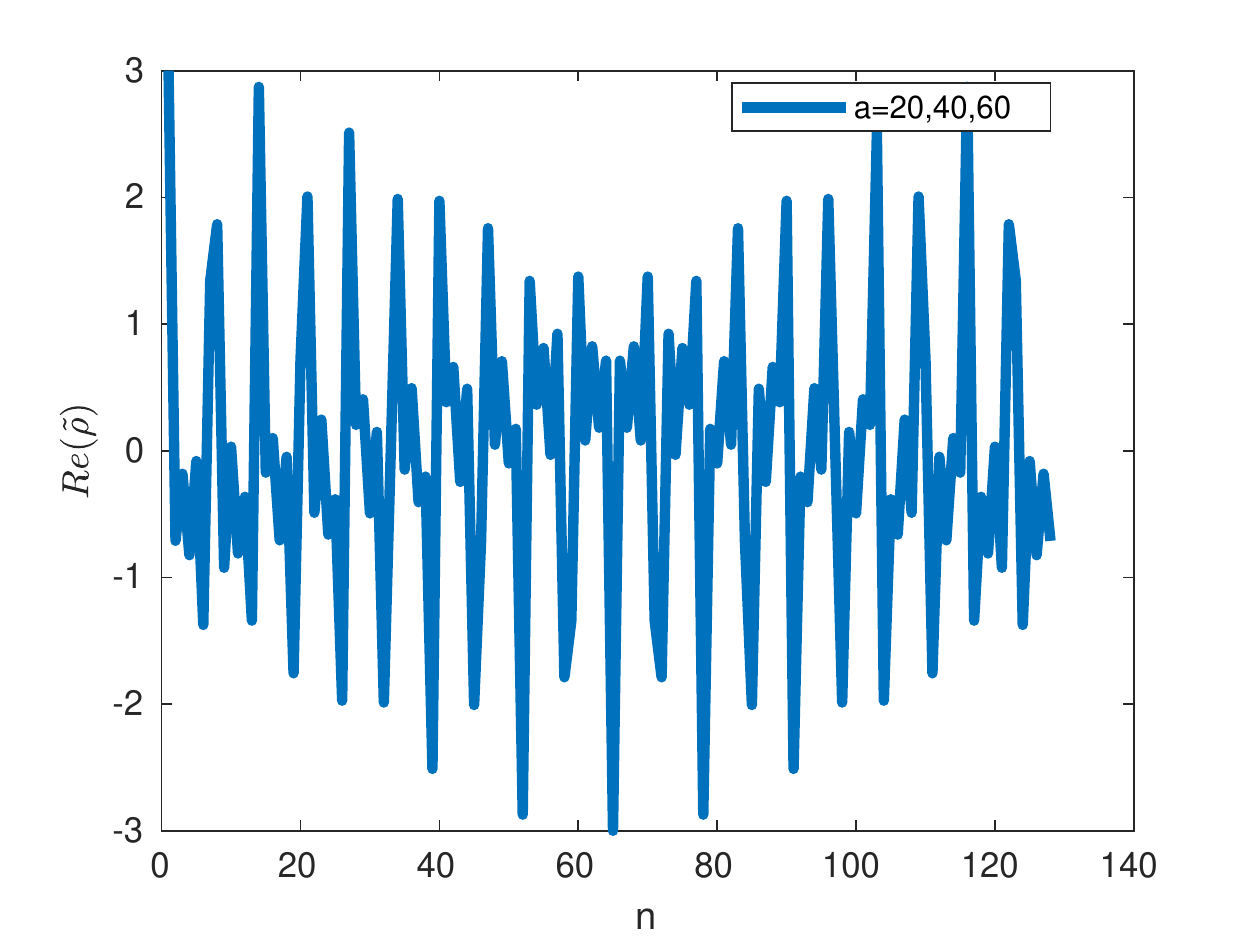}}\\
\subfloat[]{\includegraphics[width = 2.9in]{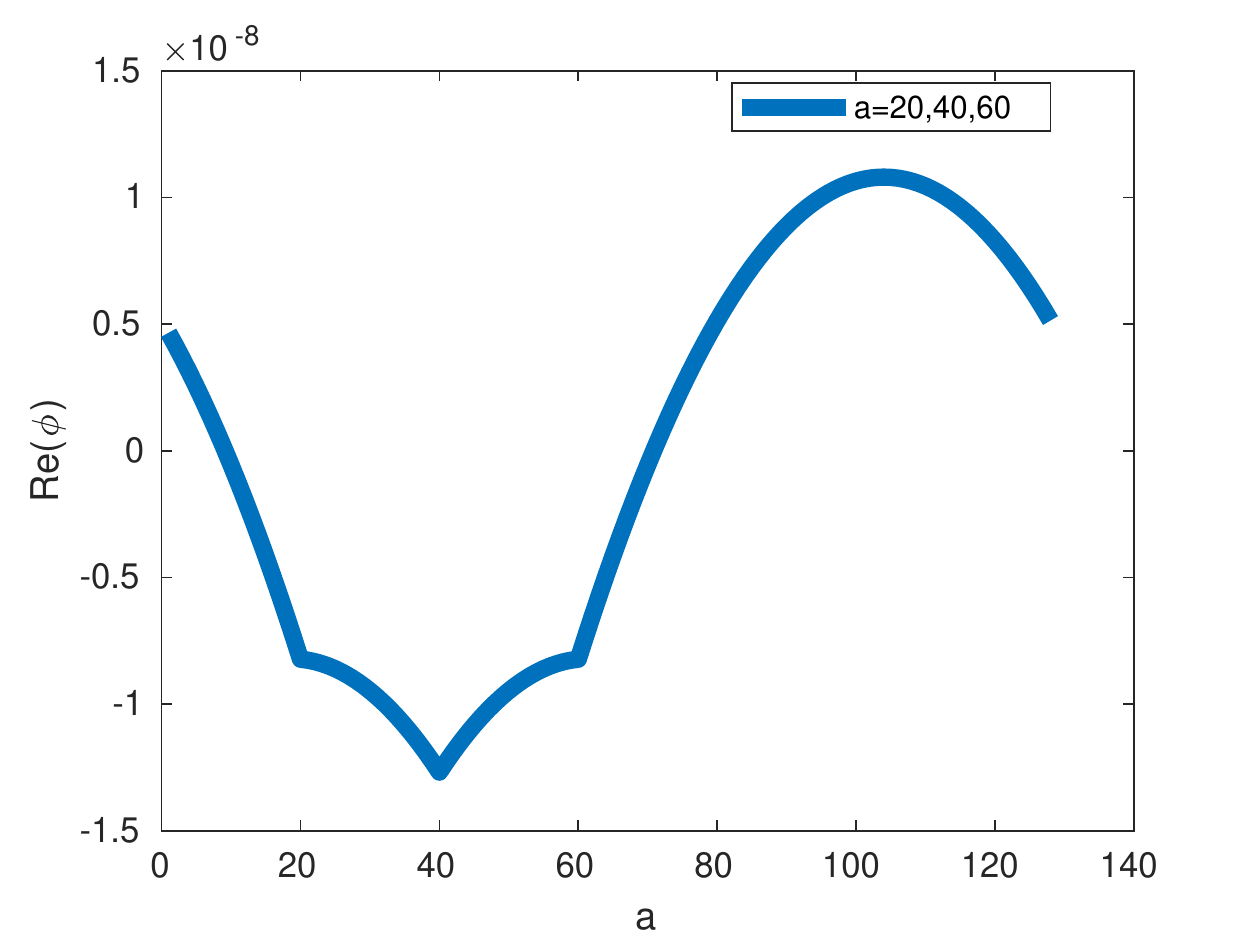}}
\subfloat[]{\includegraphics[width = 2.9in]{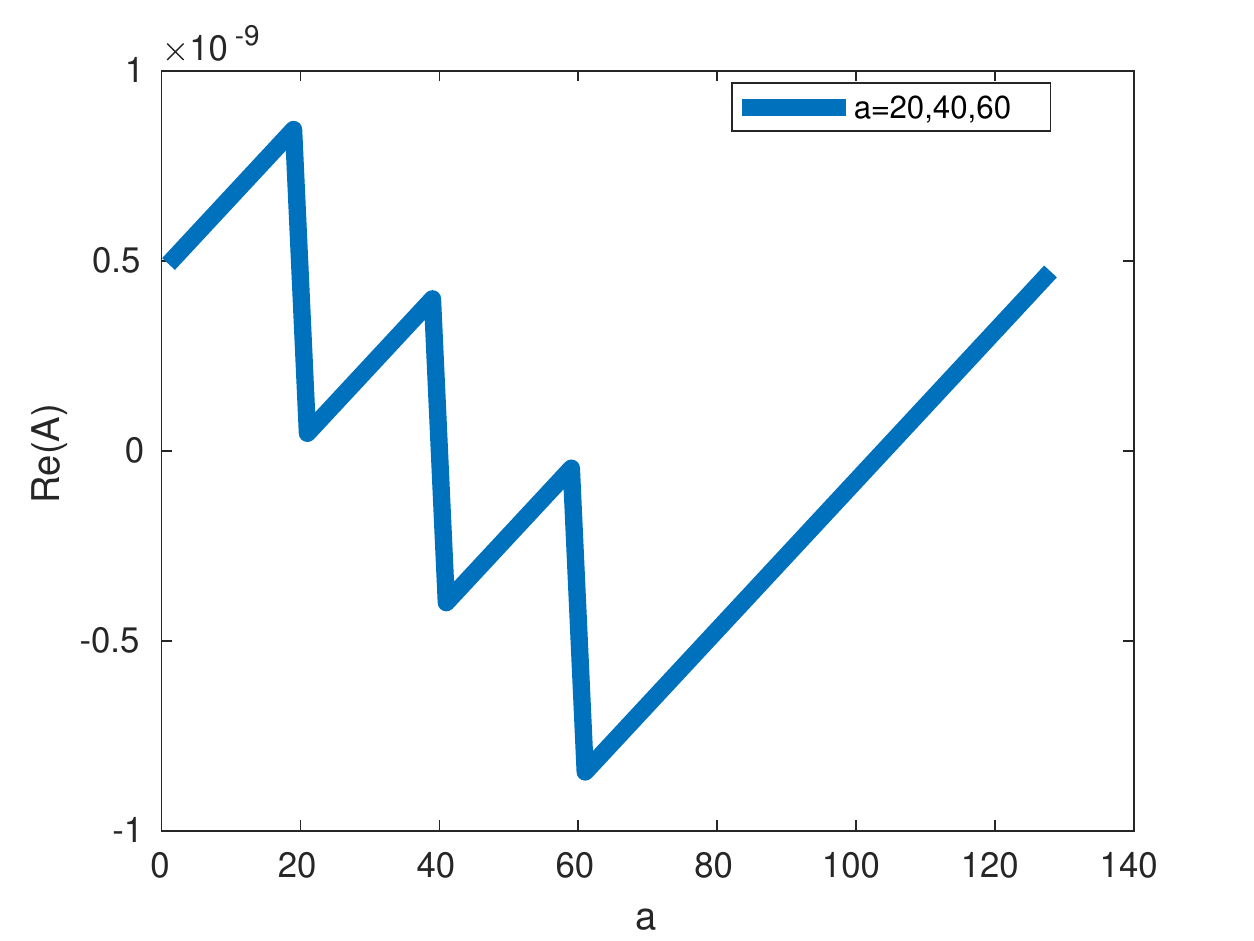}}\\
\subfloat[]{\includegraphics[width = 2.9in]{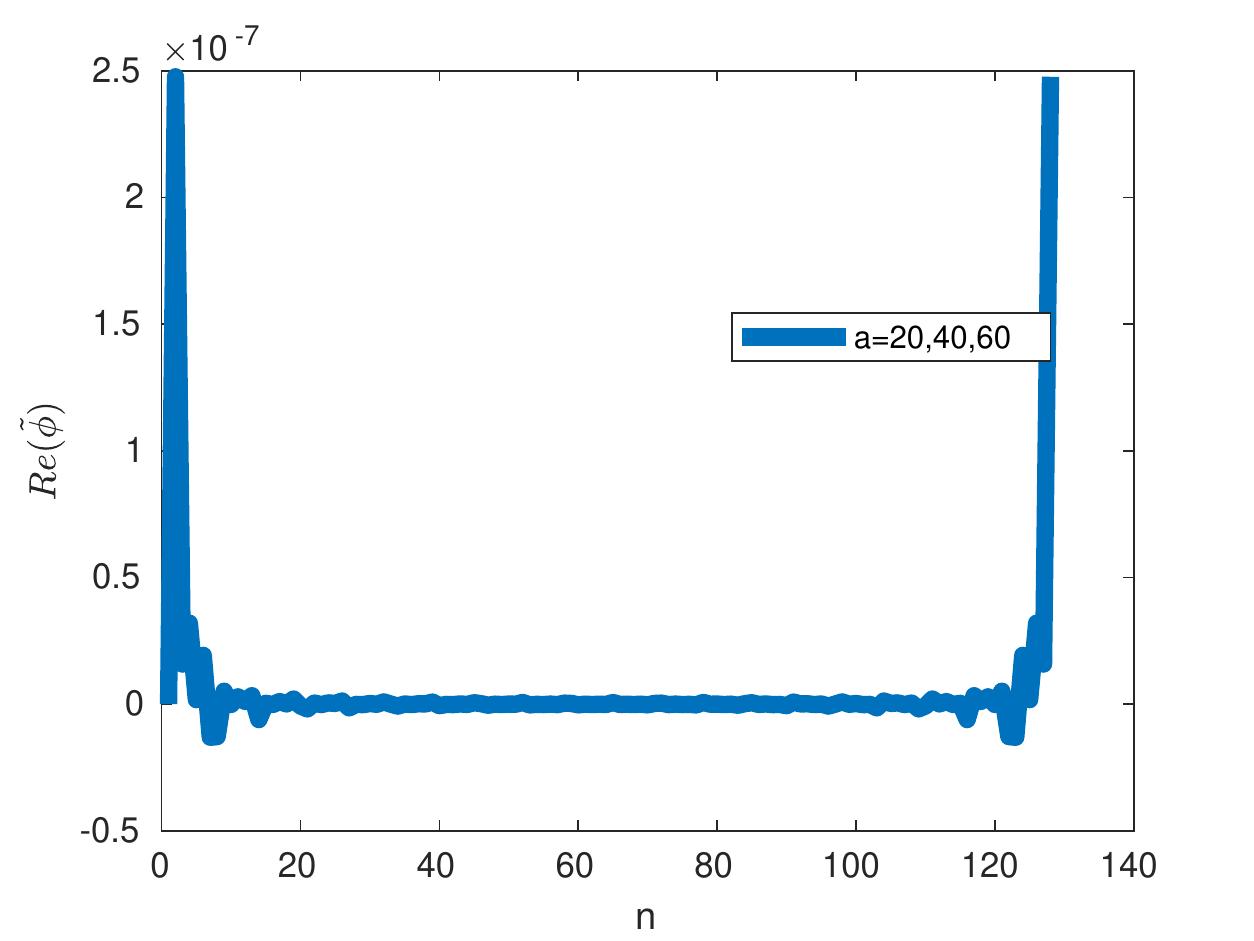}}
\subfloat[]{\includegraphics[width = 2.9in]{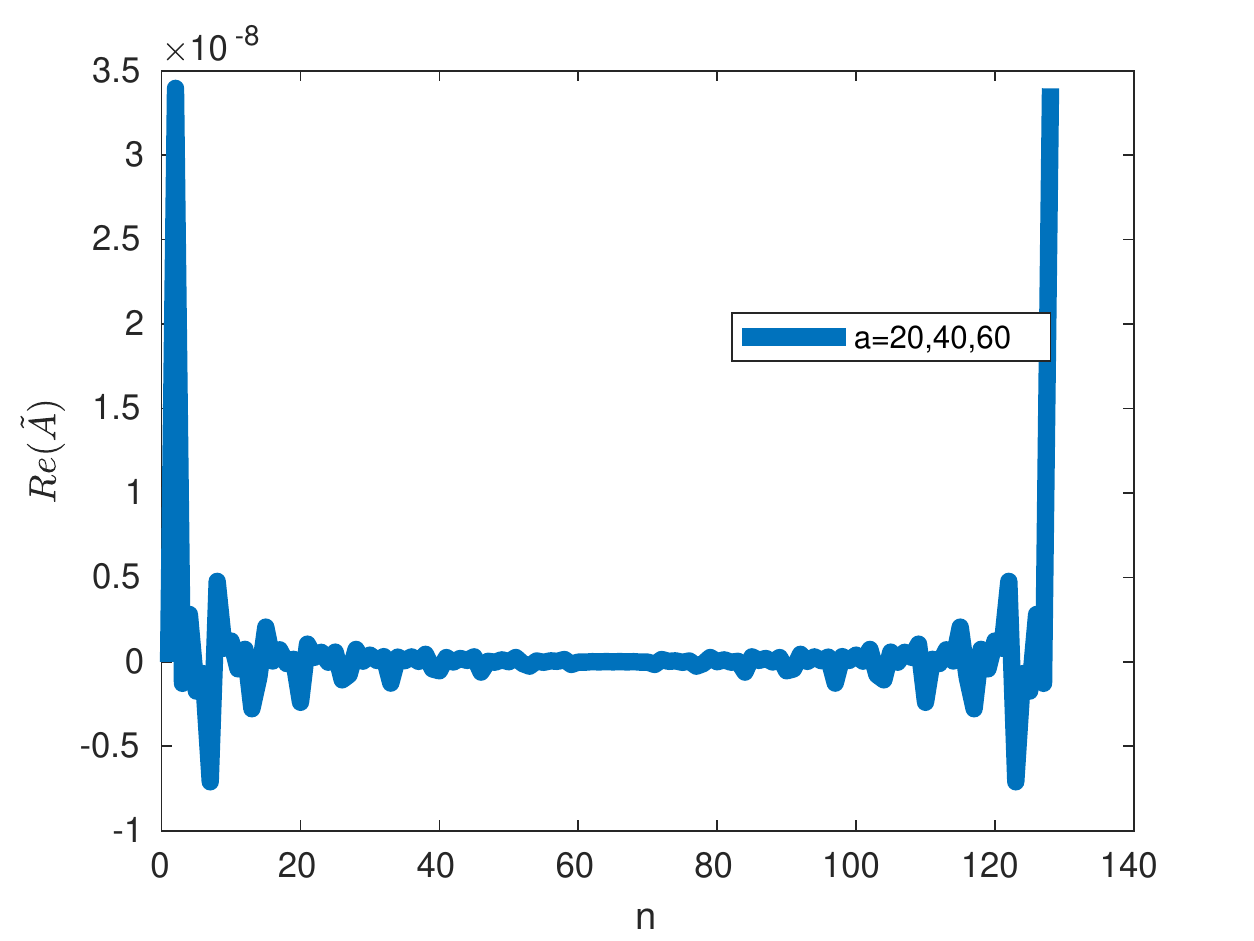}}
\caption{Real parts of all different complex parameters (three point masses at $a=20$, $a=40$ and $a=60$ are the source distribution)}
\label{fig_fr_5}
\end{figure}

\begin{figure}[ht]
\centering
\subfloat[]{\includegraphics[width = 2.9in]{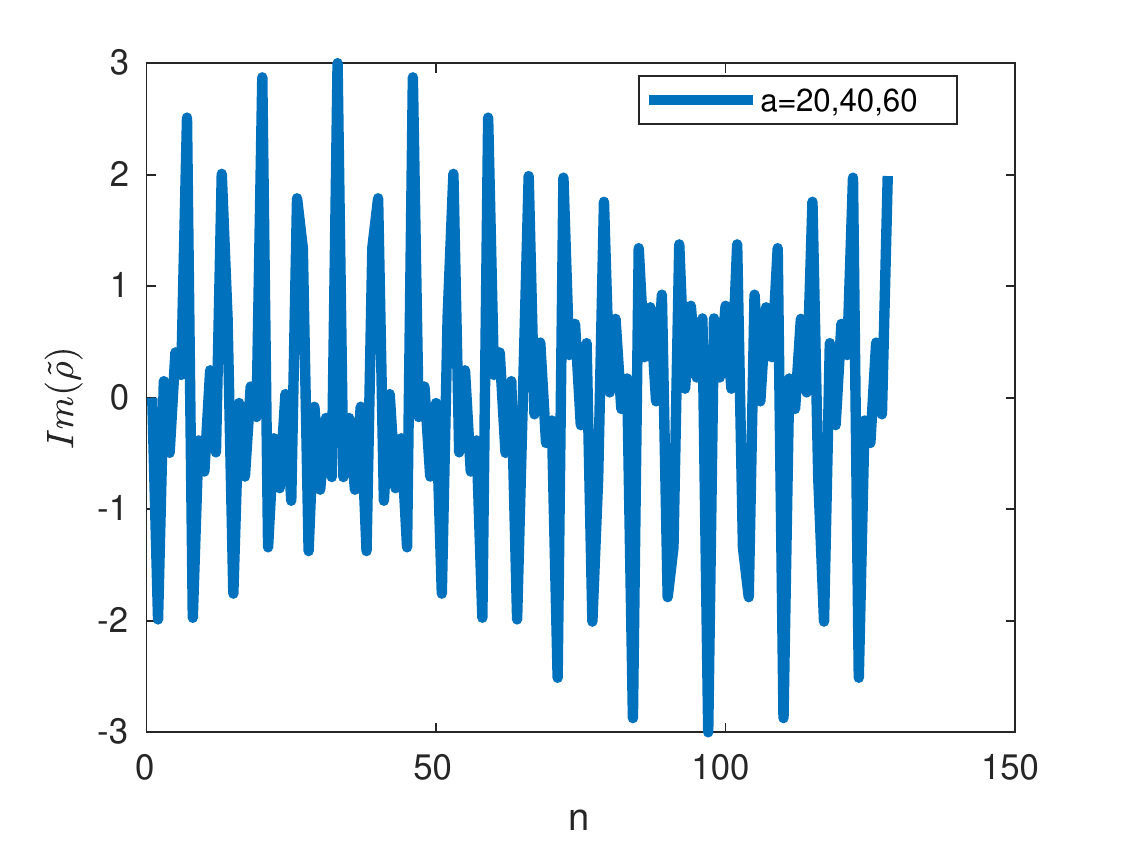}} 
\subfloat[]{\includegraphics[width = 2.9in]{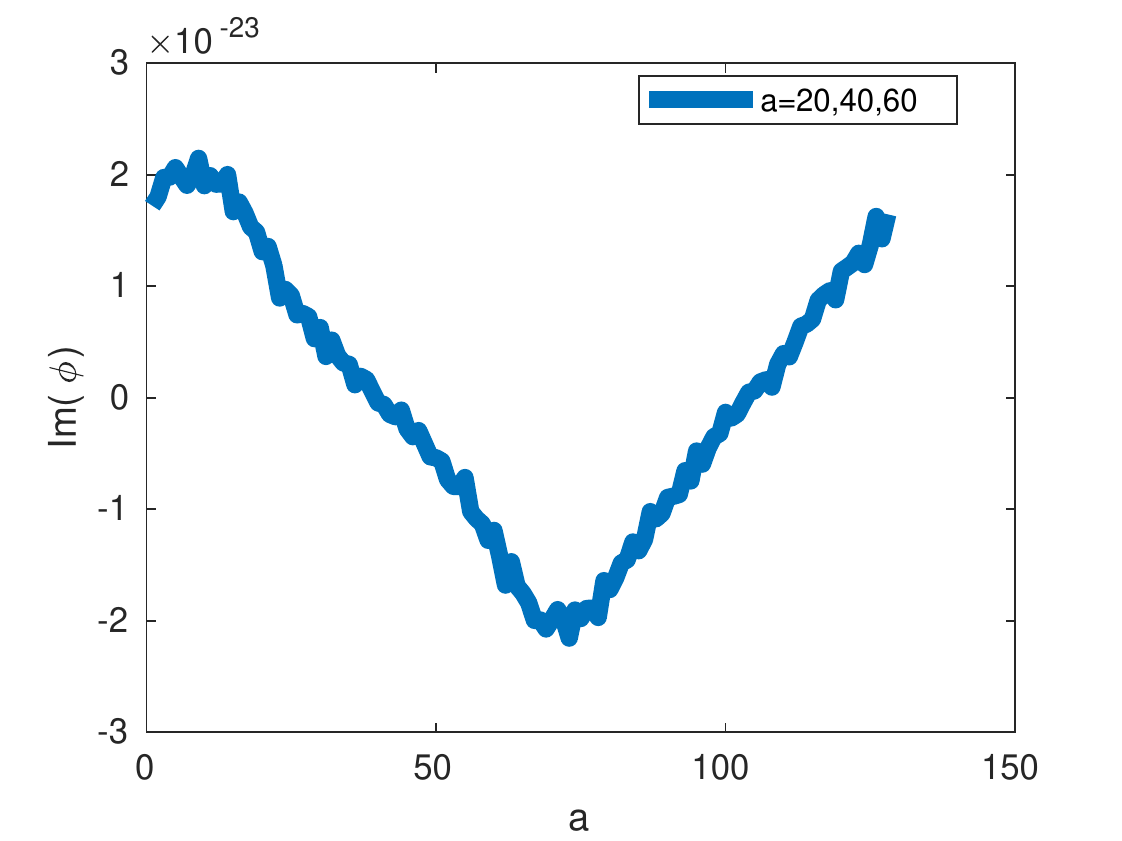}}\\
\subfloat[]{\includegraphics[width = 2.9in]{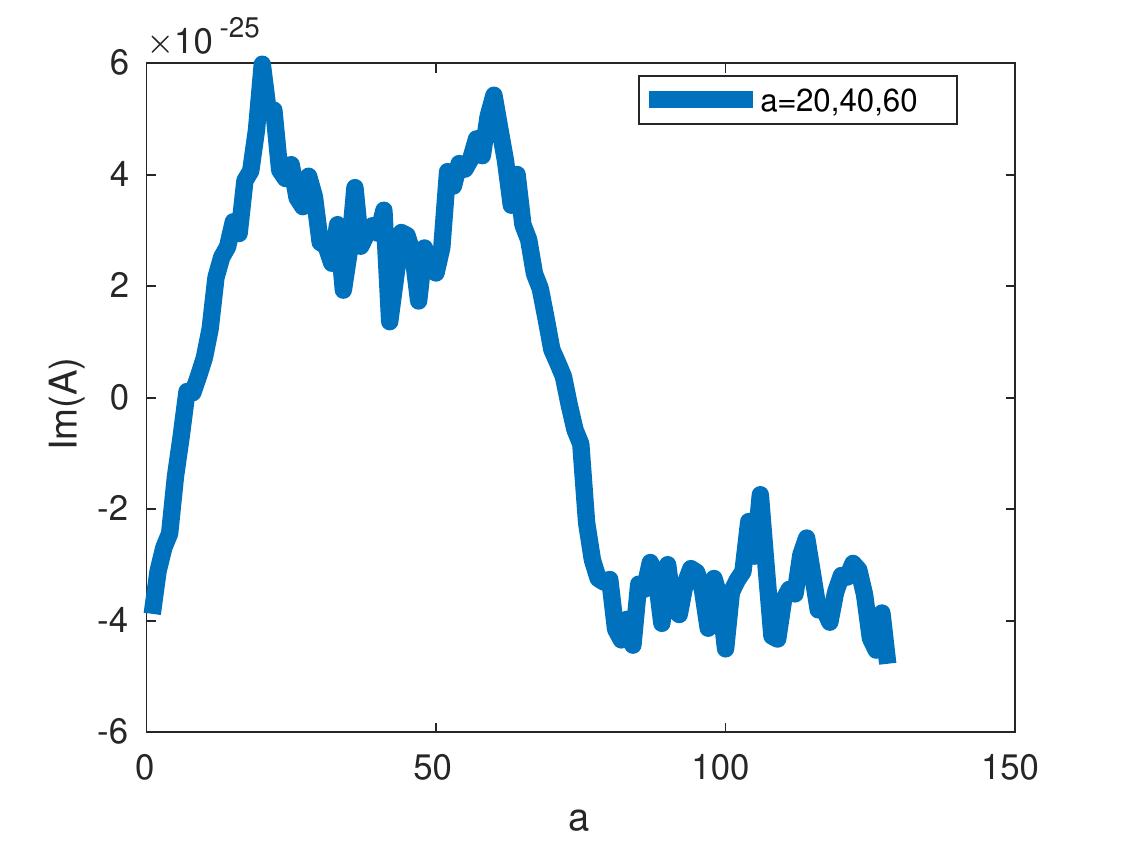}}
\subfloat[]{\includegraphics[width = 2.9in]{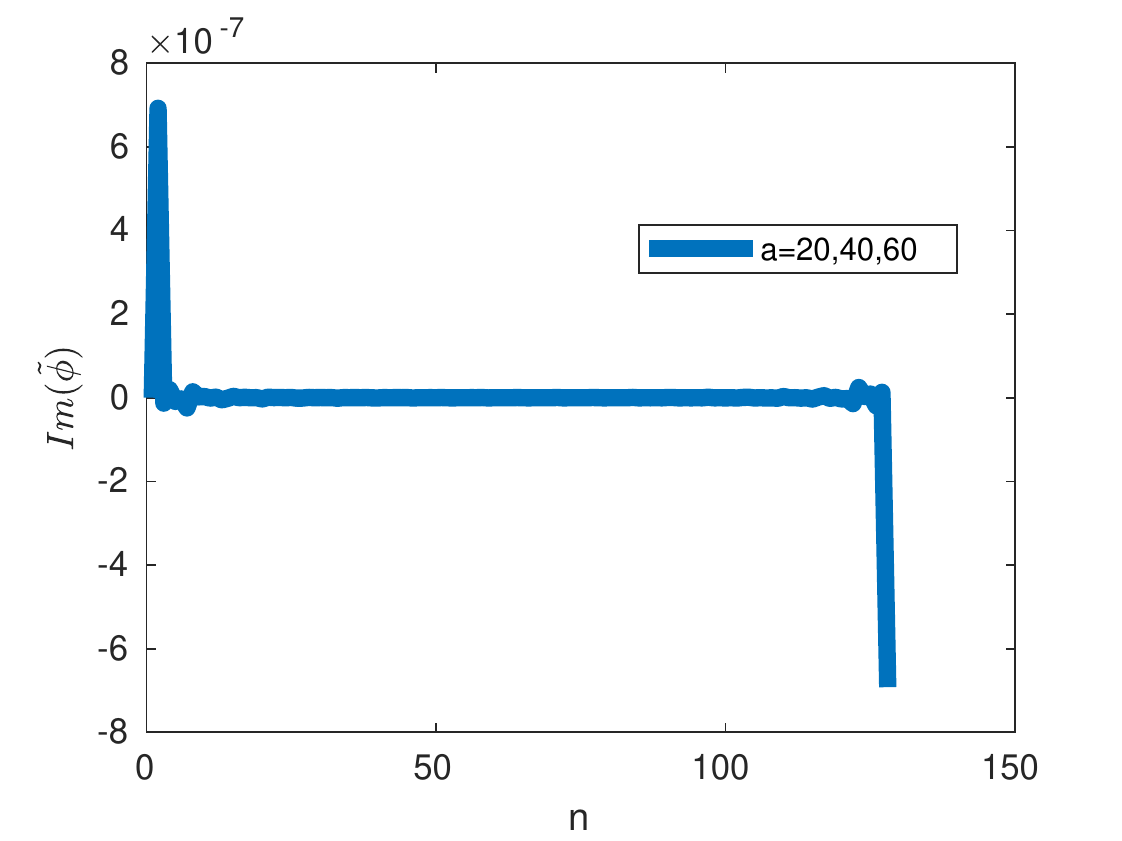}}\\
\subfloat[]{\includegraphics[width = 2.9in]{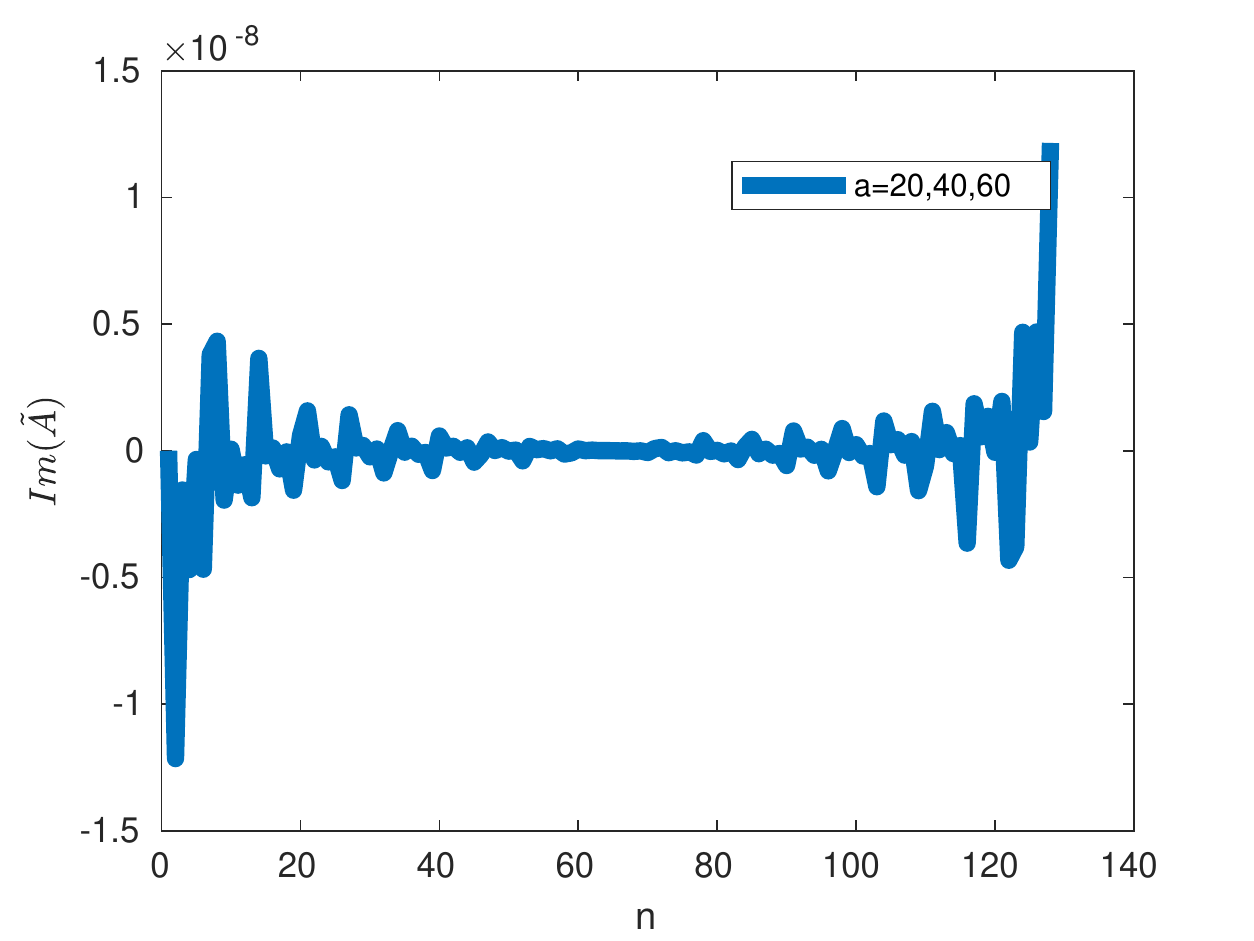}}
\caption{Imaginary parts of all different complex parameters (three point masses at $a=20$, $a=40$ and $a=60$ are the source distribution)}
\label{fig_fr_6}
\end{figure}

In figure \ref{fig_fr_5} and figure \ref{fig_fr_6} the source is three point masses situated at $a=20$, $a=40$ and $a=60$.

 \chapter{Solution of Poisson's Equation by CIC Method}

\section{Aim}
To find the acceleration value at test particle position, where in generally test particle may not be situated on a grid position. 

\section{Introduction}
In Assignment-4 we solved the acceleration and potential field for a given mass density. However the acceleration and potential field that we calculated there is on the grid. The density of particles also was distributed on the grid points only. In real physical problem the source particle and test particle positions may not be always on a grid point. In generally it can be anywhere between the grid points.

In that case we have to use Cloud in Cell (CIC) method. 
\section{Brief Theory}
If the position for any source particle between two grid points is $x$ then according to CIC method the mass of the particle will distribute among the grid points. The mass distribution will be weighted average. The closer a mass particle to a grid point, more weight will be given for that grid point. 

For a 1D case if the source particle position is $x$ and it is situated between the grid points $a=floor$ and $a=ceil$. If the mass of the particle is $m$ then

\begin{equation}
    w_1=1-|\:x-floor\:| \: \: \text{and} \:\: w_2=1-|\:ceil-x\:|
\end{equation}
and, 
\begin{equation}
    \rho_{floor}=w_1 \times m \:\:\: \text{and} \:\:\: \rho_{ceil}=w_2 \times m
\end{equation}

Now, as we have density of mass on the grid the calculation of potential field and acceleration on the grid is as the previous assignment (Assignment-4). 

If the test particle is at $x_{test}$ and situated between the grid points $a=ft$ and $a=ct$ then acceleration at $x_{test}$ is

\begin{equation}
    A\left(x_{test}\right)=w_1 \times A_{ft} + w_2 \times A_{ct}
\end{equation}

where,

\begin{equation}
    w_1=1-|\:x-ft\:| \: \: \text{and} \:\: w_2=1-|\:ct-x\:|
\end{equation}

\section{Simulation Parameters}
We simulate for a 1D grid of $L=128$. 

The source is an unit point mass at $x=64.4$

The test particle is an unit point mass at $x_{test}=10.8$

\section{Results and Plots}
The value of the acceleration at the test particle position ($x_{test}=10.8$) is $A(x_{test}) = 6.814577850552632\times 10^{-11}$ $N/kg$

\begin{figure}[ht]
\centering
\subfloat[]{\includegraphics[width = 2.9in]{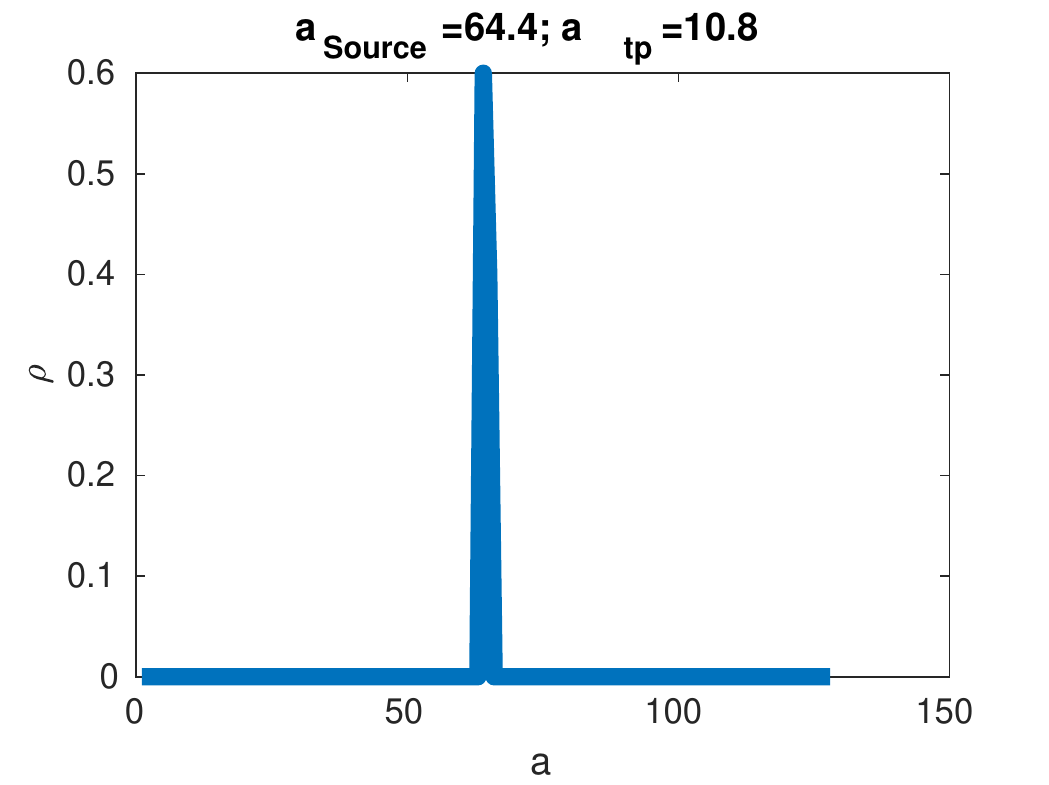}} 
\subfloat[]{\includegraphics[width = 2.9in]{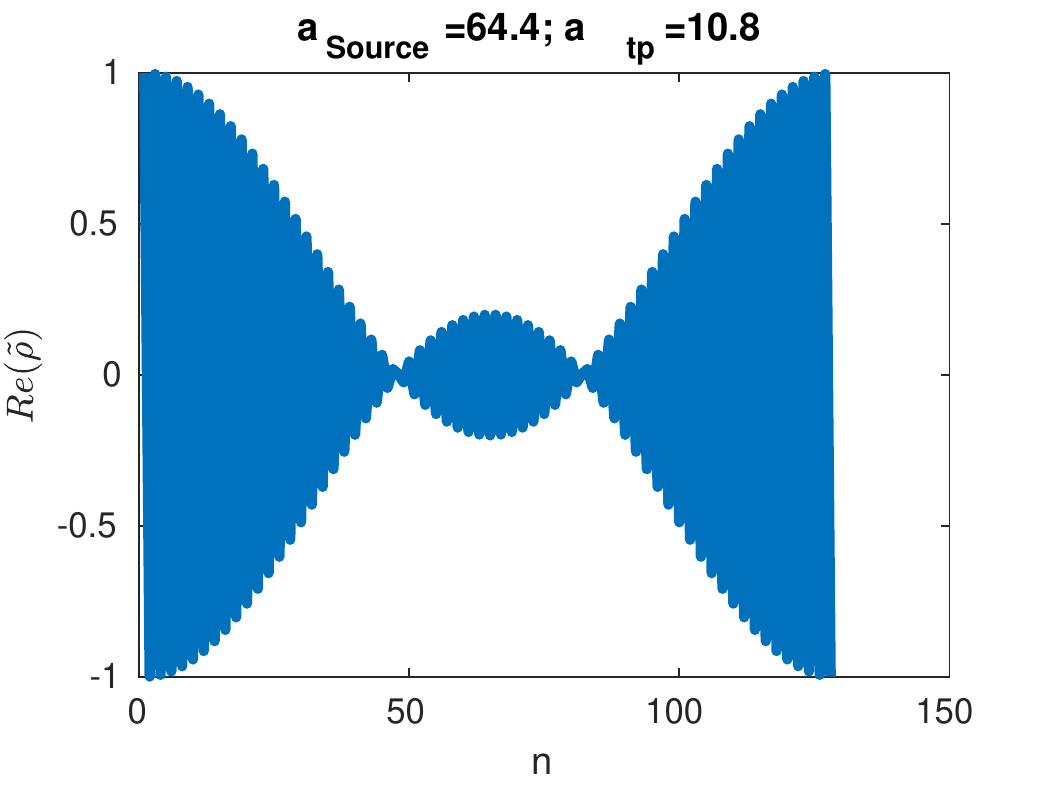}}\\
\subfloat[]{\includegraphics[width = 2.9in]{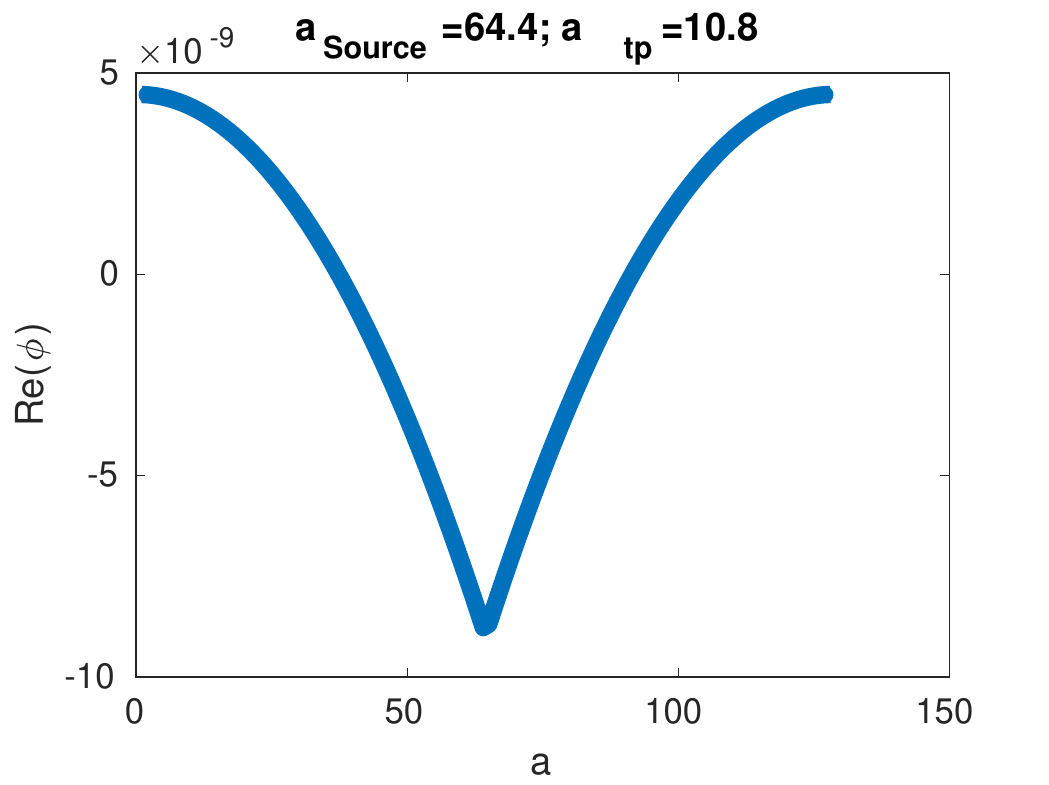}} 
\subfloat[]{\includegraphics[width = 2.9in]{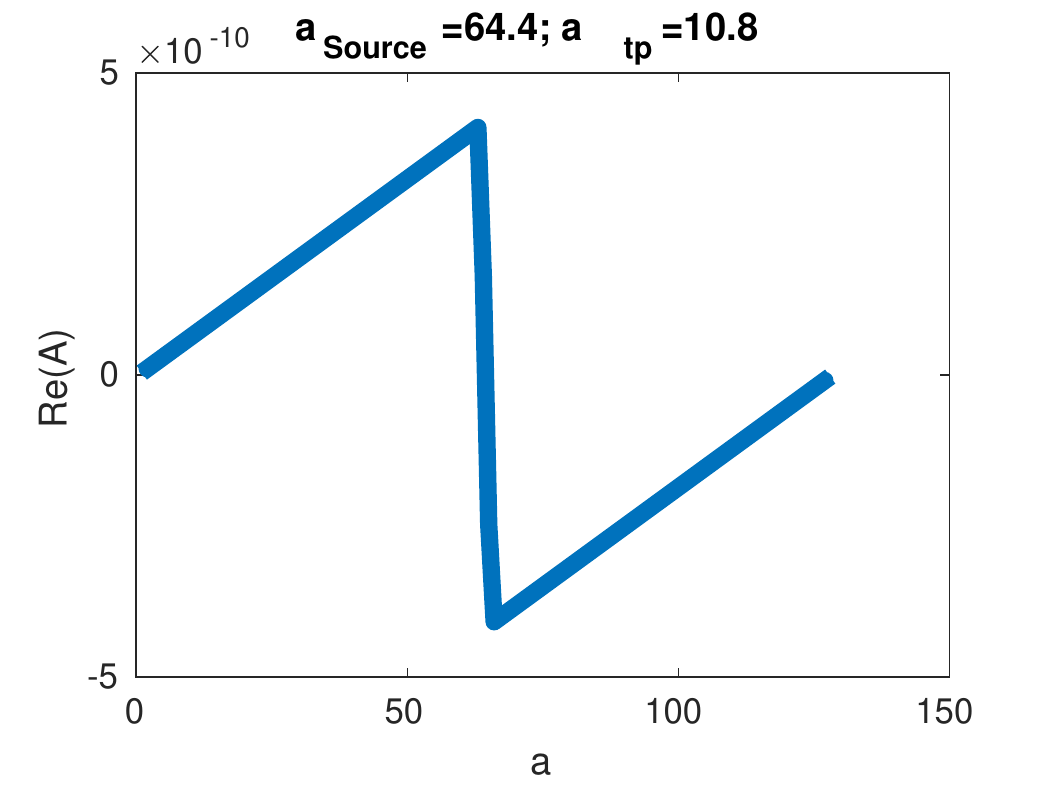}}\\ \subfloat[]{\includegraphics[width = 2.9in]{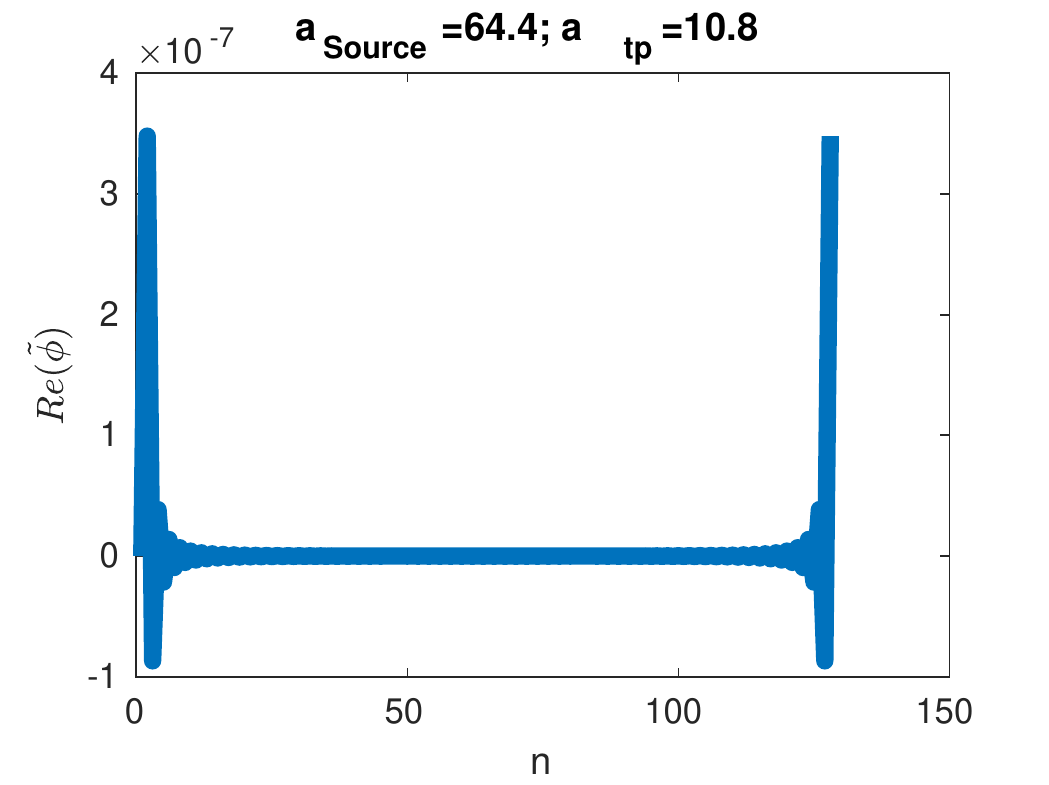}} 
\subfloat[]{\includegraphics[width = 2.9in]{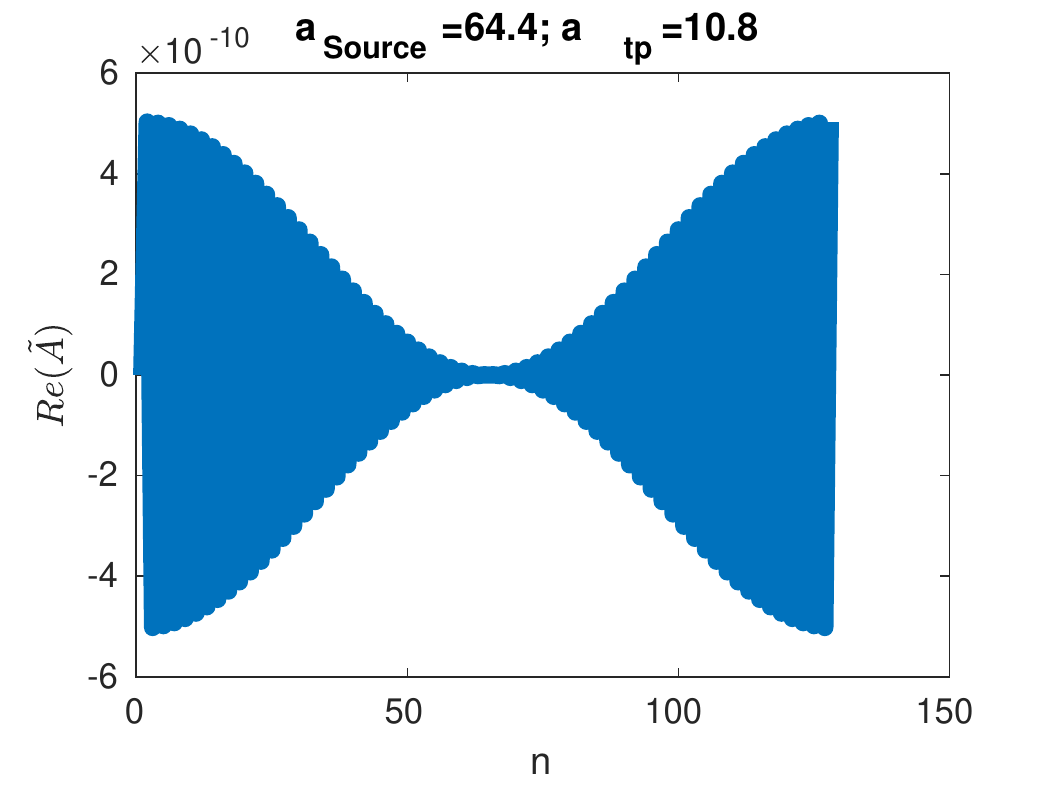}} 
\caption{Real parts of all different complex parameters (The source particle is at $x=64.4$ and test particle is at $x_{test}=10.8$)}
\label{fig_cic_1}
\end{figure}

\begin{figure}[ht]
\centering
\subfloat[]{\includegraphics[width = 2.9in]{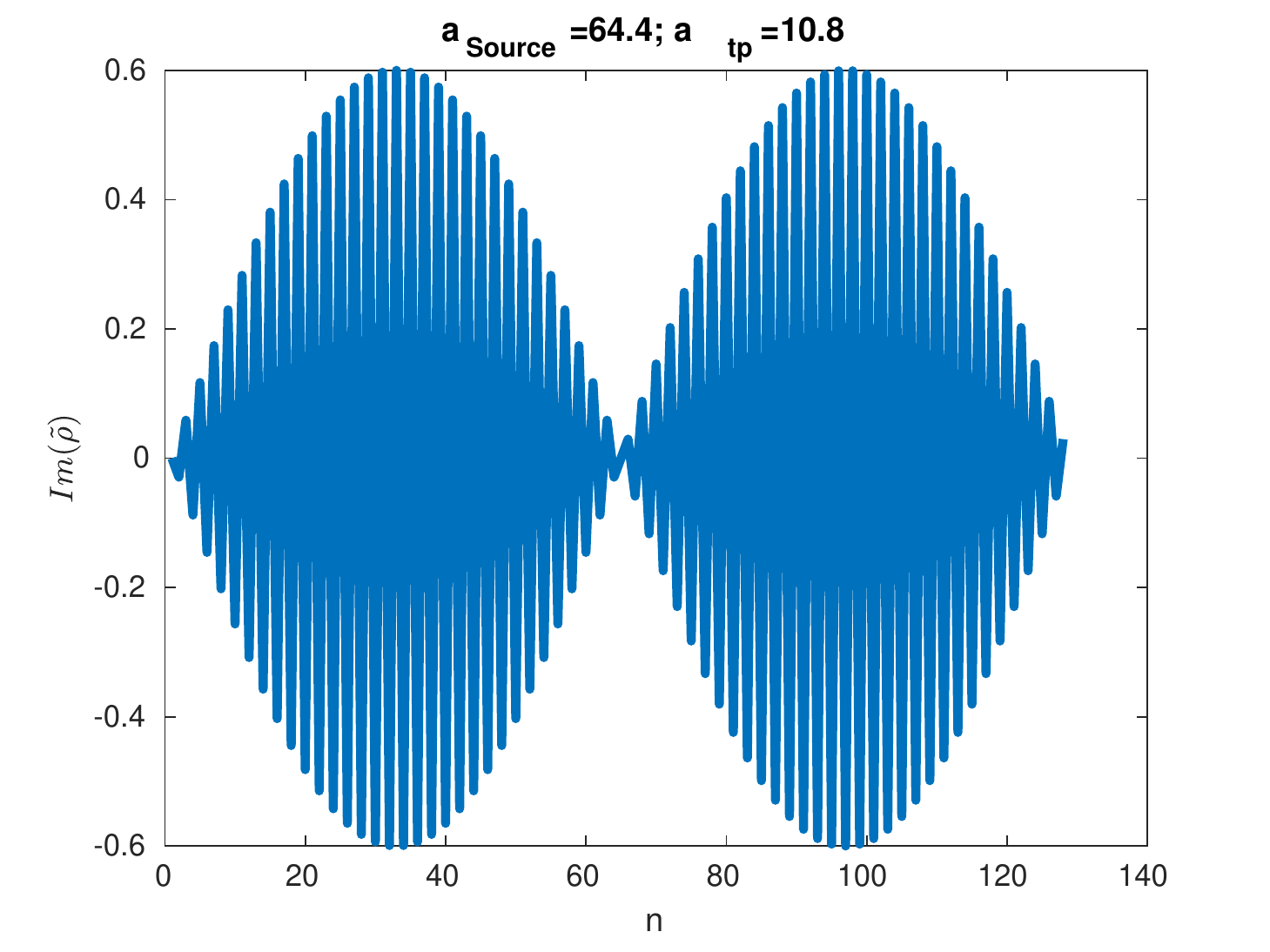}}
\subfloat[]{\includegraphics[width = 2.9in]{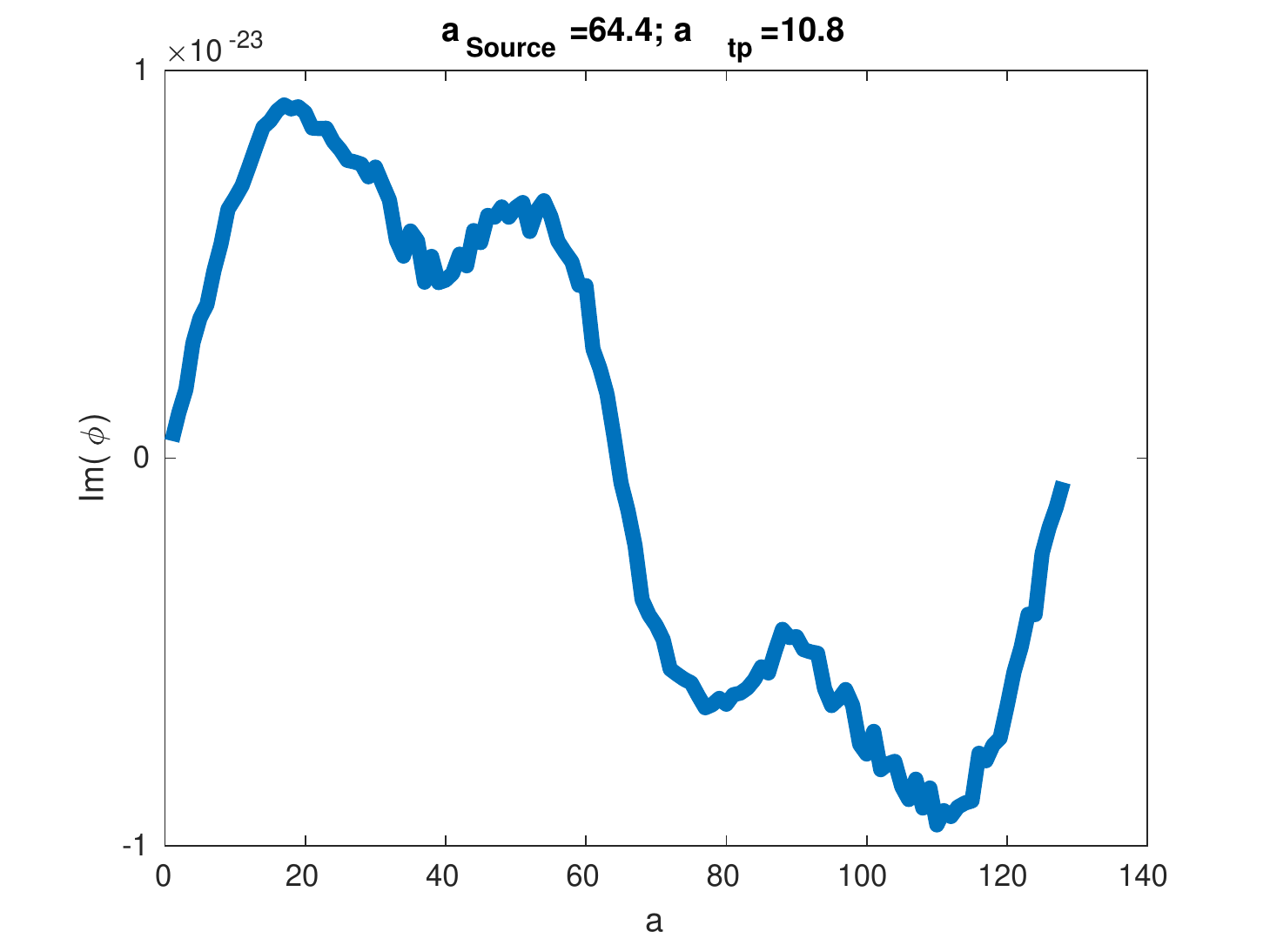}}\\
\subfloat[]{\includegraphics[width = 2.9in]{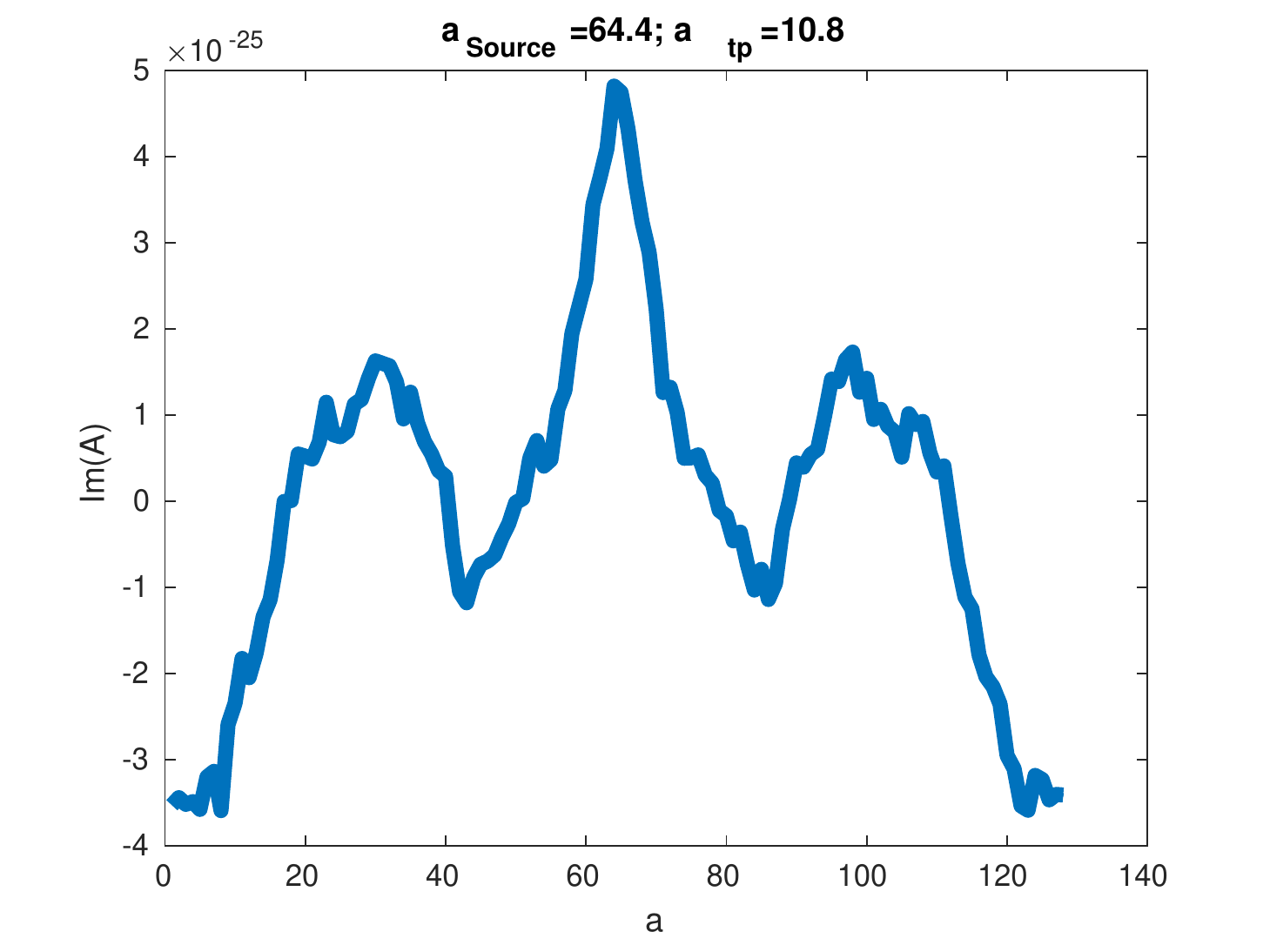}}
\subfloat[]{\includegraphics[width = 2.9in]{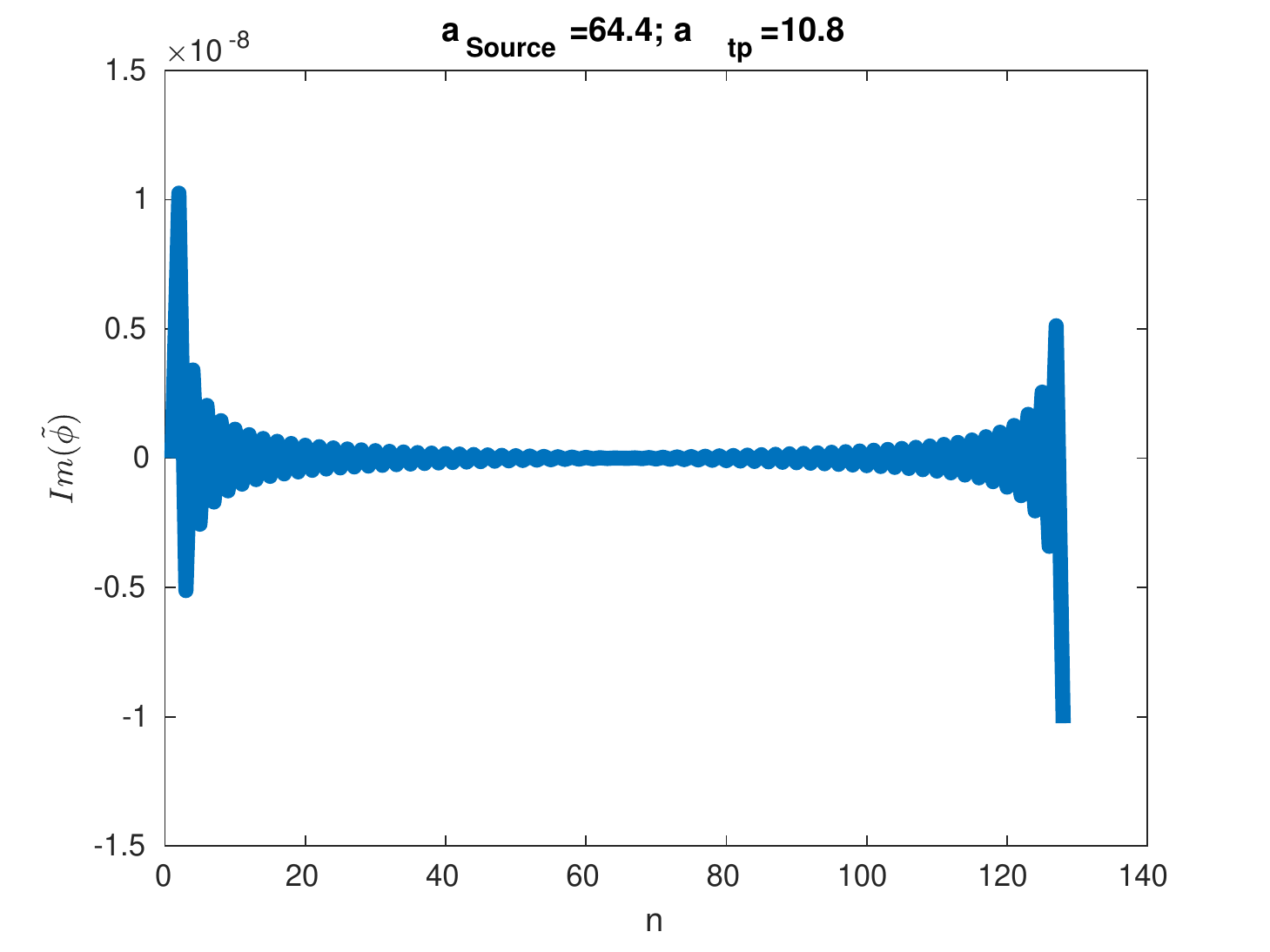}}\\
\subfloat[]{\includegraphics[width = 2.9in]{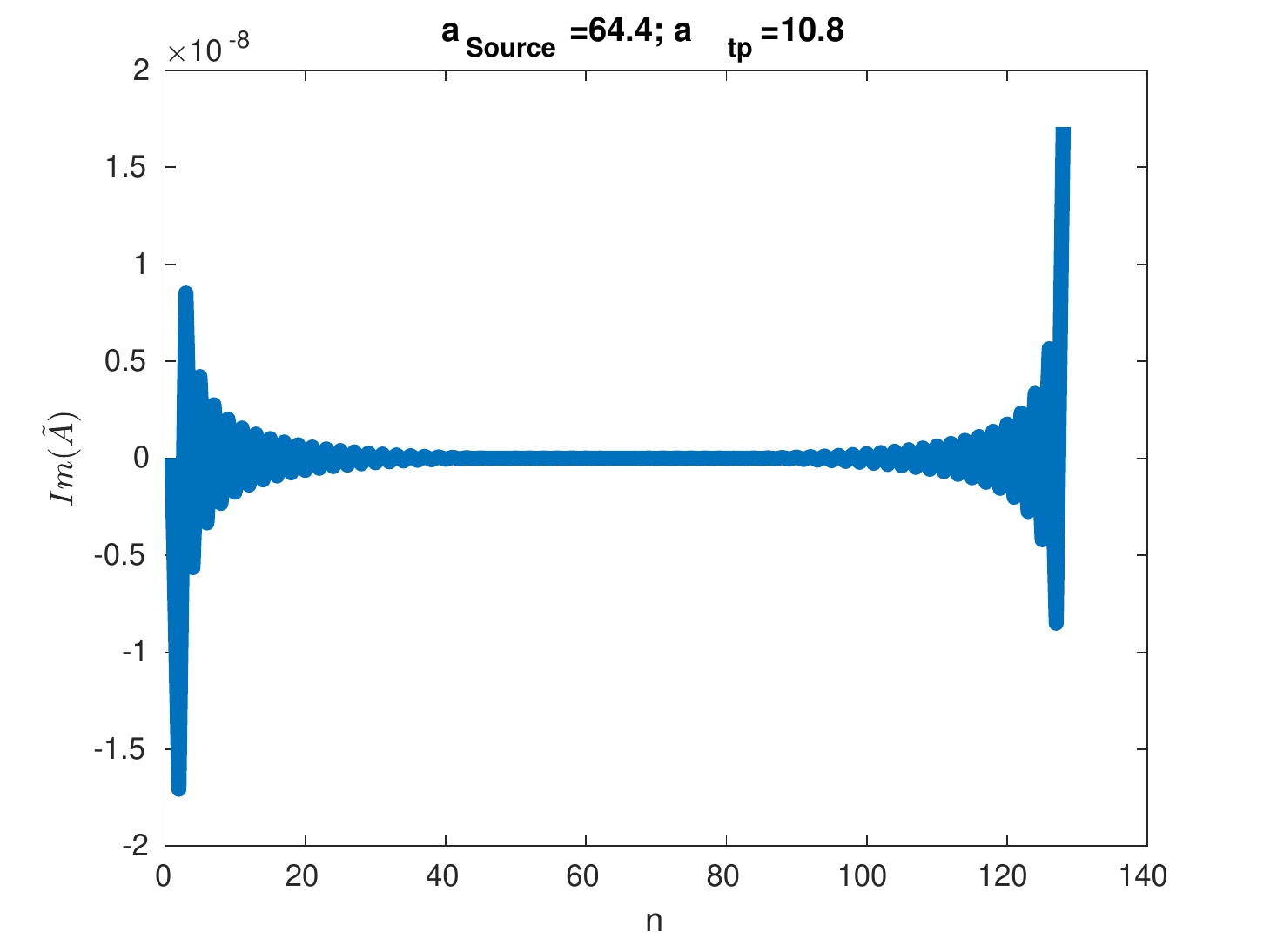}}
\caption{Imaginary parts of all different complex parameters (The source particle is at $x=64.4$ and test particle is at $x_{test}=10.8$)}
\label{fig_cic_2}
\end{figure}

For this system we plot the real and imaginary solutions of different real and Fourier space parameters in the figure \ref{fig_cic_1} (real parts of the solutions) and figure \ref{fig_cic_2} (imaginary parts of the solutions).

 \chapter{Simulation of 1D Self-Gravitating System}
\section{Aim}
To simulate a 1D distribution of particle under the influence of its own gravity. To draw the phase space snapshots over time. 

\section{Introduction}
After doing all the previous assignments we now know how to solve differential equation, we know that by Leapfrog method the solution is energy conserving. From assignment-4 we know how to solve Poisson's equation using Fourier method. From assignment-5 we know how to even solve for acceleration and potential when the test and source particles are not on grid (by CIC method). Now finally we are going to address our goal, to solve a self gravitating system. In this assignment we solve the 1D self gravtating system. 

\section{Brief Theory}
We normalize the time and space using appropriate scaling parameters. We normalize time using $t_g=\frac{a}{\sqrt{G \rho}}$. This $t_g$ is the gravitational time scale, which is the time taken by a uniformly distributed spherical body to collapse. The length $x$ is normalized using the grid spacing $L$ as a scaling length.

So, dimensionless $X=x/L$ and dimensionless $T=t/t_g$

The Hamilton's equation in dimensionless form is

\begin{equation}
    \frac{d X}{d T}=\frac{t_g P}{L m} \:\:\: \text{and} \:\:\: \frac{d P}{d T}=\frac{t^2_g f}{L m}=F \:\:\: \text{(dimensionless force)}
 \end{equation}
 
The form of dimensionless force $F$ can be calculated as

\begin{equation}
    F=4 \pi \left[- \nabla_x . (\nabla ^{-2}_x)\frac{\rho}{\Bar{\rho}}\right]
\end{equation}

$\Bar{\rho}=1$ for our simulation, but $\rho$ will evolve as the system evolve. 

\section{Simulation Parameters}
Simulation parameters are:

$L      = 100$;  total space in 1D

$dx     = 1$;  length of each segment

$N      = S/dx$;  number of grid points

$T      = 100$;  total run time

$dt     = 0.1$; time step

All the variables are normalized in this simulation.

\section{Explanation of plots}
\begin{figure}[ht]
\centering
\subfloat[$t=10$]{\includegraphics[width = 1.45in]{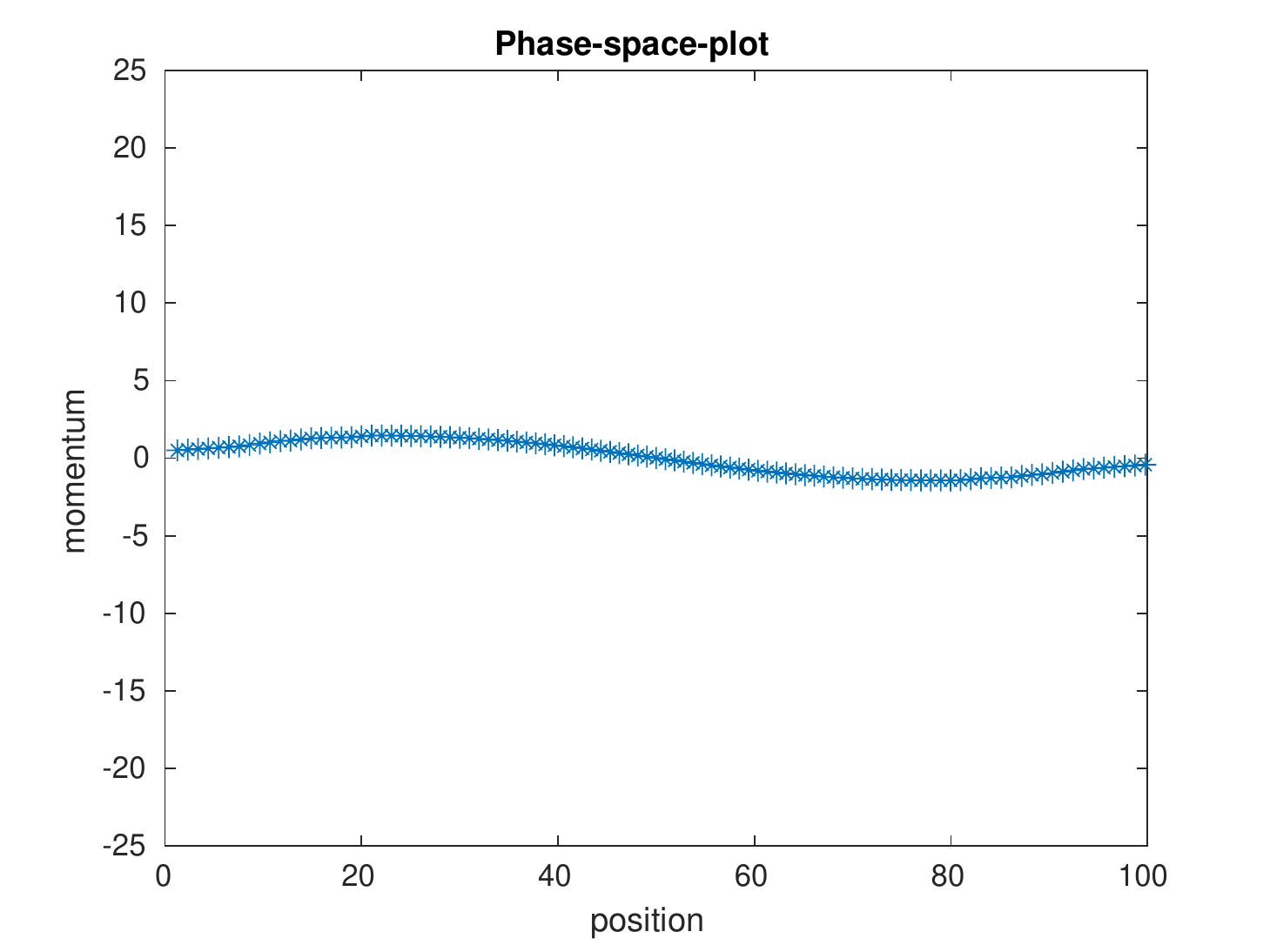}}
\subfloat[$t=20$]{\includegraphics[width = 1.45in]{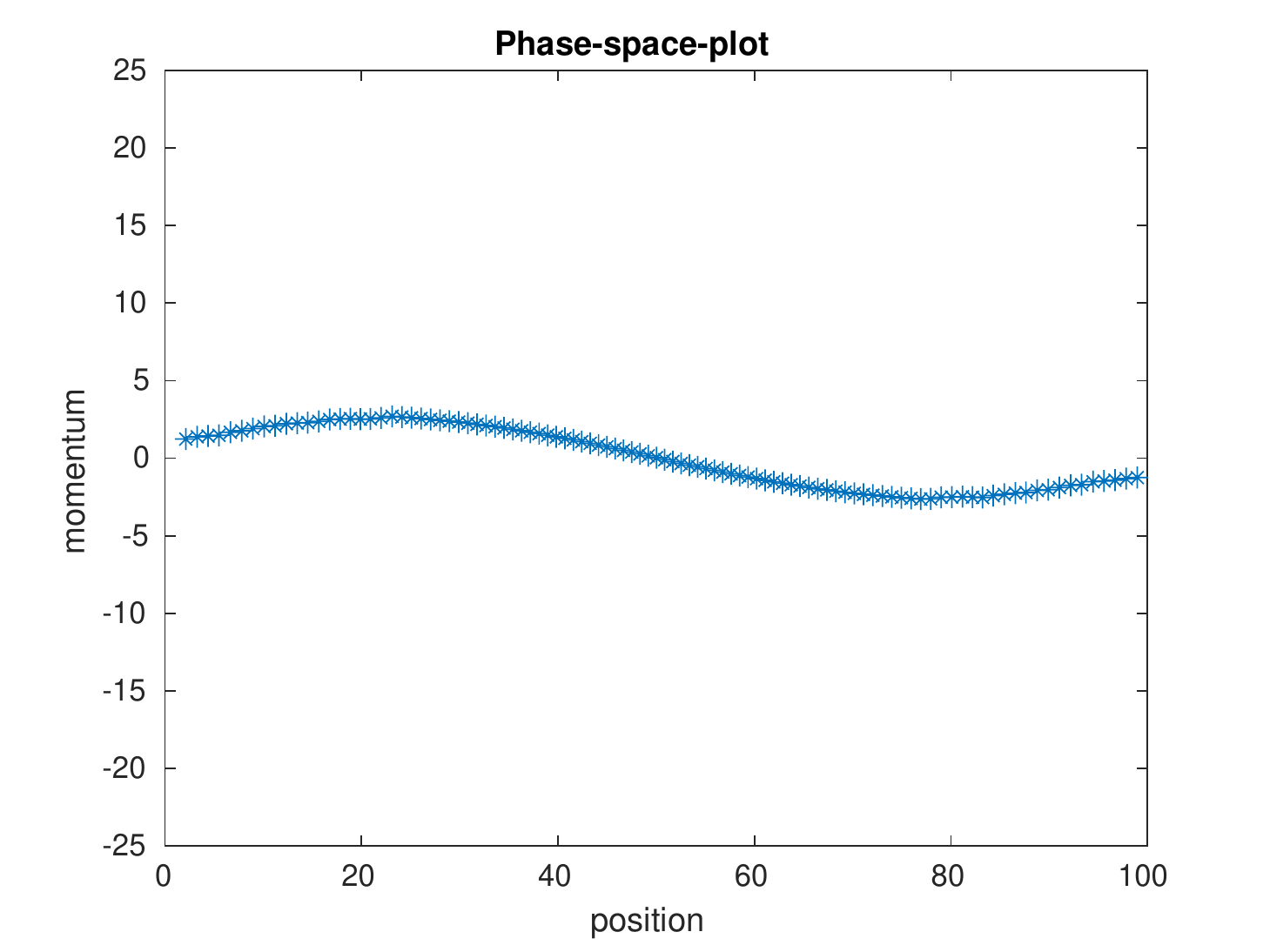}}
\subfloat[$t=30$]{\includegraphics[width = 1.45in]{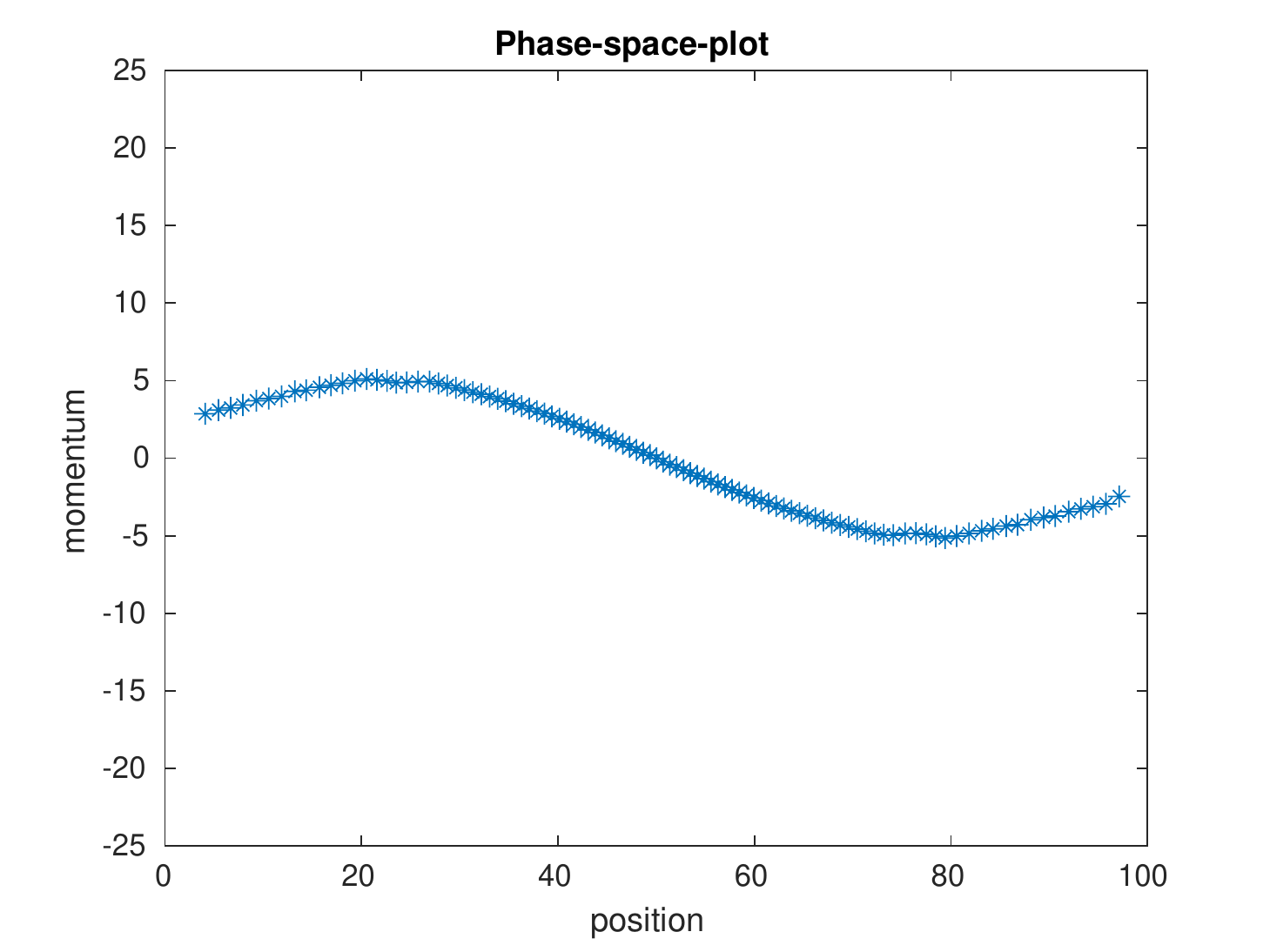}}
\subfloat[$t=40$]{\includegraphics[width = 1.45in]{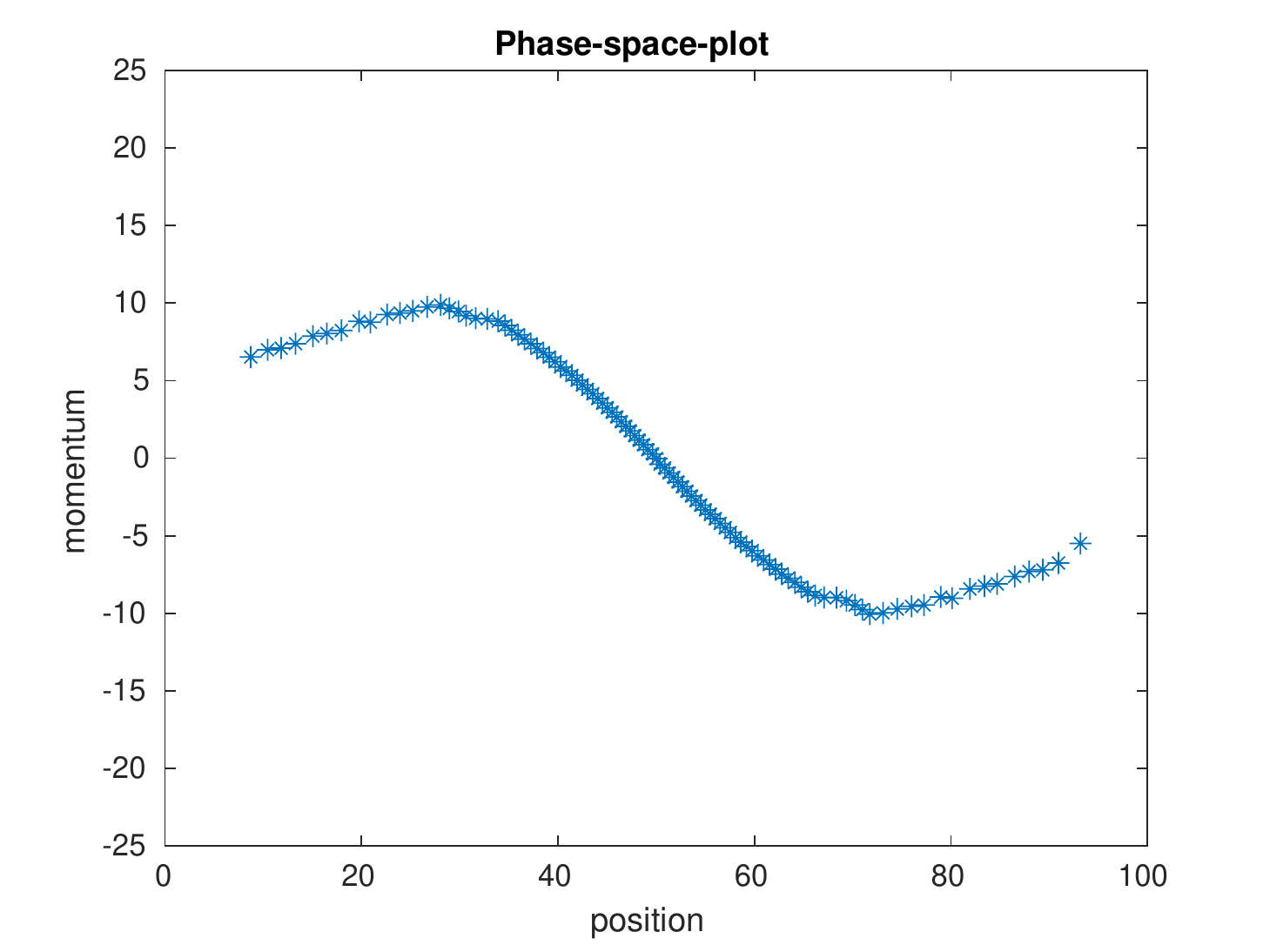}}\\
\subfloat[$t=50$]{\includegraphics[width = 1.45in]{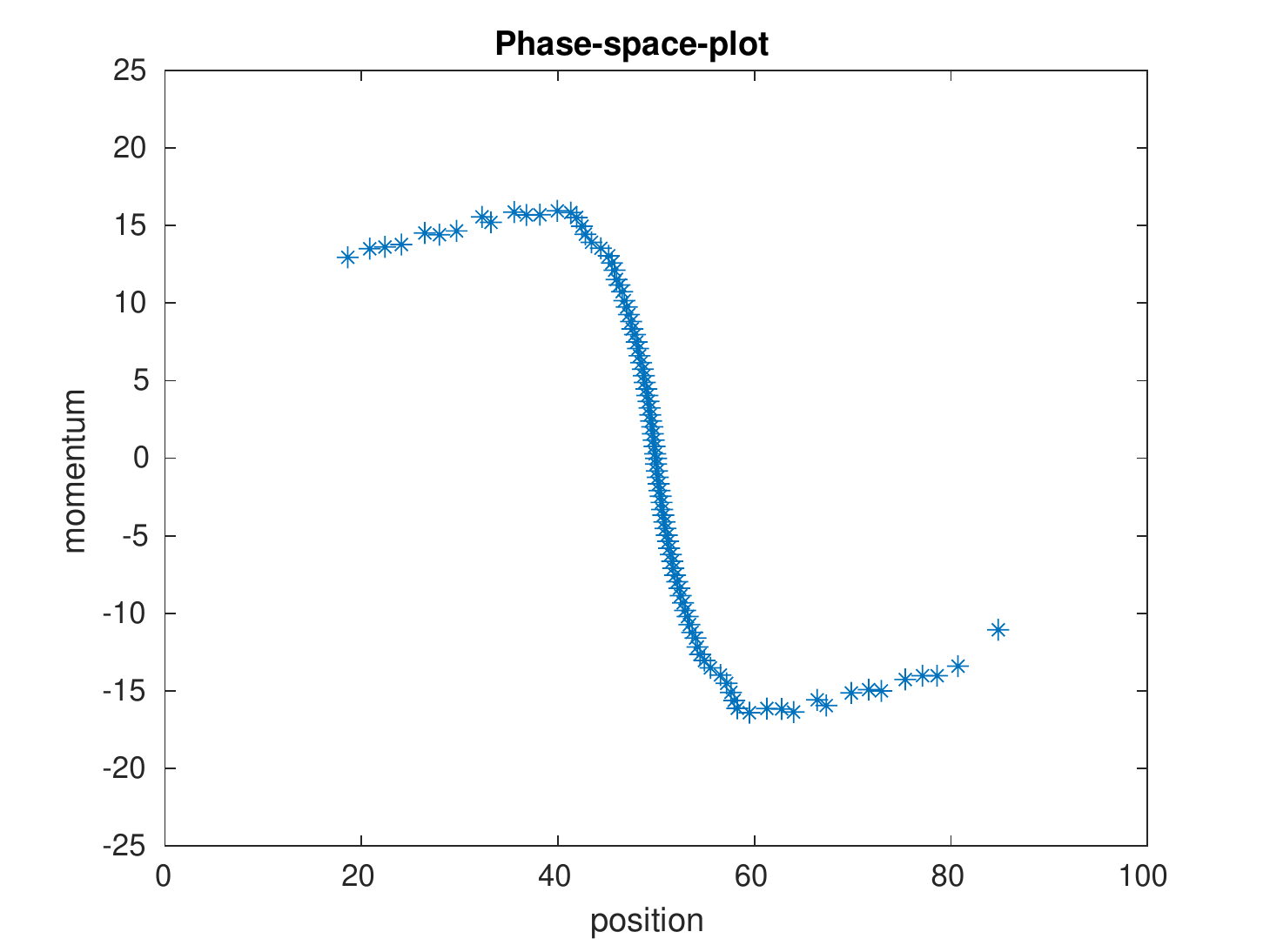}}
\subfloat[$t=60$]{\includegraphics[width = 1.45in]{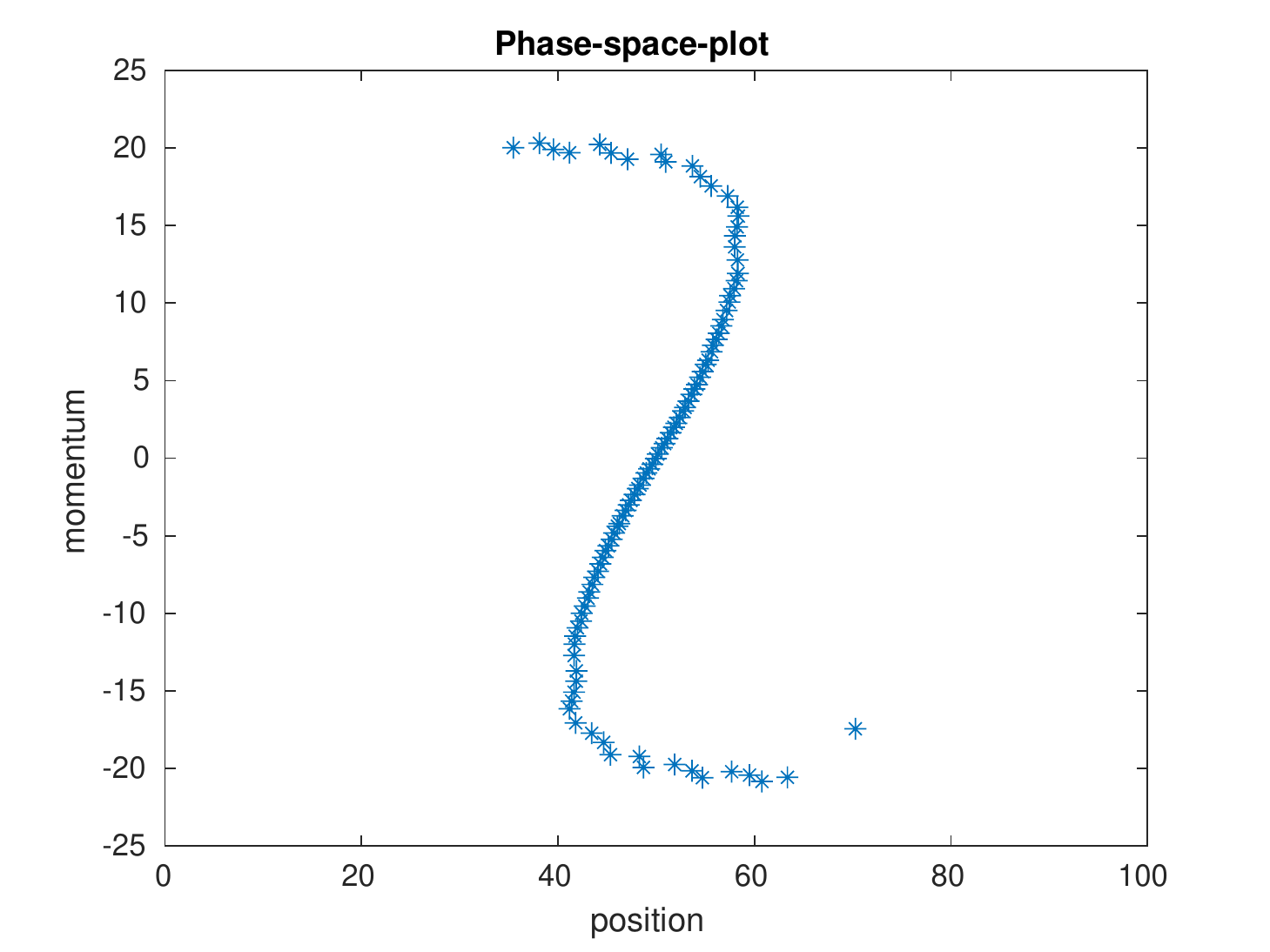}}
\subfloat[$t=70$]{\includegraphics[width = 1.45in]{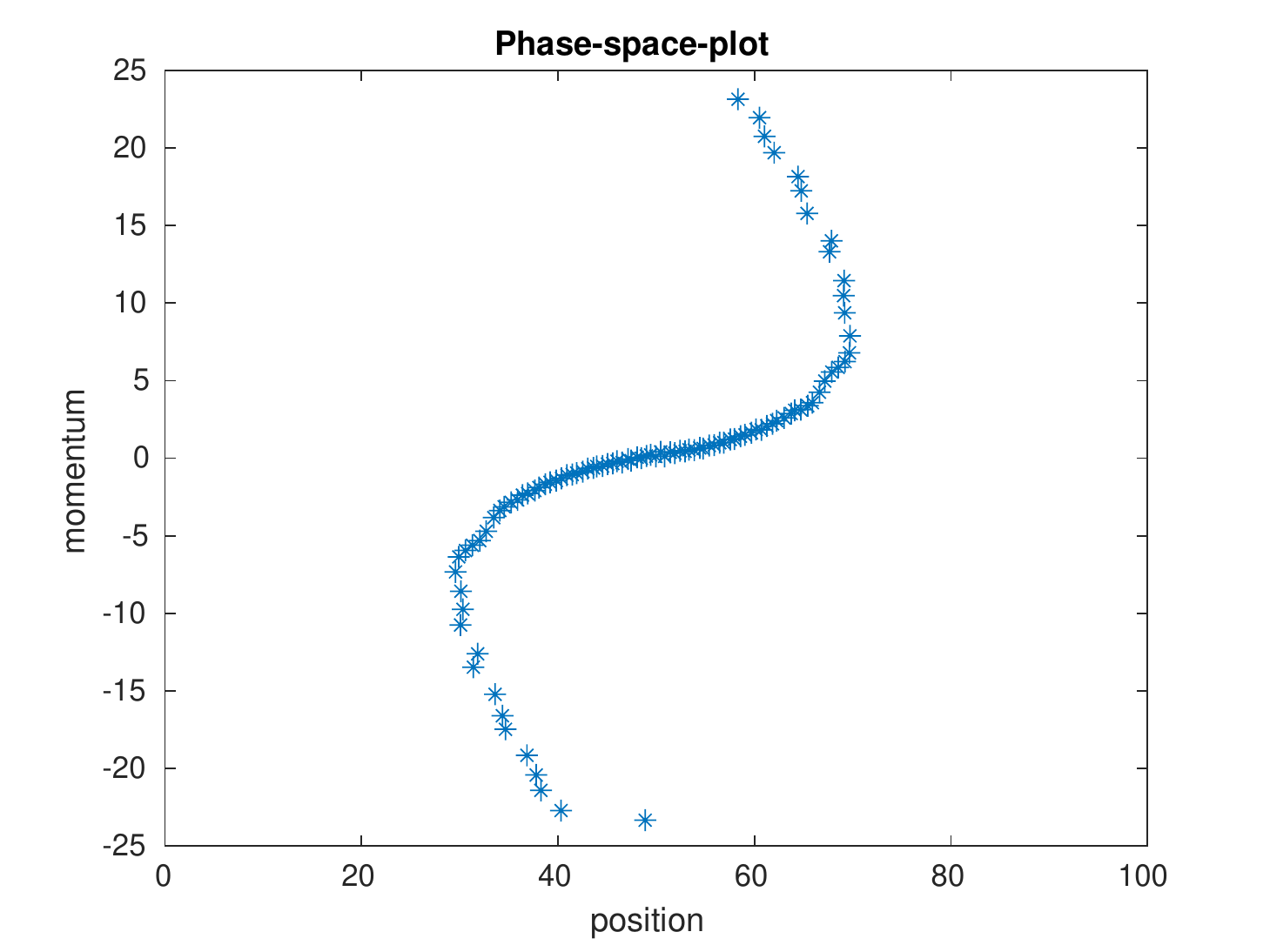}}
\subfloat[$t=80$]{\includegraphics[width = 1.45in]{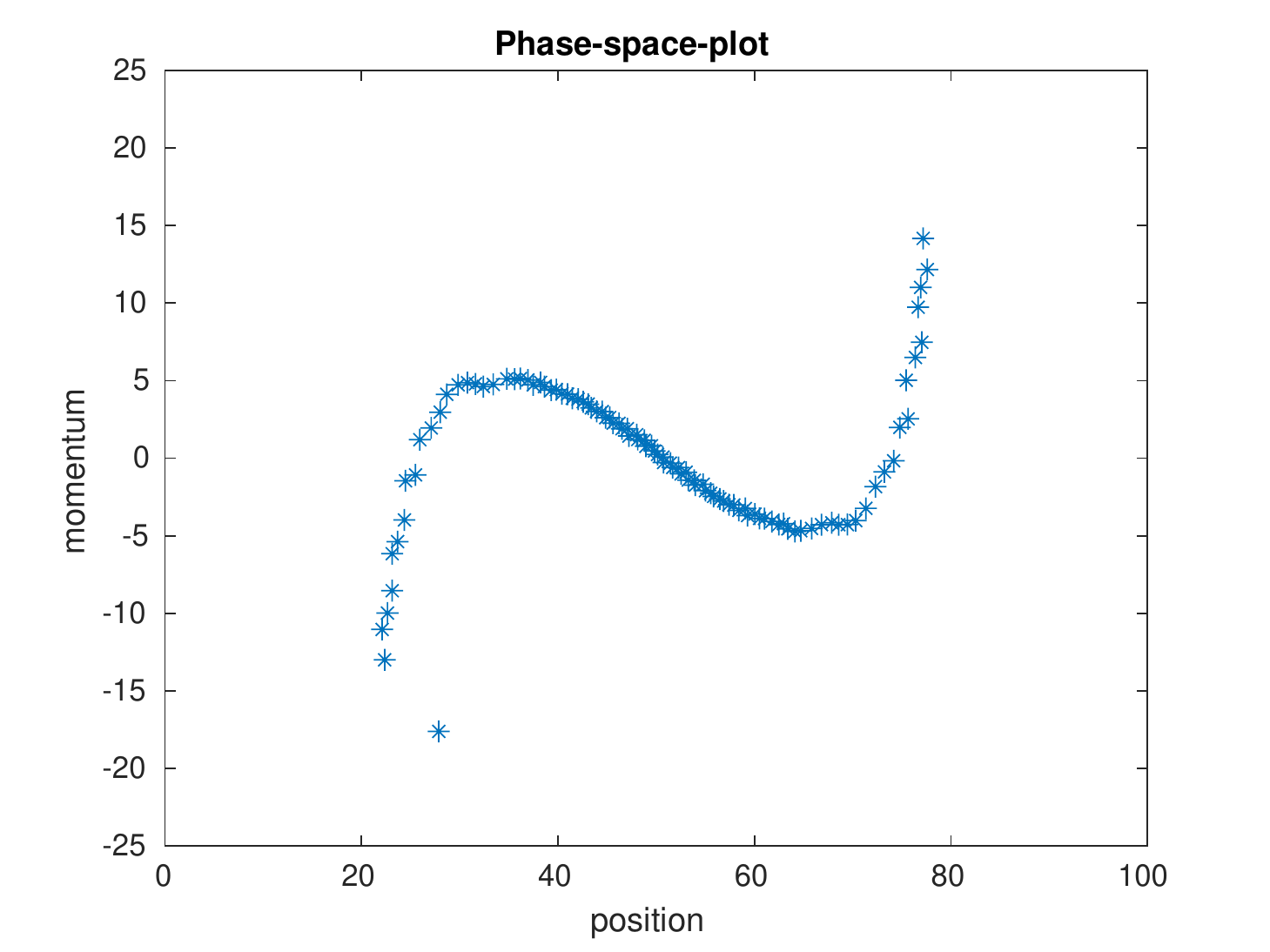}}\\
\subfloat[$t=90$]{\includegraphics[width = 1.45in]{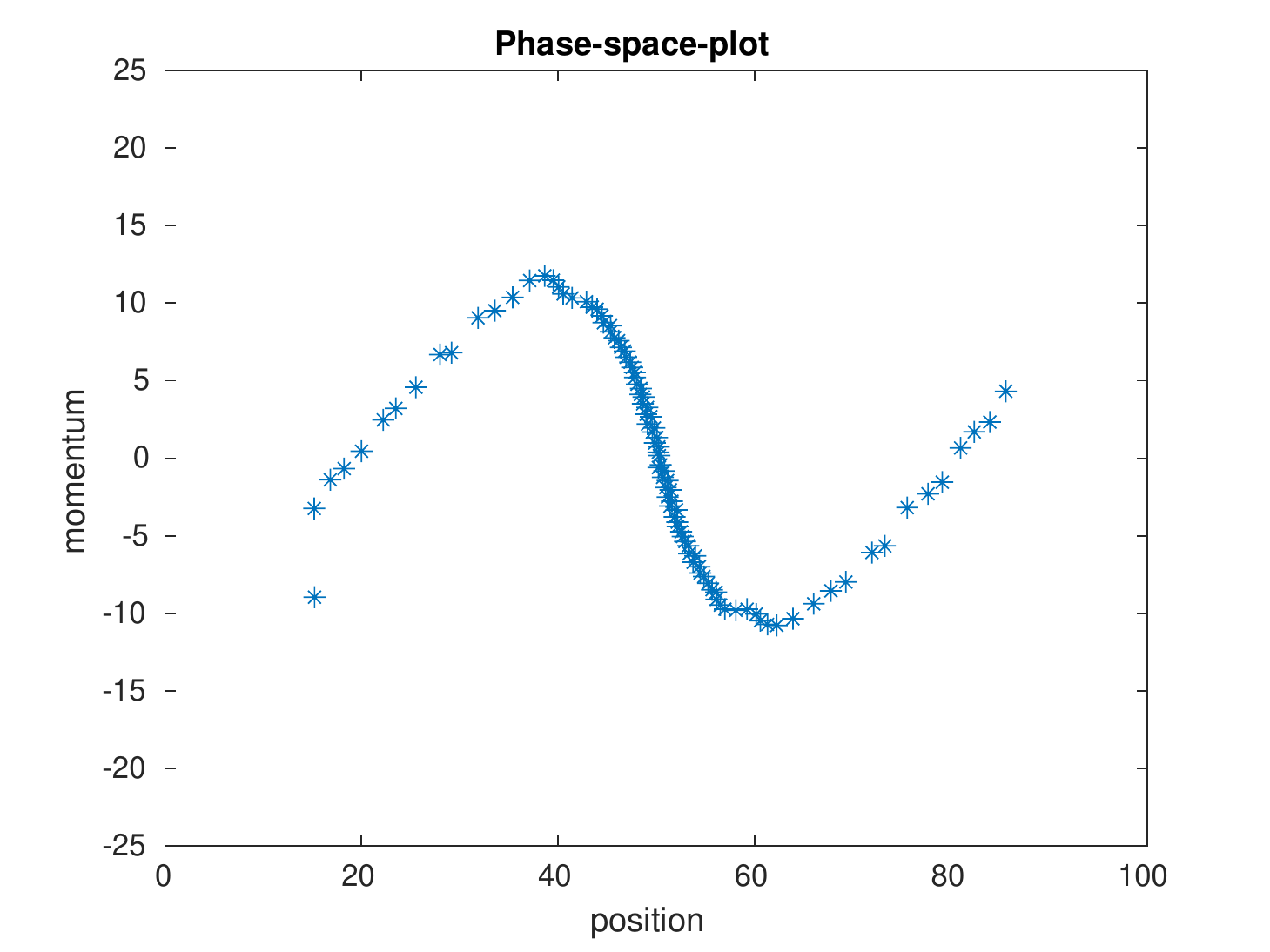}}
\subfloat[$t=100$]{\includegraphics[width = 1.45in]{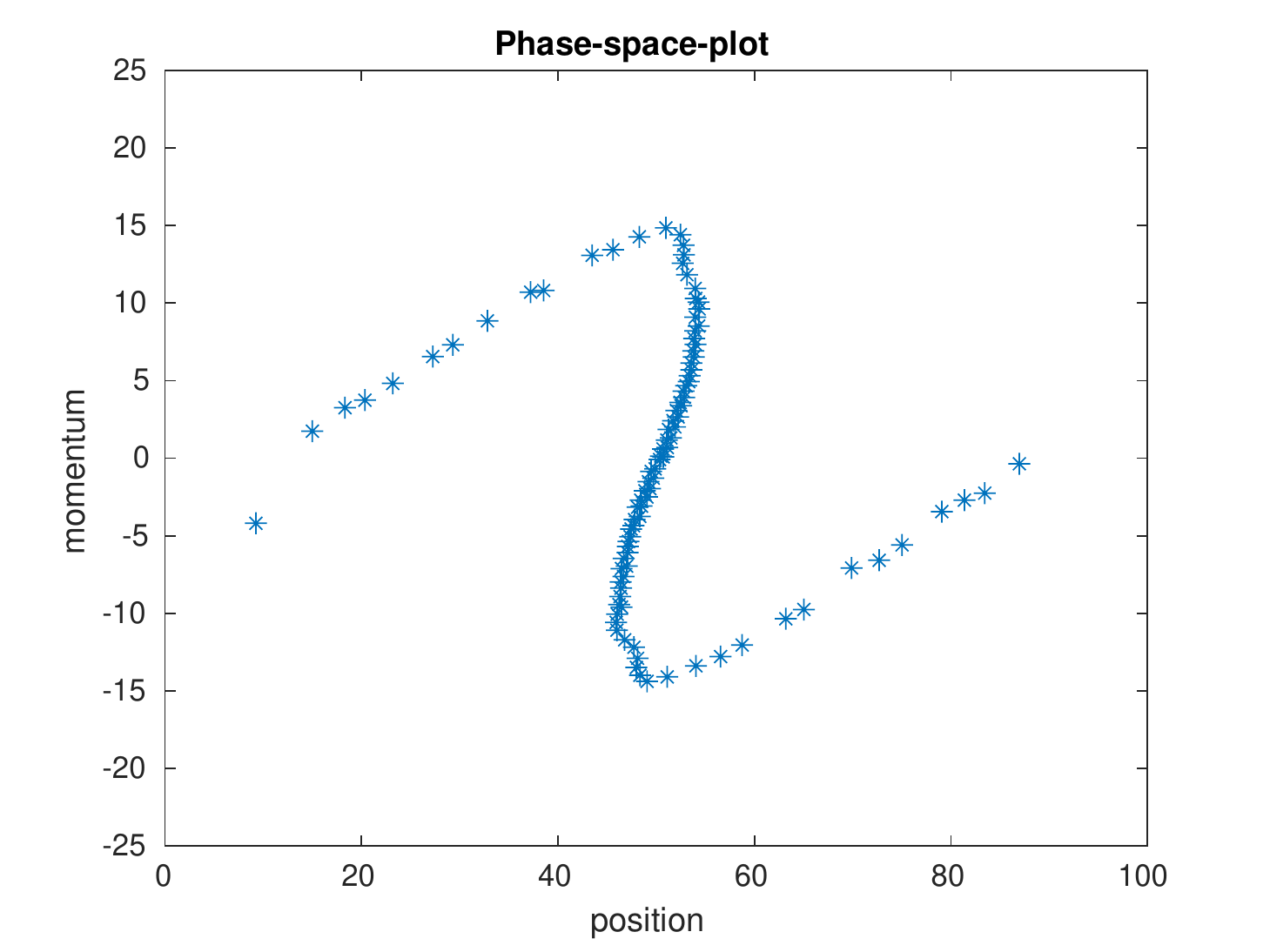}}
\subfloat[$t=110$]{\includegraphics[width = 1.45in]{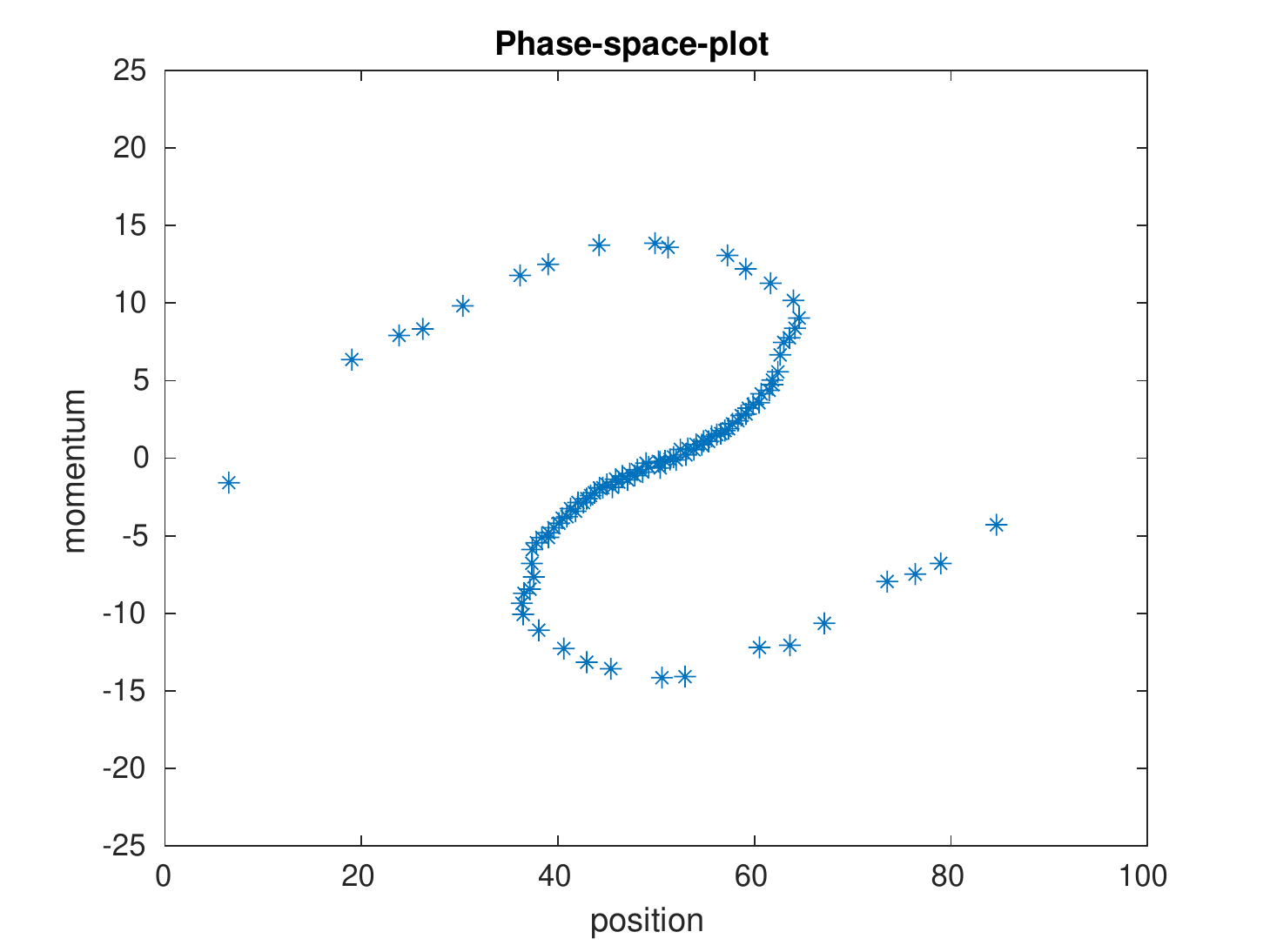}}
\subfloat[$t=120$]{\includegraphics[width = 1.45in]{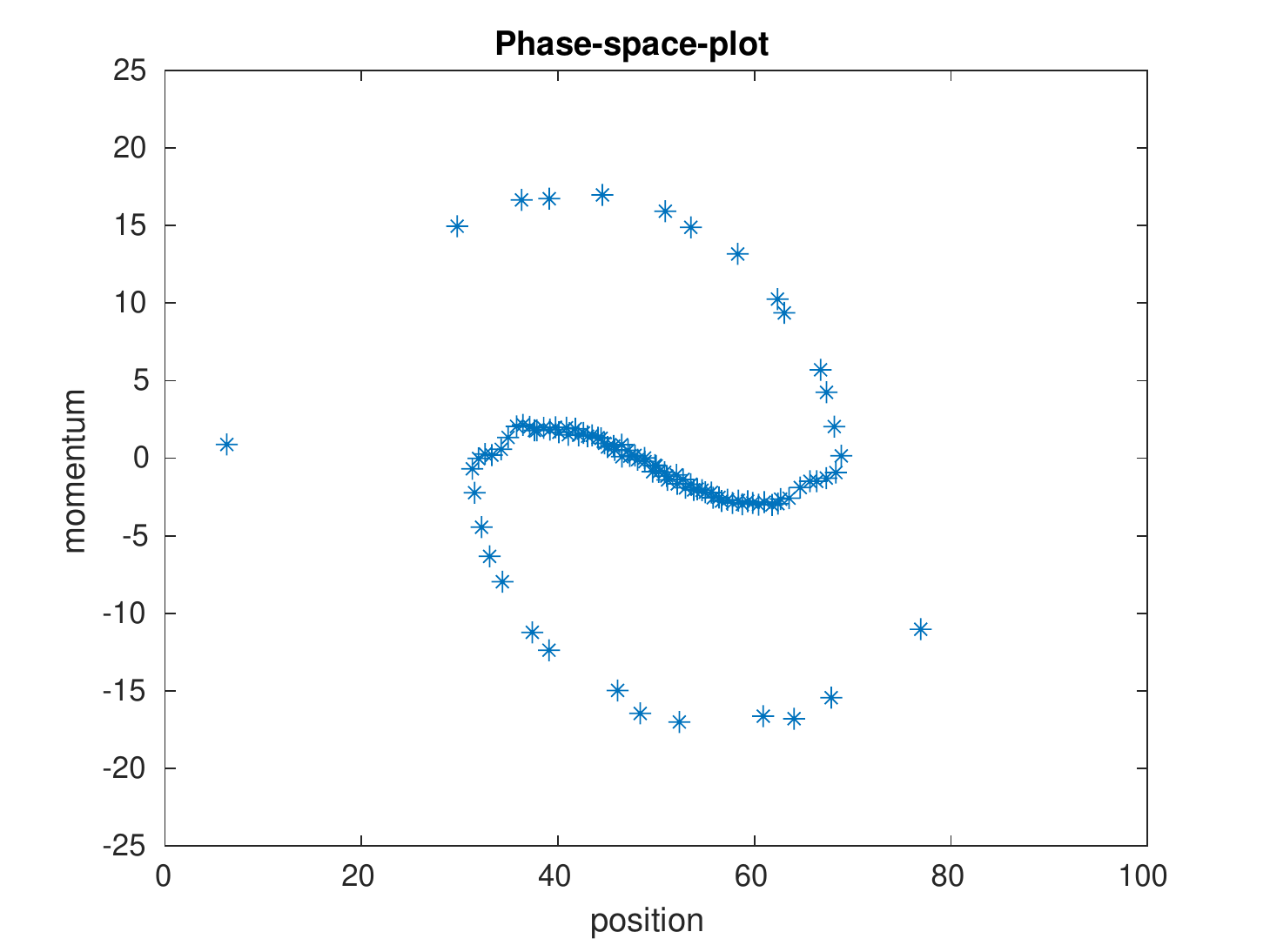}}\\
\subfloat[$t=130$]{\includegraphics[width = 1.45in]{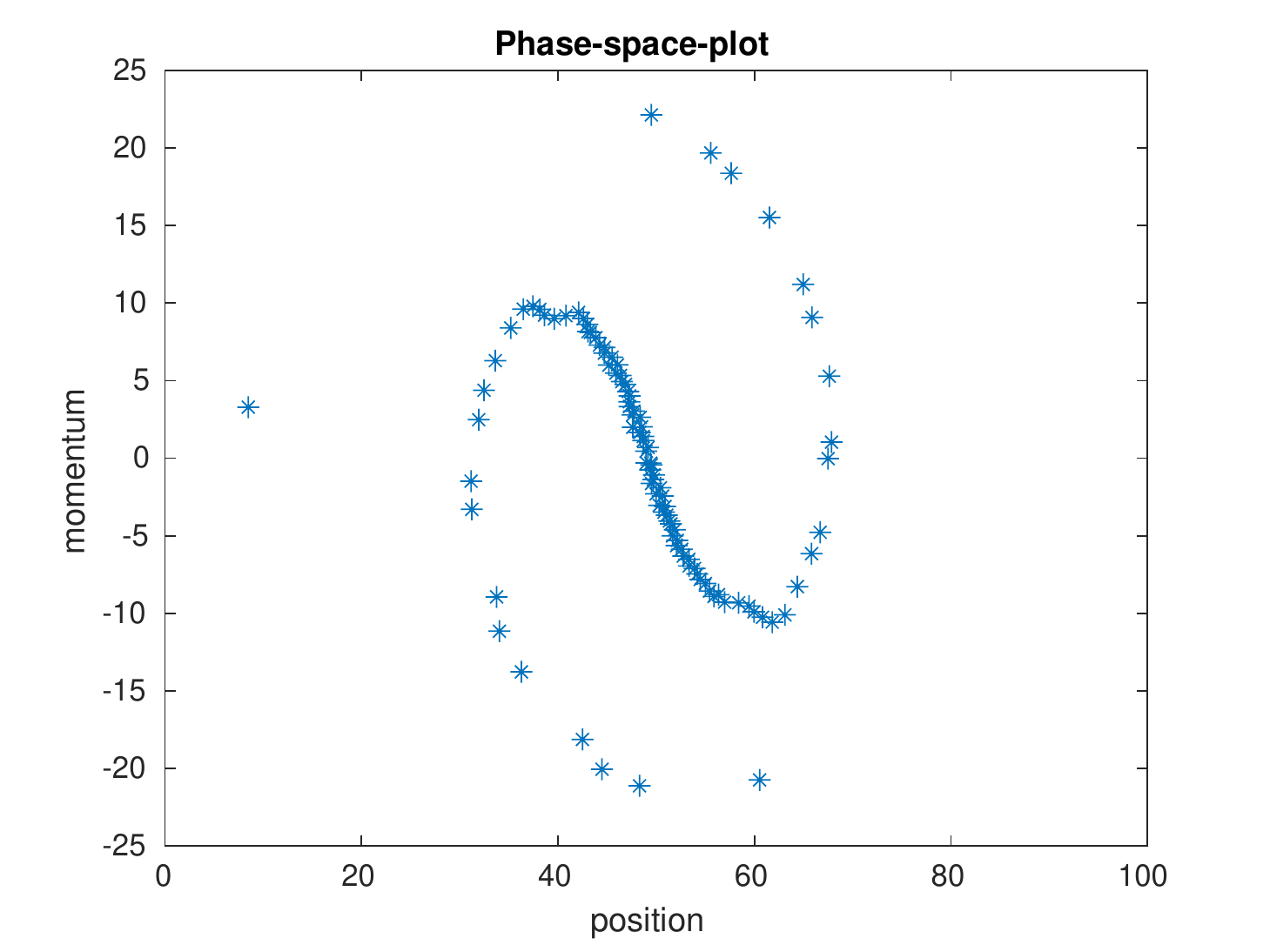}}
\subfloat[$t=140$]{\includegraphics[width = 1.45in]{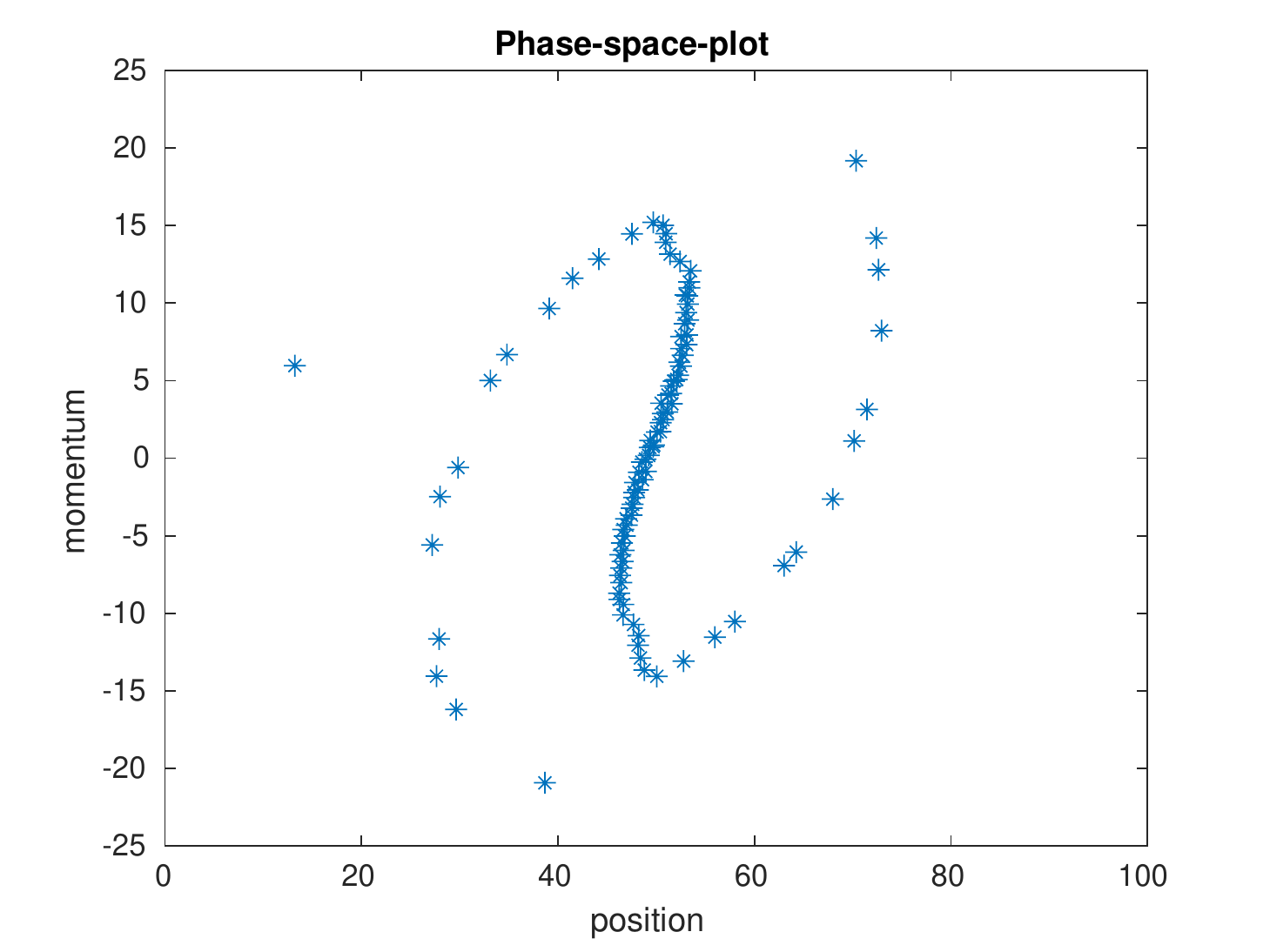}}
\subfloat[$t=150$]{\includegraphics[width = 1.45in]{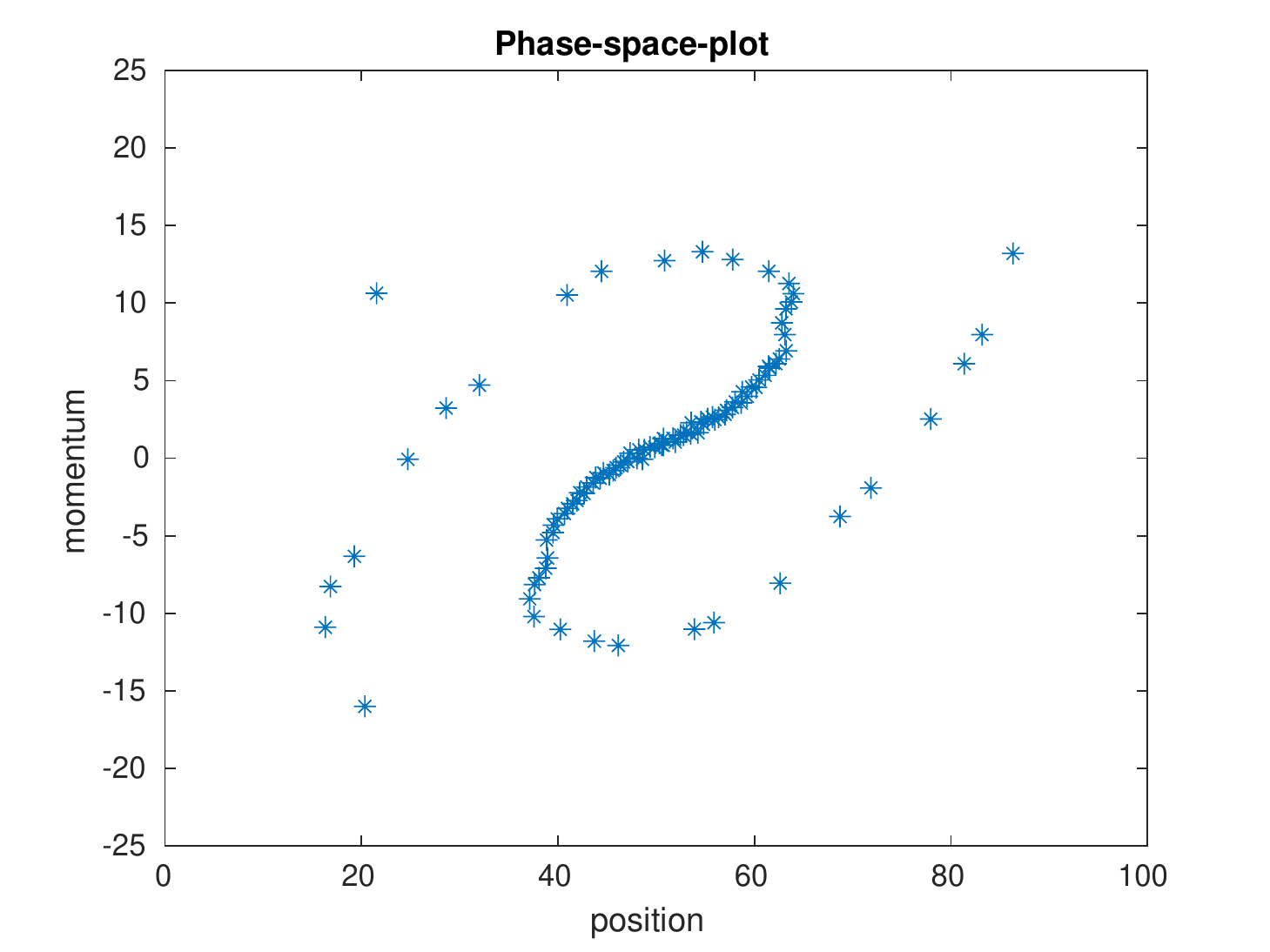}}
\subfloat[$t=160$]{\includegraphics[width = 1.45in]{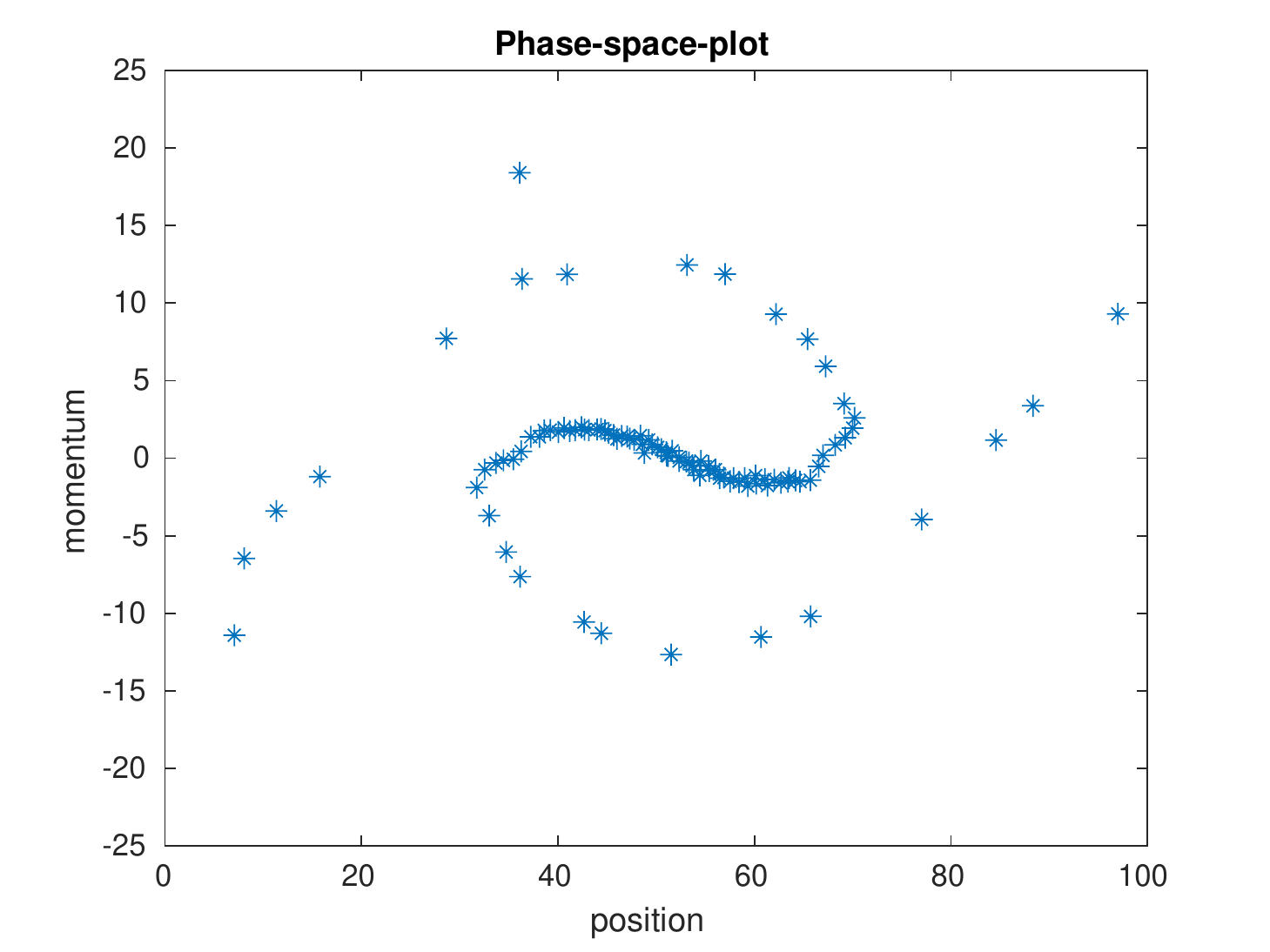}}\\
\subfloat[$t=170$]{\includegraphics[width = 1.45in]{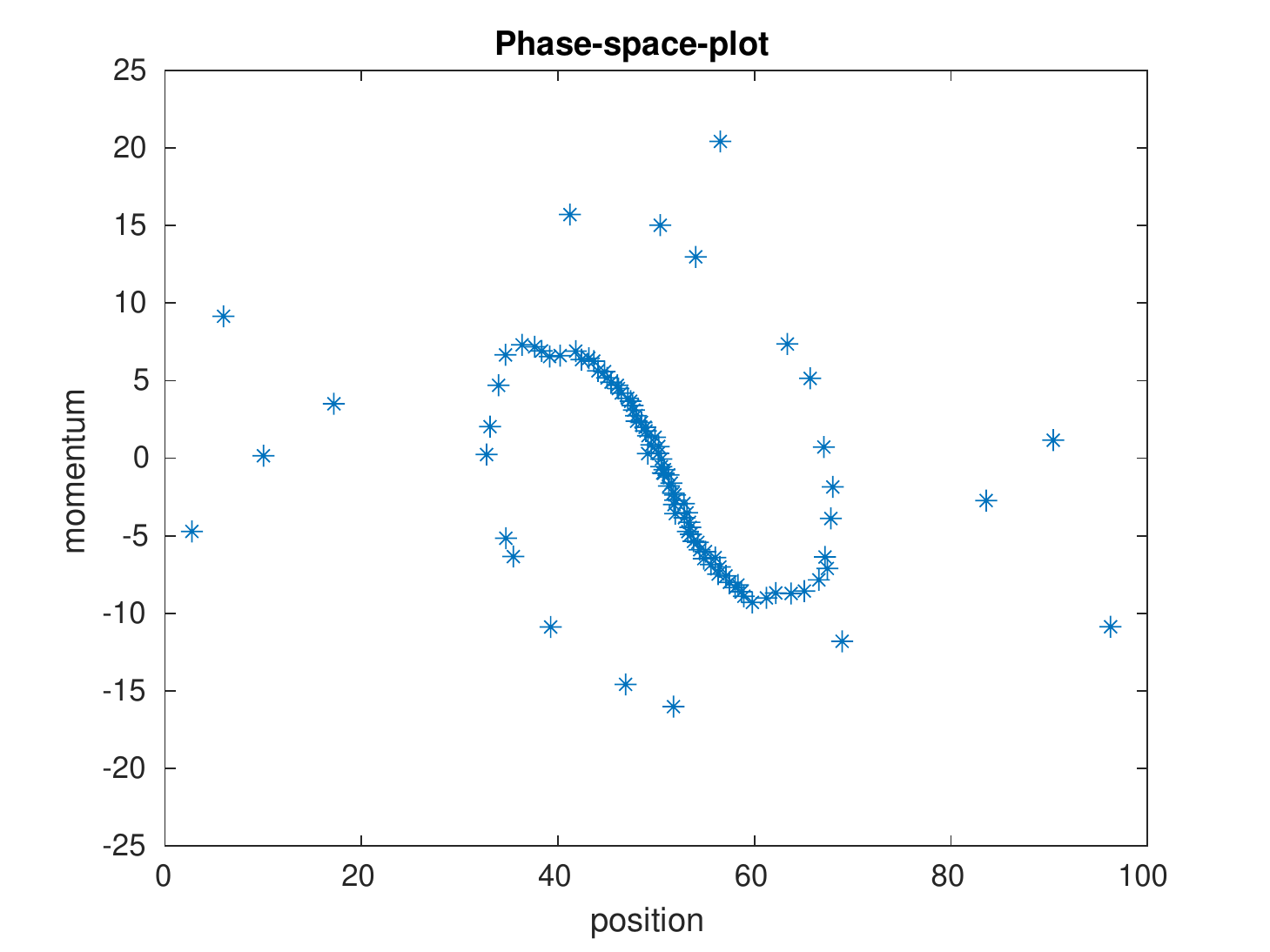}}
\subfloat[$t=180$]{\includegraphics[width = 1.45in]{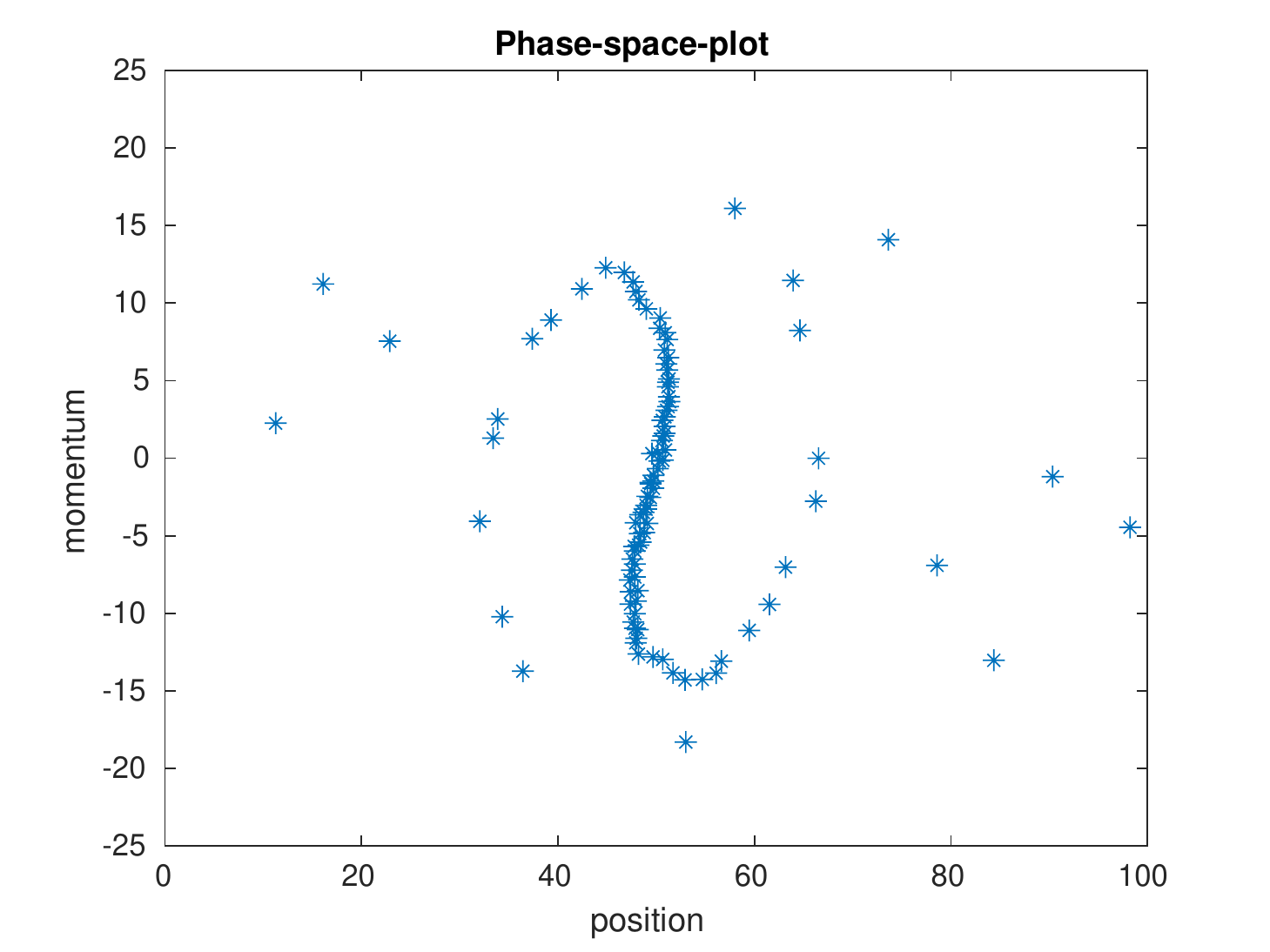}}
\subfloat[$t=190$]{\includegraphics[width = 1.45in]{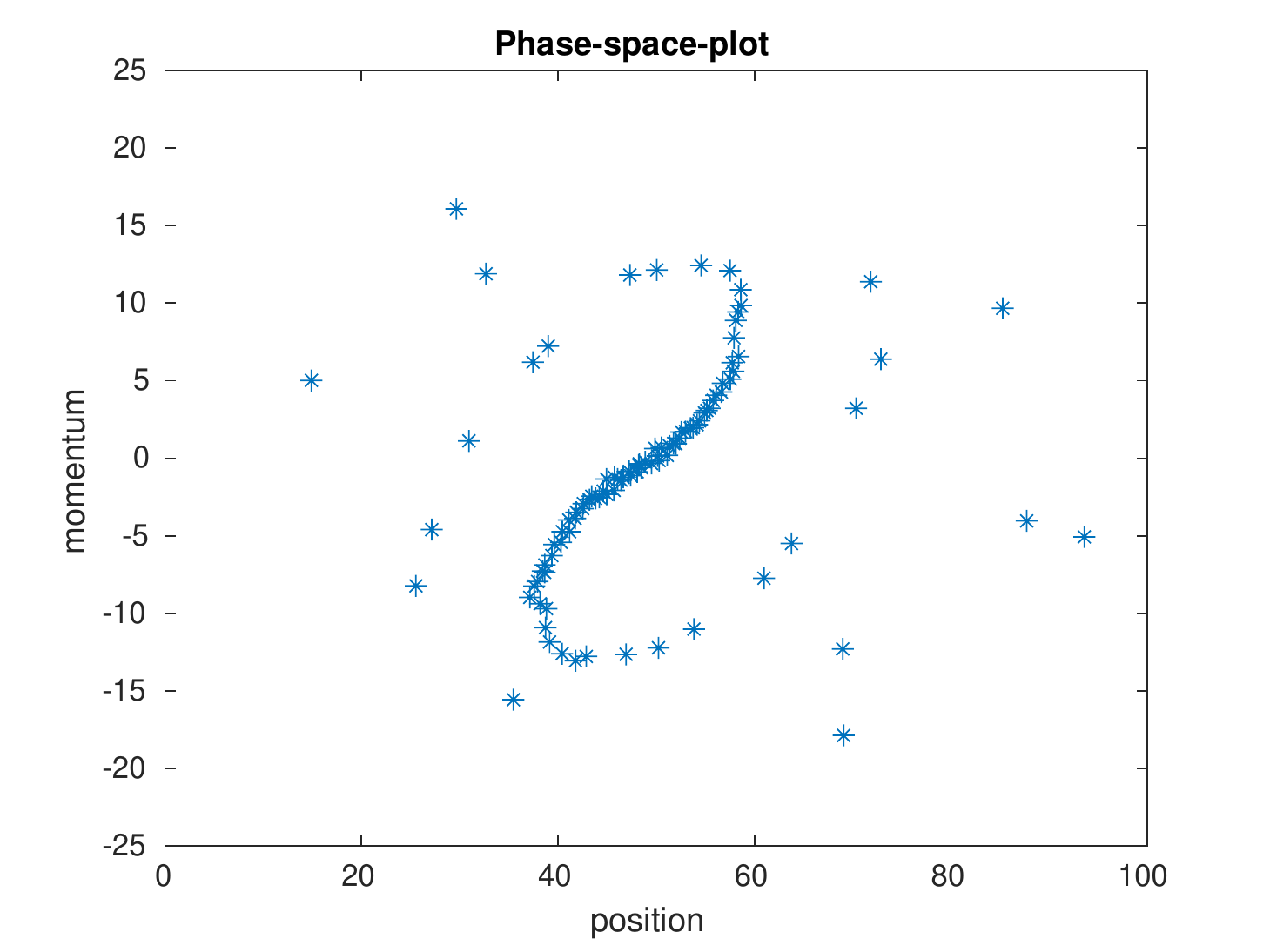}}
\subfloat[$t=200$]{\includegraphics[width = 1.45in]{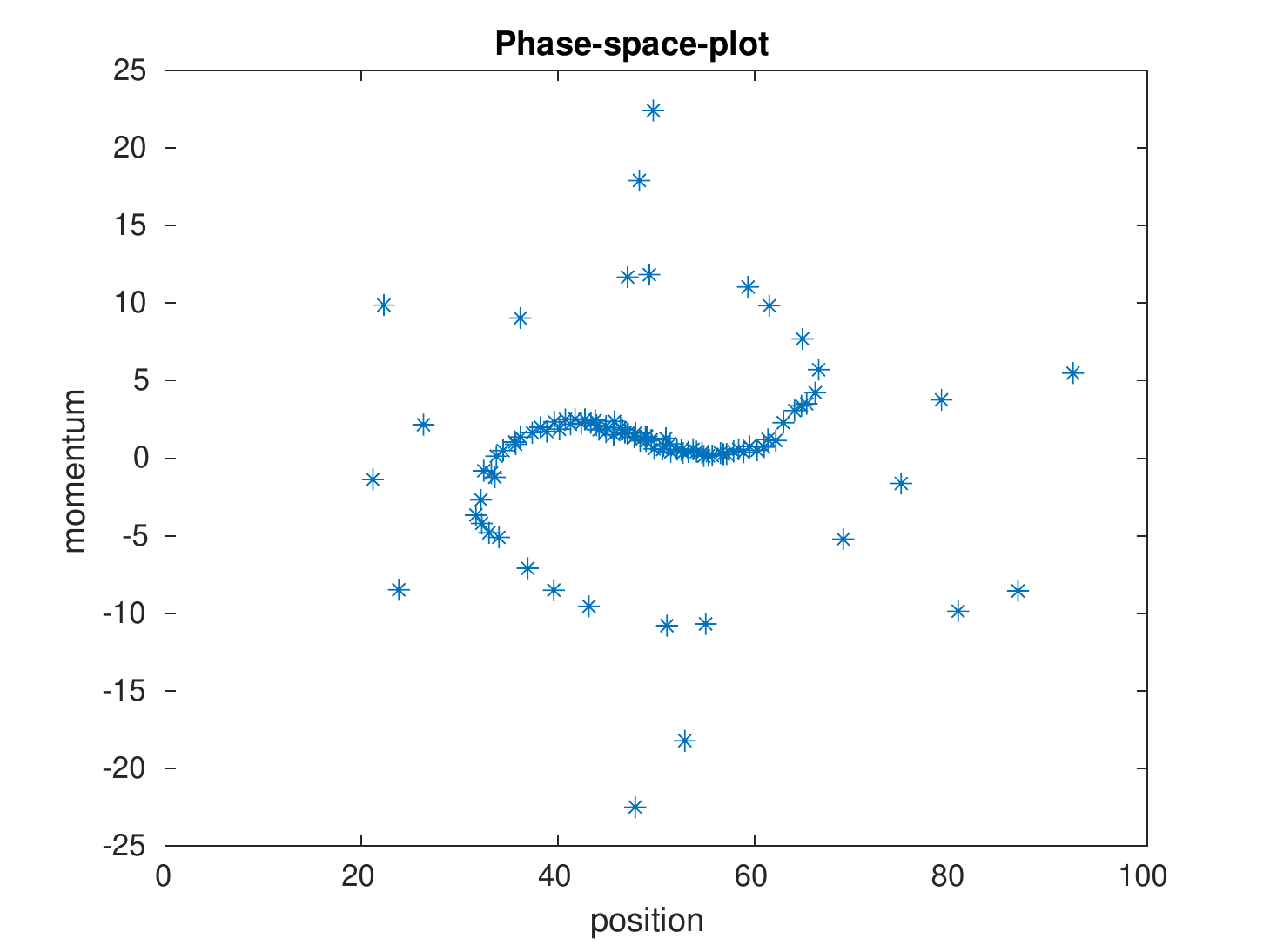}}
\caption{Phase Space snapshots for $t=10$ to $t=200$ with time separation of $10$}
\label{fig_sg_1}
\end{figure}

\begin{figure}[ht]
\centering
\subfloat[$t=610$]{\includegraphics[width = 1.45in]{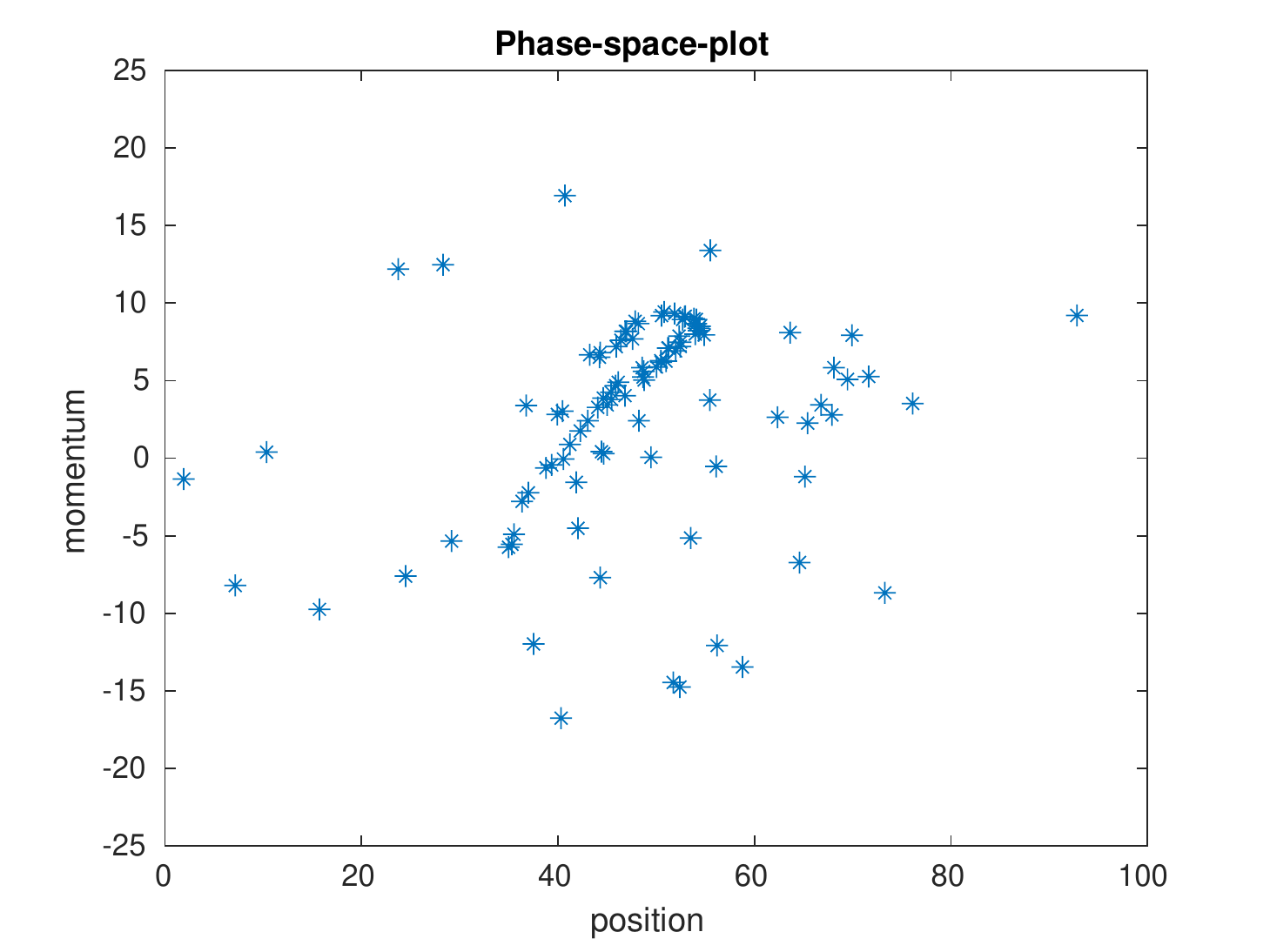}}
\subfloat[$t=620$]{\includegraphics[width = 1.45in]{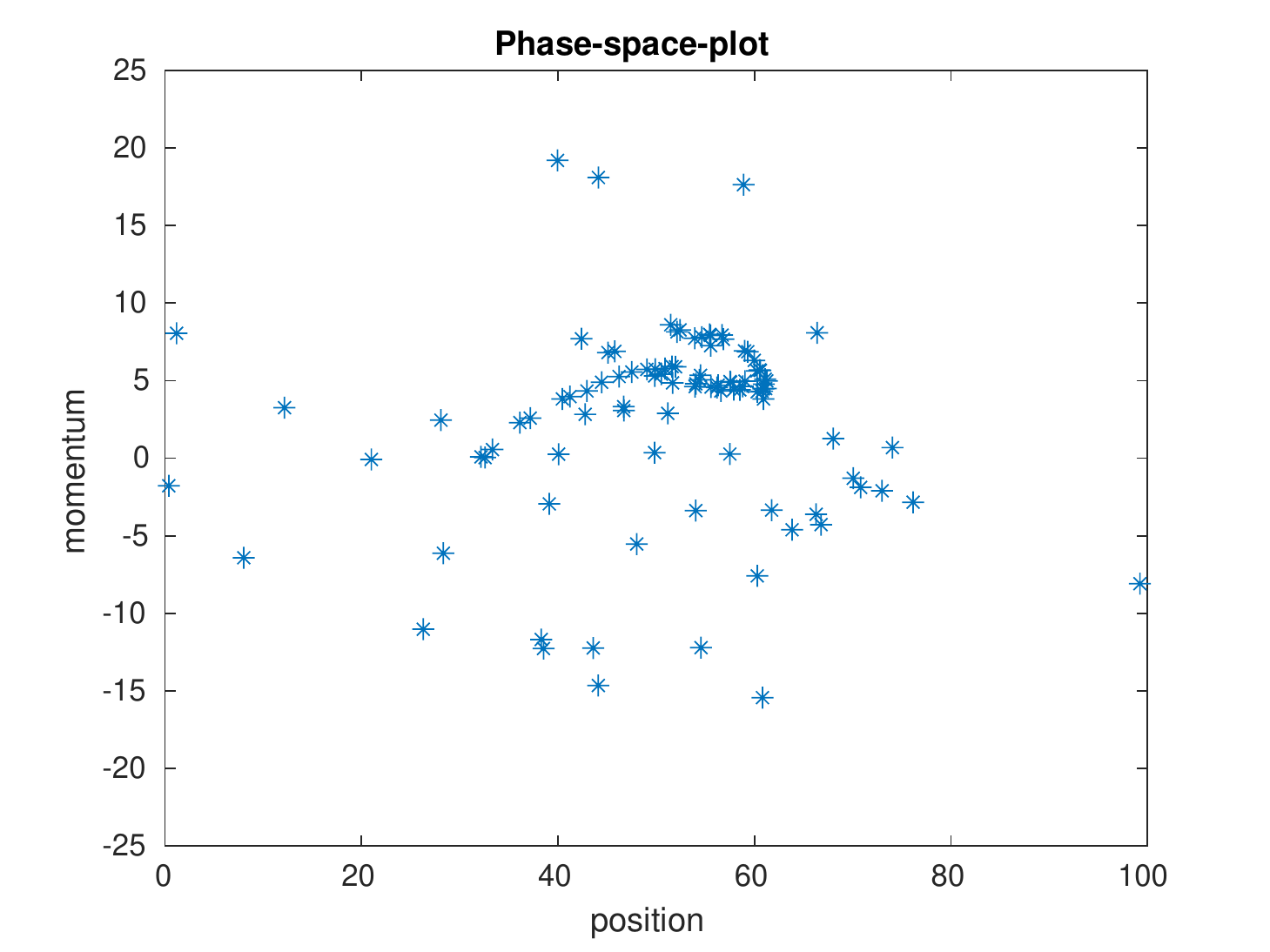}}
\subfloat[$t=630$]{\includegraphics[width = 1.45in]{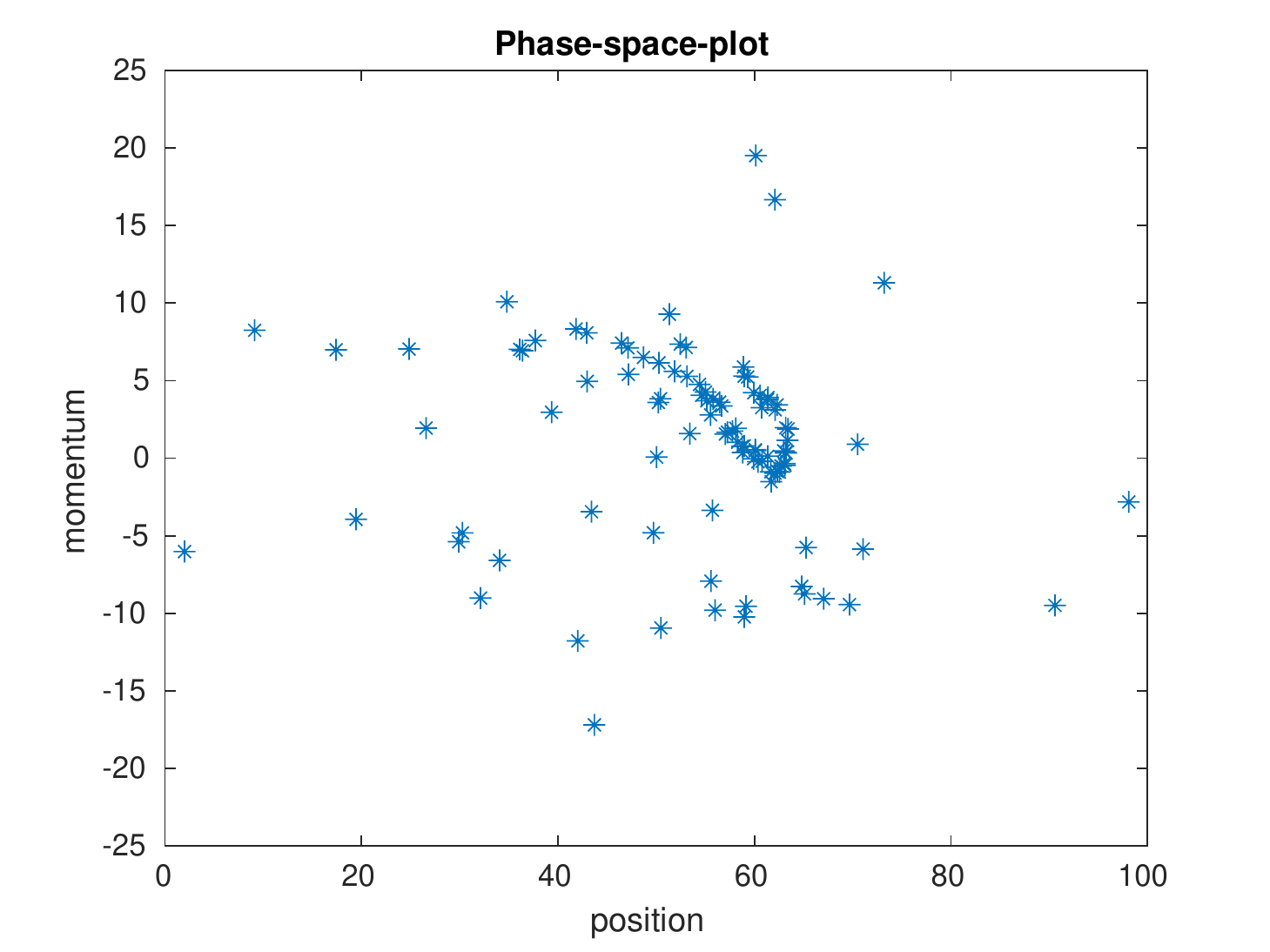}}
\subfloat[$t=640$]{\includegraphics[width = 1.45in]{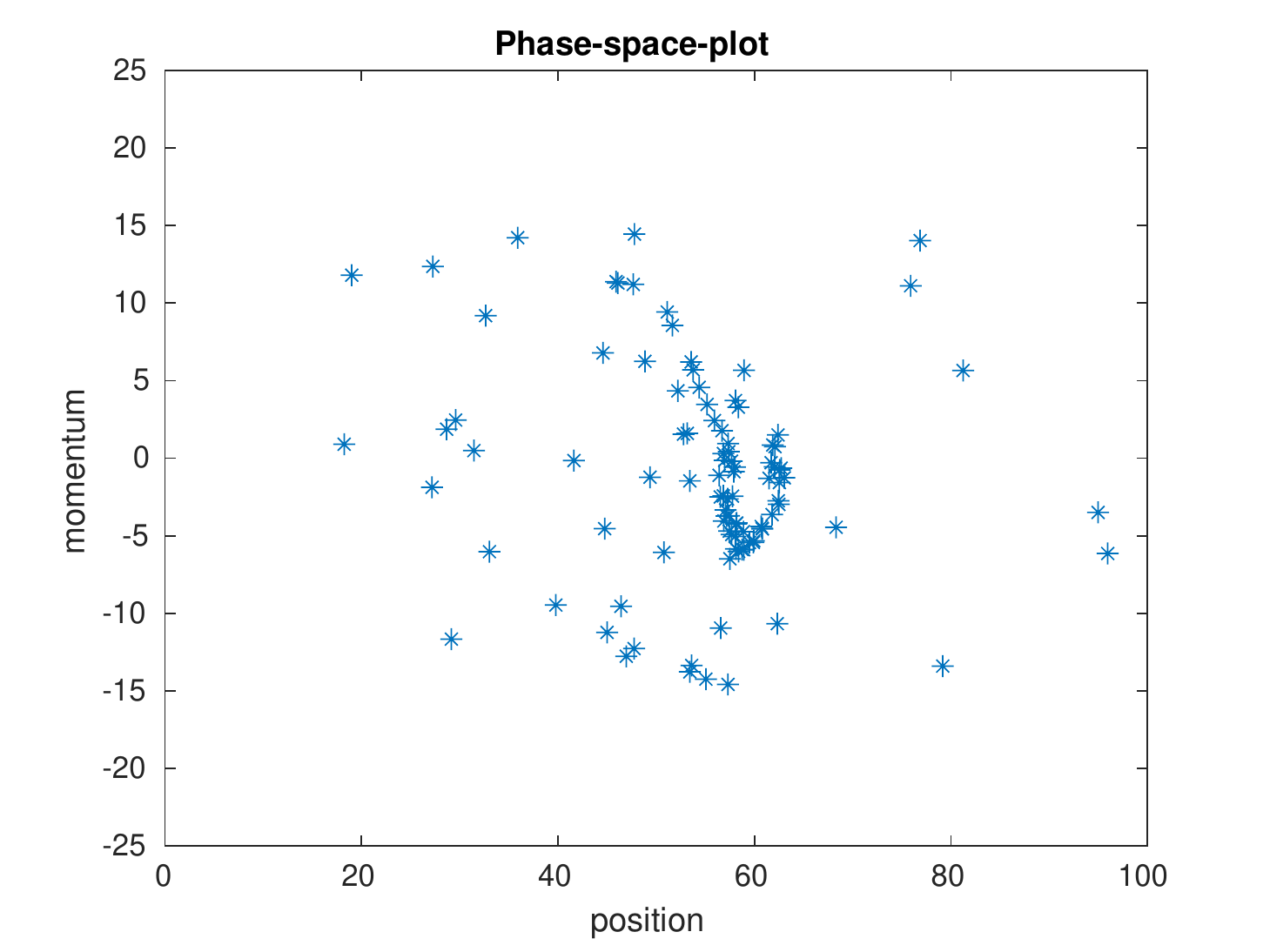}}\\
\subfloat[$t=650$]{\includegraphics[width = 1.45in]{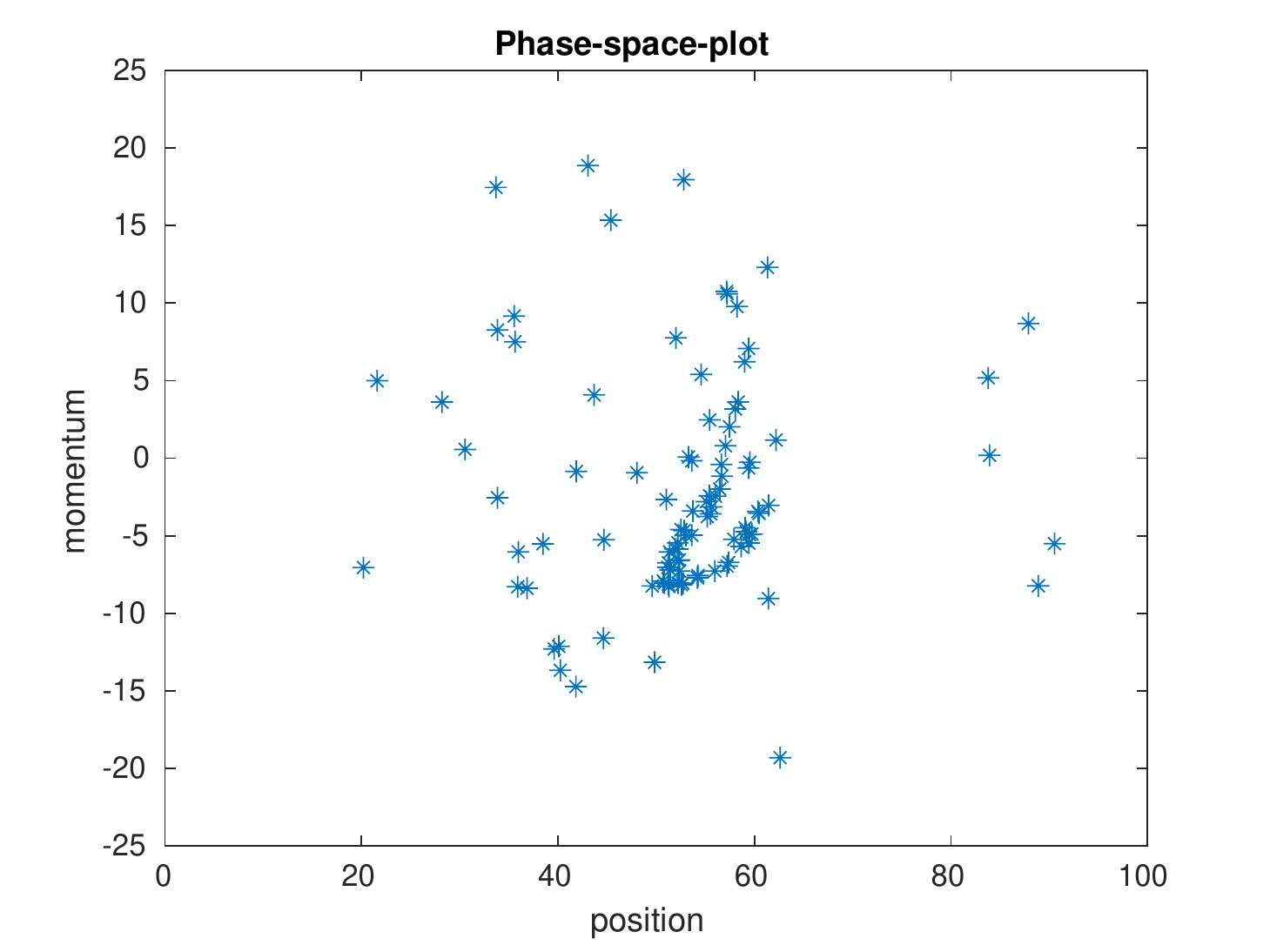}}
\subfloat[$t=660$]{\includegraphics[width = 1.45in]{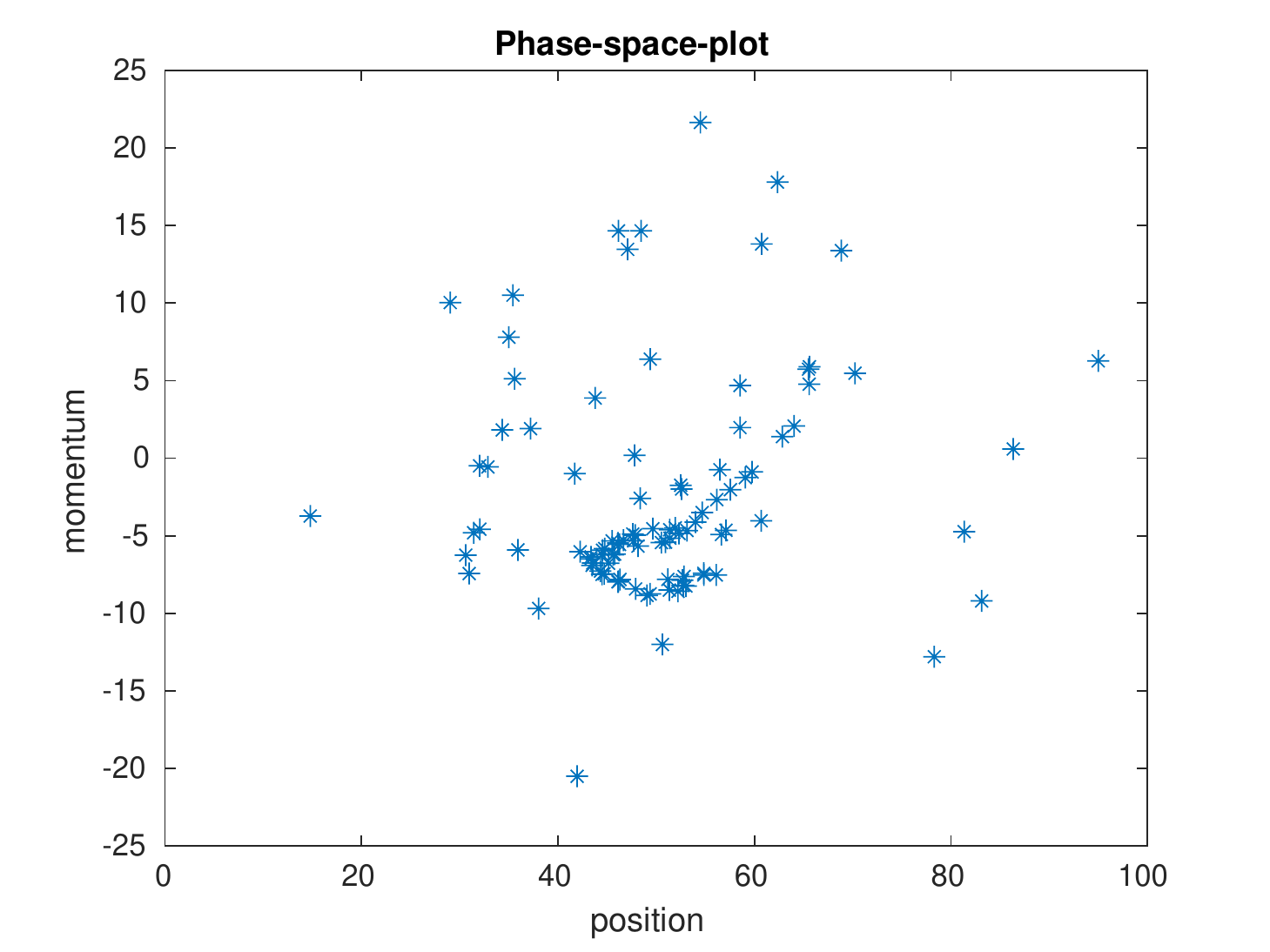}}
\subfloat[$t=670$]{\includegraphics[width = 1.45in]{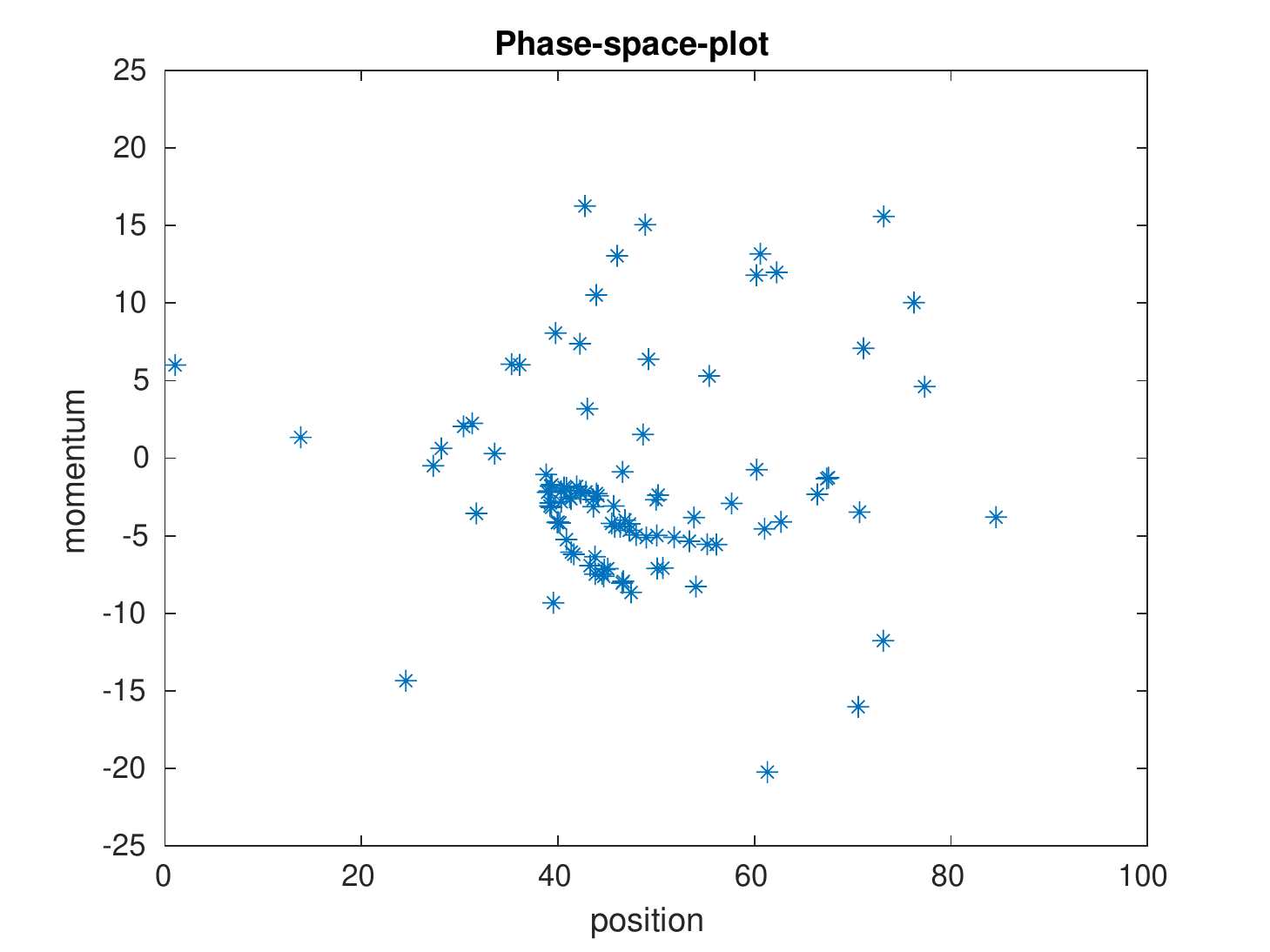}}
\subfloat[$t=680$]{\includegraphics[width = 1.45in]{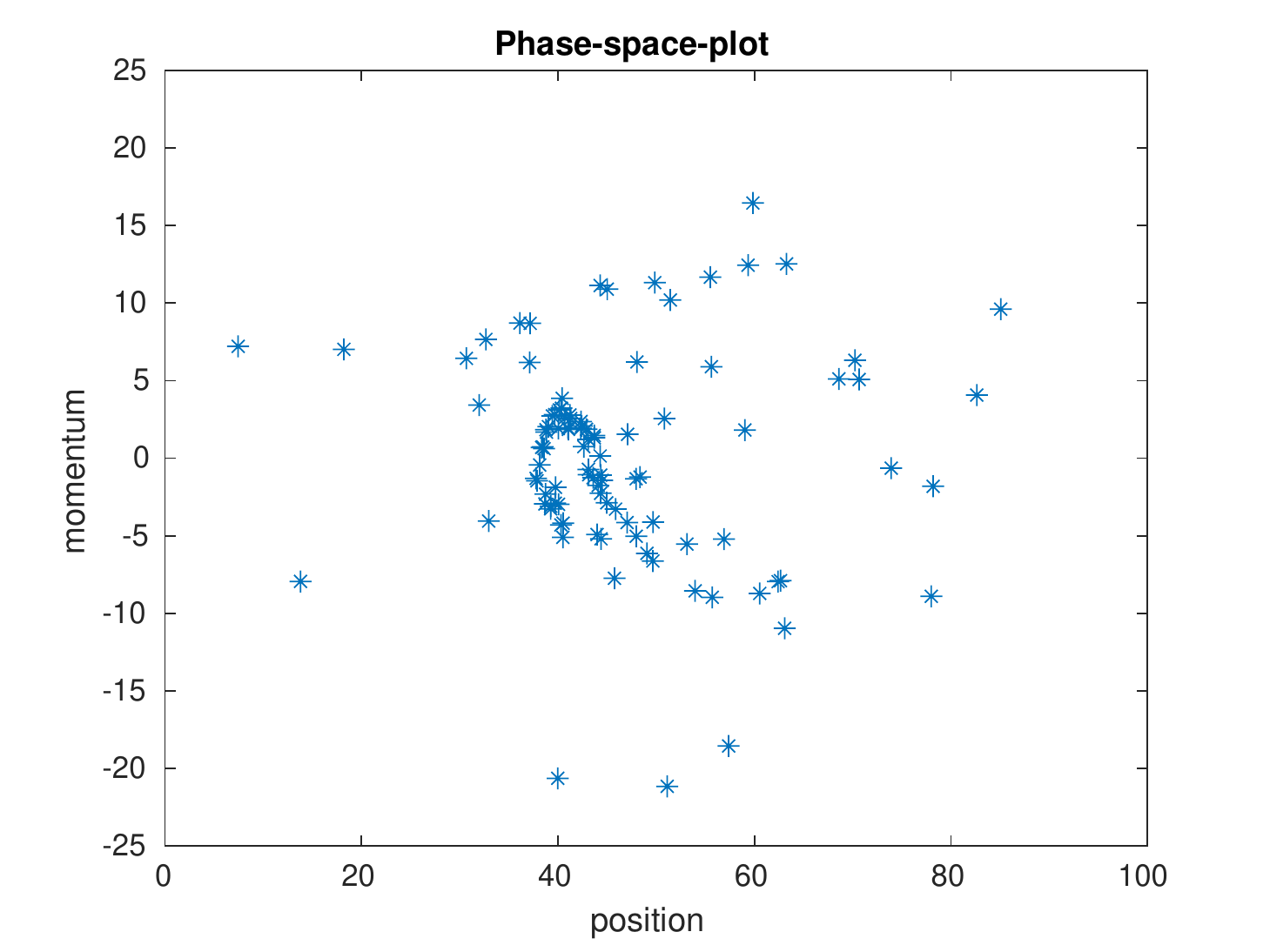}}\\
\subfloat[$t=690$]{\includegraphics[width = 1.45in]{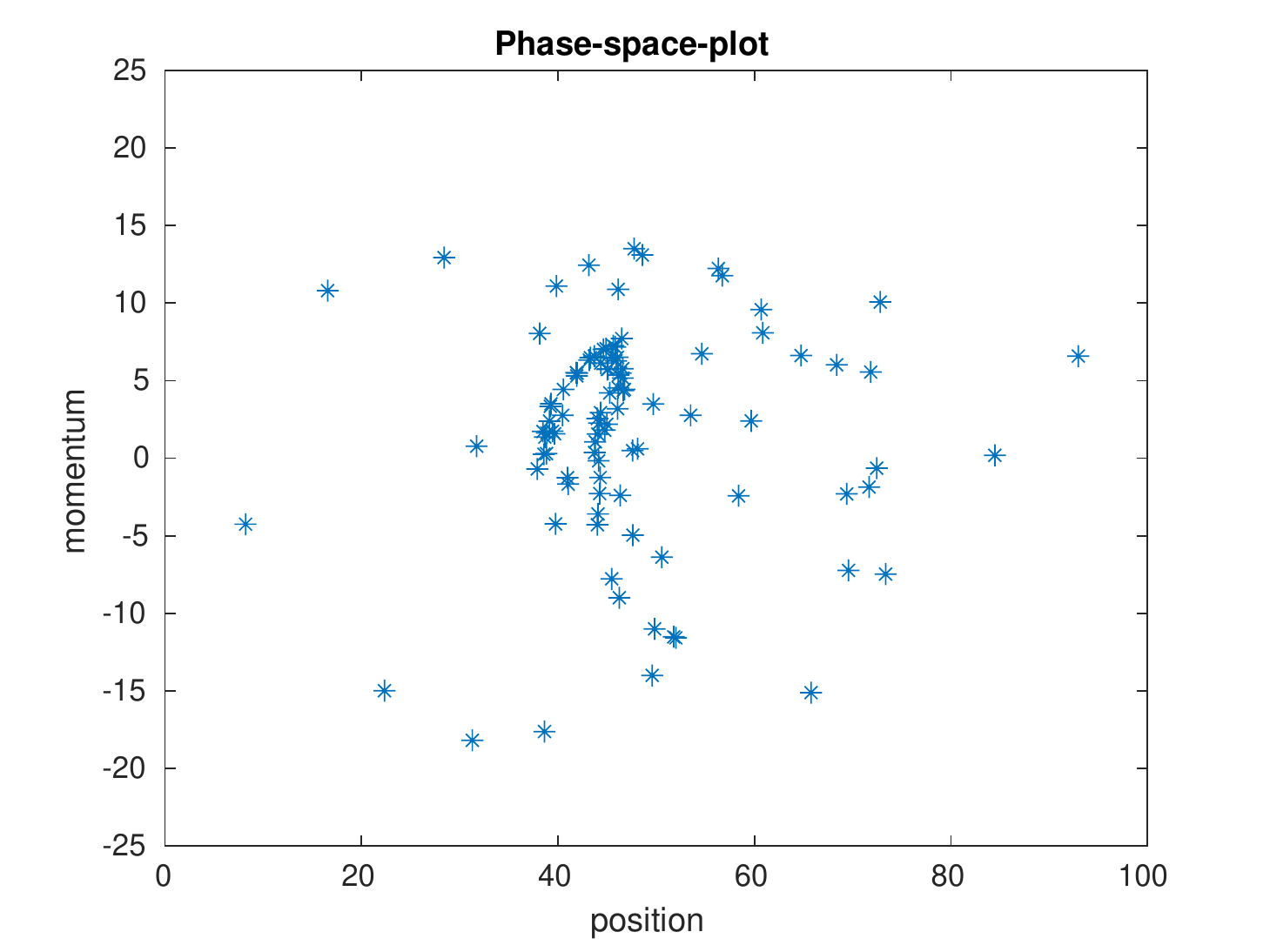}}
\subfloat[$t=700$]{\includegraphics[width = 1.45in]{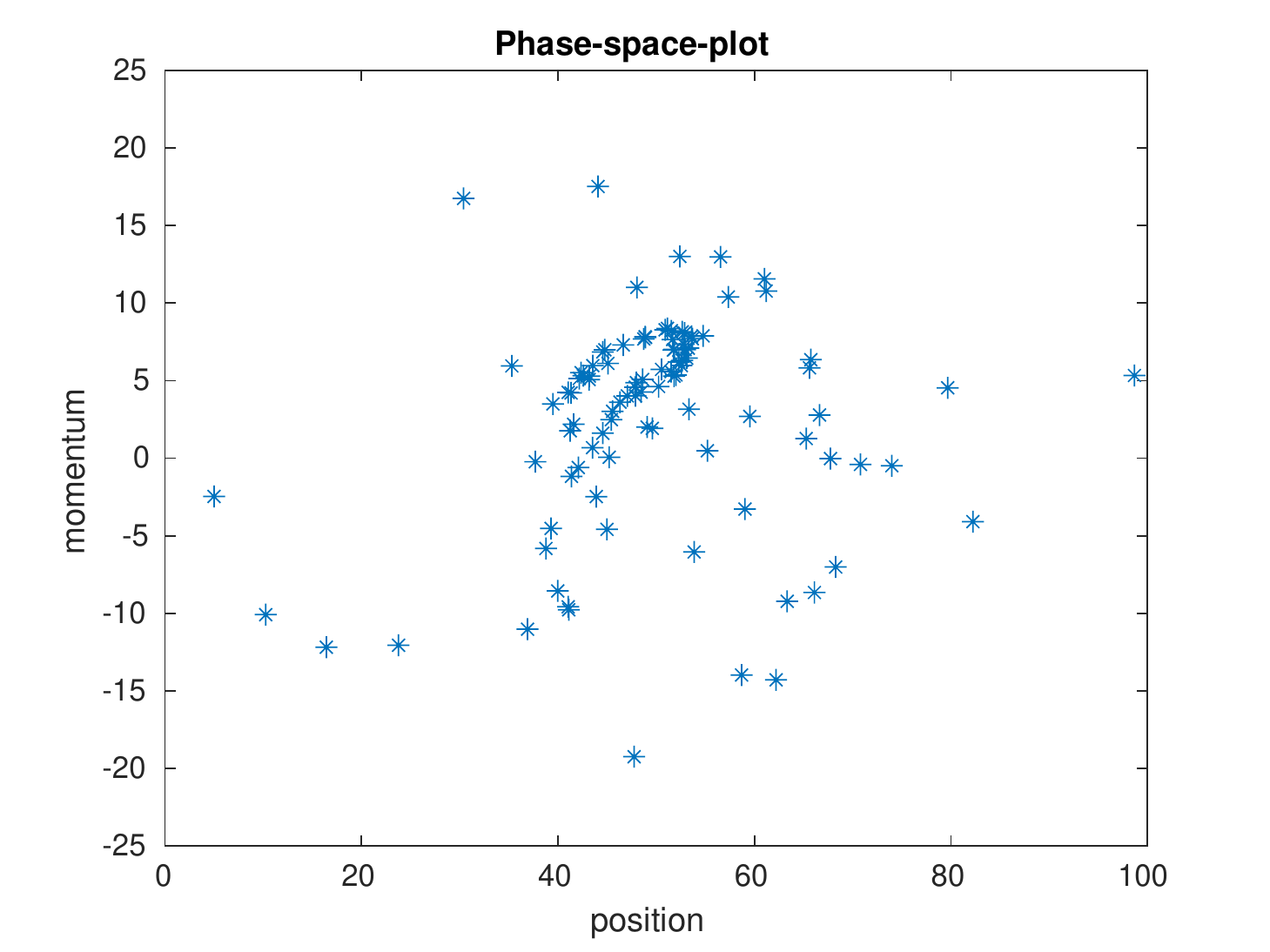}}
\subfloat[$t=710$]{\includegraphics[width = 1.45in]{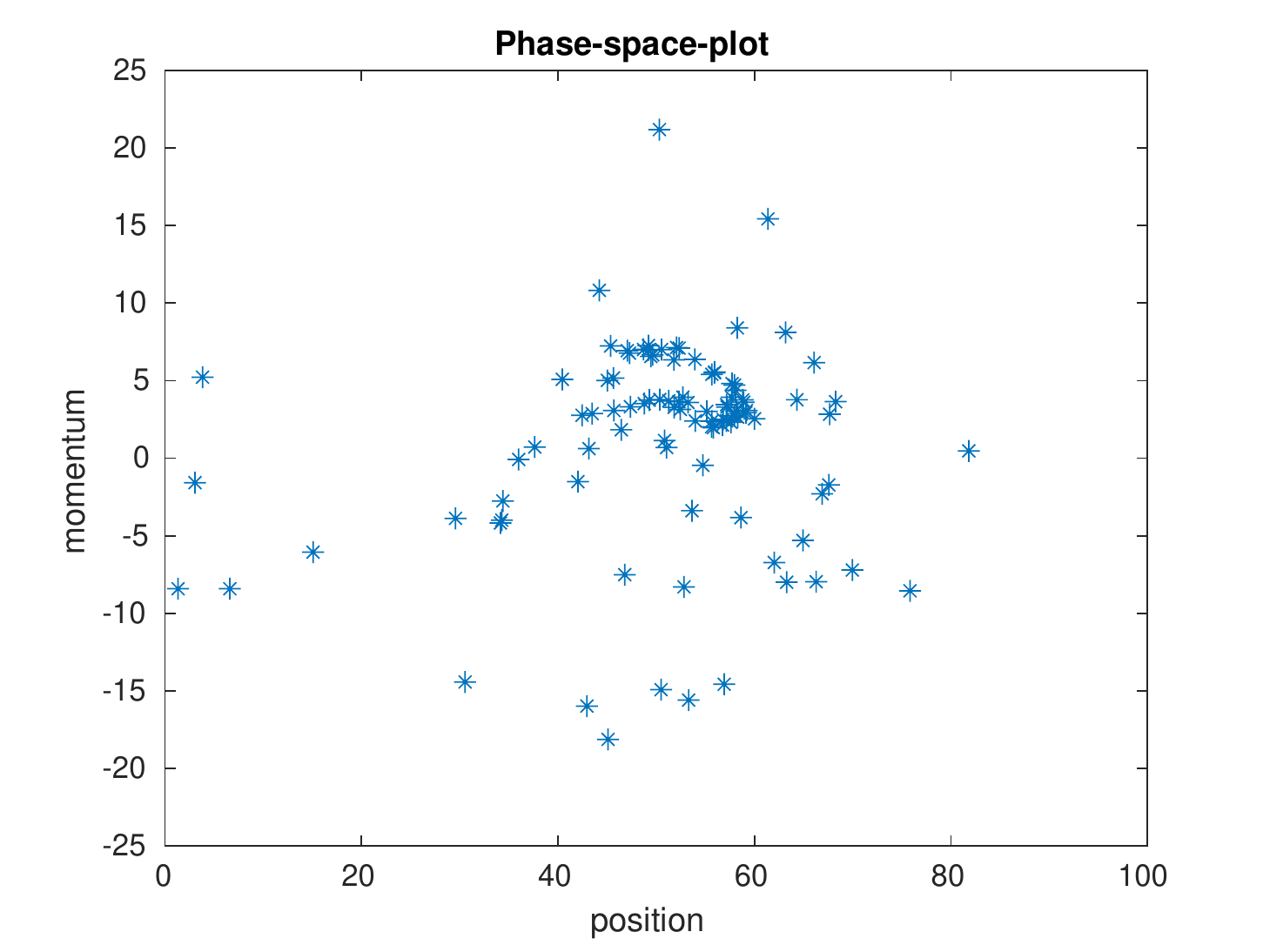}}
\subfloat[$t=720$]{\includegraphics[width = 1.45in]{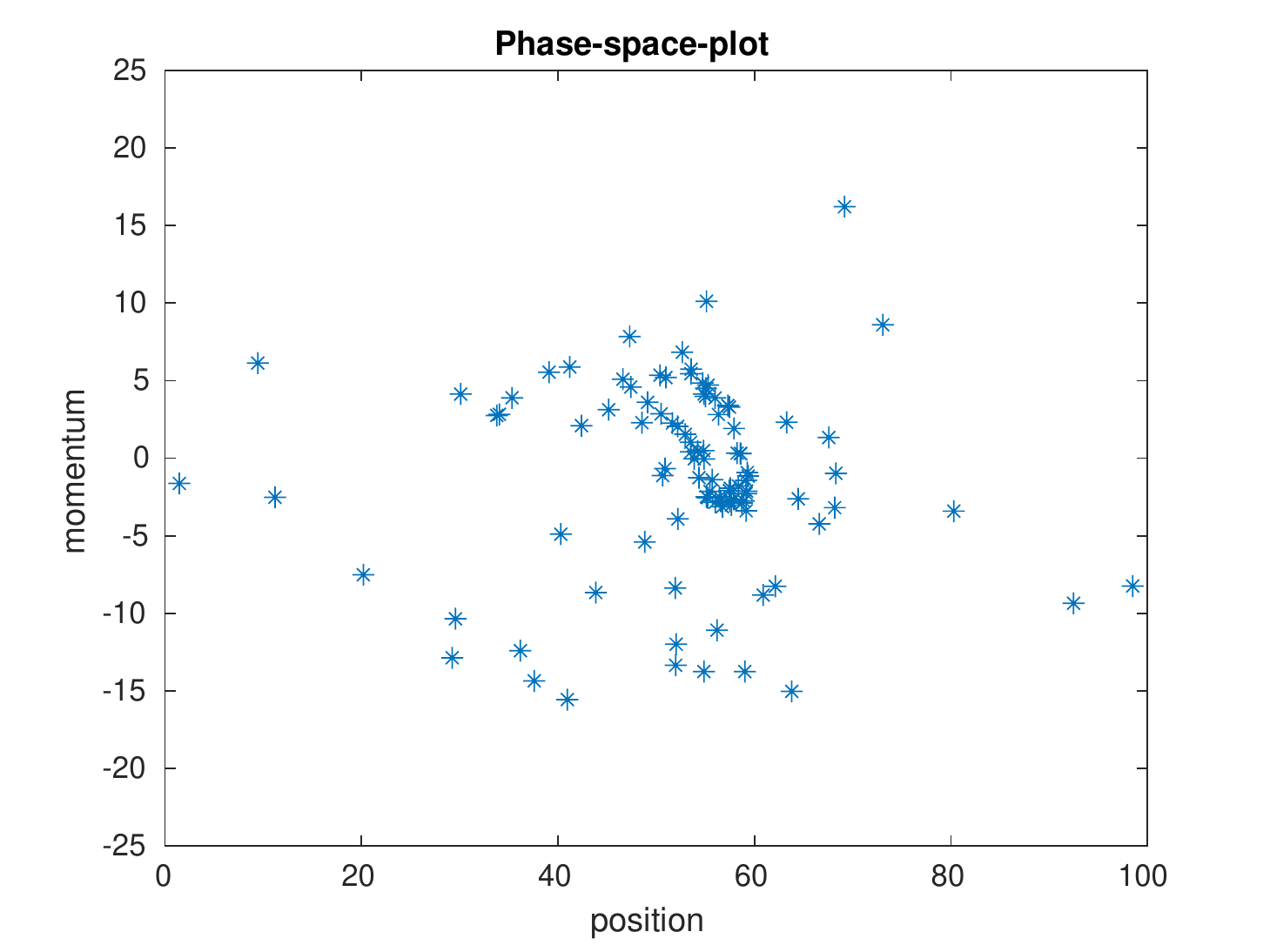}}\\
\subfloat[$t=730$]{\includegraphics[width = 1.45in]{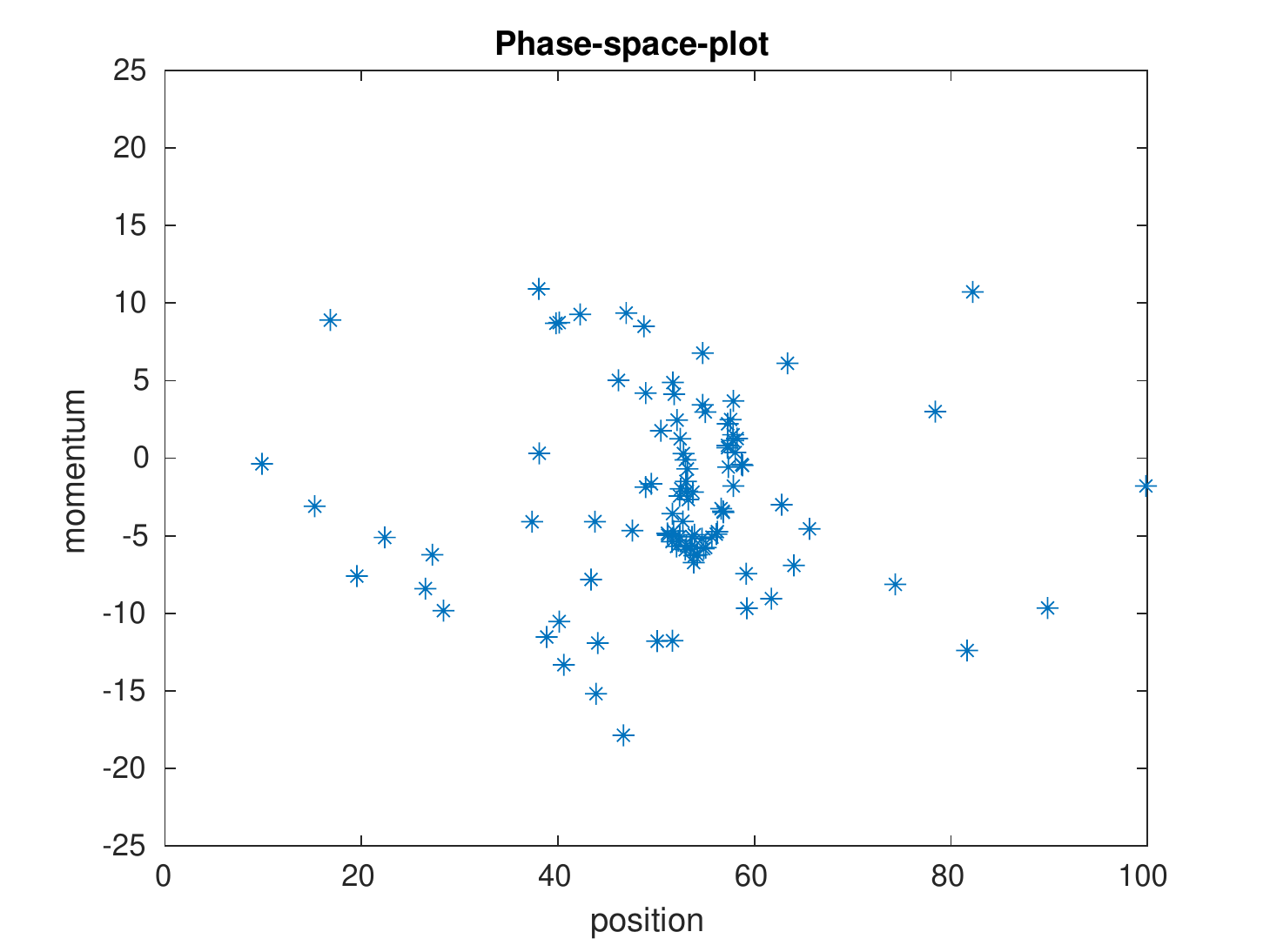}}
\subfloat[$t=740$]{\includegraphics[width = 1.45in]{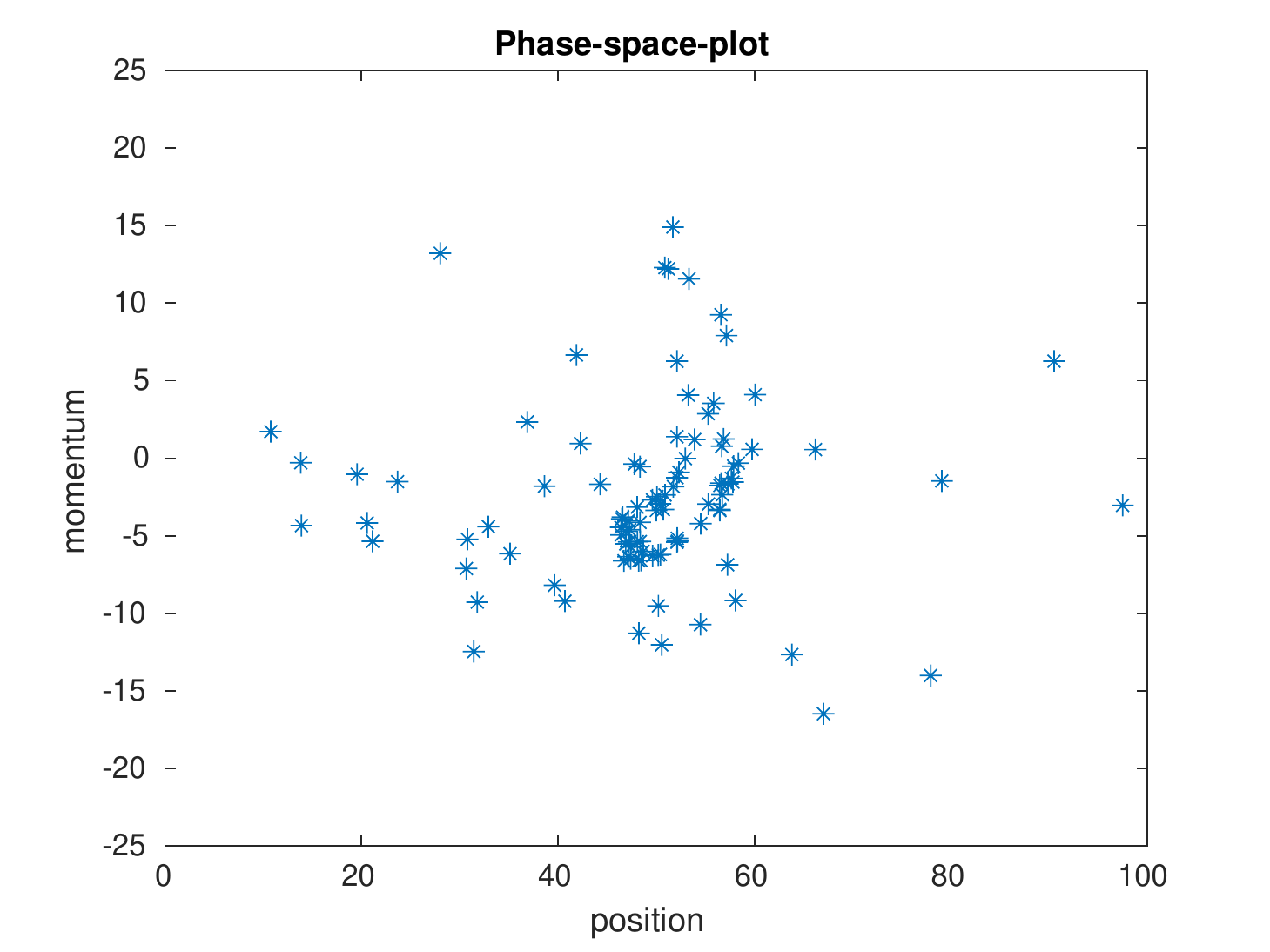}}
\subfloat[$t=750$]{\includegraphics[width = 1.45in]{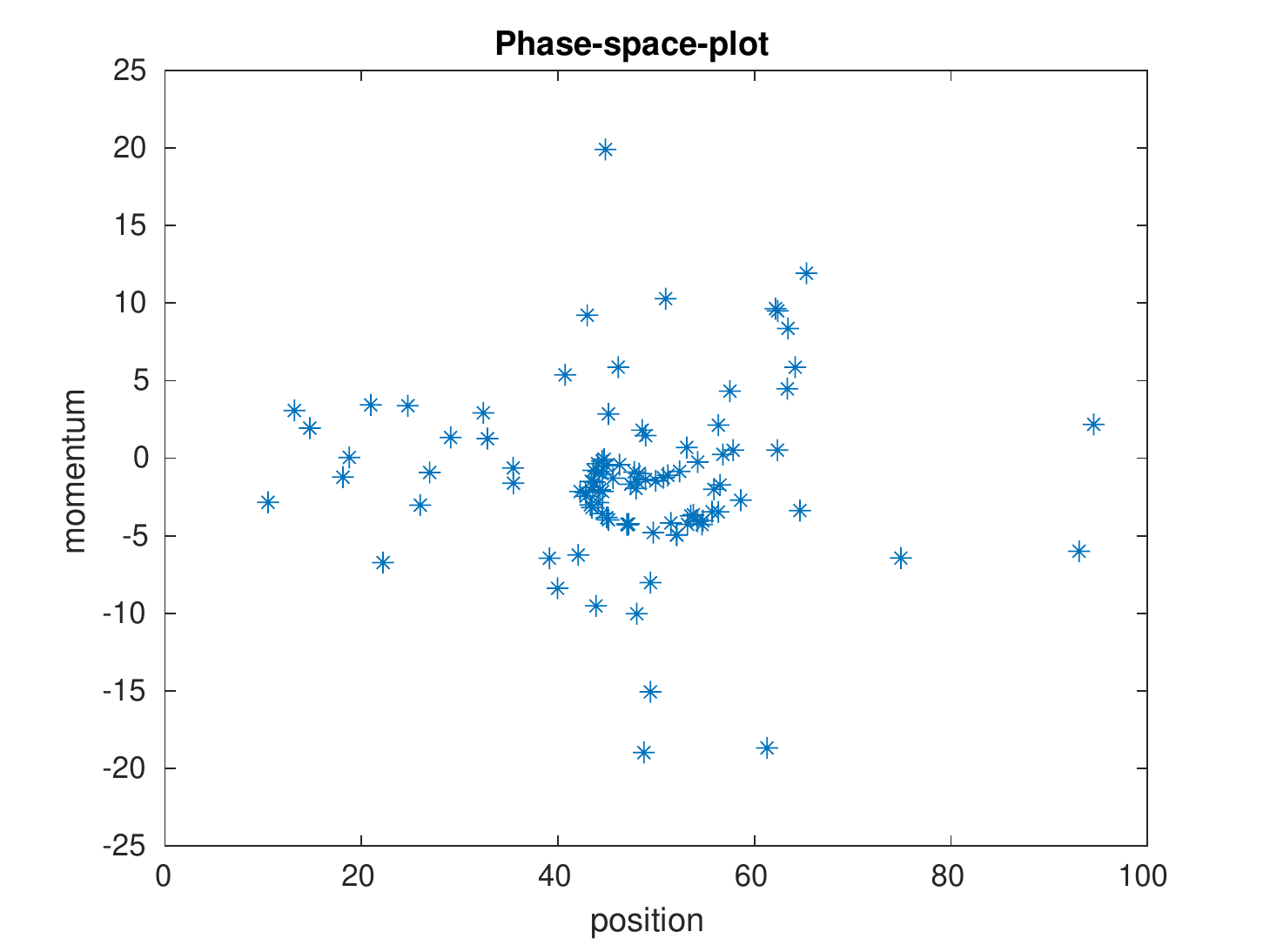}}
\subfloat[$t=760$]{\includegraphics[width = 1.45in]{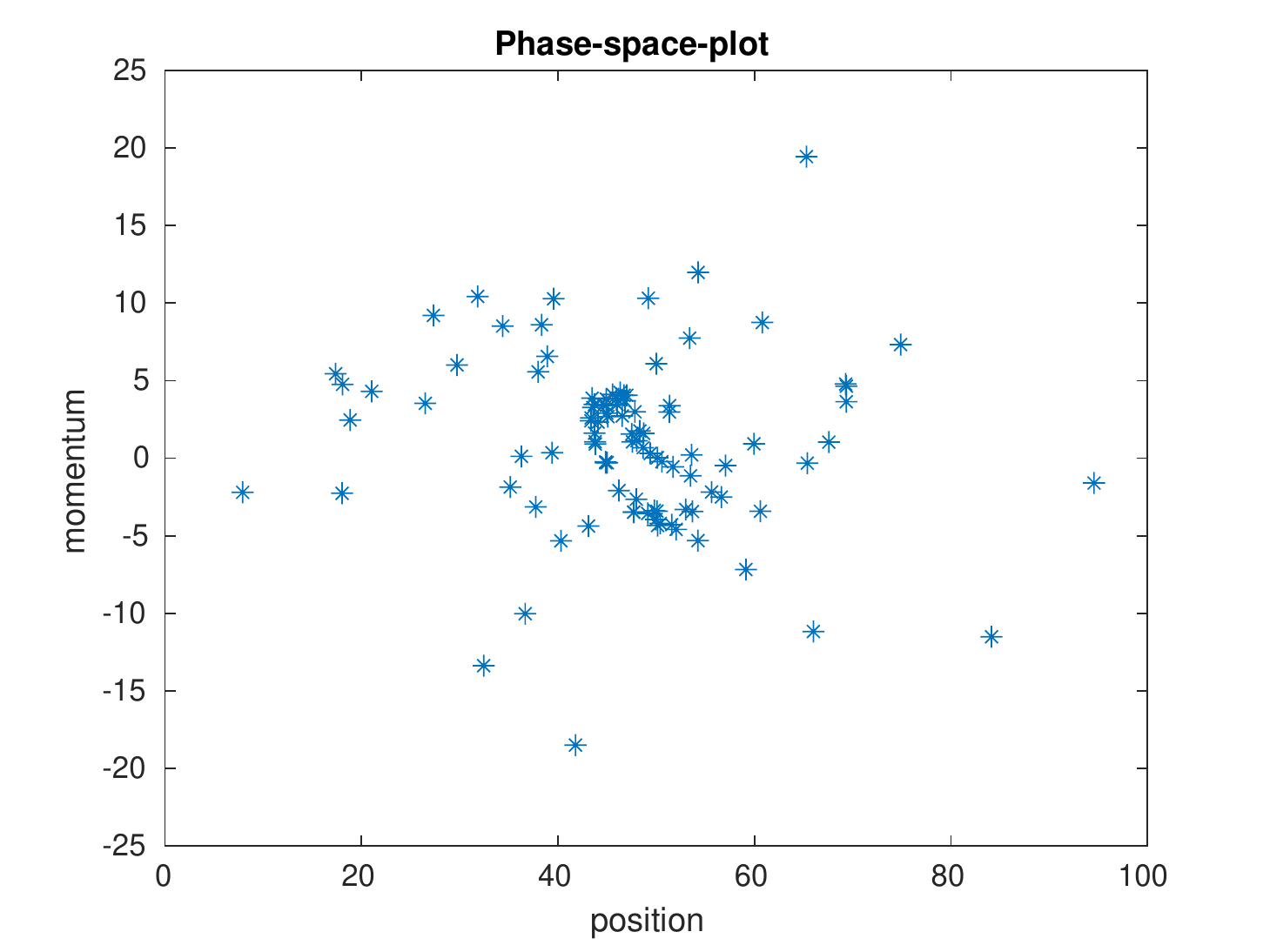}}\\
\subfloat[$t=770$]{\includegraphics[width = 1.45in]{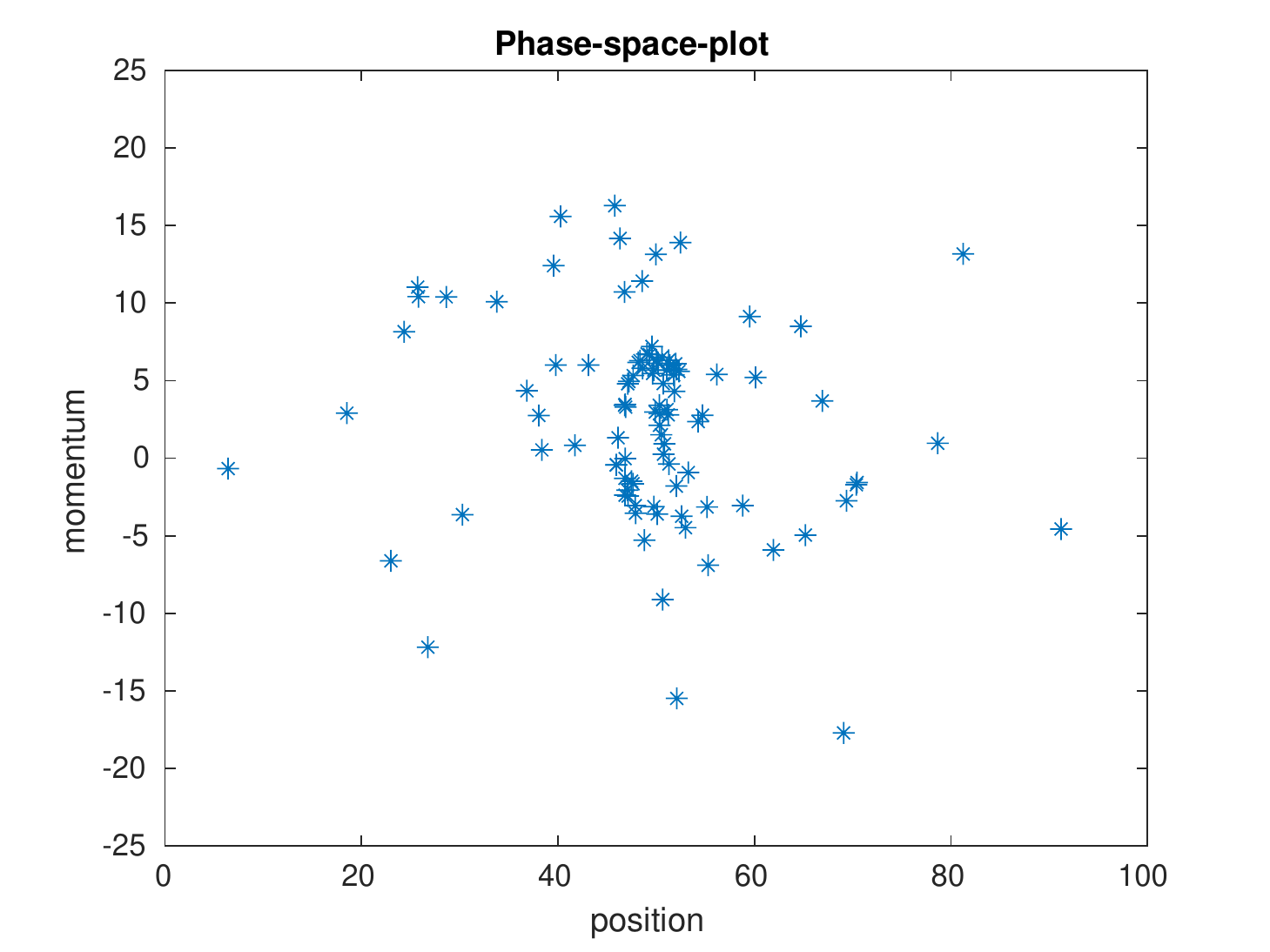}}
\subfloat[$t=780$]{\includegraphics[width = 1.45in]{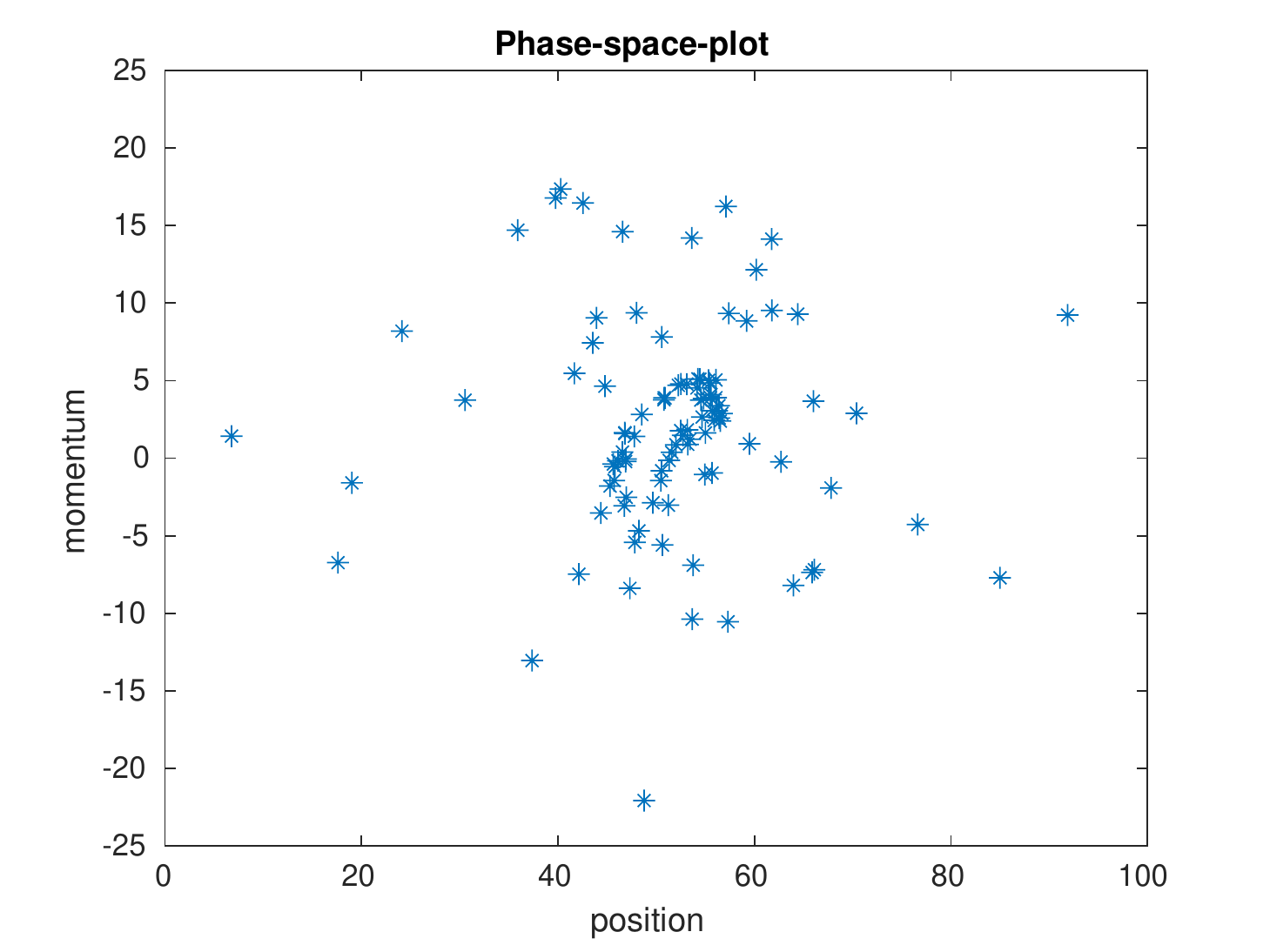}}
\subfloat[$t=790$]{\includegraphics[width = 1.45in]{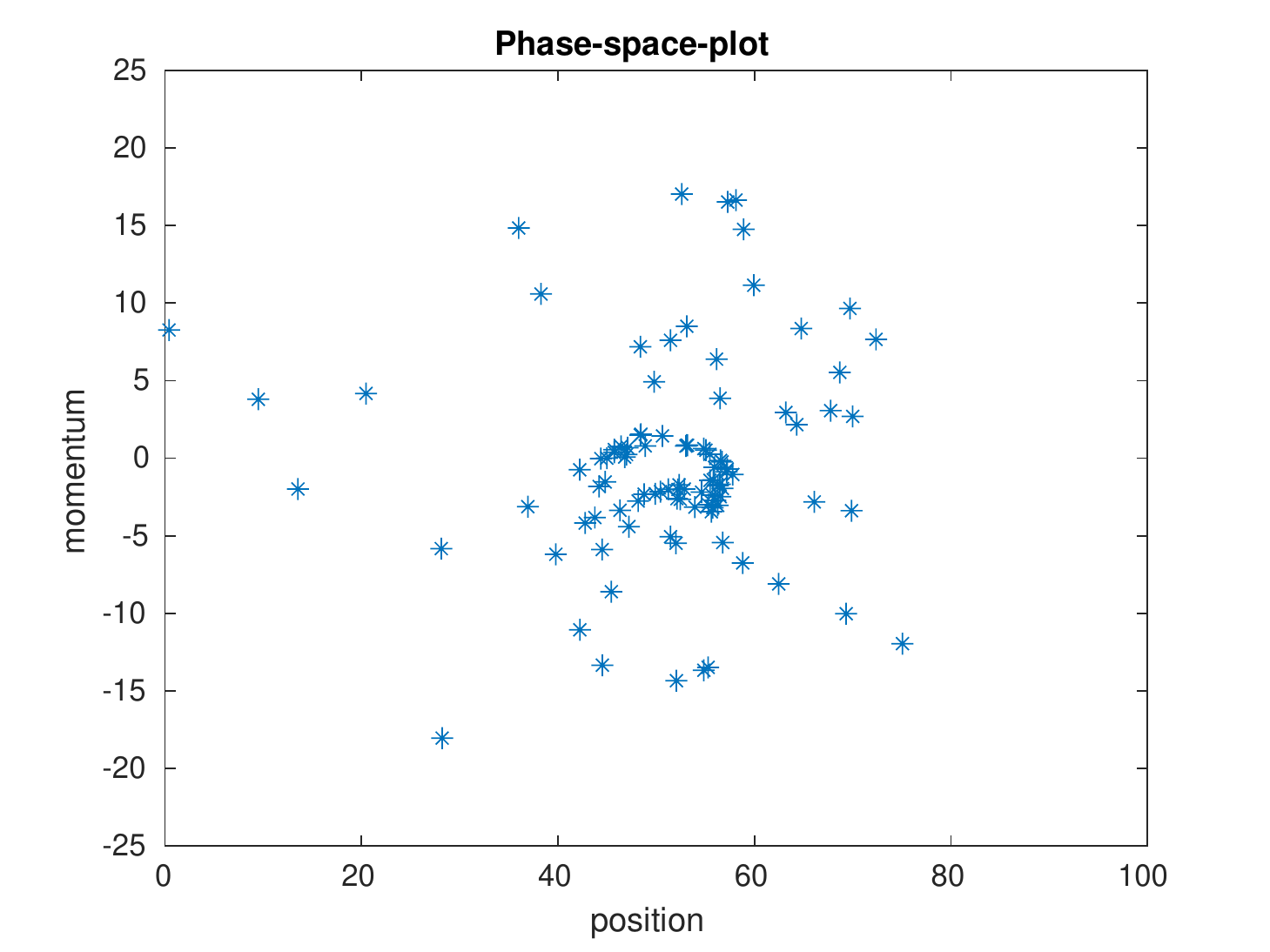}}
\subfloat[$t=800$]{\includegraphics[width = 1.45in]{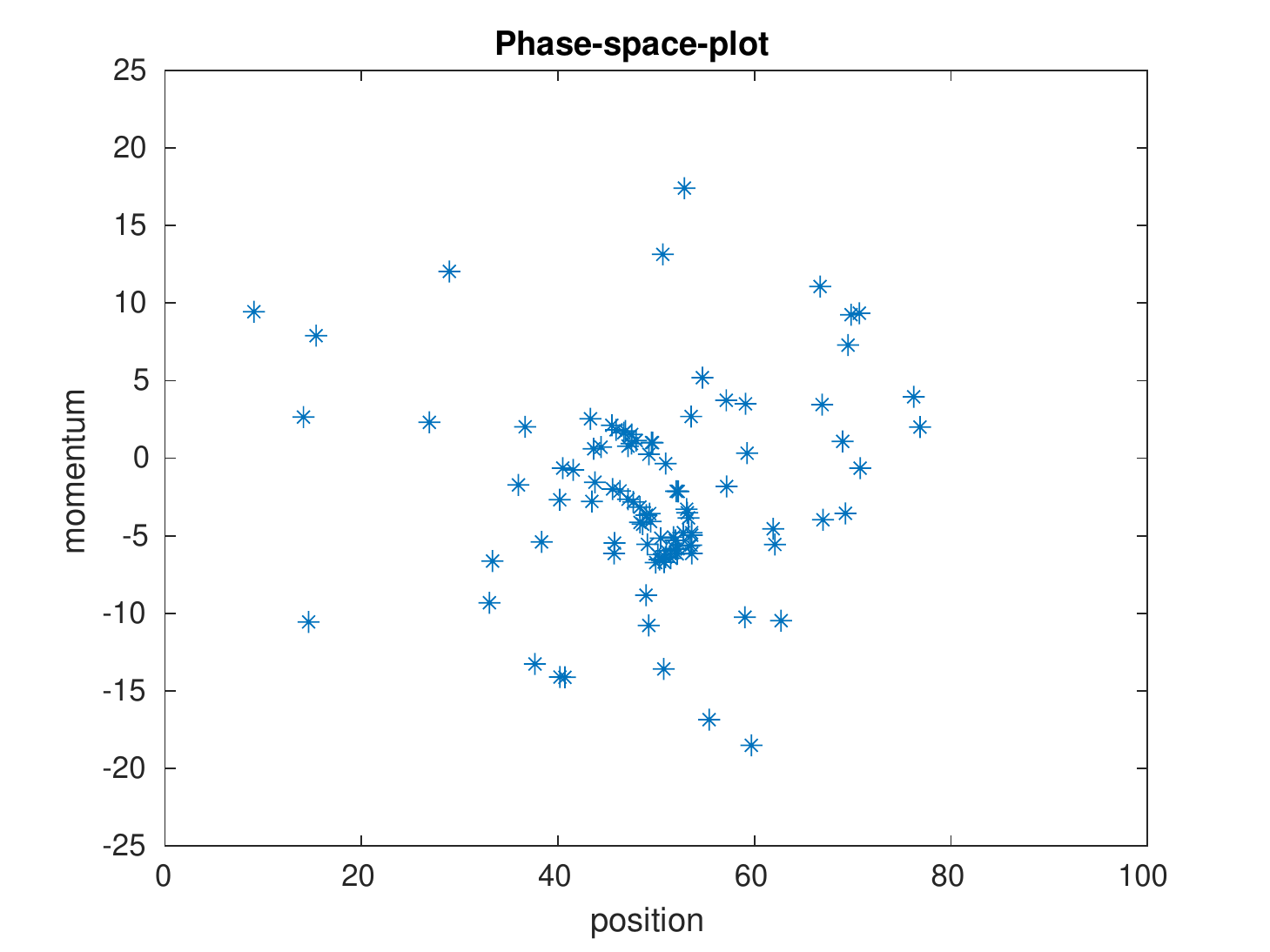}}
\caption{Phase Space snapshots for $t=610$ to $t=800$ with time separation of $10$}
\label{fig_sg_4}
\end{figure}

All the plots in figure \ref{fig_sg_1} to figure \ref{fig_sg_4} are the phase space snapshots for time $t=10$ to $t=800$ with an interval of $t=10$. We start with particles uniformly distributed over a 1D line with zero momentum. As time pass particles will get momentum due to gravitational attraction and try to collapse (cluster) at the centre. We use periodic boundary condition here. 

\section{Video of the simulation}
A video of the simulation done in this assignment can be found in this link \url{https://youtu.be/tzBqdSGEpTQ}. However with this assignment PDF file the video is also attached.

\begin{figure}[ht]
\centering
\subfloat[$t=210$]{\includegraphics[width = 1.45in]{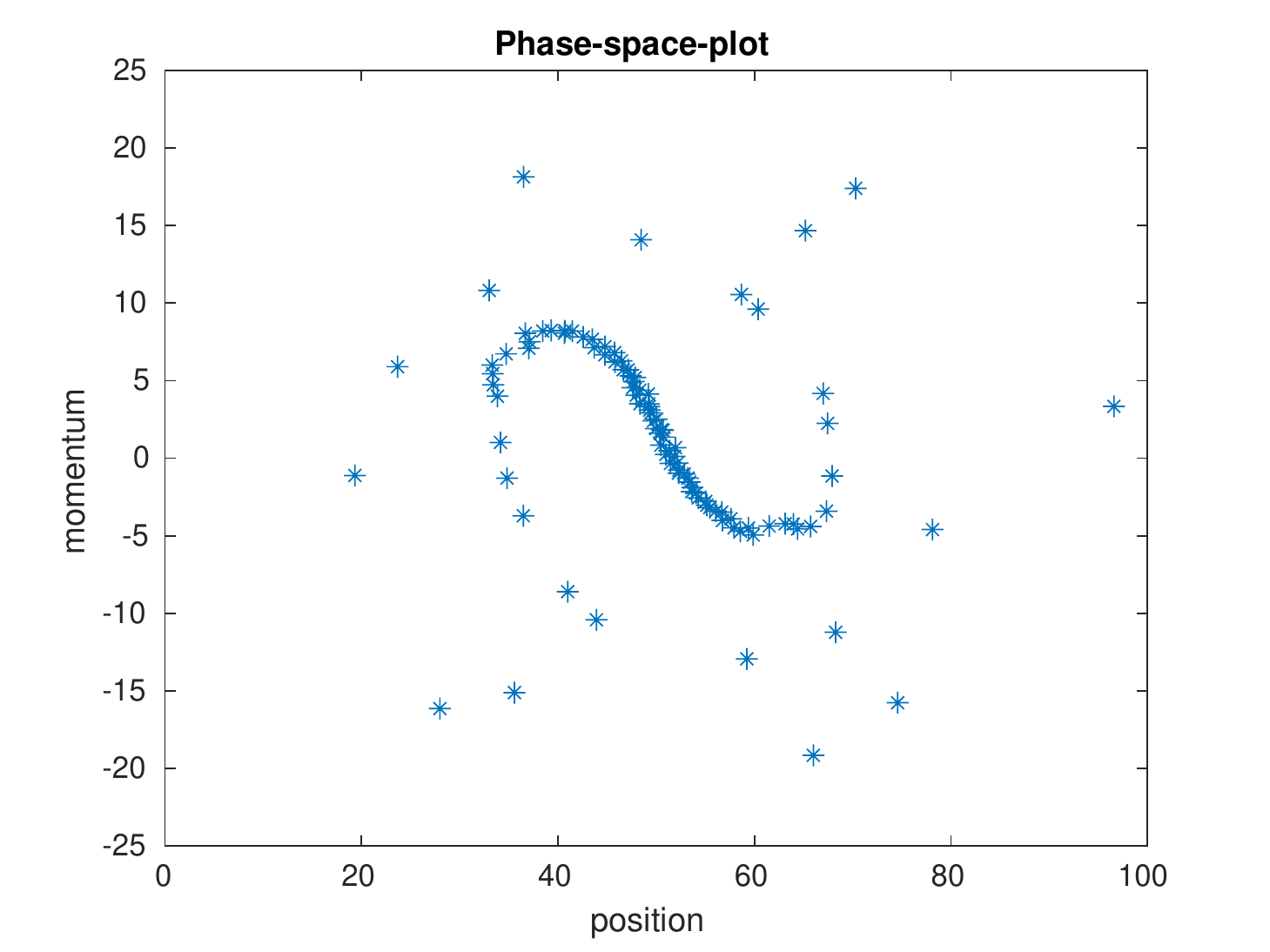}}
\subfloat[$t=220$]{\includegraphics[width = 1.45in]{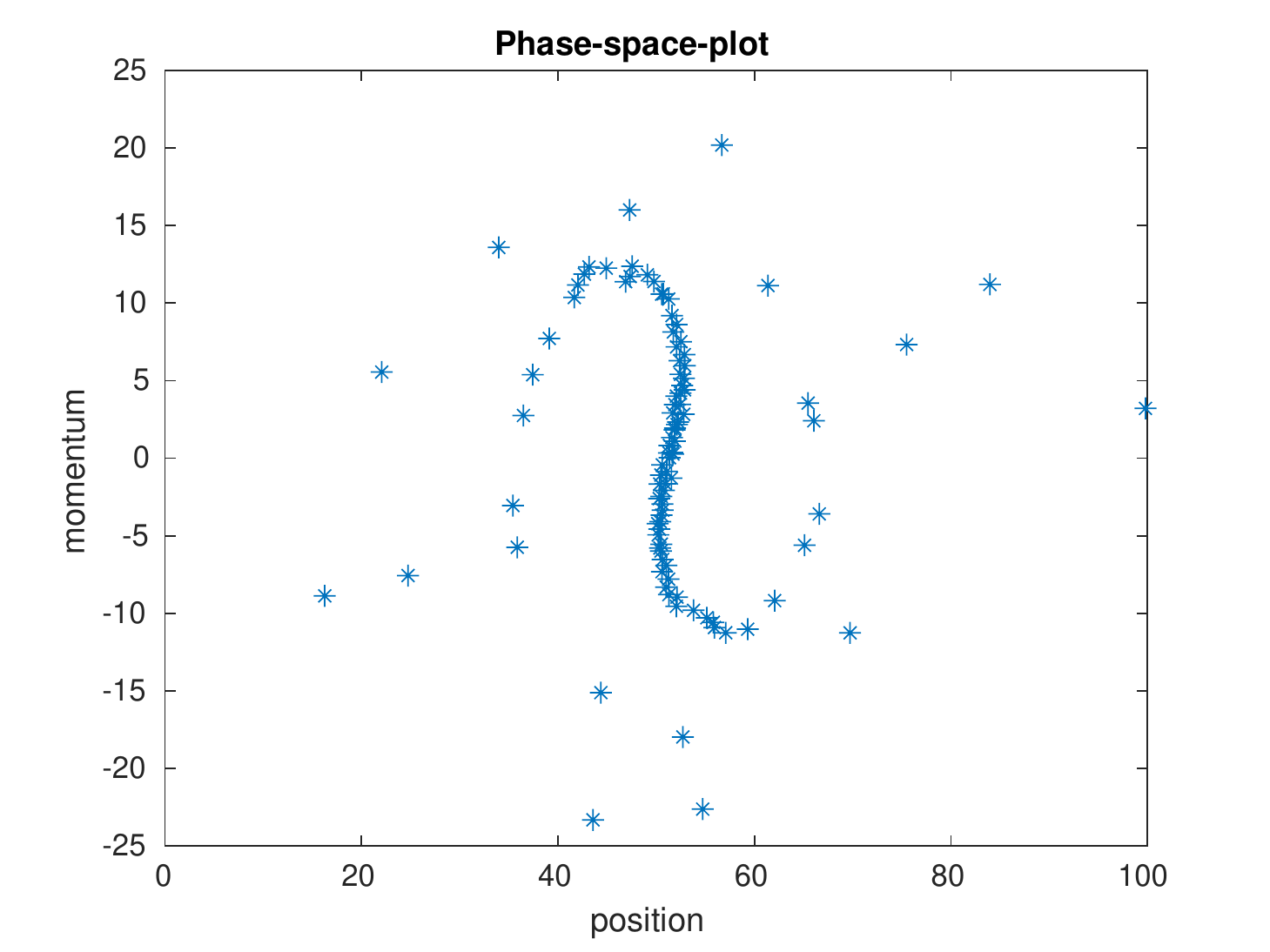}}
\subfloat[$t=230$]{\includegraphics[width = 1.45in]{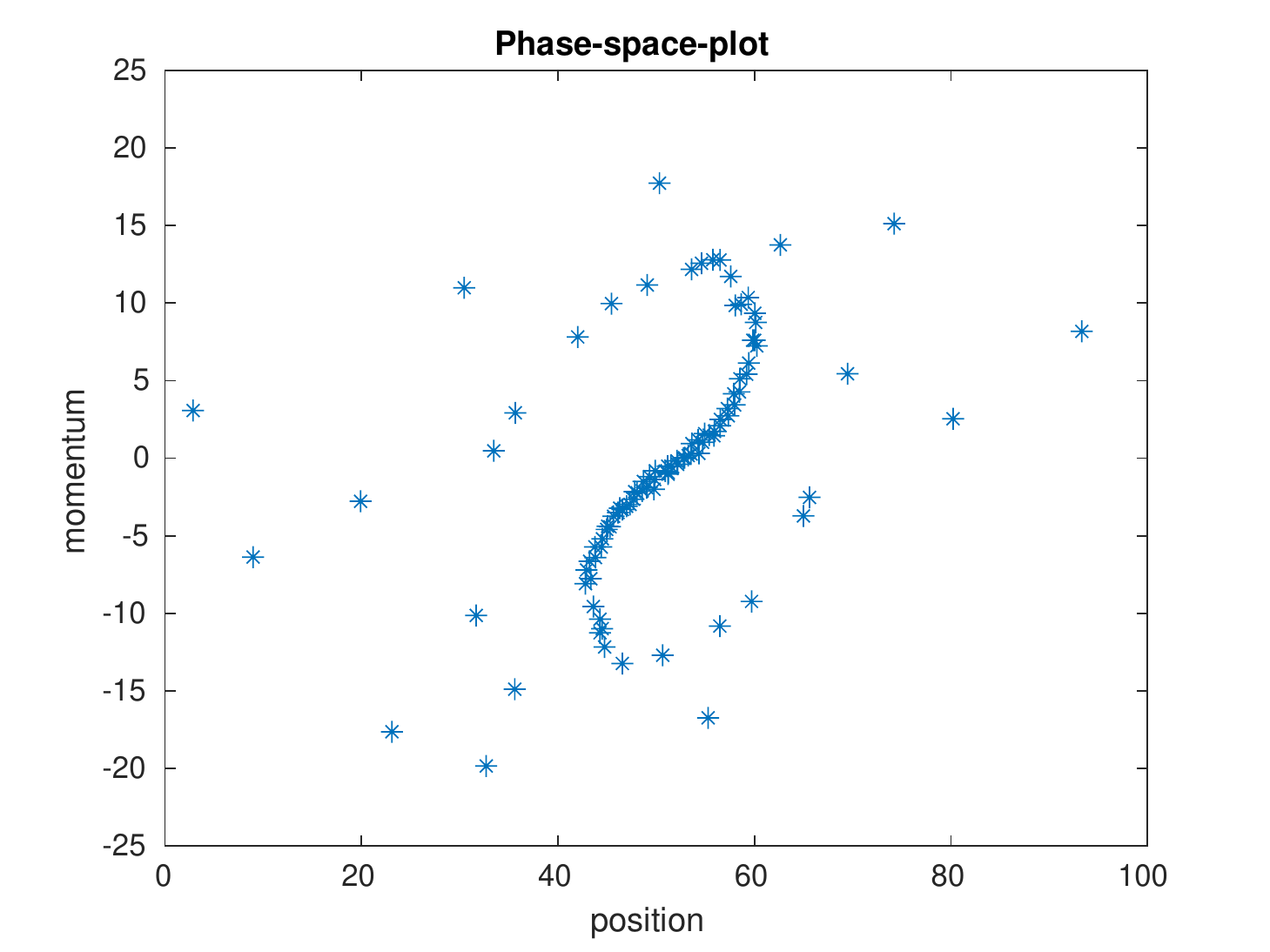}}
\subfloat[$t=240$]{\includegraphics[width = 1.45in]{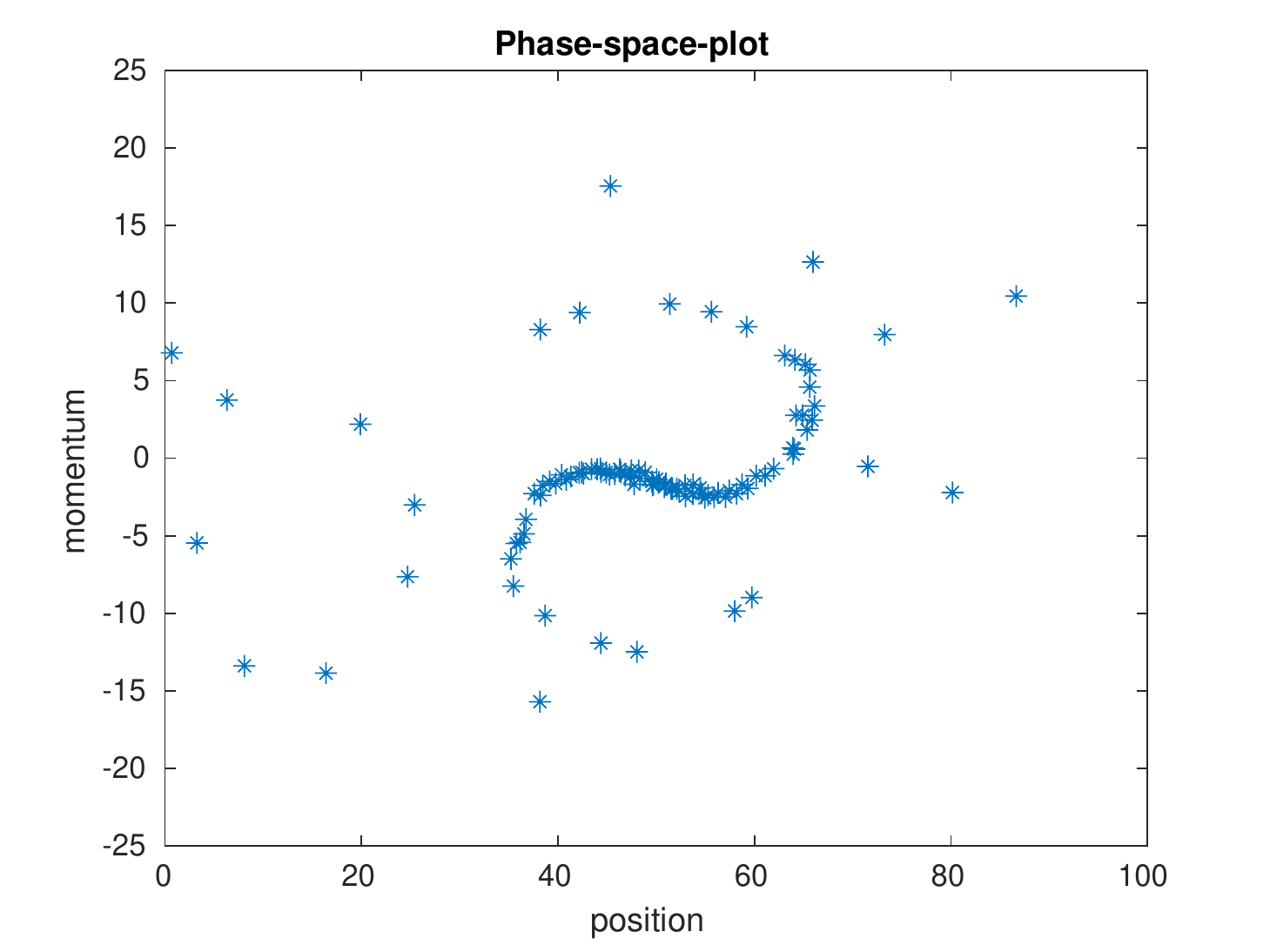}}\\
\subfloat[$t=250$]{\includegraphics[width = 1.45in]{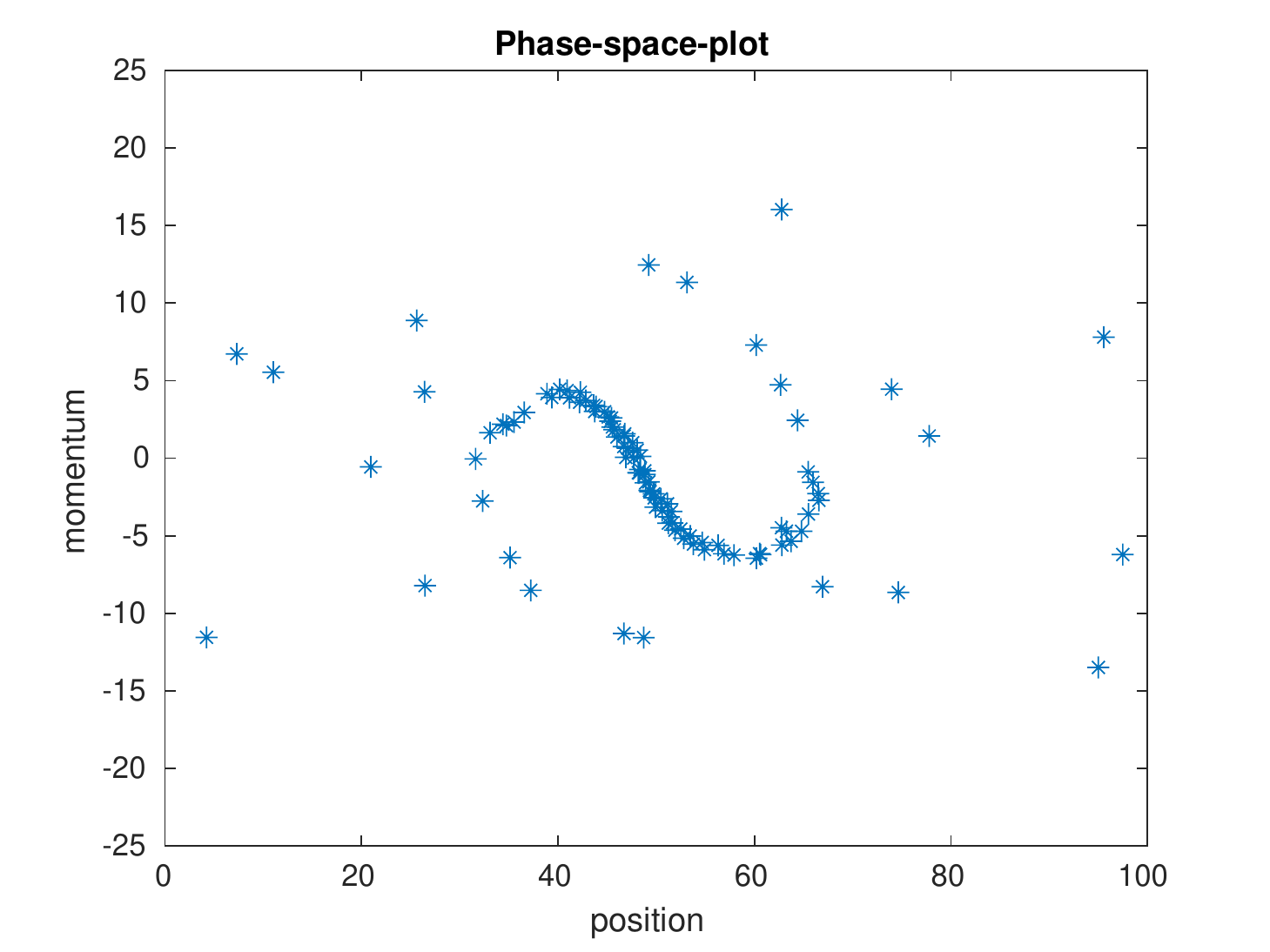}}
\subfloat[$t=260$]{\includegraphics[width = 1.45in]{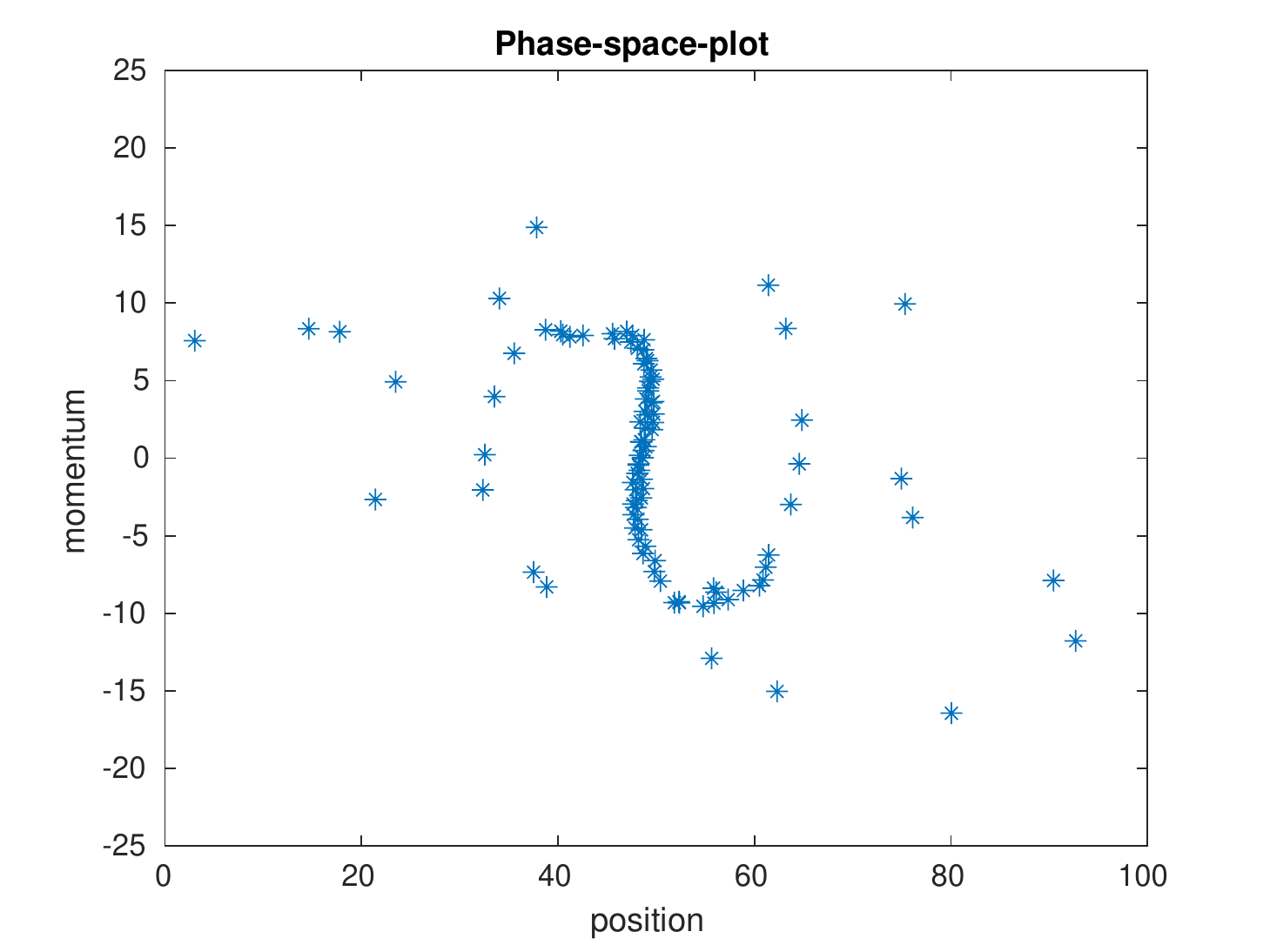}}
\subfloat[$t=270$]{\includegraphics[width = 1.45in]{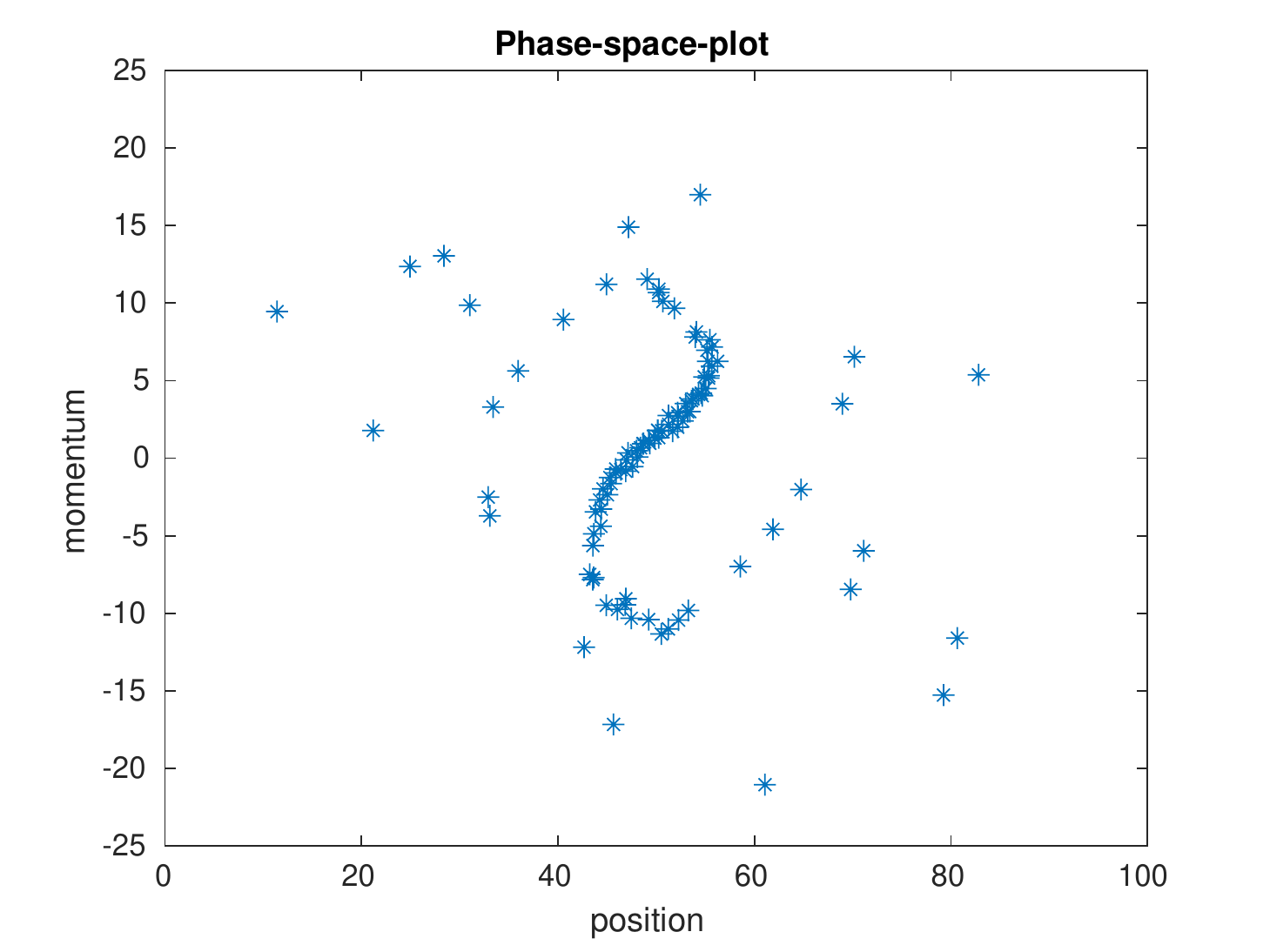}}
\subfloat[$t=280$]{\includegraphics[width = 1.45in]{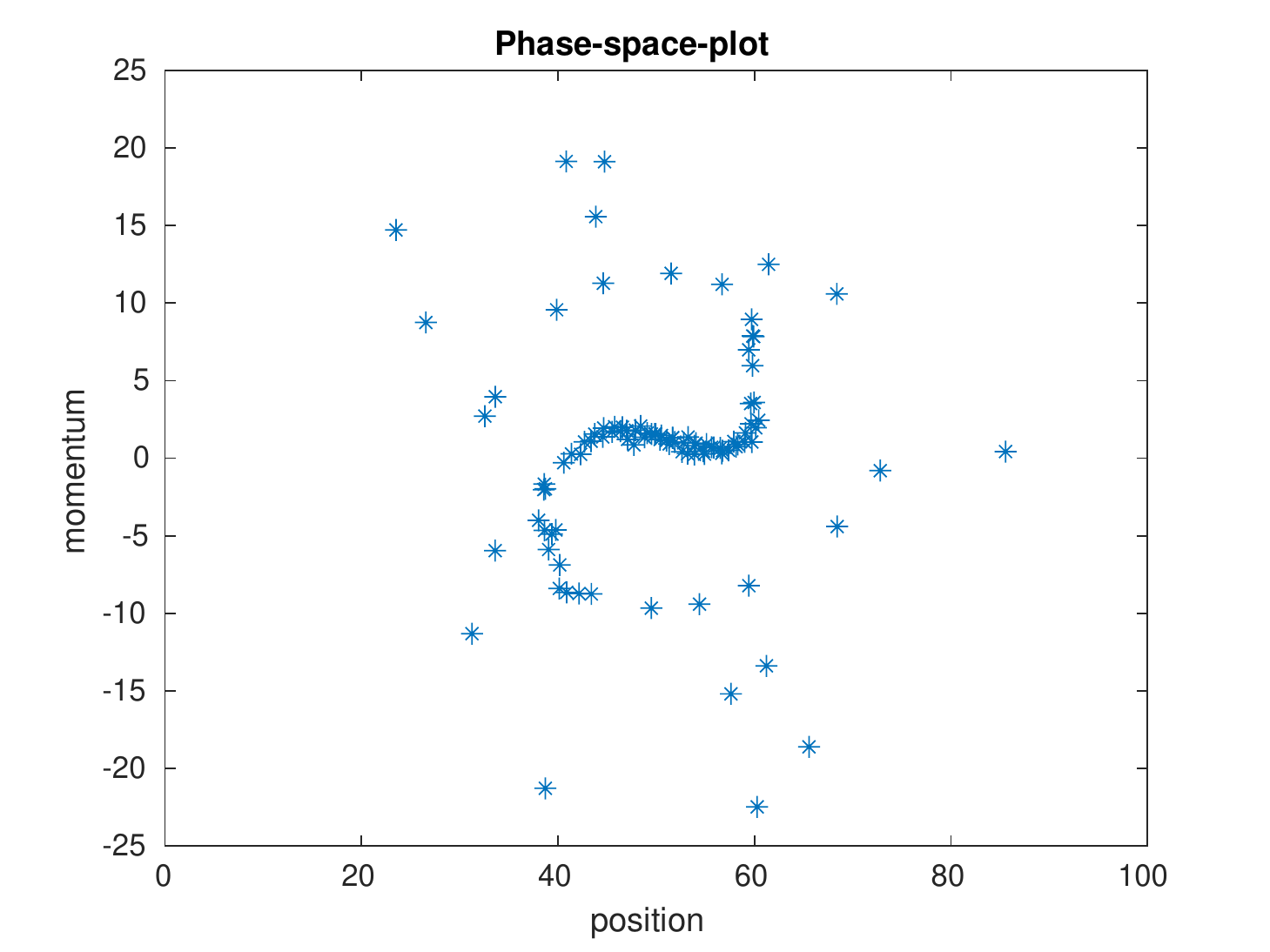}}\\
\subfloat[$t=290$]{\includegraphics[width = 1.45in]{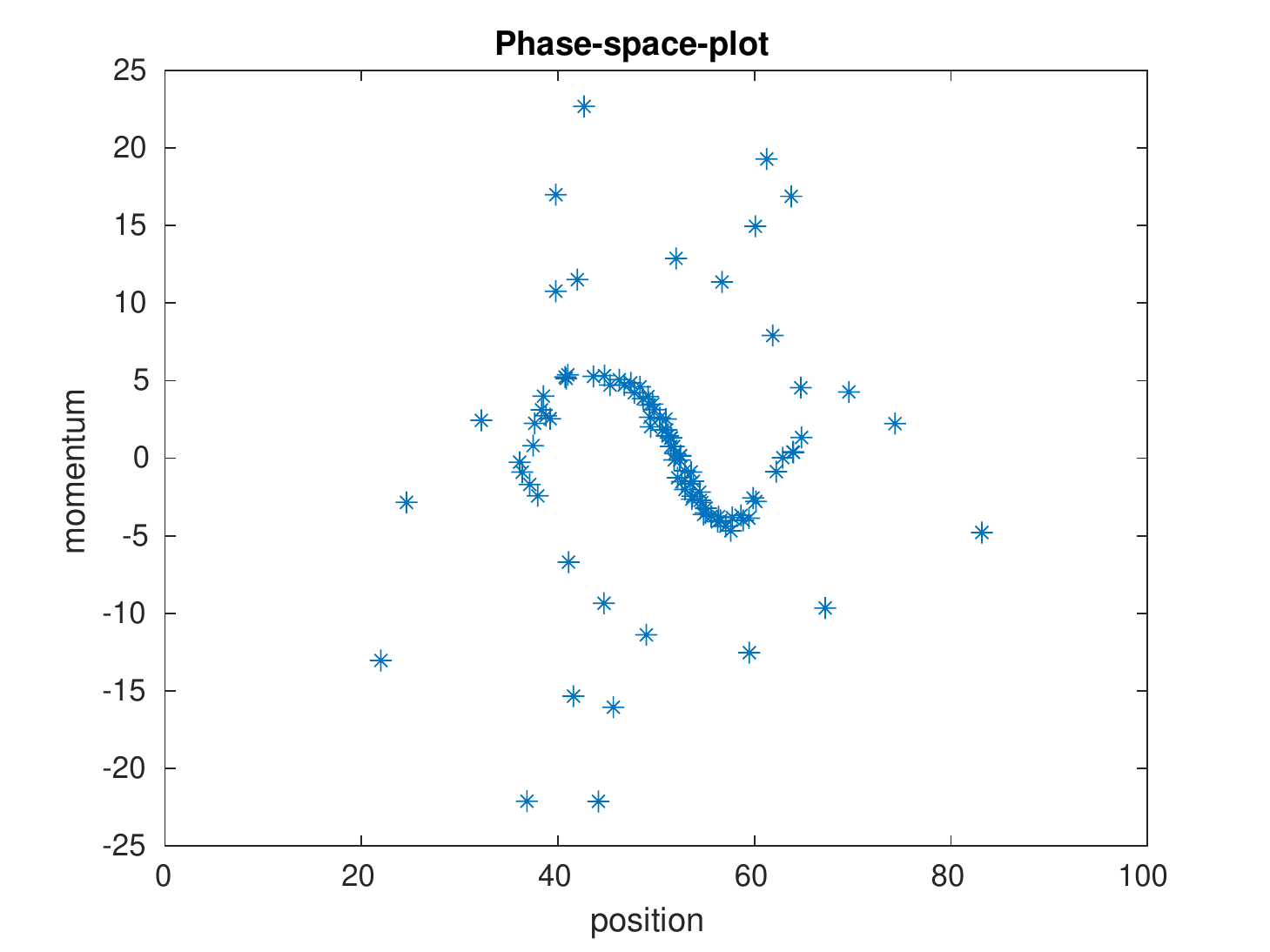}}
\subfloat[$t=300$]{\includegraphics[width = 1.45in]{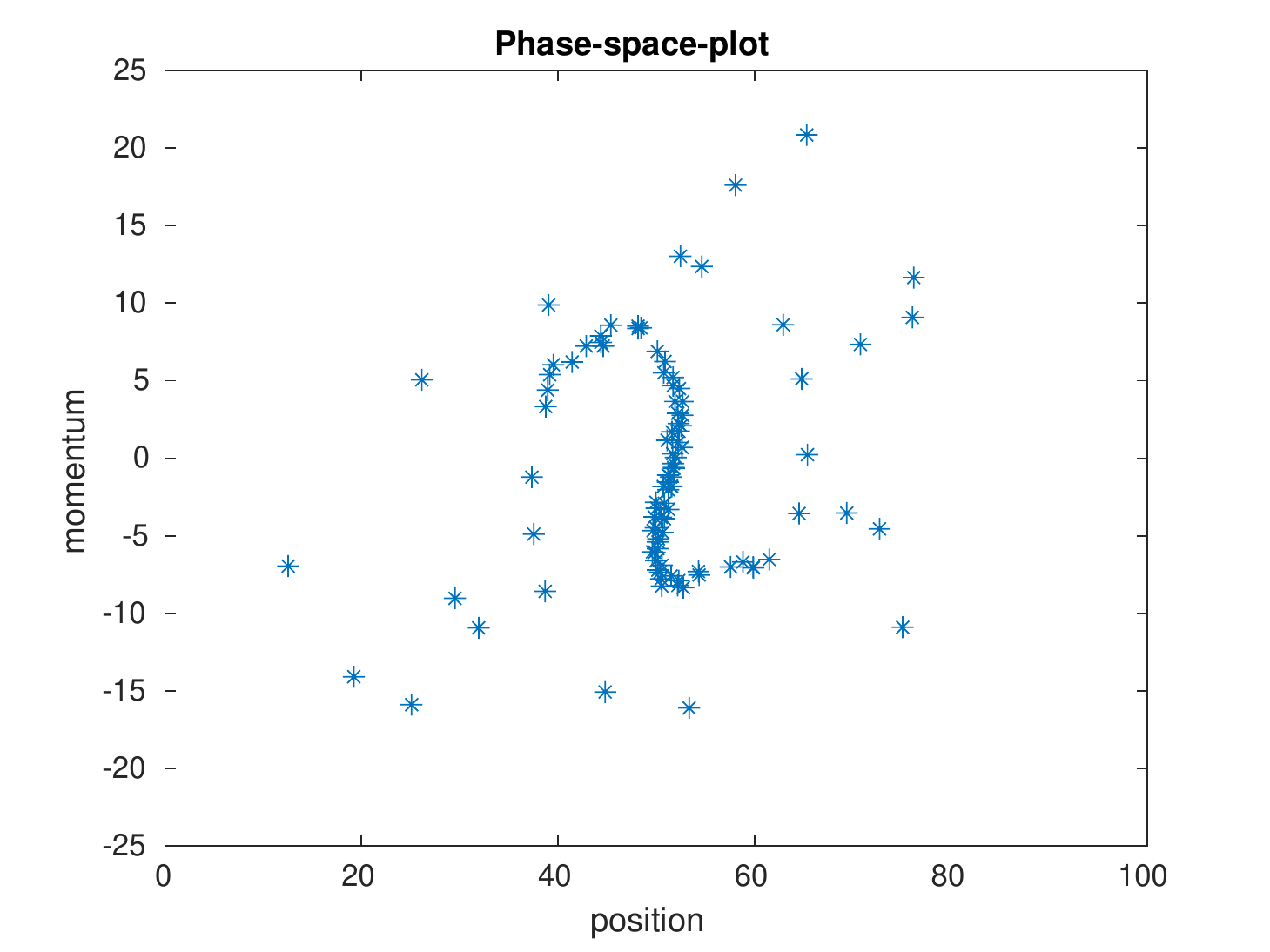}}
\subfloat[$t=310$]{\includegraphics[width = 1.45in]{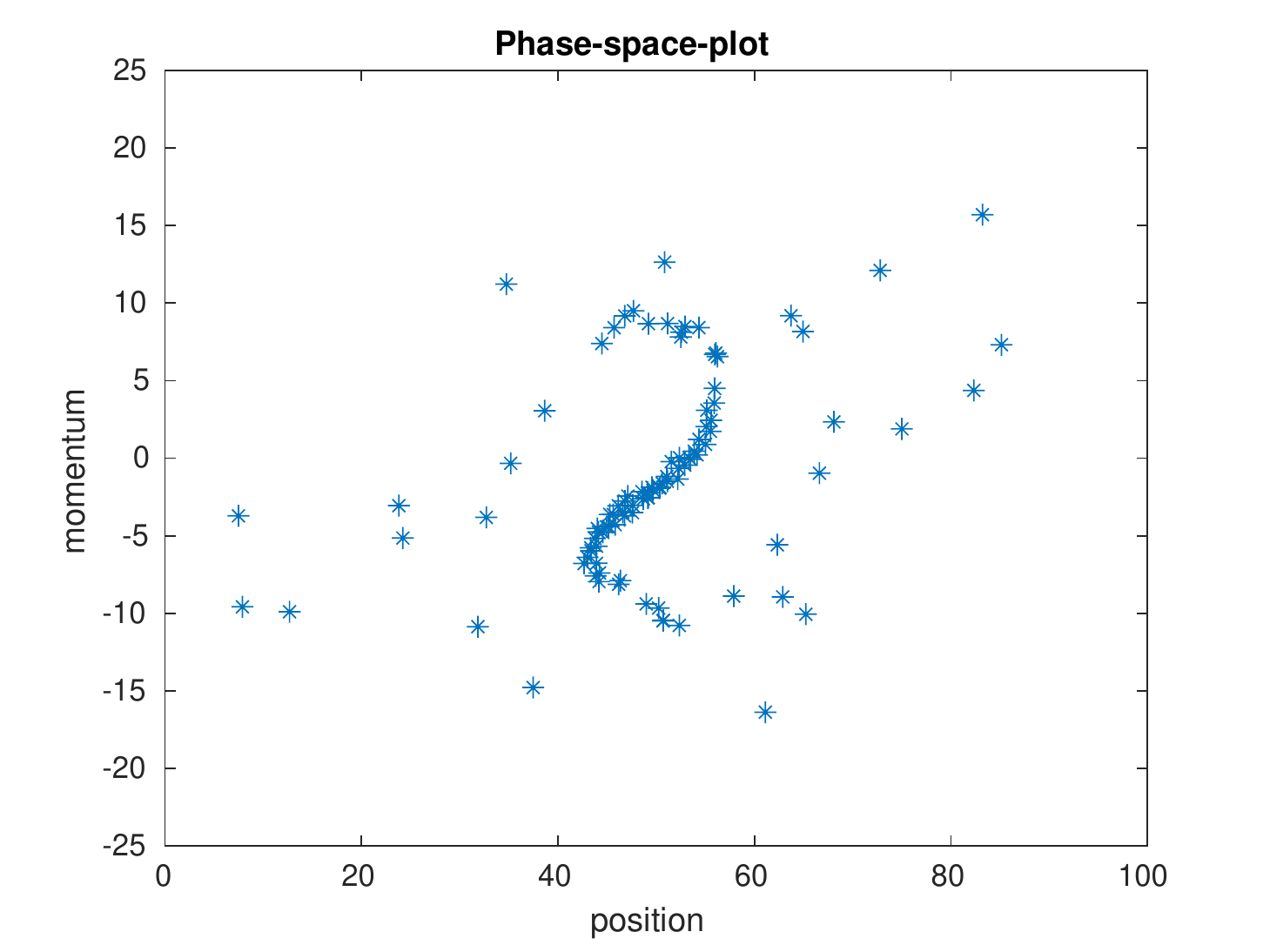}}
\subfloat[$t=320$]{\includegraphics[width = 1.45in]{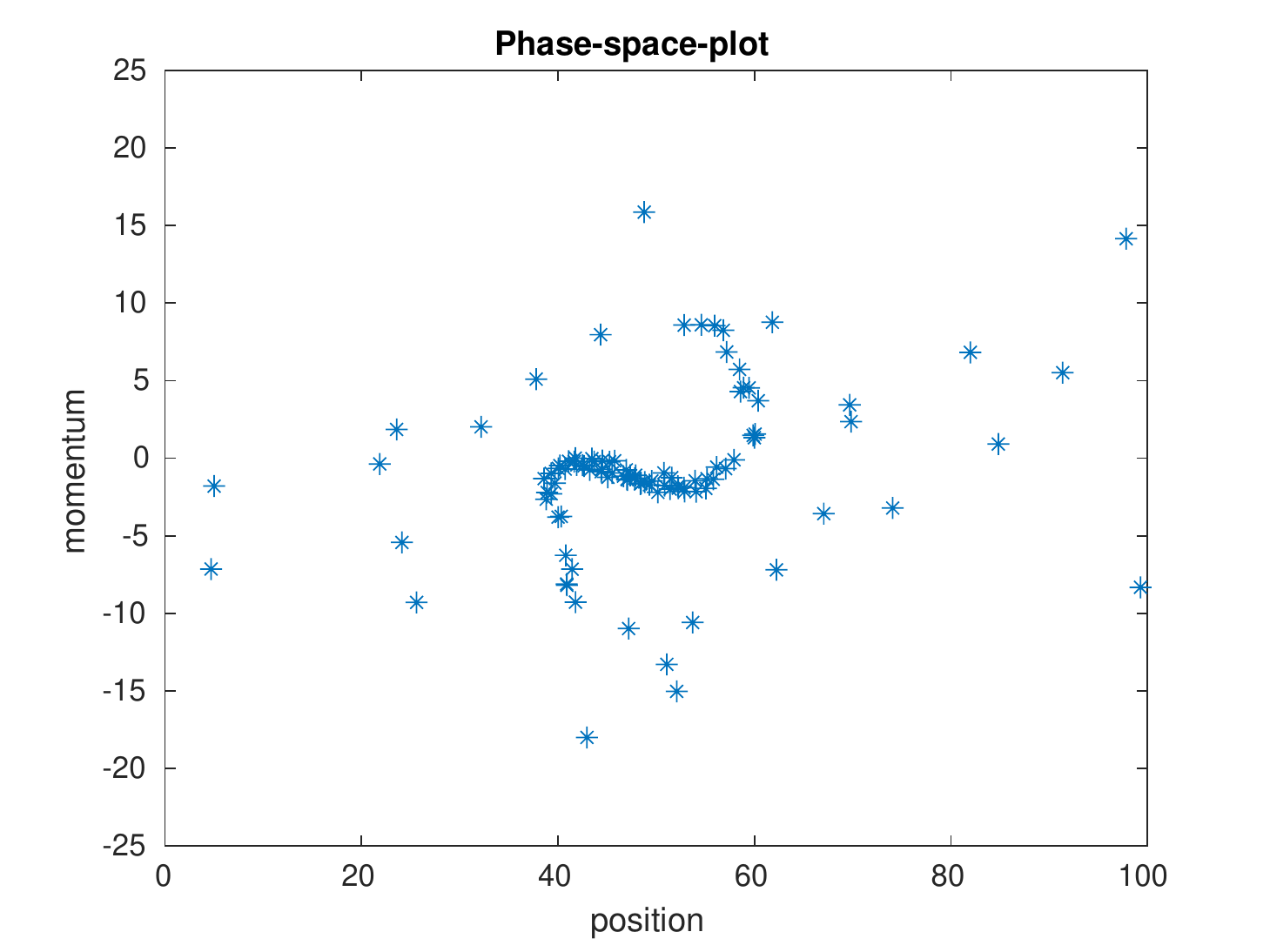}}\\
\subfloat[$t=330$]{\includegraphics[width = 1.45in]{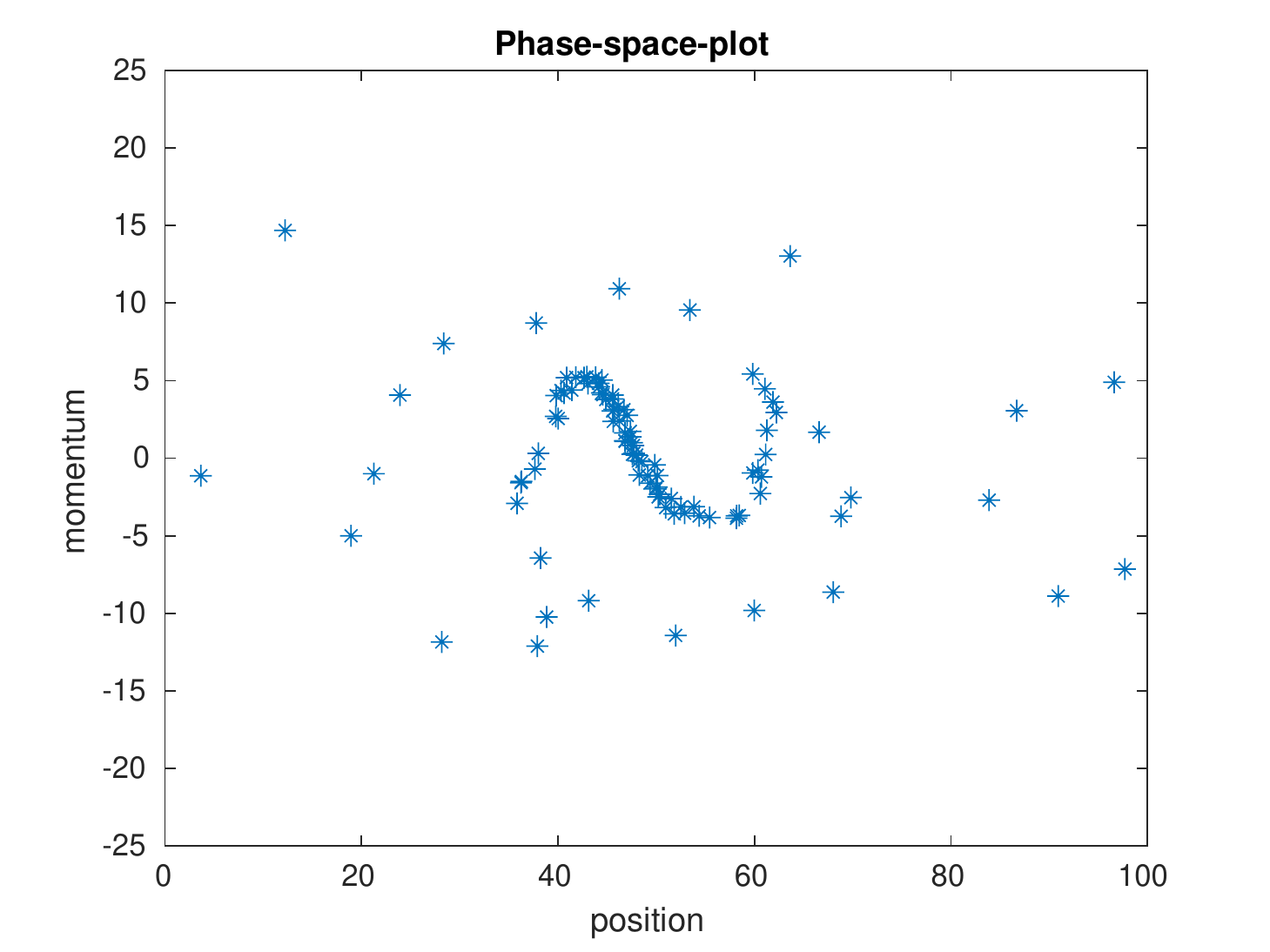}}
\subfloat[$t=340$]{\includegraphics[width = 1.45in]{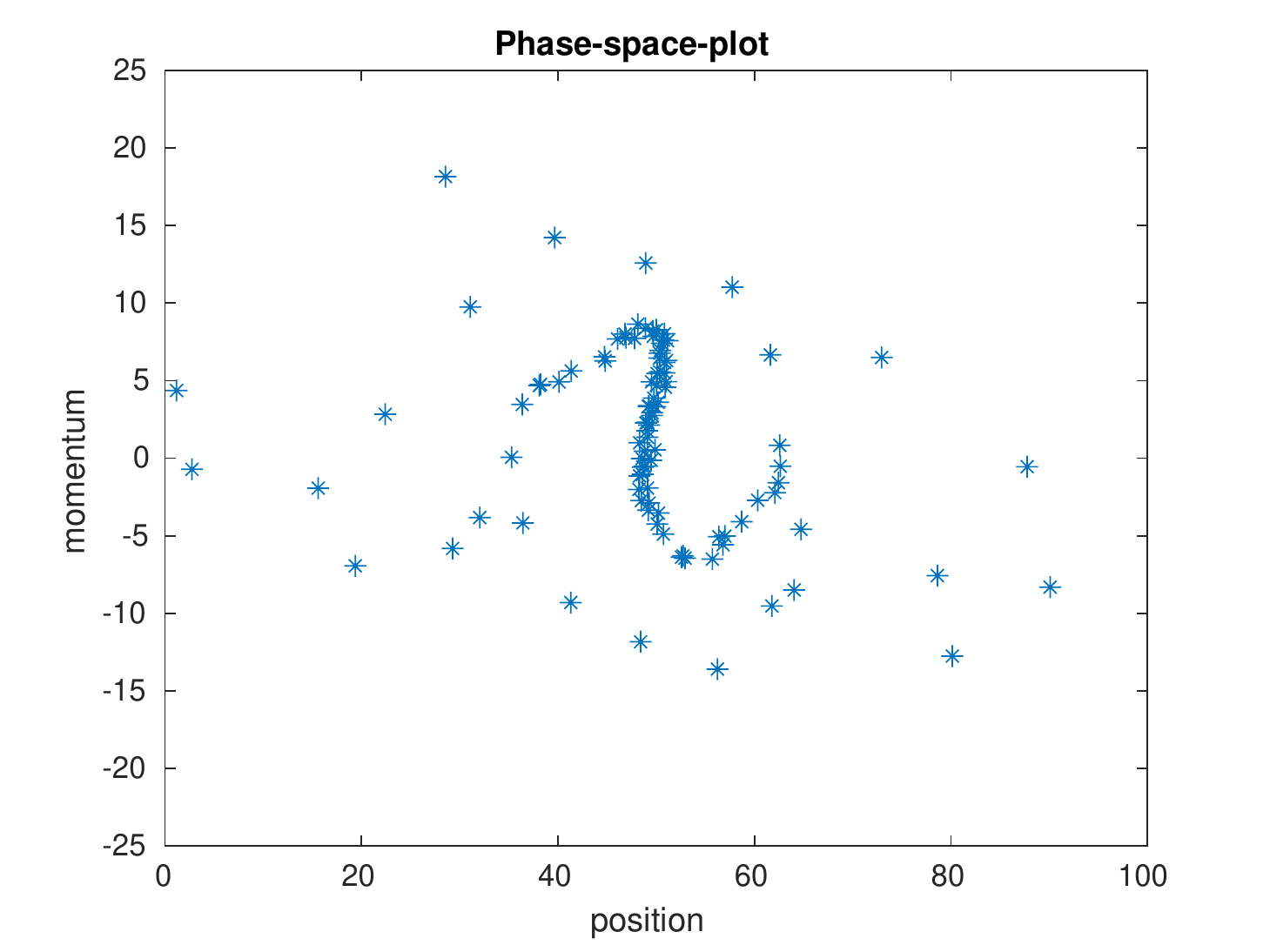}}
\subfloat[$t=350$]{\includegraphics[width = 1.45in]{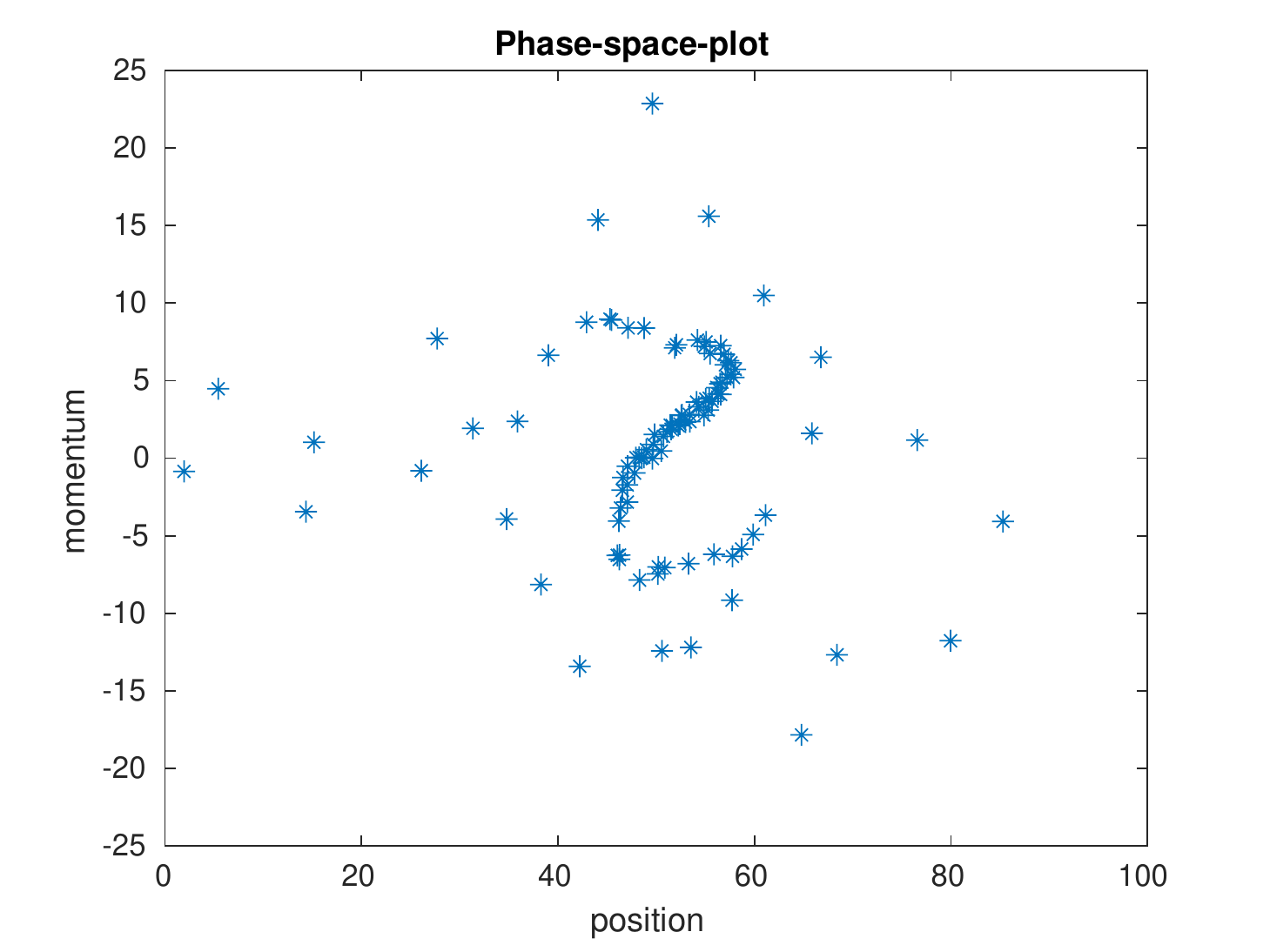}}
\subfloat[$t=360$]{\includegraphics[width = 1.45in]{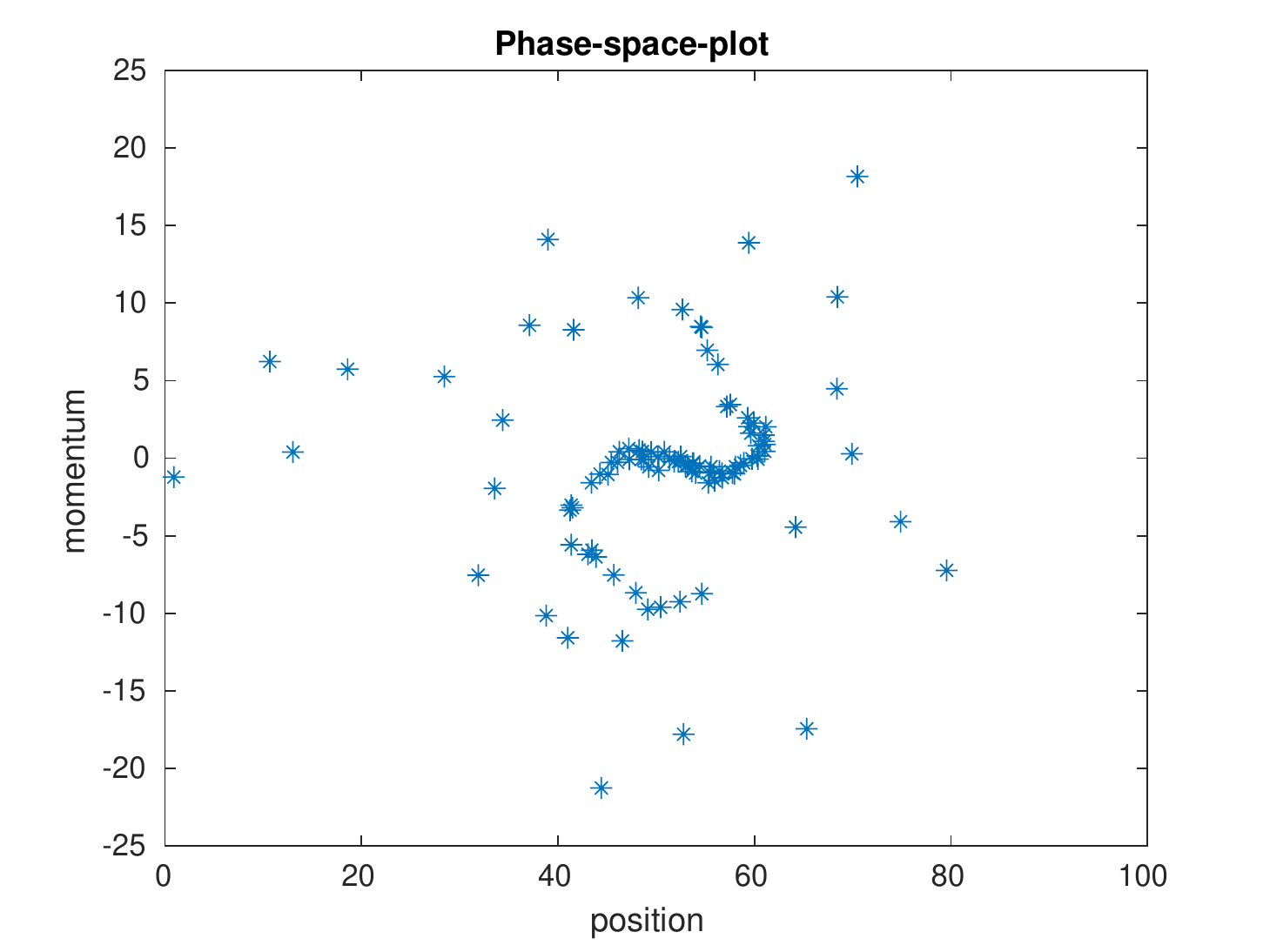}}\\
\subfloat[$t=370$]{\includegraphics[width = 1.45in]{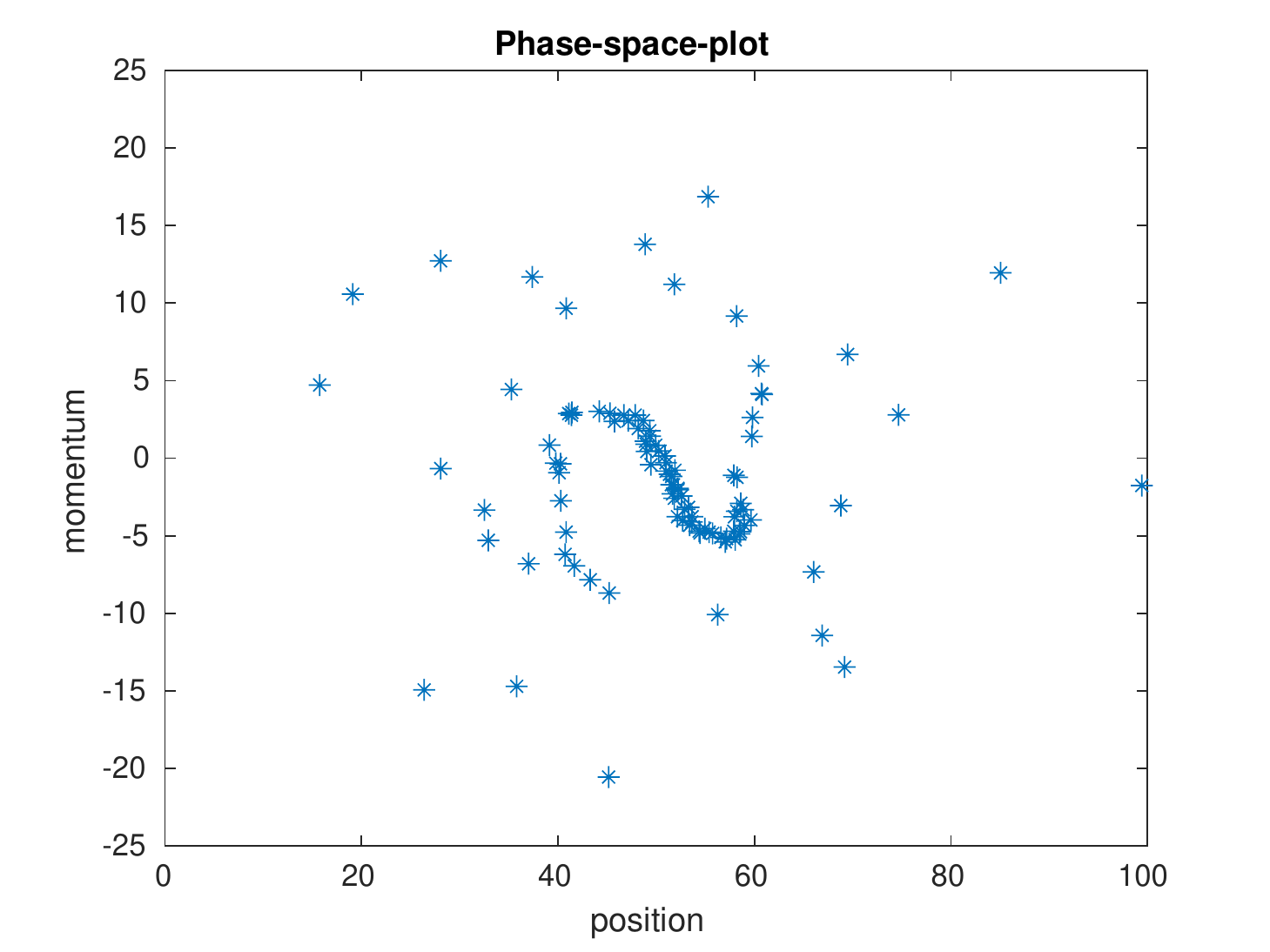}}
\subfloat[$t=380$]{\includegraphics[width = 1.45in]{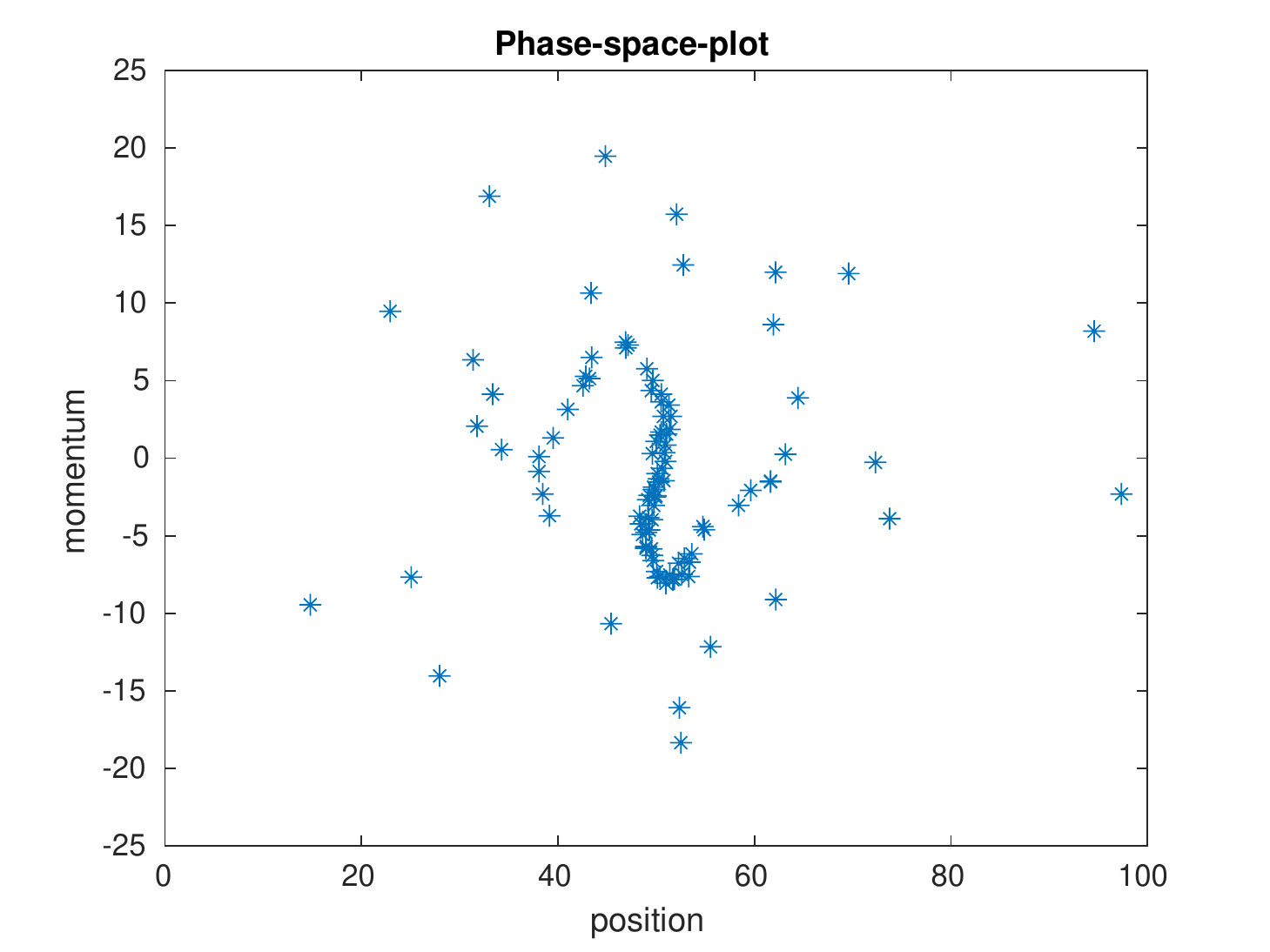}}
\subfloat[$t=390$]{\includegraphics[width = 1.45in]{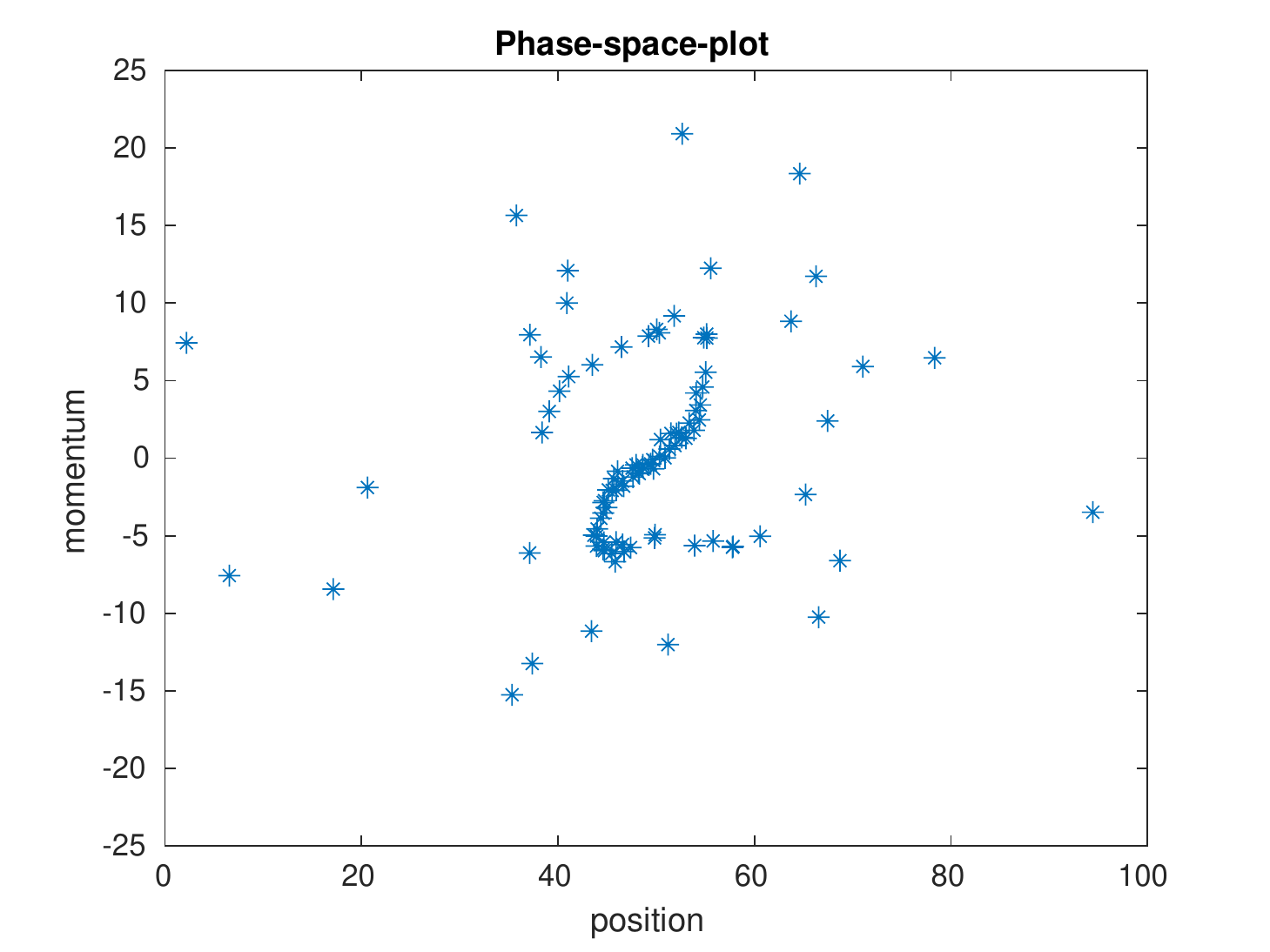}}
\subfloat[$t=400$]{\includegraphics[width = 1.45in]{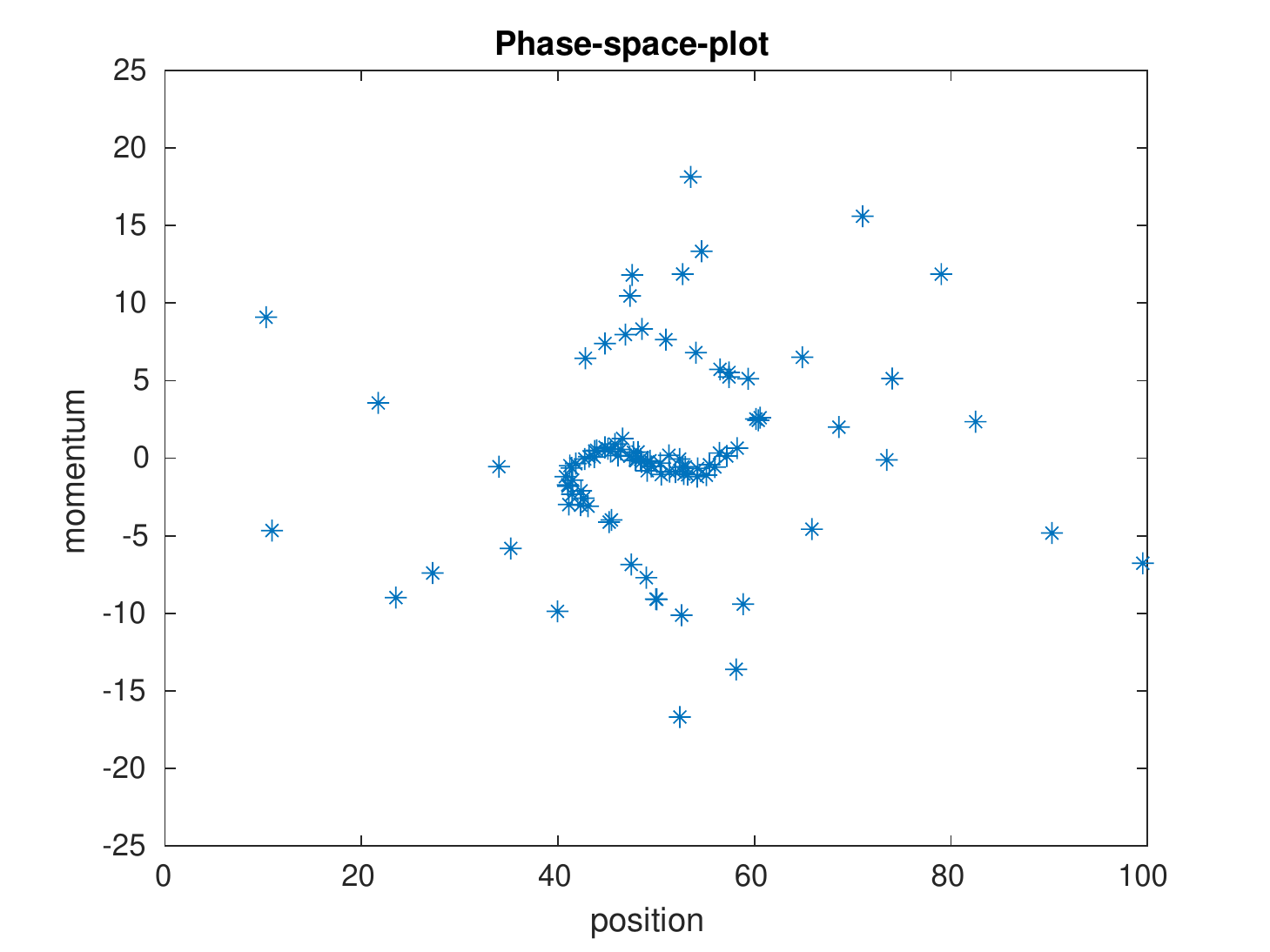}}
\caption{Phase Space snapshots for $t=210$ to $t=400$ with time separation of $10$}
\label{fig_sg_2}
\end{figure}

\begin{figure}[ht]
\centering
\subfloat[$t=410$]{\includegraphics[width = 1.45in]{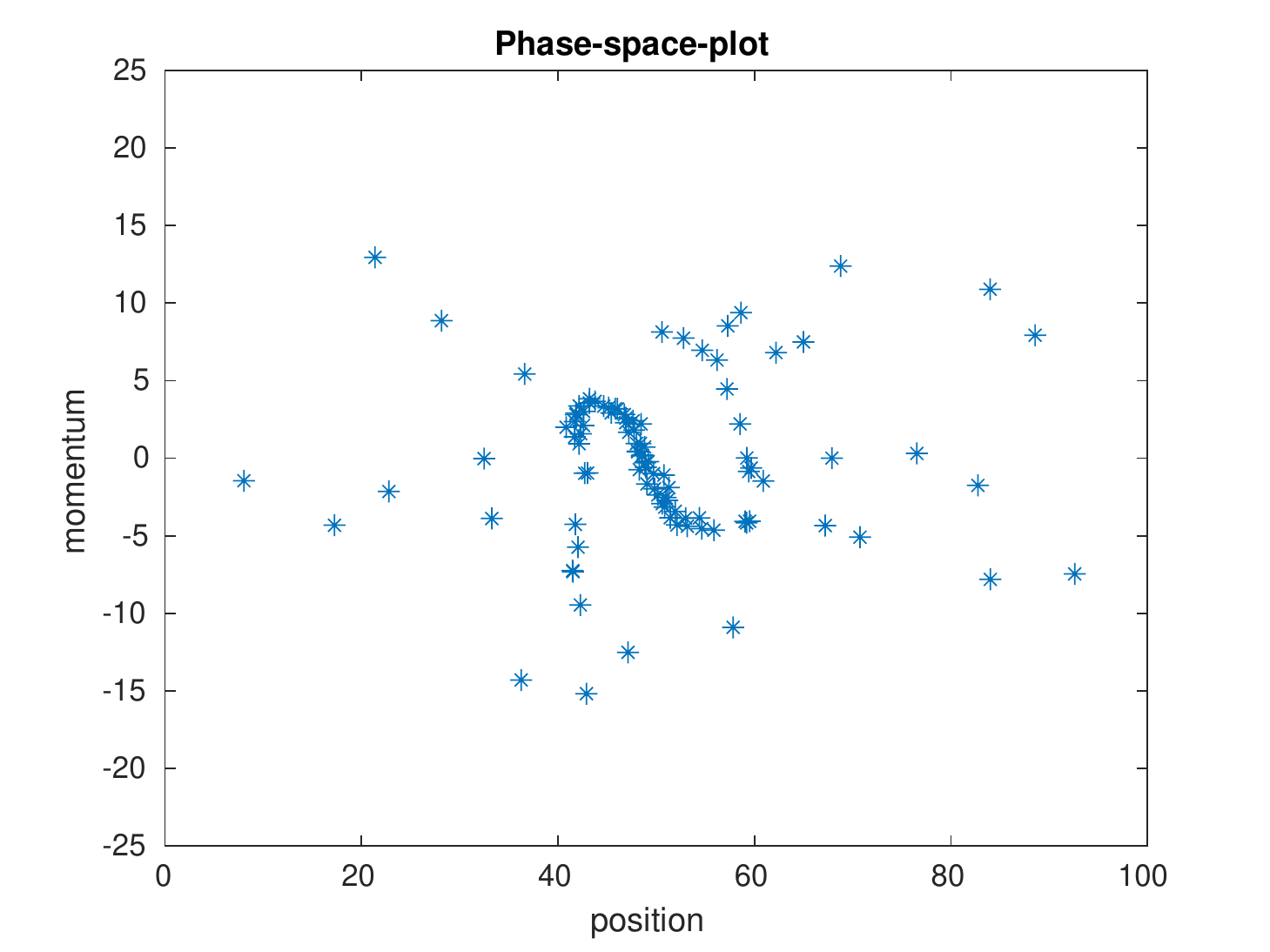}}
\subfloat[$t=420$]{\includegraphics[width = 1.45in]{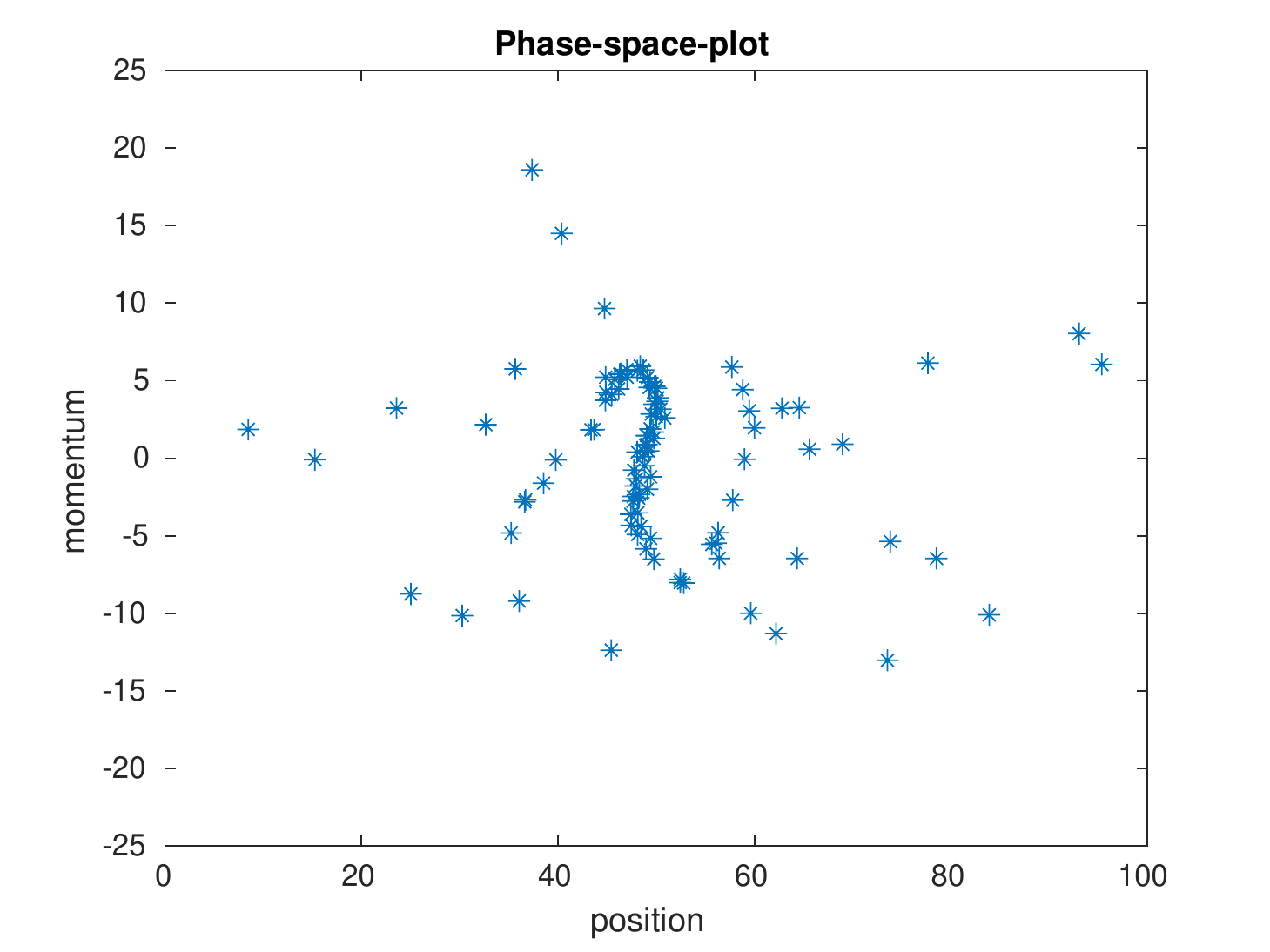}}
\subfloat[$t=430$]{\includegraphics[width = 1.45in]{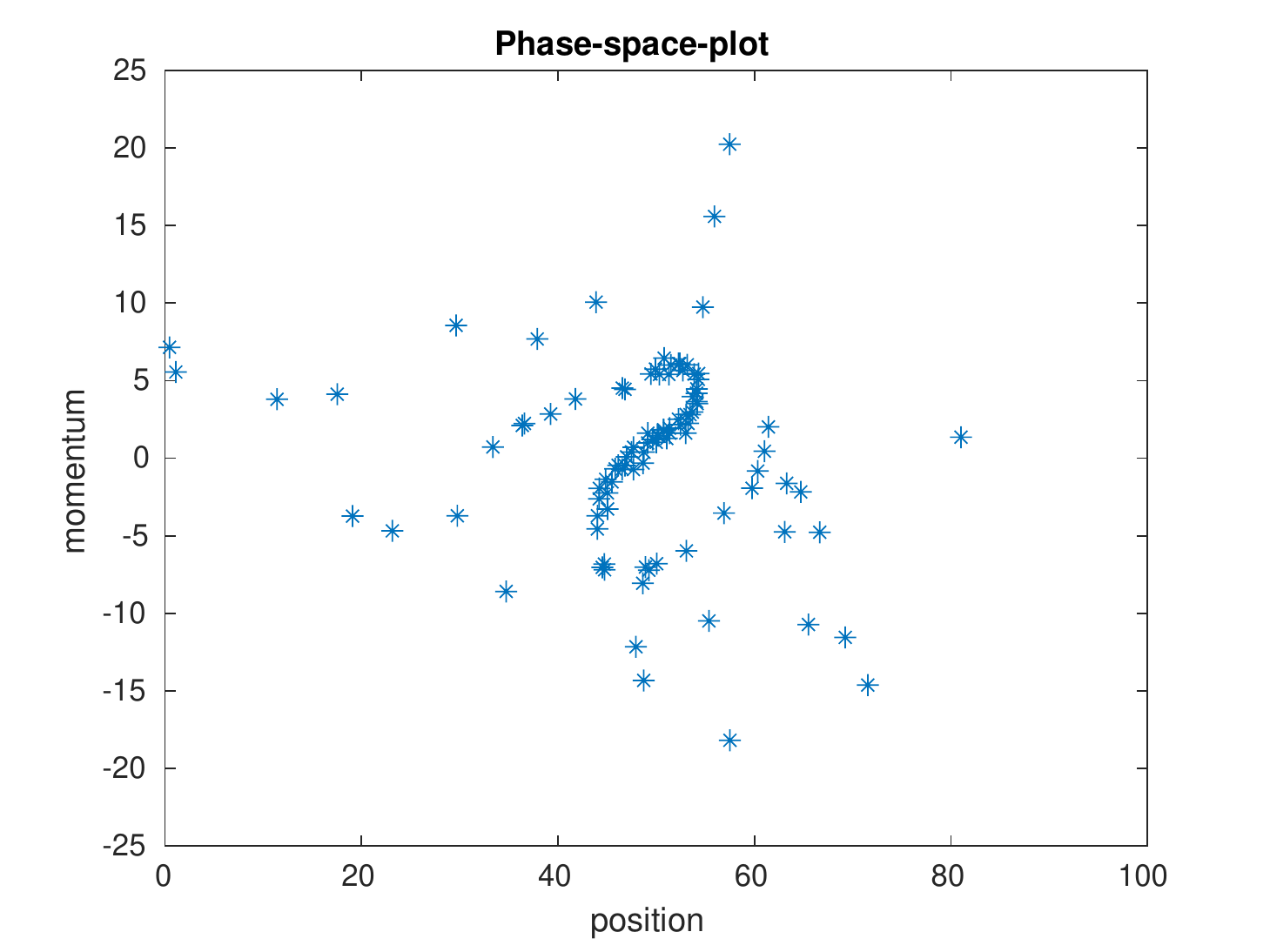}}
\subfloat[$t=440$]{\includegraphics[width = 1.45in]{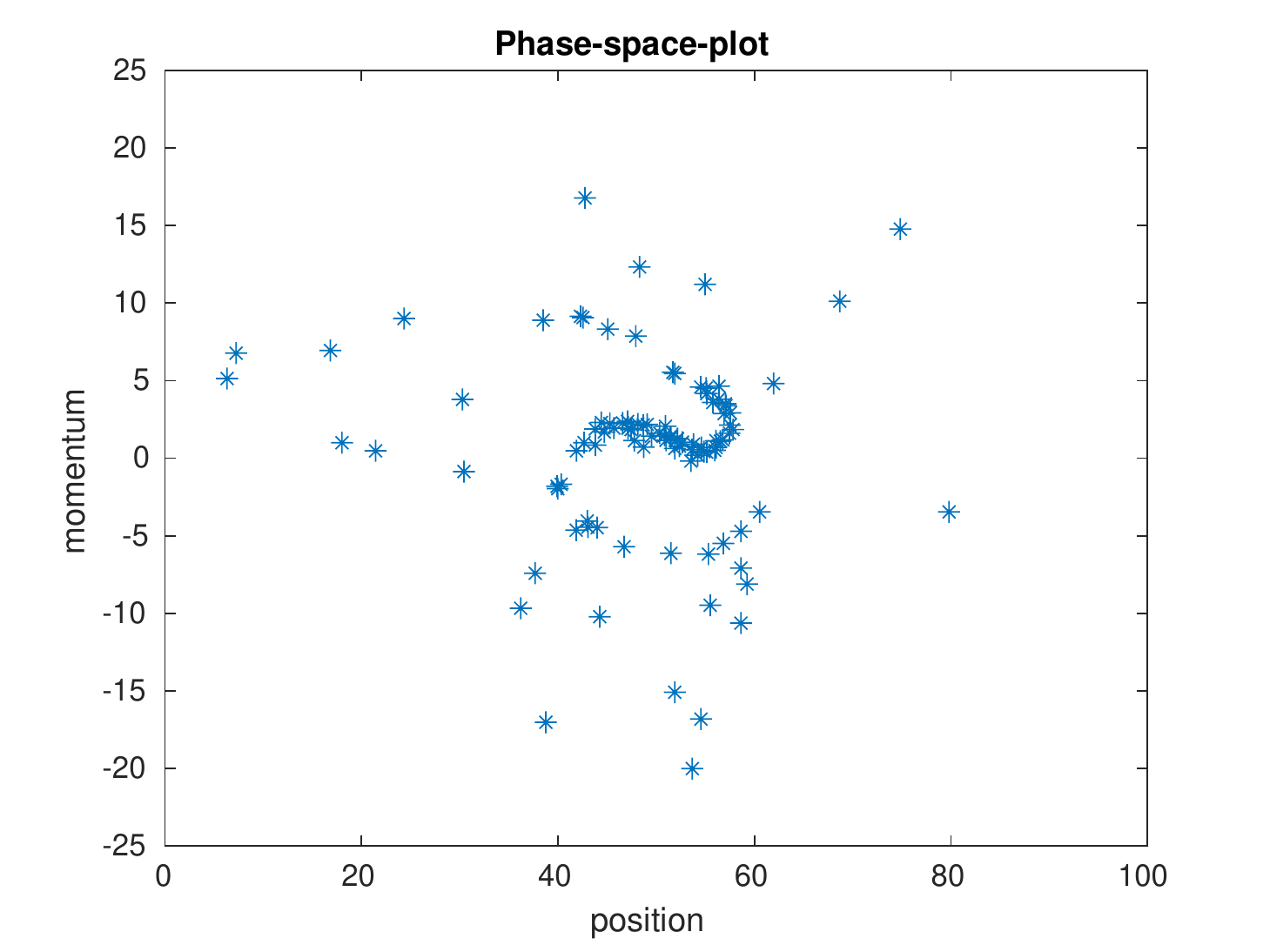}}\\
\subfloat[$t=450$]{\includegraphics[width = 1.45in]{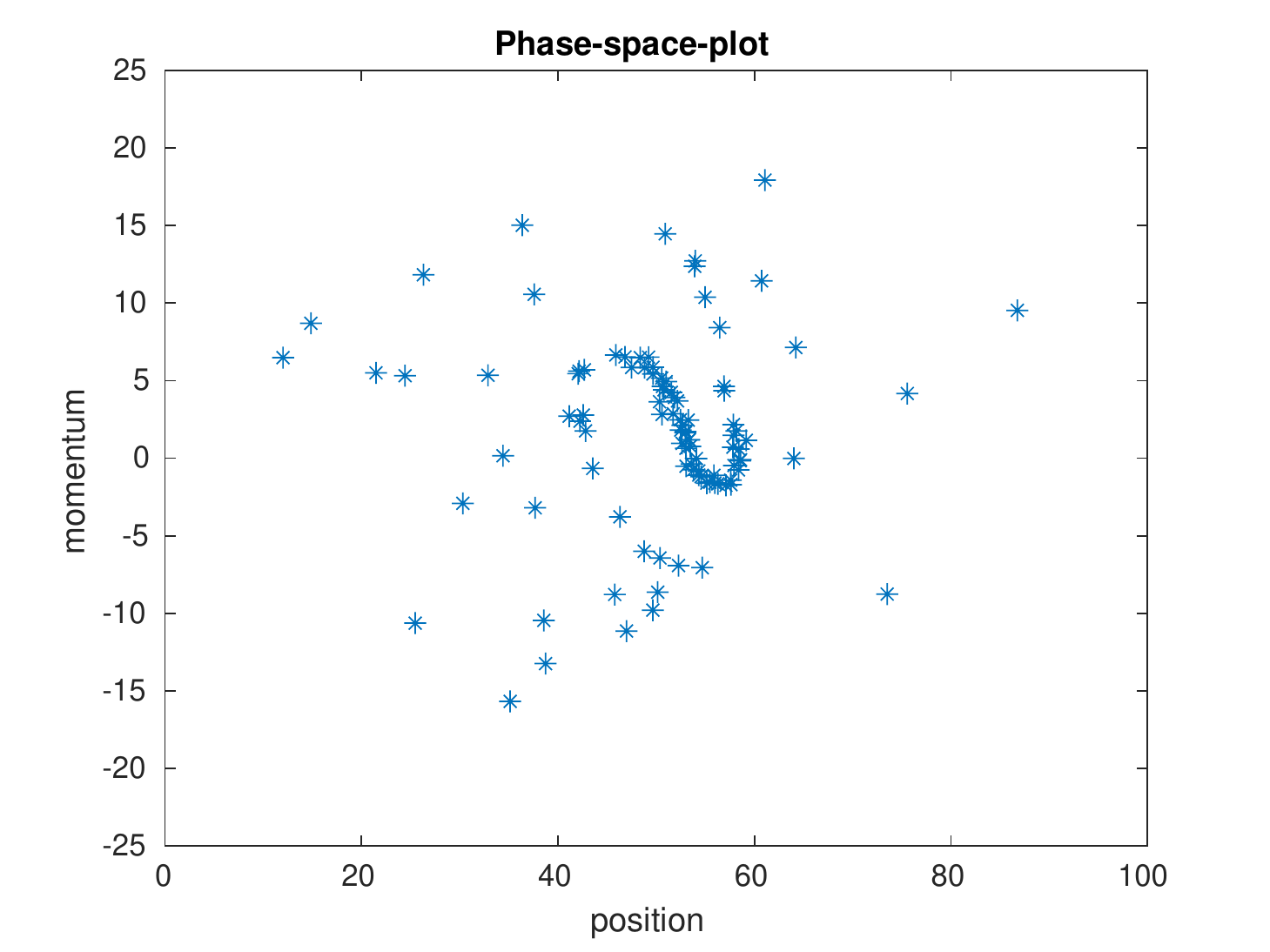}}
\subfloat[$t=460$]{\includegraphics[width = 1.45in]{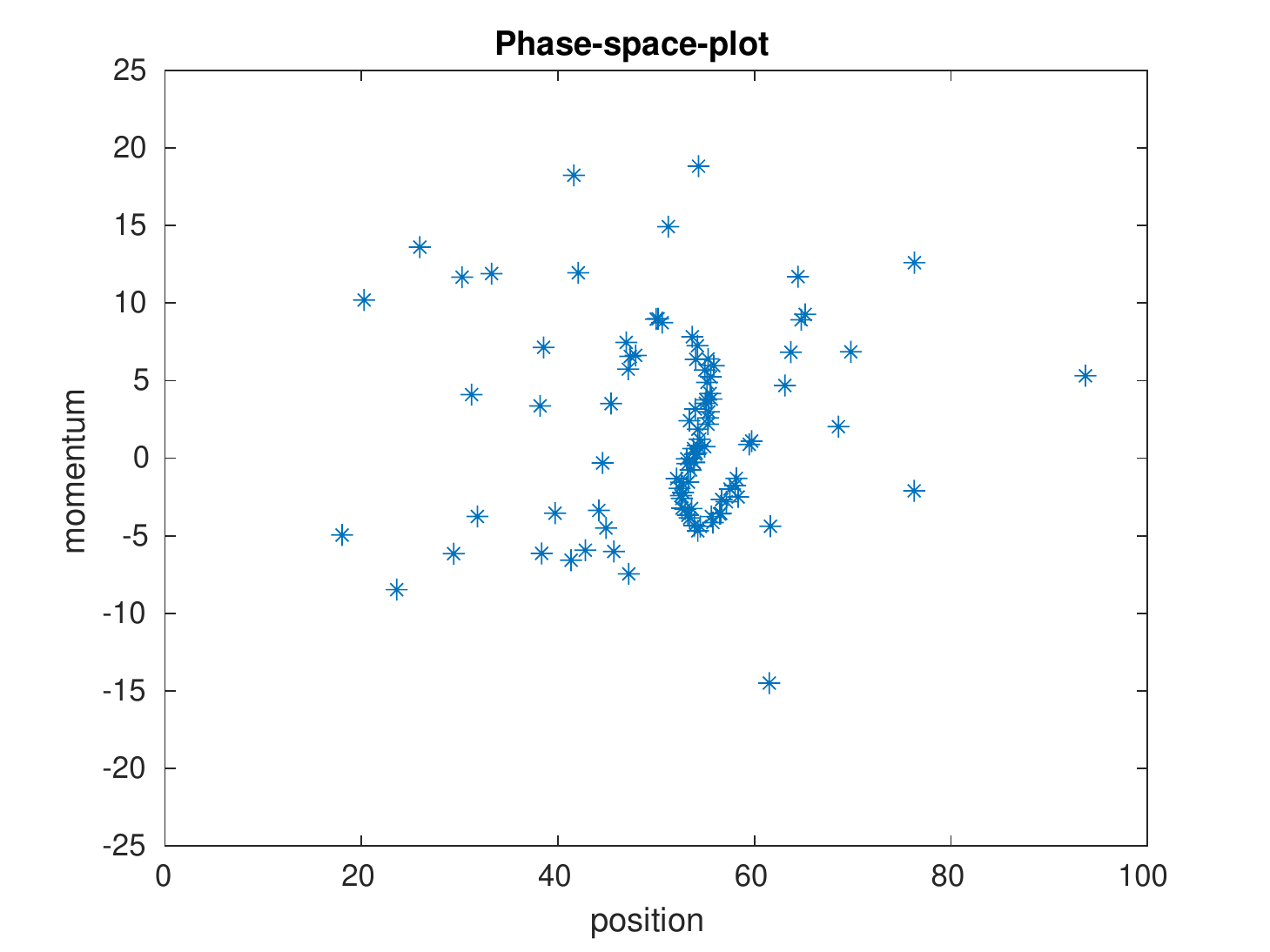}}
\subfloat[$t=470$]{\includegraphics[width = 1.45in]{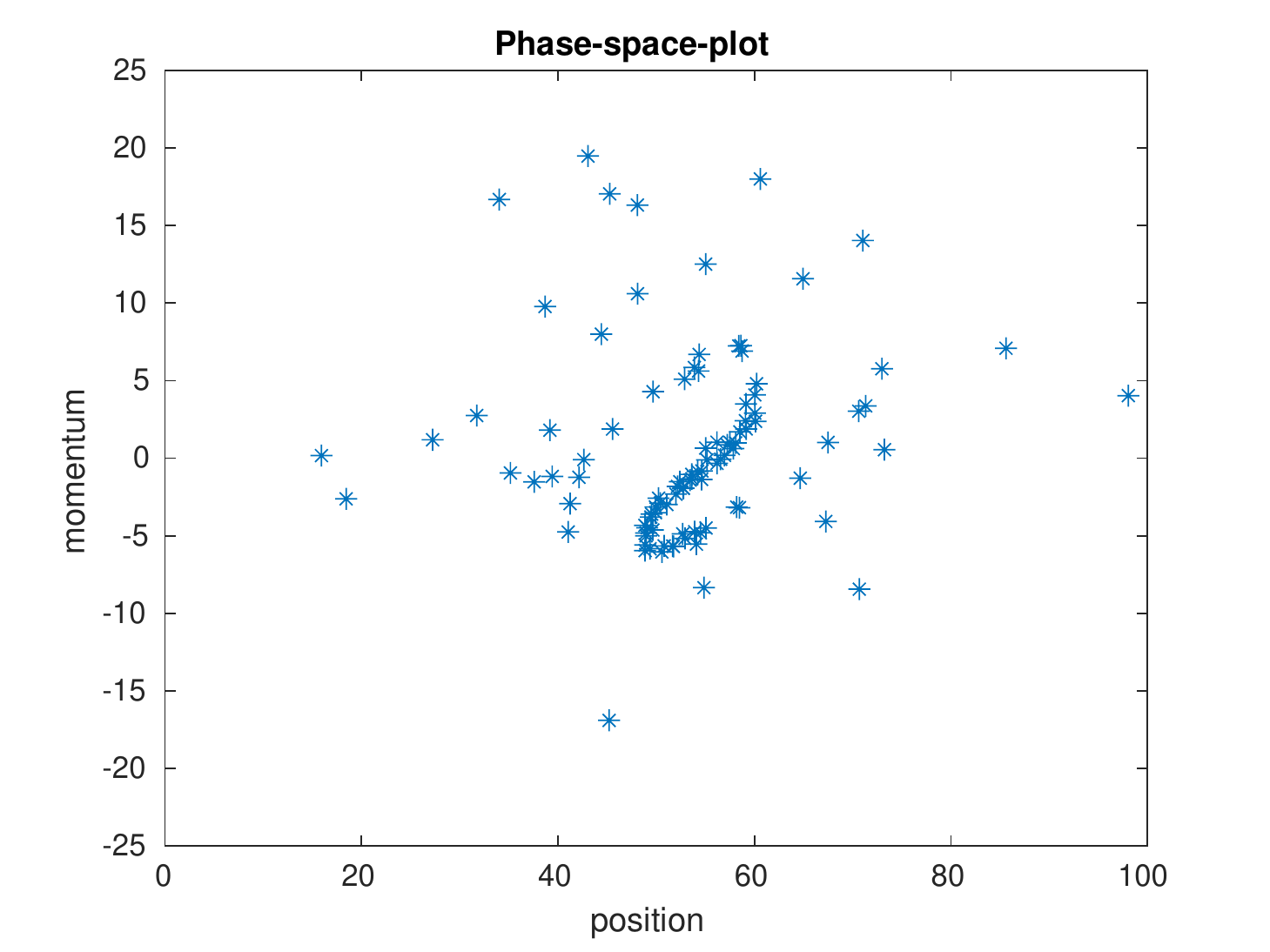}}
\subfloat[$t=480$]{\includegraphics[width = 1.45in]{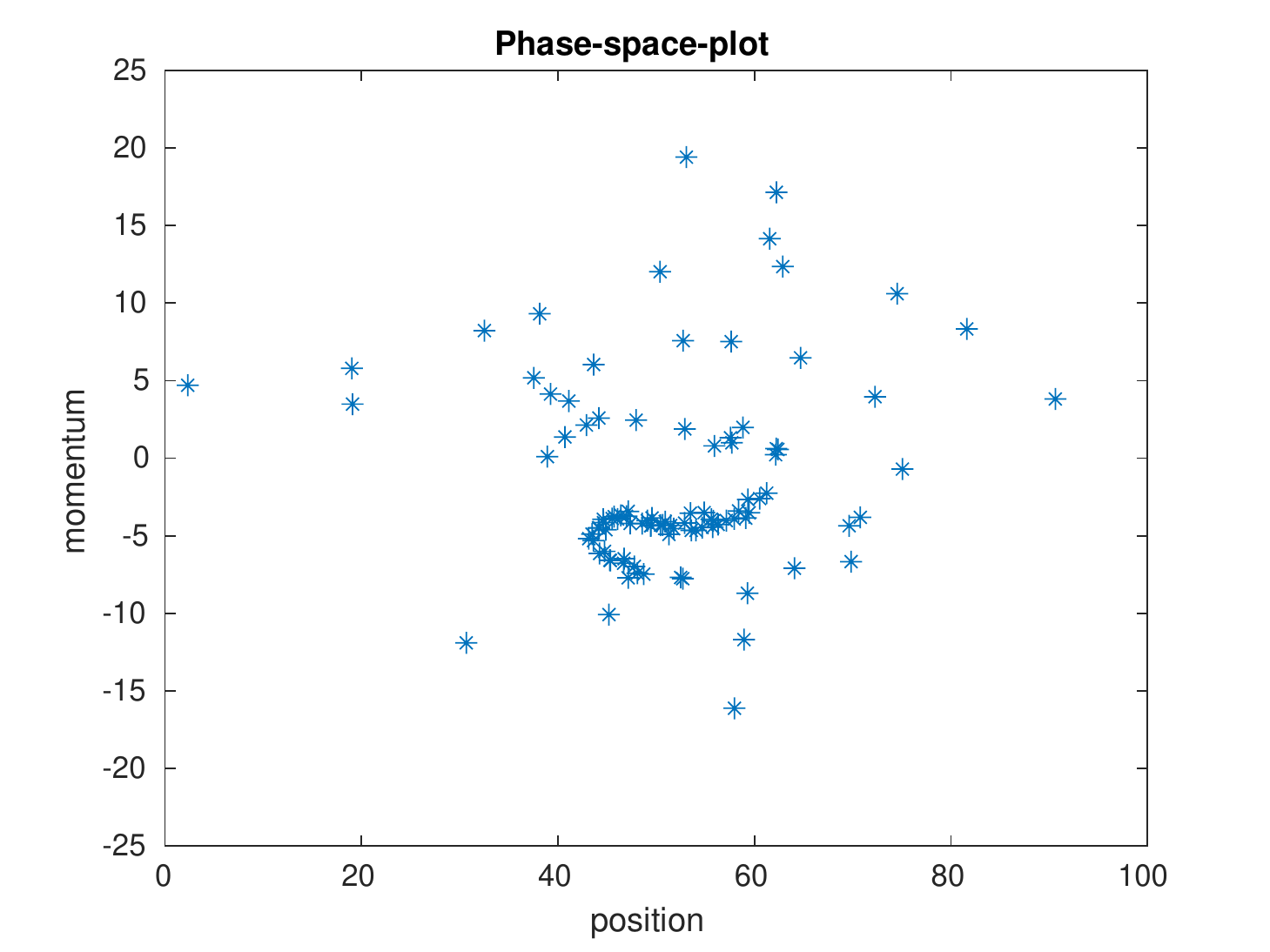}}\\
\subfloat[$t=490$]{\includegraphics[width = 1.45in]{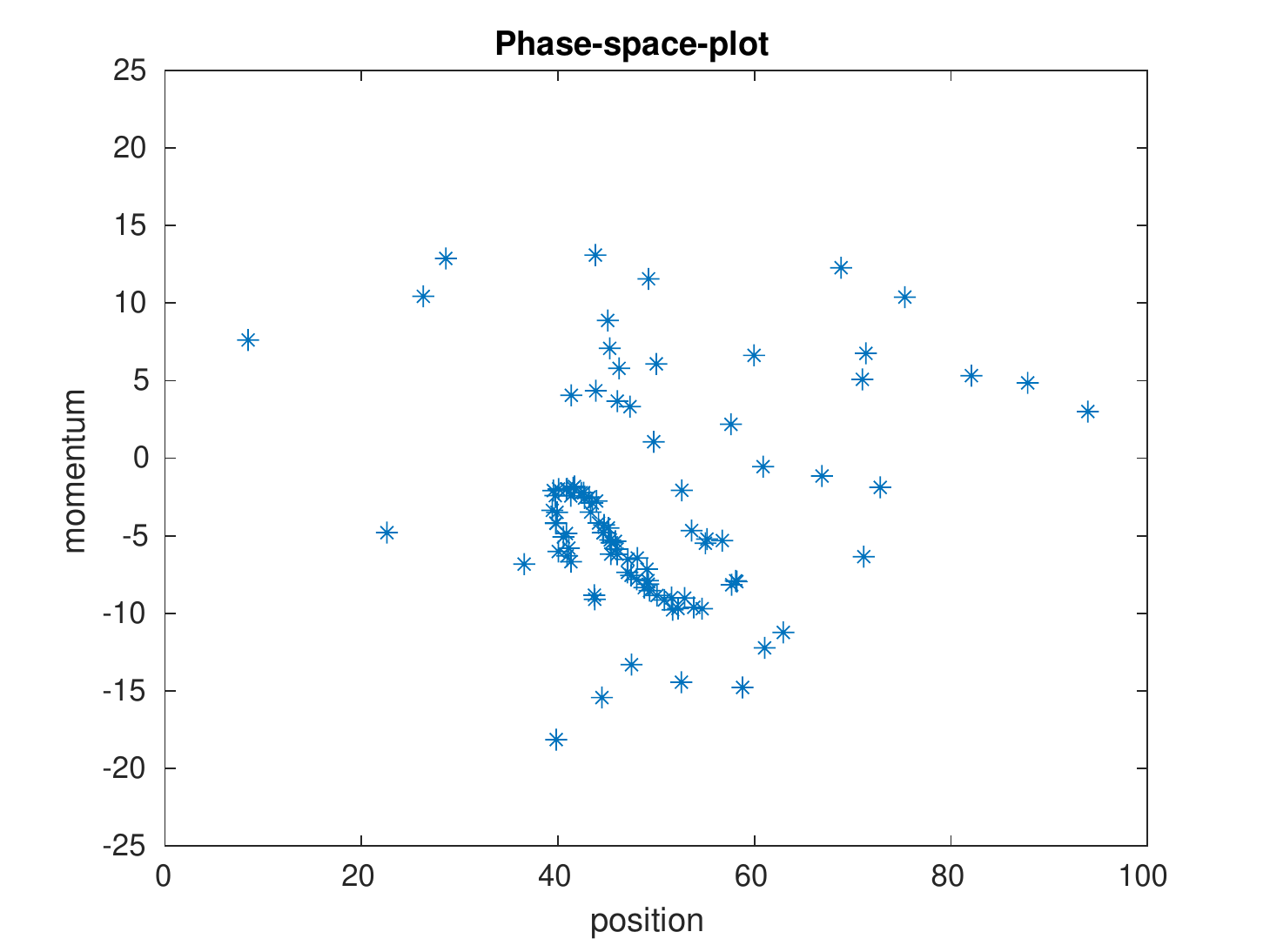}}
\subfloat[$t=500$]{\includegraphics[width = 1.45in]{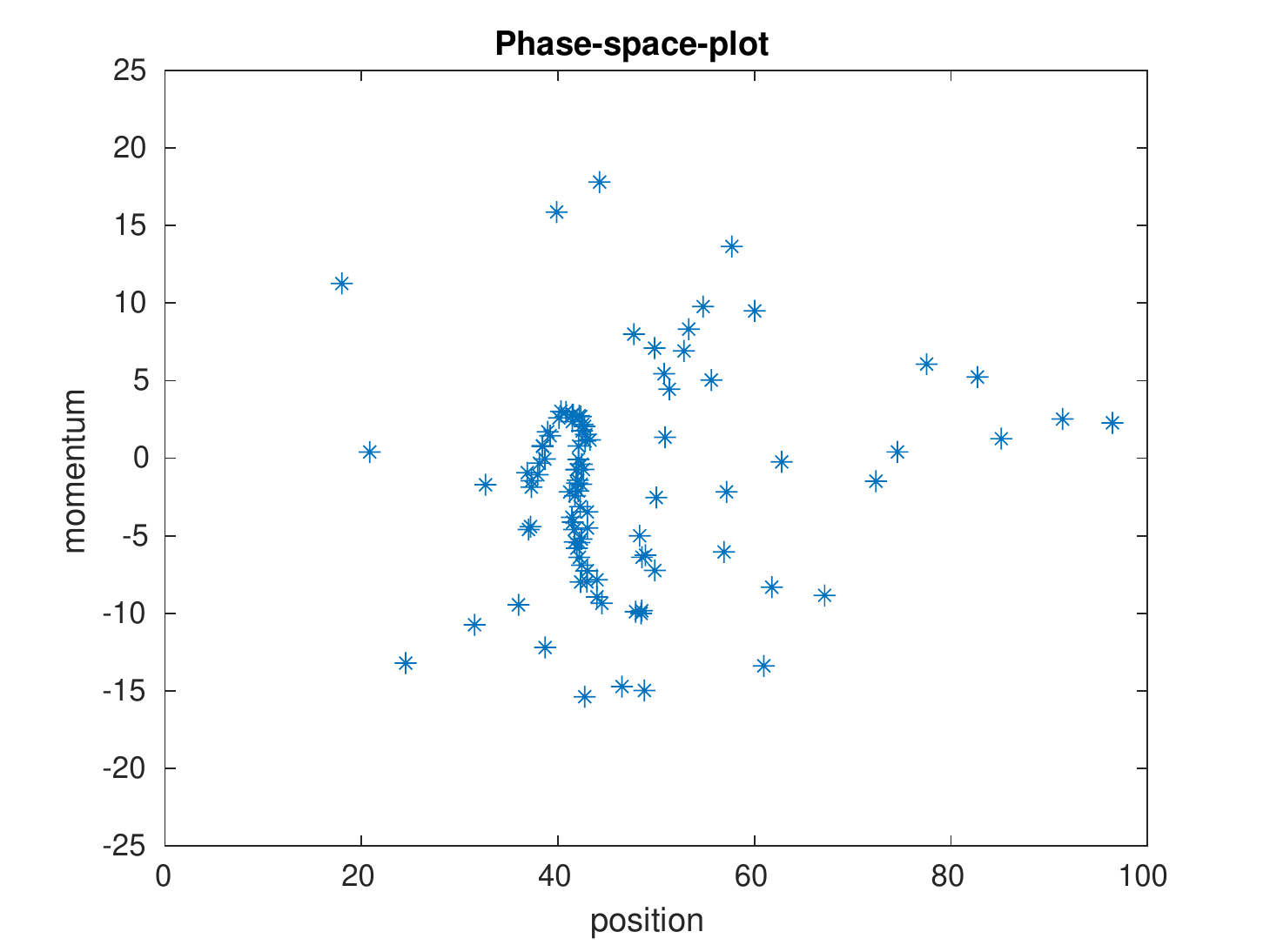}}
\subfloat[$t=510$]{\includegraphics[width = 1.45in]{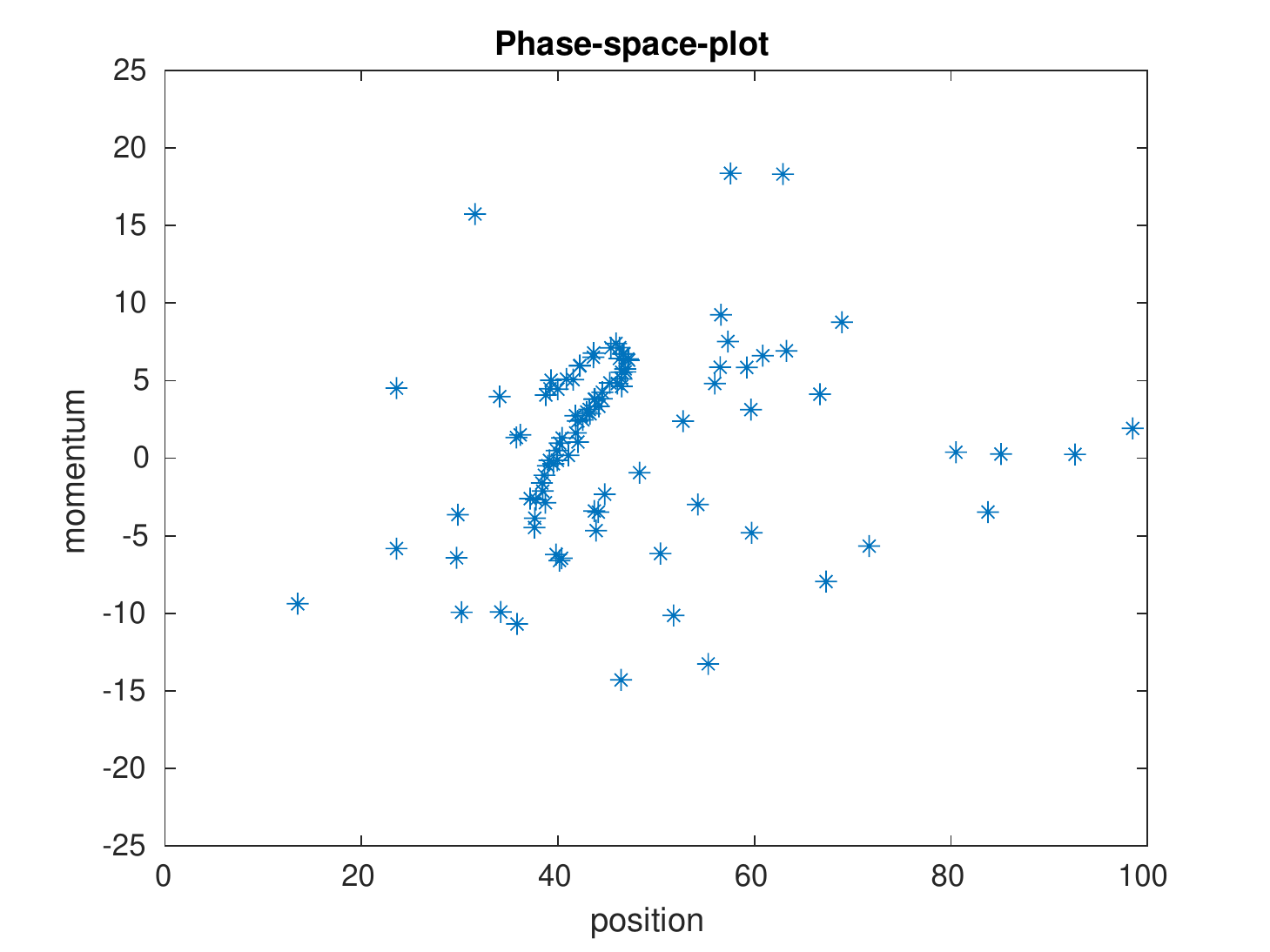}}
\subfloat[$t=520$]{\includegraphics[width = 1.45in]{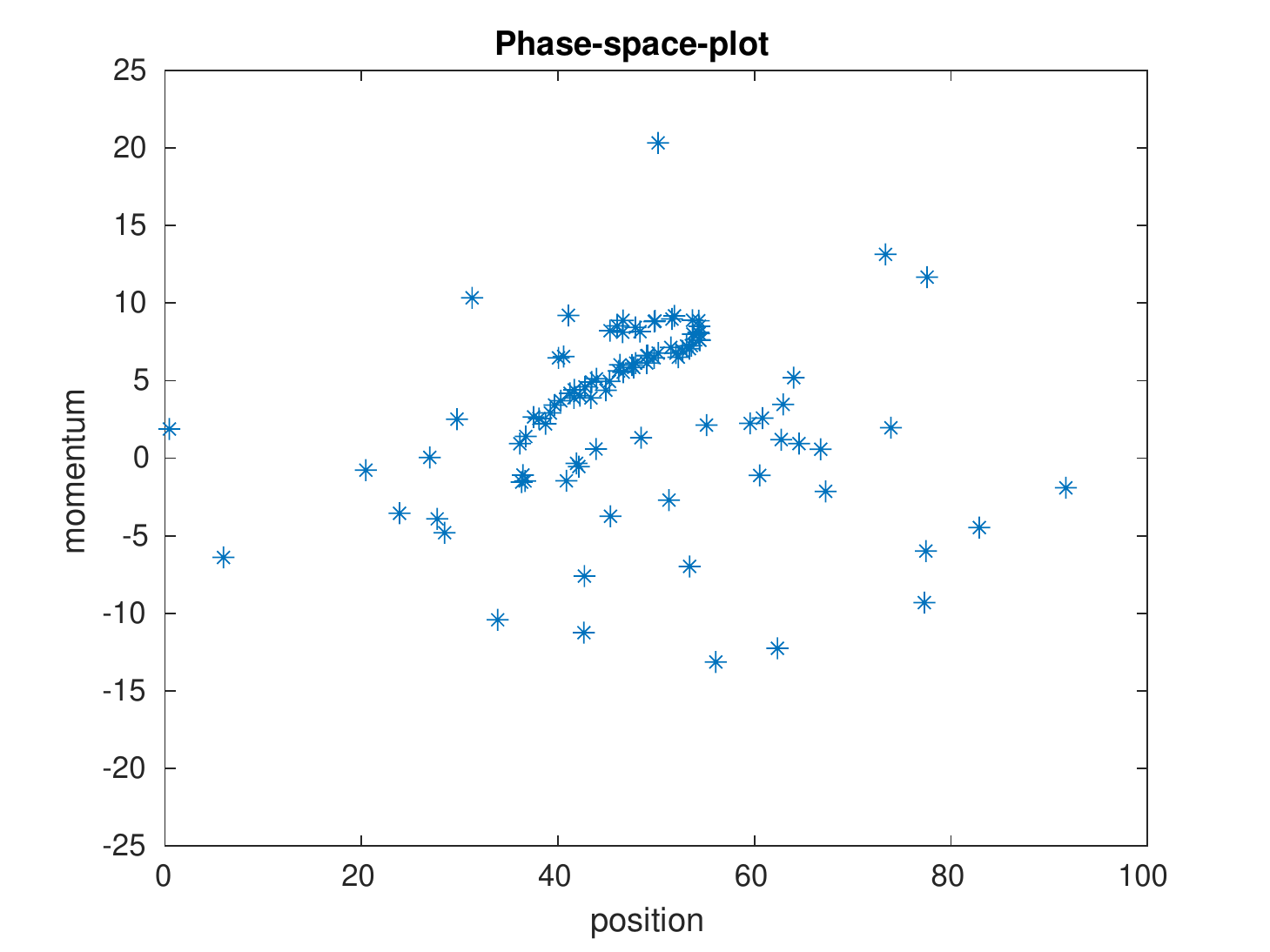}}\\
\subfloat[$t=530$]{\includegraphics[width = 1.45in]{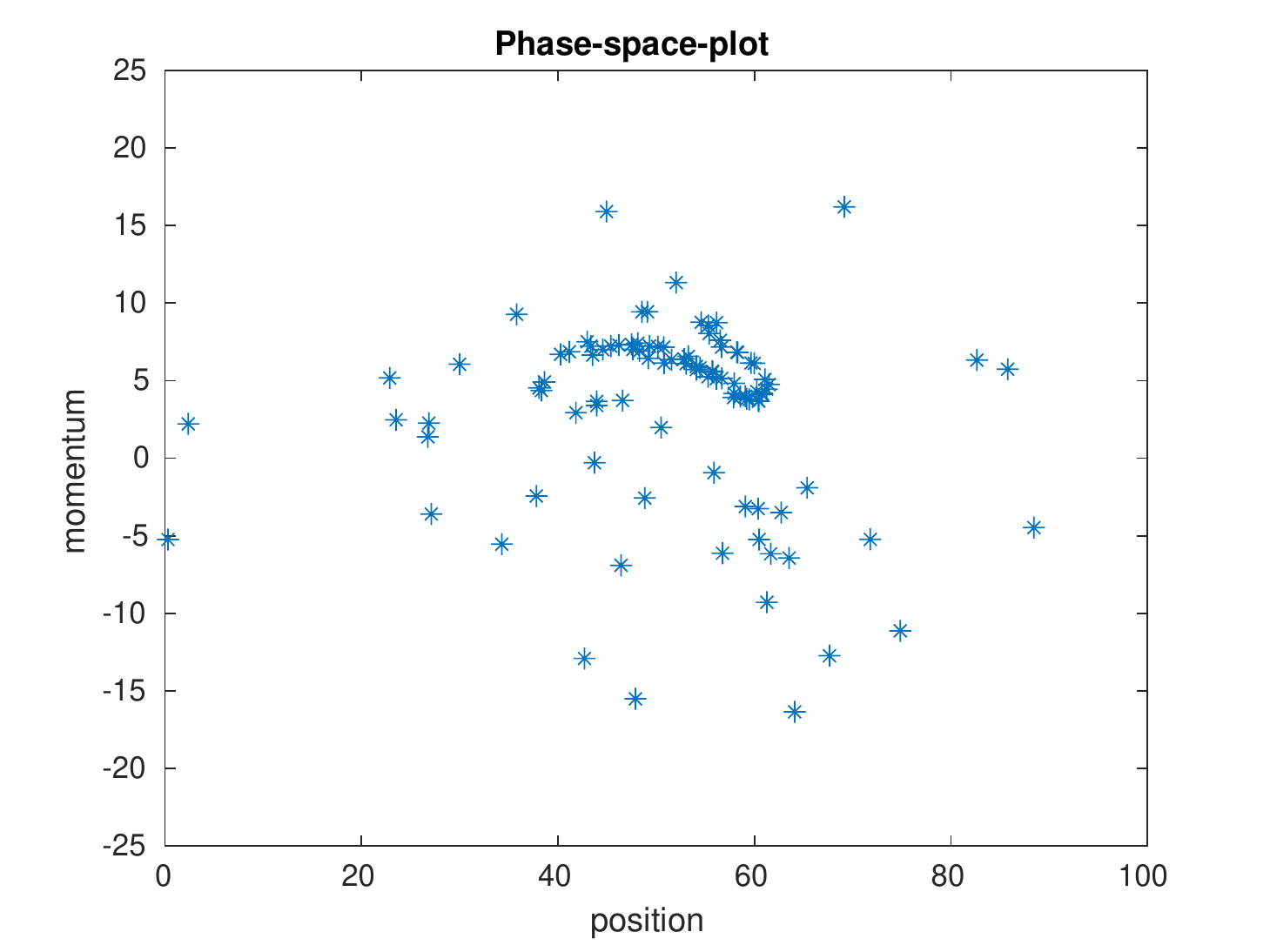}}
\subfloat[$t=540$]{\includegraphics[width = 1.45in]{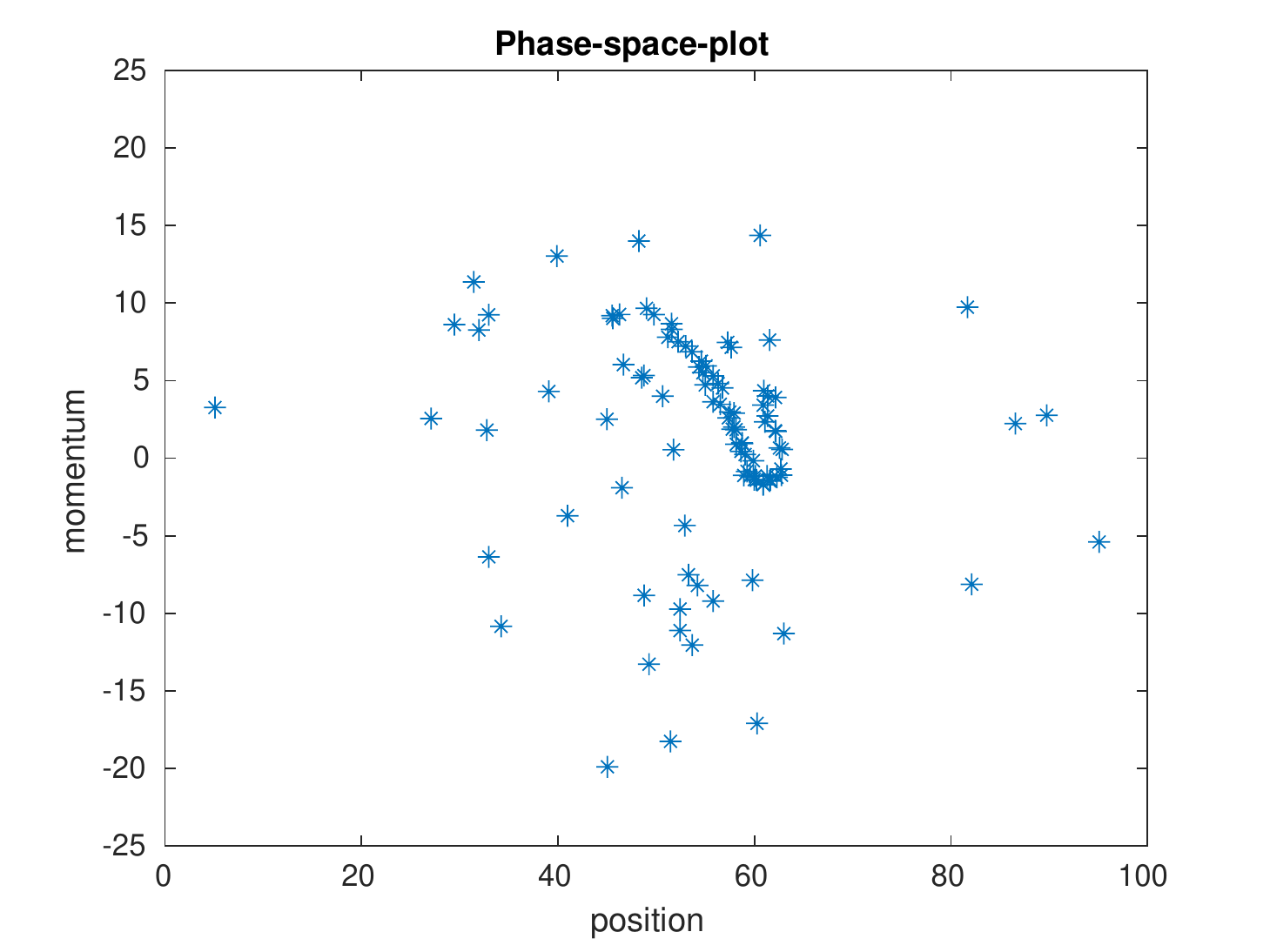}}
\subfloat[$t=550$]{\includegraphics[width = 1.45in]{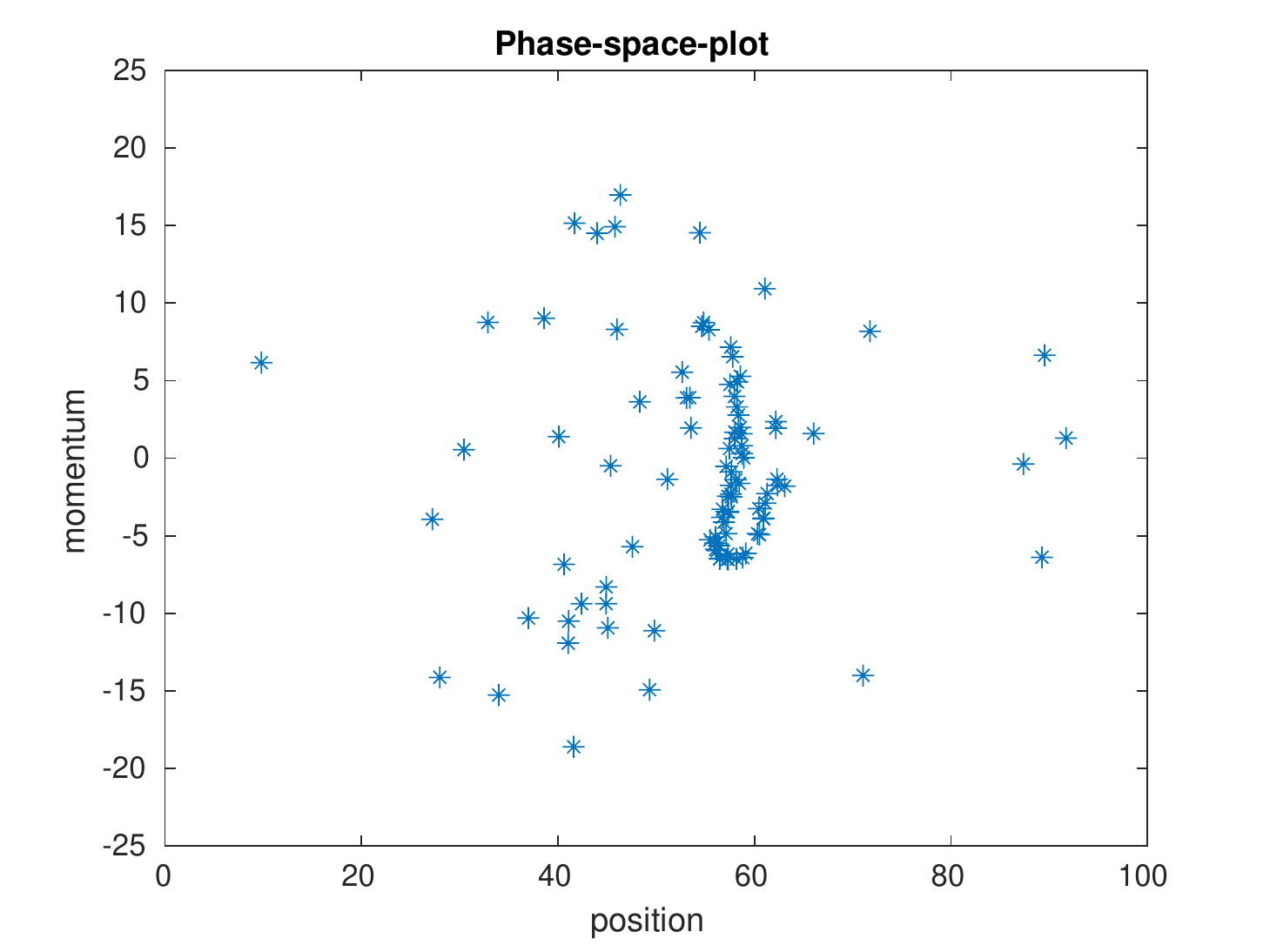}}
\subfloat[$t=560$]{\includegraphics[width = 1.45in]{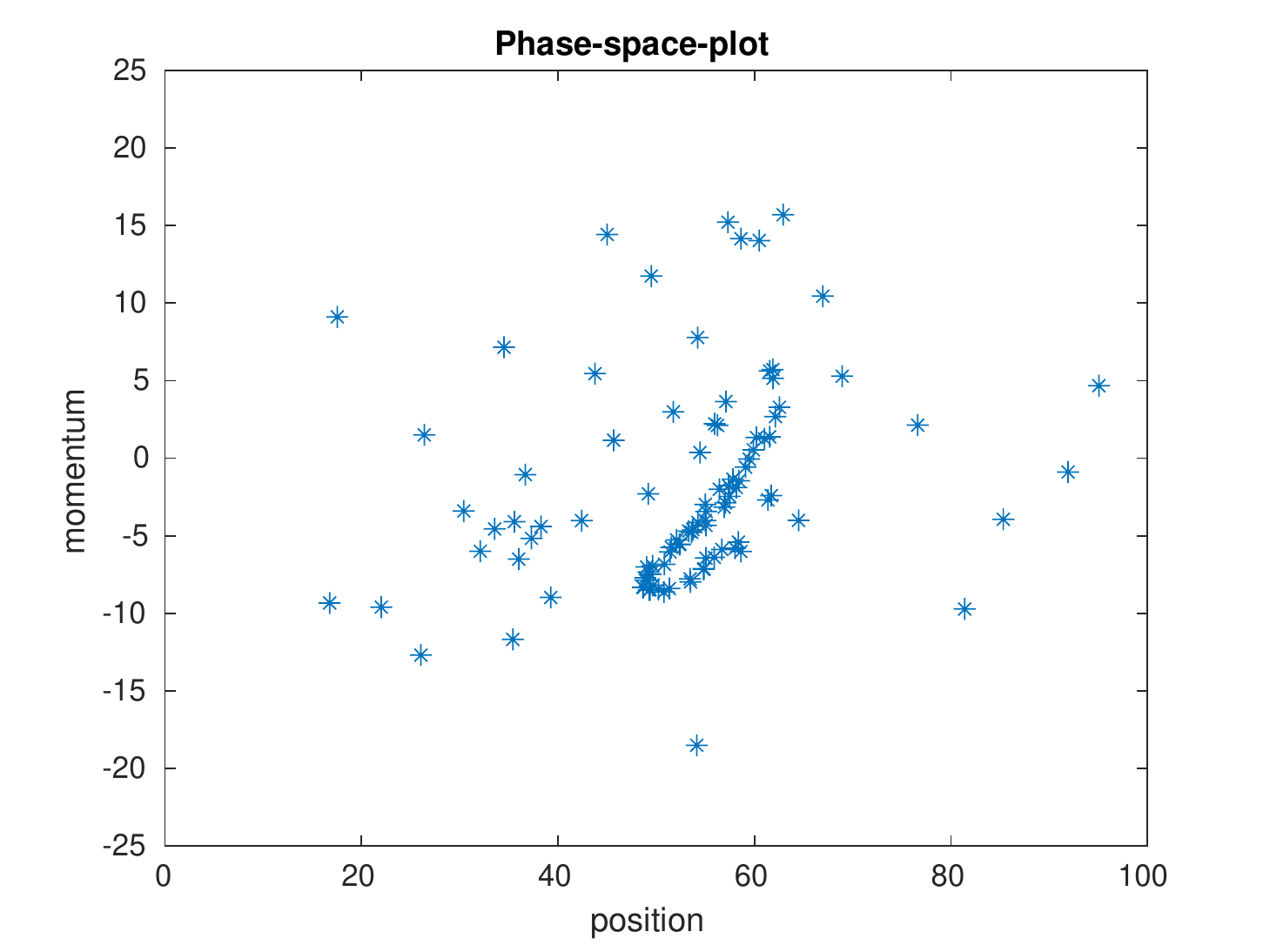}}\\
\subfloat[$t=570$]{\includegraphics[width = 1.45in]{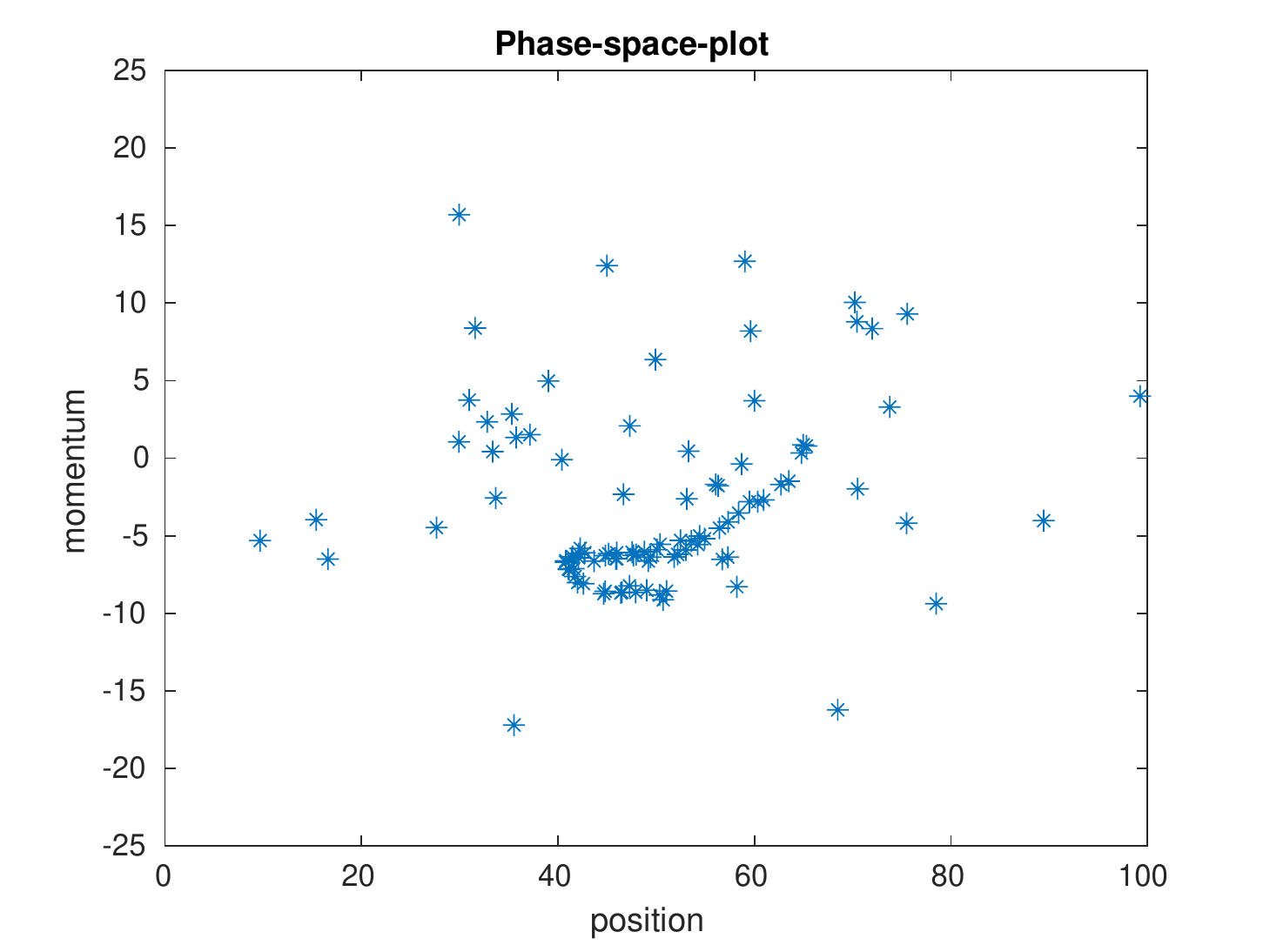}}
\subfloat[$t=580$]{\includegraphics[width = 1.45in]{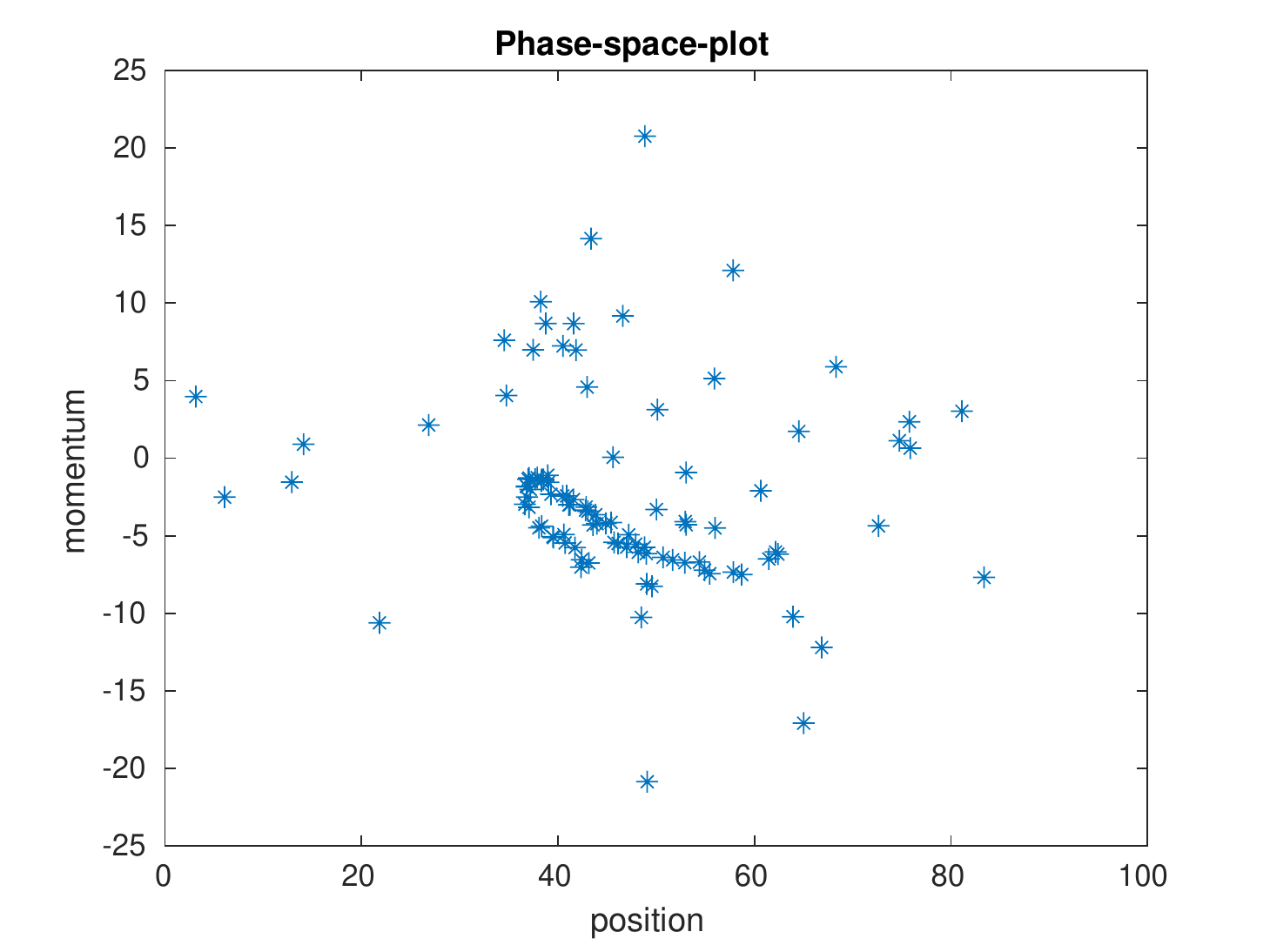}}
\subfloat[$t=590$]{\includegraphics[width = 1.45in]{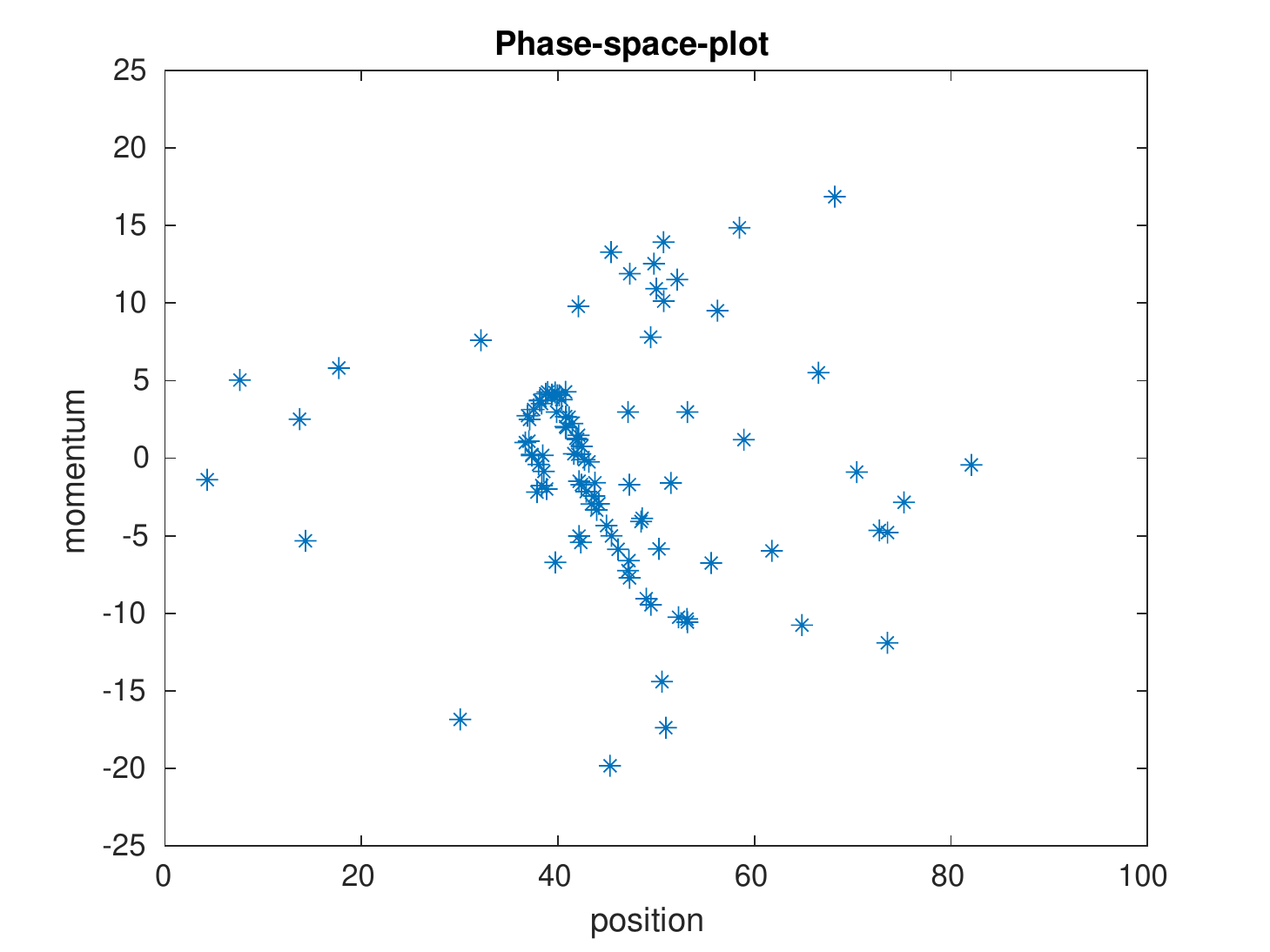}}
\subfloat[$t=600$]{\includegraphics[width = 1.45in]{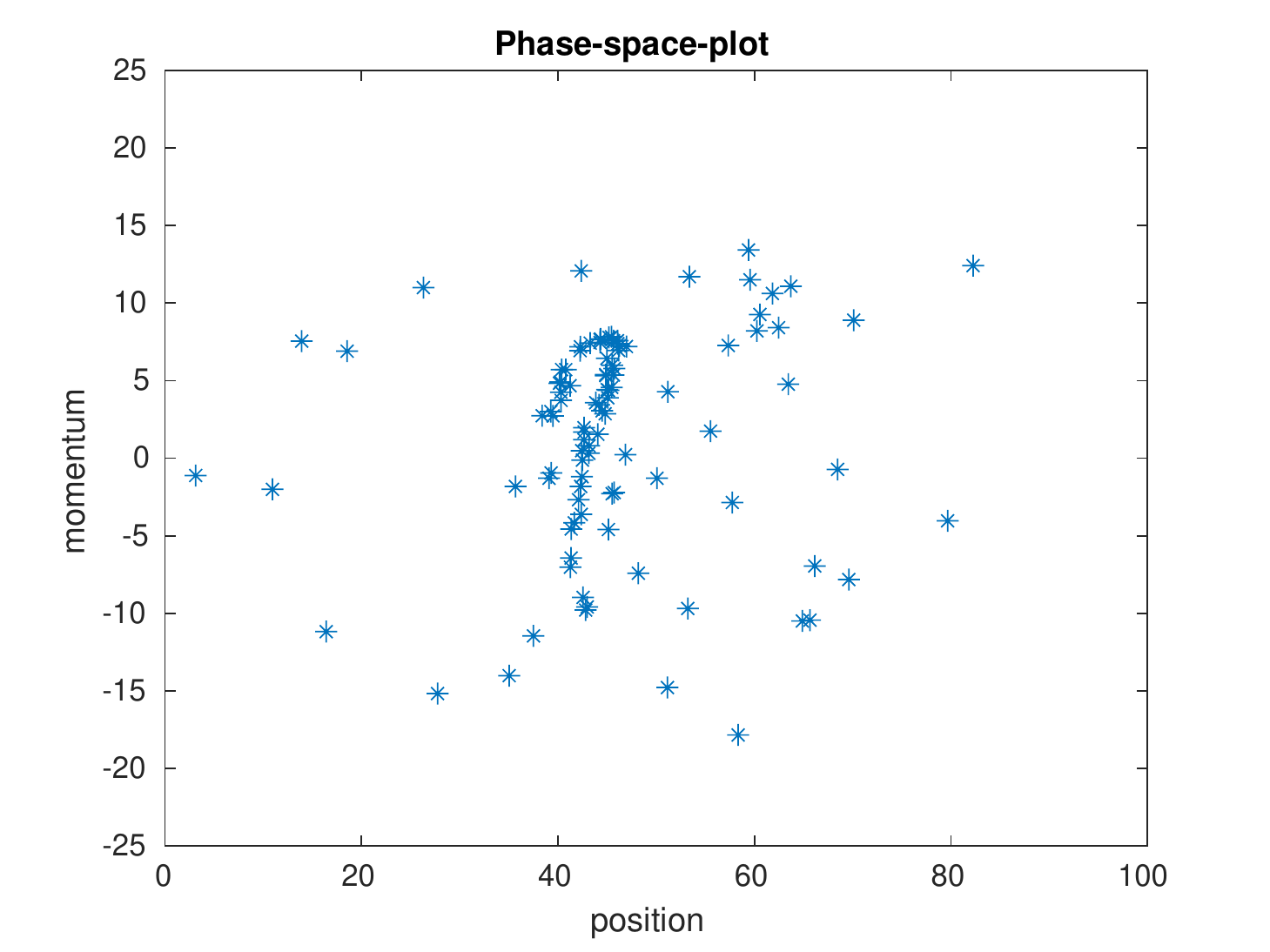}}
\caption{Phase Space snapshots for $t=410$ to $t=600$ with time separation of $10$}
\label{fig_sg_3}
\end{figure}

 %Bibliography
 
 \addcontentsline{toc}{chapter}{Bibliography}
 \nocite*{}
 \bibliographystyle{agsm}
 \bibliography{references}
 
\end{document}